\documentclass[aps,notitlepage,nofootinbib,longbibliography,
twocolumn,prx,
superscriptaddress]{revtex4-2}
\usepackage{amsmath}
\usepackage{amsthm}
\usepackage{amssymb}
\usepackage{slashed}
\usepackage{graphicx}
\usepackage[usenames,dvipsnames]{xcolor}

\usepackage[normalem]{ulem}

\definecolor{midblue}{rgb}{0.05,0.15,0.8}
\definecolor{darkblue}{rgb}{0.1,0.2,0.6} \definecolor{darkred}{rgb}{0.8,0.1,0.2}
 \definecolor{bluepurple}{rgb}{0.3,0,0.7}
  \definecolor{bluepurple2}{rgb}{0.06,0,0.6}
\definecolor{salmon}{rgb}{1,0.57,0.64}
\usepackage{bm}
\usepackage[colorlinks,citecolor=blue,linkcolor=bluepurple2,urlcolor=bluepurple2]{hyperref}
\usepackage{dsfont}
\usepackage{verbatim}
\usepackage{changepage}
\usepackage{array}
\usepackage{makecell}

\newcommand{\bit}{\begin{itemize}}
\newcommand{\eit}{\end{itemize}}

\newcommand{\f}{\frac}
\renewcommand{\>}{\right\rangle}
\newcommand{\<}{\left\langle}
\newcommand{\ba}{\begin{align}}
\newcommand{\ea}{\end{align}}
\newcommand{\be}{\begin{equation}}
\newcommand{\ee}{\end{equation}}
\newcommand{\bi}{\begin{itemize}}
\newcommand{\ei}{\end{itemize}}
\newcommand{\lf}{\left(}
\newcommand{\ri}{\right)}
\newcommand{\dd}{\mathrm{d}}

\newcommand{\Tr}{\operatorname{Tr}}
\newcommand{\tr}{\operatorname{tr}}

\newcommand{\red}{\color{red}}

\newcommand{\er}{Erd\H{o}s-R\'{e}nyi \ }

\newcommand{\bra}[1]{\< #1 \right|}
\newcommand{\ket}[1]{\left| #1 \>}

\renewcommand{\vec}[1]{\boldsymbol{\mathbf{#1}}}

\newcommand{\pperc}{P_\textrm{perc}}
\newcommand{\Ztyp}{Z^\text{typ}}

\begin{document}

\title{Measurement and entanglement phase transitions in all-to-all quantum circuits, \\ on quantum trees, and in Landau-Ginsburg theory}

\author{Adam Nahum}
\affiliation{Rudolf Peierls Centre for Theoretical Physics, Clarendon Laboratory, Oxford University, Parks Road, Oxford OX1 3PU, United Kingdom}
\author{Sthitadhi Roy}
\affiliation{Rudolf Peierls Centre for Theoretical Physics, Clarendon Laboratory, Oxford University, Parks Road, Oxford OX1 3PU, United Kingdom}
\affiliation{Physical and Theoretical Chemistry, Oxford University, South Parks Road, Oxford OX1 3QZ, United Kingdom}
\author{Brian Skinner}
\affiliation{Department of Physics, Ohio State University, Columbus, OH 43210, USA}
\author{Jonathan Ruhman}
\affiliation{Department of Physics, Bar-Ilan University, 52900, Ramat Gan, Israel}
\affiliation{Center for Quantum Entanglement Science and Technology, Bar-Ilan University, 52900, Ramat Gan, Israel}

\date{\today}

\begin{abstract}
A quantum many-body system whose dynamics includes local measurements at a nonzero rate can be in distinct dynamical phases, with differing entanglement properties.
We introduce theoretical approaches to  measurement-induced phase transitions (MPT)
and also to entanglement transitions in random tensor networks. 
Many of our results are for ``all-to-all'' quantum circuits with unitaries and measurements, in which any qubit can couple to any other, and related settings where some of the complications of low-dimensional models are reduced.
We also propose field theory descriptions for spatially local systems of any finite dimensionality.
To build intuition, we first solve the simplest ``minimal cut'' toy model for entanglement dynamics in all-to-all circuits, finding scaling forms and exponents within this approximation.
We then show that certain all-to-all measurement circuits allow exact results by exploiting local tree-like structure in the circuit geometry. For this reason, we make a detour to give general universal results for entanglement phase transitions in a class of random tree tensor networks { with bond dimension 2}, making a connection with the classical theory of directed polymers on a tree. 
We then compare these results with numerics in all-to-all circuits, both for the MPT and for the simpler ``Forced Measurement Phase Transition'' (FMPT).  We characterize the two different phases in all-to-all circuits using observables that are sensitive to the amount of information that is propagated between the initial and final time. We demonstrate signatures of the two phases that can be understood from simple models.
Finally we propose Landau-Ginsburg-Wilson-like field theories for the measurement phase transition, the forced measurement phase transition, and for entanglement transitions in random tensor networks. 
This analysis shows a surprising difference between the measurement phase transition and the other cases. 
We discuss variants of the measurement problem with additional structure { (for example free-fermion structure),} and questions for the future.
\end{abstract}

\maketitle

\tableofcontents

\section{Introduction}
\label{sec:intro}

A quantum system whose unitary dynamics is interspersed with repeated measurements follows a random trajectory through Hilbert space \cite{basche1995direct,gleyzes2007quantum,vijay2011observation,robledo2011high,minev2019catch}, determined both by the unitary part of the dynamics and by the sequence of measurement outcomes. 
In the many-body case this random dynamics admits  a ``measurement phase transition''  (MPT) 
between two qualitatively different, stable dynamical phases, with distinct entanglement properties \cite{skinner2019measurement,FisherZeno,CNPS,li2019measurement,szyniszewski2019entanglement,choi2020quantum,gullans2019dynamical, bao2020theory,jian2020measurement,li2020conformal,zabalo2020critical,gullans2020scalable,tang2020measurement,fuji2020measurement,lunt2020measurement,szyniszewski2020universality,turkeshi2020measurement,van2020entanglement,fan2020self,shtanko2020classical,vijay2020measurement,li2020statistical}. 
For definiteness, consider a system of many spins in a pure state,  evolving under a quantum circuit that includes both entangling two-spin unitary gates
and  measurements, which are made at random times at a finite rate per spin. 
Informally, sufficiently frequent  measurements yield a ``disentangling'' phase: 
in this phase, the state at a given time is weakly entangled, and is fully specified by the outcomes of  a relatively recent set of measurements.
(The limiting case of this disentangling dynamics is where all the spins are measured simultaneously, leaving the system in a product state that can be read off from the measurement outcomes.)
But when the frequency of measurements falls below a critical threshold, the dynamics enters an entangling phase \cite{skinner2019measurement,FisherZeno}. 
In this phase the dynamics produces states with extensive entanglement, which retain quantum information from much earlier times. 
If the initial state is mixed, rather than pure, then it will rapidly be purified \cite{gullans2019dynamical} by the repeated measurements in the disentangling phase, but not in the entangling phase.

The simplest toy model for the MPT  arises from thinking about the connectivity of the spacetime diagram of the quantum circuit, viewed as a tensor network \cite{skinner2019measurement}. 
In this representation a measurement event is a break in the worldline of a spin, across which quantum information cannot be transmitted. 
When measurements become sufficiently frequent, the circuit falls apart into disconnected pieces, implying that entanglement in the final state is short ranged and there is no transmission of quantum information, from the initial to the final state, over long timescales.

The existence of the MPT poses  several types of questions.
Viewing the circuit as a quantum information processor, the MPT is  a transition in the properties of a randomly generated error-correcting code \cite{choi2020quantum, fan2020self,Aharonov}, 
the structure of which { can be optimal in a certain sense \cite{gullans2019dynamical}; understanding the MPT may lead to useful insights into fault-tolerant quantum computation~\cite{Gottesman2009fault-tolerant}.  

The transition also has consequences} for the computational difficulty, for a \textit{classical} computer, of simulating various types of   open or monitored quantum systems 
\cite{skinner2019measurement,Bonnes2014superoperators, napp2019efficient, plenio1998quantum, daley2014quantum}. 
{ As a simplified thought experiment, we may imagine that we are given the sequence of measurement outcomes obtained in an experimental run (as well as the information about the Hamiltonian, and the intial state), 
and asked to calculate  the quantum state of the system following these measurements.
If the dynamics is in the disentangling phase,  an efficient matrix-product or tensor network representation of the evolving state may be possible, 
while if the dynamics is in the entangling phase, the computation may be  intractable. In this sense the MPT can function as an ``epistemological'' phase transition in which the quantum wavefunction becomes essentially unknowable.

Philosophically, we may also} wonder what the existence of two {phases} implies about how to distinguish  dynamical processes that are intrinsically quantum from those that are effectively classical.  For example, in both of the phases separated by the MPT, quantum correlations between local observables are ``weak'', but for  different reasons. 
In the disentangled phase, a local operator is correlated only with a few others nearby.
In the entangled phase, it may have nontrivial correlations, but these are detectable only by highly nonlocal, ``scrambled'' operators, and hidden from local ones.
 Only close to the transition point does the system escape both mechanisms, allowing nontrivial correlations for local operators \cite{skinner2019measurement,li2019measurement, gullans2020scalable, bao2020theory, zabalo2020critical, li2019measurement, li2020conformal, jian2020measurement, fan2020self}.
Yet another key question is how to probe the MPT experimentally \cite{bao2020theory,gullans2020scalable}. This question is nontrivial: for example, a naive approach leads to a severe sampling problem (due to the need to compare measurements in distinct experimental runs that have the same measurement outcomes).

Another way of looking at the MPT is as a 
problem in statistical mechanics and critical phenomena \cite{skinner2019measurement,gullans2019dynamical, jian2020measurement, bao2020theory,FisherZeno,li2019measurement,li2020statistical,li2020conformal,shtanko2020classical,zabalo2020critical,szyniszewski2019entanglement,turkeshi2020measurement,fan2020self, zhou2018emergent}. 
Open questions abound, both about the nature of the phases and about the critical point separating them.
Many variants of the measurement transition can be imagined; how do we sort them into universality classes?
Are there simplifying limits where exact results are possible?
Are there useful continuum field theories for the MPT and related problems, that allow us to apply the tools of the renormalization group?

This statistical mechanics problem is closely connected to an entanglement transition that takes place in random tensor networks \cite{hayden2016holographic,vasseur2019entanglement,jian2020measurement} (we will explore the similarities and differences further here) and the same questions apply in that setting. 
These problems are challenging  partly because of the need to average over randomness: 
either intrinsic randomness in the definition of the dynamics (for example if we consider dynamics using a random quantum circuit) 
or simply the inevitable quantum-mechanical randomness in measurement outcomes.

 \begin{figure}[b]
    \includegraphics[width=0.48\linewidth]{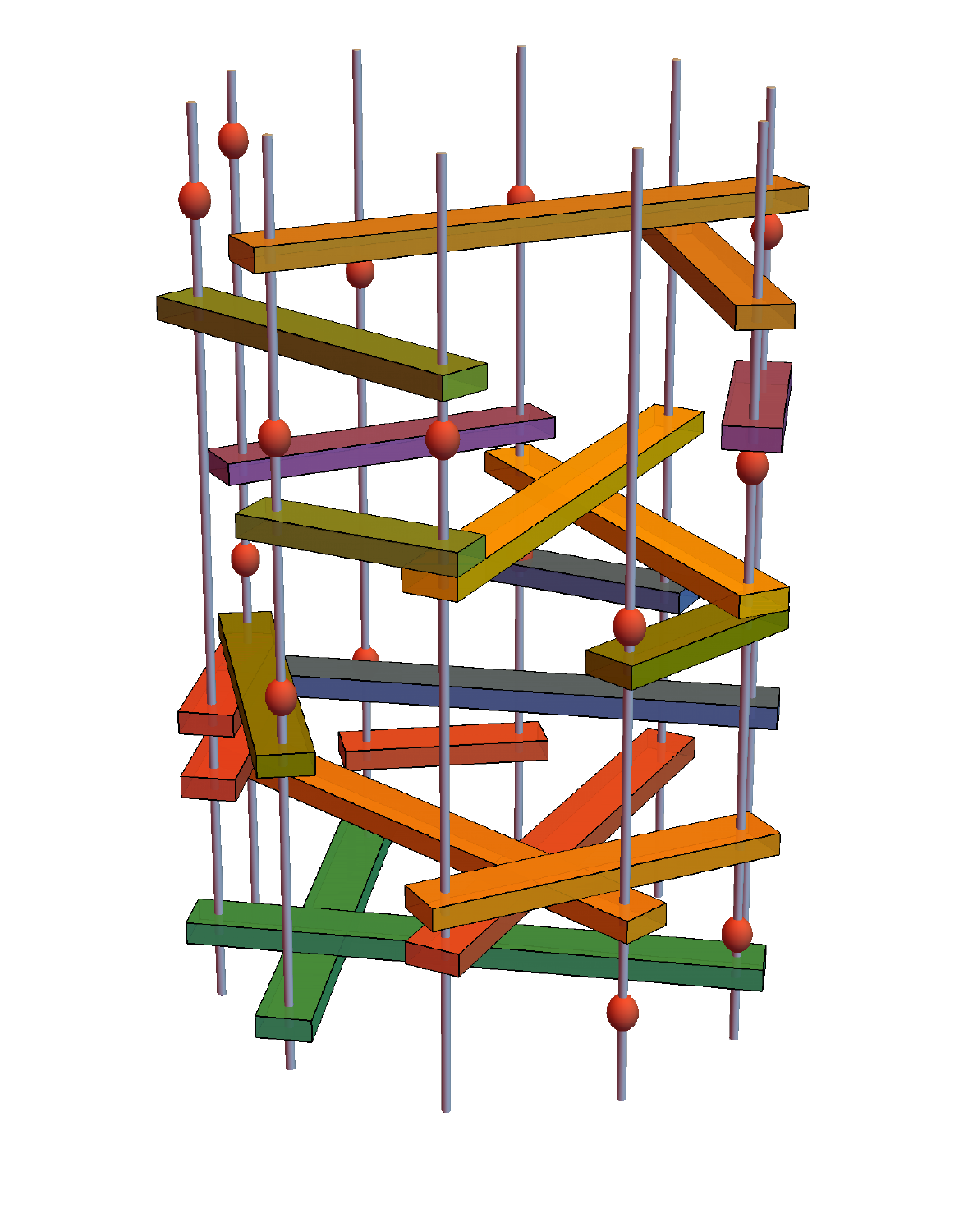}
    \includegraphics[width=0.48\linewidth]{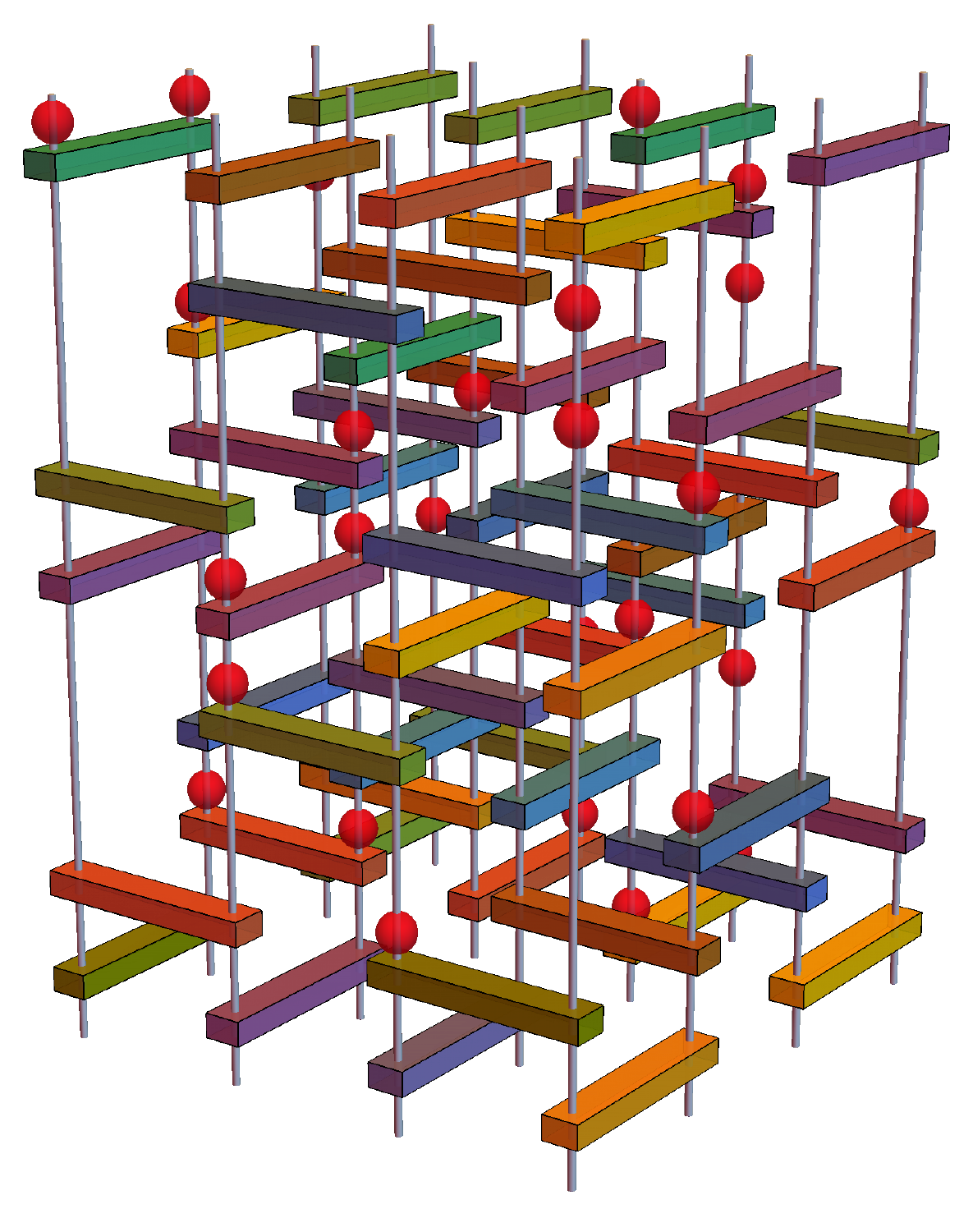}
    \caption{Random circuits with measurements.
    Vertical lines are world lines of individual spins, with time running vertically. 
   The blocks connecting different world lines are independently random unitary gates. 
   These are interspersed by projective measurements, represented by the red blobs. 
   The left panel shows an all-to-all 2-local circuit whereas the right panel shows a regular 2+1D circuit.}
    \label{fig:circuit-cartoon}
\end{figure}
 
Our focus in this paper is on circuits built from generic unitary gates (for example Haar-random gates). 
An alternative profitable direction is to study circuits made from Clifford unitaries \cite{FisherZeno,li2020conformal,gullans2020scalable,gullans2019dynamical,nahum2020entanglement,
sang2020measurement,lavasani2020measurement,ippoliti2020entanglement,lang2020entanglement,
turkeshi2020measurement}.  Clifford circuits are efficiently classically simulable, which has allowed direct tests of conformal invariance at the MPT in 1+1D 
\cite{li2019measurement, li2020conformal}  and simulations in 2+1D \cite{turkeshi2020measurement}. In general, the universality class of the MPT is expected to differ for Clifford versus generic unitaries (see e.g. Ref.~\cite{zabalo2020critical}), though many features of the stable phases are similar.

This paper is a journey  through several approaches to the MPT, and also to the closely related ``forced'' measurement phase transition { (FMPT, defined in Sec.~\ref{sec:overviewmodels} below)}, and to the entanglement transition in various types of random tensor networks (RTN).  Our aim is to find settings in which exact results can be obtained for the transition, as well as to clarify the properties of the two phases.  We examine several different tools and settings, but the unifying feature is that we consider measurement and entanglement transitions in situations where the complications arising from low-dimensional spatial structure are reduced.

\begin{figure}[t]
    \centering
    \includegraphics[width=\linewidth]{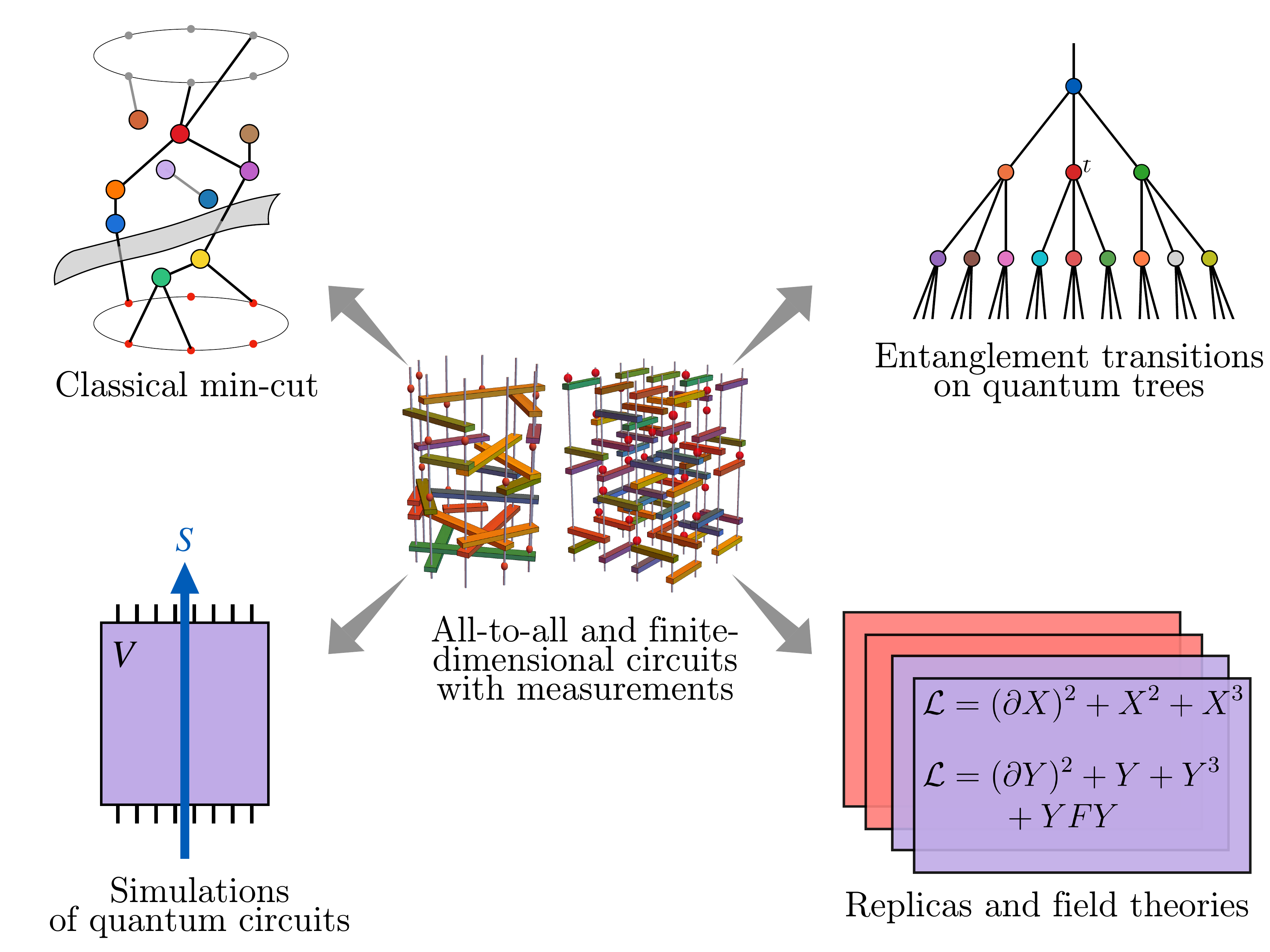}
    \caption{Some of the approaches in this paper. 
    To gain intuition, we start by solving a classical toy model for the transition in the all-to-all circuit, which gives scaling forms for the ``minimal cut'' cost that determines the zeroth R\'enyi entropy (Sec.~\ref{sec:classical}).
    In Sec.~\ref{sec:quantumtree} we turn to the truly ``quantum'' problem,  obtaining exact results for random tree tensor networks. These can be applied to the true quantum transition in all-to-all   circuits with ``forced'' measurements. 
    Sec.~\ref{sec:simulationscircuits} simulates all-to-all measurement circuits, using operator entanglement of the time evolution operator, and the convergence with time of two initially orthogonal states, to diagnose the preservation of information over an exponentially long timescale that is a hallmark of the entangling phase. Sec.~\ref{sec:landautheory} develops analytical approaches to the MPT and to entanglement transitions based on the replica trick, clarifying the properties of the two phases and suggesting candidate field theories for the critical points in various settings.}
    \label{fig:summary}
\end{figure}

Much of this paper is concerned with circuits with all-to-all couplings between qubits, i.e.\ with no fixed spatial geometry, which we study using analytical arguments and numeric simulations. (Various types of all-to-all circuit have also been discussed recently in Refs.~\cite{gullans2019dynamical, vijay2020measurement}.) These circuits are in turn closely related to tree tensor networks, for which we give exact results, including the first exact identification of an entanglement transition in a generic system with finite bond dimension.

Turning to models { a finite number of spatial dimensions}, we discuss and extend tools based on mappings to effective ``lattice magnets'' \cite{hayden2016holographic, nahum2018operator,zhou2018emergent,vasseur2019entanglement,jian2020measurement,bao2020theory,zhou2019entanglement,hunter2019unitary,liu2020entanglement}, involving a replica limit \cite{zhou2018emergent,vasseur2019entanglement,jian2020measurement,bao2020theory}, which capture the properties of the two phases, and in principle the critical point.  We suggest alternative ways of thinking about these effective models,  making connections with ideas from disordered magnetism:
{ in particular we suggest a construction of  order parameters for the MPT and for entanglement transitions random tensor networks, based on overlap of Feynman trajectories.}
A key outstanding question is the existence of effective field theories for the MPT. Here we
propose --- speculatively --- two  Landau-Ginsburg theories,  one for the MPT and one for both the FMPT  and the RTN. 

The cartoon in Fig.~\ref{fig:circuit-cartoon} contrasts an all-to-all measurement circuit and a circuit with a fixed spatial geometry. In this figure, time runs vertically, and each worldline represents a spin/qubit. 
Unitary gates are applied between randomly chosen spins at random times, and projective measurements are applied to randomly chosen spins. 
All-to-all coupling is perhaps the simplest setting for the MPT.
Since the distinction between area and volume law breaks down in the all-to-all case 
(as also in the limit of infinite dimensions), it is natural to focus instead on the transmission of information between initial and final times. 
Here we characterize this transmission via the operator entanglement \cite{zanardi_entanglement_2001,prosen_operator_2007,pizorn_operator_2009,dubail_entanglement_2016,zhou_operator_2017,jonay2018coarse} of the nonunitary time evolution operator, defined below. 
This quantity has a simple interpretation in terms of the surface tension of the ``entanglement membrane'' in the effective replica description, which we discuss. An even simpler heuristic picture for it comes from the  classical toy model, in terms of the minimal cut that separates the top of the circuit from the bottom.

We apply all the approaches mentioned above 
(tree approximations, simulations, replica field theories)
in the setting of generic quantum circuits for spin-1/2, as well as  related random tensor networks, giving results for scaling properties in the entangled phase and close to the critical point. We also study a solvable ``classical'' limit of the problem.
Our main approaches are illustrated in Fig.~\ref{fig:summary}, and the ensuing section, Sec.~\ref{sec:overview}, gives an overview of our results. In closing this Introduction, however, let us briefly clarify the logic of our four-pronged approach to understanding the MPT and its relatives.

Before tackling the ``true'' quantum circuit problem, we find it instructive to first solve the classical toy model mentioned above, in the particular setting of all-to-all circuits (Sec.~\ref{sec:classical}). In this model the entanglement is described in terms of a ``minimal cut'' through a circuit in which worldlines have been broken by measurement. The minimal cut becomes an exact description of the MPT in certain limits, but in general it does \emph{not} capture either the location of the critical point or the true critical scaling of the quantum problem. Nonetheless, the minimal cut problem yields some useful lessons for the full quantum problem. Most prominently, it captures key qualitative features of the two phases, including the appearance of an exponentially long timescale for survival of quantum information within the entangled phase. Solving the minimal cut problem also makes clear certain crucial concepts for understanding the MPT in all-to-all circuits, including the local tree structure of the circuit and the relevance of crossover scaling phenomena.

The fact that all-to-all circuits have a local tree structure { then} motivates us to study entanglement transitions in quantum trees (Sec.\ \ref{sec:quantumtree}).  In this setting we are able to obtain the exact location of the entanglement transition (and exact critical properties) for a { tree tensor network that is relevant to dynamics with Haar-random} gates.  This result may be useful for further investigations: studies of the MPT in systems with generic unitaries are often hampered by the restriction of numerics to small sizes, which make it  difficult to accurately pinpoint entanglement transitions. Moreover, we argue that the critical measurement rate that we identify in the quantum tree is also the exact result for the full all-to-all quantum circuit with forced (postselected) measurements.

Armed with the understanding gained from the minimal-cut and quantum tree problems, we turn our attention to direct numerical simulations of the quantum circuit (Sec.~\ref{sec:simulationscircuits}). The results we obtain are consistent with the critical scaling forms suggested by the previous approaches, and highlight the emergence of an exponentially long timescale associated with information transmission through the circuit in the entangling phase.

Finally, in Sec.\ \ref{sec:landautheory} we discuss mappings of the MPT and of random tensor networks to effective lattice  models for a ``pairing field'', and we discuss how to coarse-grain such models. 
We construct the simplest candidate Lagrangians that are consistent with the replica symmetry and describe some of their features. 
We also touch on free fermions subject to measurement \cite{cao2018entanglement}, which do not show the same kind of transition between weakly and strongly entangled phases but do show transitions of a different type \cite{nahum2020entanglement,chen2020emergent,alberton2020trajectory}.  We contrast these { free systems (which have a continuous, rather than discrete, replica symmetry)} with generic models, and we discuss some other variants of the MPT.

\section{Overview}
\label{sec:overview}

\subsection{Models}
\label{sec:overviewmodels}

Our starting point is a dynamical process in which a large number $N$ 
of spin-1/2s undergoes unitary evolution punctuated by projective single-spin measurements: Fig.~\ref{fig:circuit-cartoon}, Left. 
(Circuits with both unitaries and measurements have been  referred to as ``monitored'' or ``hybrid'' quantum circuits.)
The spins are ``all-to-all'' coupled, meaning that unitary gates may be applied between any two spins in the system.
These gates are applied at a uniform rate between randomly chosen pairs of spins, and are themselves drawn independently from a random ensemble (e.g. the Haar ensemble).
Measurements, which are made  in the $Z$-basis,  are also applied at a uniform rate to randomly chosen spins. 
The only  parameter is $r\in [0,1]$, which determines the relative rate of measurements and unitaries: 
in a unit interval of time there are on average $r N$ measurements and $(1-r)N$ unitary operations.

We  distinguish between two possibilities for the projective measurements, which we refer to as  ``measurements'' and ``forced measurements'', respectively.  
(Correspondingly we refer to the ``measurement phase transition'', or MPT, and ``forced measurement phase transition'', or FMPT.) 
The outcomes of ``measurements'' are determined as usual by the Born rule, 
based on the state of the system at the time of measurement. 
By contrast the probability of a given outcome for a ``forced measurement'' is independent of the state.
We will take it to be $1/2$ for both of the two possible outcomes, $\uparrow$ and $\downarrow$ ---
but in fact, for the ensembles of random unitaries we consider, 
it is completely equivalent to take all the measurement outcomes to be $\uparrow$. 
We can think of the FMPT as pertaining to a protocol in which we run (exponentially) many samples, discarding all those except those that yield the desired (``postselected'') sequence of outcomes. 

To formalize the distinction between MPT and FMPT,  define $V_{\vec m}$ to be the nonunitary time evolution operator represented by a given realization of the circuit. 
This operator is the  product of unitaries and projection operators: 
we have  labelled it by a given sequence $\vec m$ of outcomes for the measurement events: for example $\vec m = (\uparrow, \downarrow, \ldots, \uparrow)$.
($V_{\vec{m}}$ also depends on the total time $t$,  locations and times of the unitaries and measurements, and the specific random unitaries in the circuit realization, but we leave these dependencies implicit.)
For the MPT, and for a given sequence of unitaries and measurement locations,
 the probability of a sequence of measurement \textit{outcomes} $\vec m$ is 
\be\label{eq:overviewMPTprob}
P(\vec{m}) = \bra{\psi(0)} V_{\vec{m}}^\dag V_{\vec{m}}^{\phantom{\dag}} \ket{\psi(0)},
\ee
where $\ket{\psi(0)}$ is the initial state. For the FMPT it is
\be\label{eq:overviewFMPTprob}
P(\vec{m}) = 2^{-|\vec{m}|},
\ee
where $|\vec{m}|$ is the number of measurements in a given realization of the circuit. In both cases, the time evolution of a pure state is
\be\label{eq:overviewevolution}
\ket{\psi(t)} =  \f{V_{\vec{m}} \ket{\psi(0)}}{\left| V_{\vec{m}} \ket{\psi (0)} \right |}.
\ee
It is occasionally useful to generalize the circuit to a variable number of spin states $q$ for each site.  In particular, the limit of large $q$ is one way to motivate the classical problem we describe below.

Having started with the models above, we will be led to consider some other related problems. 
These models will be introduced as we need them. 
Sec.~\ref{sec:quantumtree} considers a class of tree tensor networks, one example of which is closely related to the FMPT case above. 
Sec.~\ref{sec:landautheory} addresses both circuits and tensor network models in a finite number of dimensions, in which we do have a sense of spatial locality.

\subsection{Detecting the entangling phase}
\label{sec:overviewdetecting}

Before turning to the critical properties, we discuss the more basic issue of how to distinguish the two phases.

The entanglement transition can be identified with the vanishing of an effective \textit{surface tension} for a 
 membrane-like object in spacetime, as we discuss below. 
In the classical toy model, this membrane is a minimal cut through the circuit \cite{skinner2019measurement}.
In a more precise picture, it is a domain wall in an effective statistical mechanics problem  (see following sections).
The surface tension of this membrane/domain wall is positive in the entangling phase~\cite{skinner2019measurement}.

\begin{figure}
    \centering
    \includegraphics[width=\linewidth]{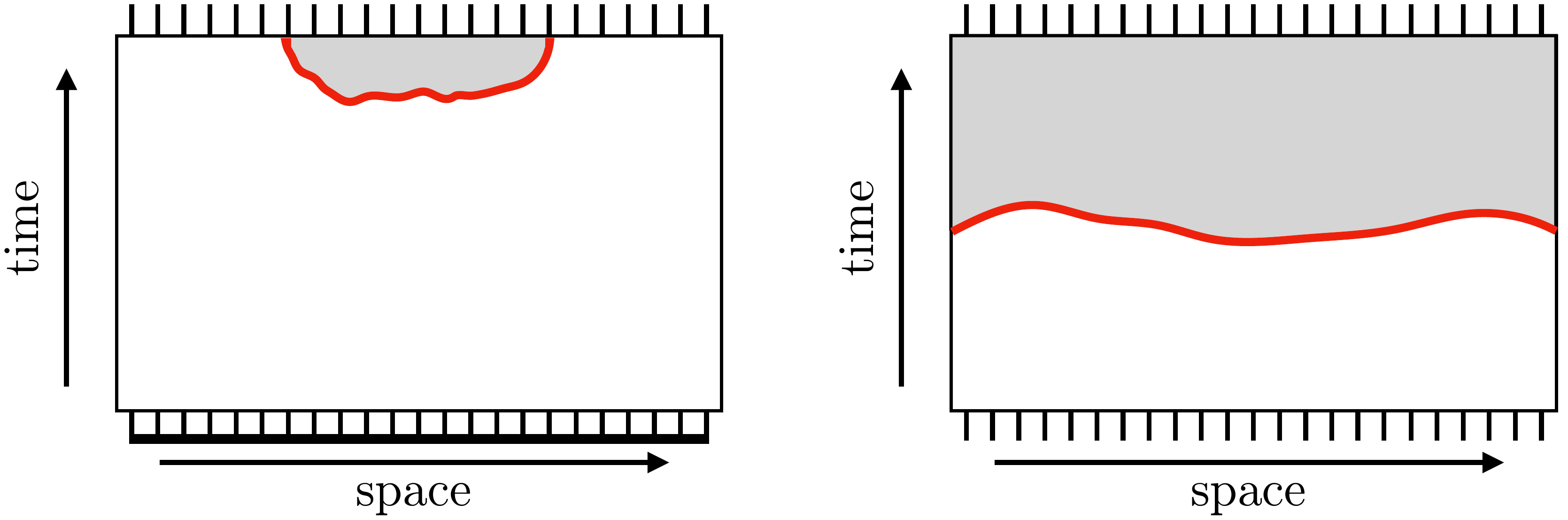}
    \caption{The entanglement entropy of states and operators can be described in terms of the surface tension of an effective membrane/domain wall. 
    Left: interpretation of entanglement entropy of a spatial subregion, in the entangling phase, as the free energy of an anchored membrane.  Right: operator entanglement of the nonunitary  circuit (in the entangling phase)
    in terms of a ``horizontal'' membrane (see also Ref.~\cite{bao2020theory}).}
    \label{fig:membrane}
\end{figure}

In finite dimensions, the vanishing of this surface tension, which we denote $s_n(r)$,\footnote{In general the membrane tension can depend on the R\'enyi index $n$ \cite{zhou2018emergent}.
It can also depend on the orientation of the membrane \cite{jonay2018coarse}, but here we are interested only in membranes that are ``horizontal'' on large scales.} implies a vanishing of the entanglement entropy density of the \textit{states} produced by the dynamics at late time.
This density is  the coefficient of the volume law for the entanglement entropy of a spatial subregion, and is given by the surface tension $s_n(r)$. This is because the subregion's entropy maps to the free energy of a membrane that is anchored on the boundary of region $A$ on the final time surface: see Fig.~\ref{fig:membrane} for a schematic in 1+1D.

In the all-to-all circuit there is no distinction between areas and volumes (as in the limit of high dimensions), so the naive attempt to define an entropy density using the entropy of a spatial subregion is contaminated by trivial short range entanglement.\footnote{I.e.\ entanglement that can be removed with a shallow-depth circuit.}
Instead it is simpler to consider the entanglement properties of the \textit{operator} that implements the time evolution itself. 
The operator entanglement \cite{zanardi_entanglement_2001,prosen_operator_2007,pizorn_operator_2009,dubail_entanglement_2016,zhou_operator_2017},  defined below, is a measure of the amount of quantum information transmitted from the initial to the final time by the nonunitary evolution operator $V_{\vec{m}}$.
In the membrane  picture, this operator entanglement is equal to the free energy of a ``horizontal'' membrane that completely traverses the system \cite{bao2020theory}, 
as shown in Fig.~\ref{fig:membrane}. This observable also detects the vanishing surface tension $s_n(r)$ for the domain wall, as detailed below, but it does not require us to specify a spatial subregion.

Gullans and Huse proposed in Ref.~\cite{gullans2019dynamical}  to think about dynamics with measurements in terms of the entropy of a state that starts out as maximally mixed,
and is gradually purified by the dynamics.
(The entangling phase is then a ``mixed’’ phase, where the state remains mixed for a long time,
and the disentangling phase is a ``pure’’ phase where the state
 is rapidly purified.)
This mixed state entropy is in fact equal to the operator entanglement of the nonunitary evolution operator.
Ref.~\cite{gullans2019dynamical} noted the exponentially long timescale for the survival of quantum information in the entangled phase (and  plateaus in various observables), which will play an important role below. See also the recent Refs.~\cite{li2020statistical, fidkowski2020dynamical}.

Formally, the $n$th operator entanglement entropy of the circuit, denoted  $S_n$ throughout this paper, may be defined via the singular value decomposition of the nonunitary time evolution operator $V_{\vec{m}}$:
\be \label{eq:overviewsingval}
V_{\vec{m}} \propto  \sum_{j=1}^{2^N} \lambda_j |j\rangle_t \langle j |_0 \,,
\ee
where $\{|j\rangle_0\}$ and $\{|j\rangle_t\}$ are bases corresponding to the initial and final time. Normalizing the $\lambda_j$ so ${\sum\lambda_j^2=1}$,
\be\label{eq:overviewSndefn}
S_n ={1\over 1-n}\,\ln \bigg(  \sum_{j} \, \lambda_j^{ 2n} \bigg).
\ee 
For the unitary case ($r=0$), $S_n = N \ln 2$ is maximal at all times. 
For positive $r$, and for asymptotically late times, a single term dominates Eq.~\ref{eq:overviewsingval}, meaning essentially that all initial states are projected onto the same final state --- i.e. the final state can be read off from measurement outcomes $\vec{m}$ (and the structure of the circuit) without knowledge of which initial state was fed in.

We will also discuss another observable for quantifying the transmission of quantum information from initial to final times, which is more numerically tractable: this is the overlap between two initially orthogonal states, both subjected to the same ${V_{\vec{m}}}$. (In the entangling phase, initially orthogonal states remain orthogonal for a long time.)

We will characterize the operator entanglement in the classical toy model (Sec.~\ref{sec:classical}), in numerical simulations (Sec.~\ref{sec:simulationscircuits}), using the replica trick (Sec.~\ref{sec:replicatimescale}), and with a crude toy model based on multiplying random matrices (Appendix \ref{app:multiplyingmatrices}). 
The following basic points hold in all of these approaches.

First, in the entangling phase a nonzero density can be associated with the operator entanglement:
\be\label{eq:snfirstdefn}
s_n(r) \equiv \lim_{t\rightarrow \infty} \lim_{N\rightarrow \infty} \f{S_n(r, N, t)}{N}. 
\ee
We think of this quantity as the information transmitted per spin, or in the membrane picture as the surface tension for a ``horizontal'' membrane.
$s_n(r)$ is positive in the entangling phase, and vanishes continuously,  for all $n\geq 1$, as the critical measurement rate $r_c$ is approached from below.

As for almost any product of many random matrices, we expect that if $N$ and $r$ are fixed, then at sufficiently late times one of the singular values dominates the others and $S_n$ decays exponentially in time. But if $s_n(r)$ is positive, this exponential decay does not set in until a time $\tau(r,N)$ that is exponentially large in $N$. We may define
\be
a(r) = \lim_{N\rightarrow \infty} \f{\ln\, \tau(r,N)}{N}.
\ee
Close to the transition, at $r\lesssim r_c$ where $s_n(r)$ is small,
\be
a(r) \sim s_n(r)
\ee
(up to an order 1 constant of proportionality). On times $t$ satisfying $\ln t\ll \ln \tau$, the entanglement deviates logarithmically from the ``plateau'' value dictated by  Eq.~\ref{eq:snfirstdefn}. 
For example in one regime,
\be
S_n(r,N,t) \simeq s_n(r) N - \ln t.
\ee
This formula has also been obtained in various limits in Refs.~\cite{gullans2019dynamical,li2020statistical}: 
in particular Li and Fisher in Ref.~\cite{li2020statistical} give a discussion very similar to that in Sec.~\ref{sec:replicatimescale}, in terms of domain walls in an effective quasi-1D model. In this interpretation $\ln t$ is the translational entropy of a domain wall.  More generally, randomness and other effects can modify the nature of the subleading term above slightly, depending on the time regime.

In contrast to the above, the information transmitted per spin, ${\lim_{N\rightarrow\infty} \f{1}{N} S_n(r,N,t)}$, decays exponentially with $t$ in the disentangled phase. 

We now give an overview of our approaches to critical properties of these circuits and related models, considering each approach in turn (summarized in Fig.~\ref{fig:summary}).
The reader may obtain the key points of each approach from the corresponding Overview section.
We also highlight some points that are not yet resolved, and places where our arguments rely on conjectures that could be tested further.

\subsection{Min-cut toy model}
\label{sec:overviewmincut}

Before attempting an exact treatment of the true quantum transition in the spin-1/2 circuit, 
we consider a limit 
(we will sometimes refer to this as the ``classical'' limit)
in which the entanglement transition becomes a simple geometric problem involving a random  graph. 
This graph is defined such that its edges represent the time evolution of each spin, 
which can be severed by measurement, 
and its nodes represent interactions (applied gates) between spins. 
The analog of the operator entanglement entropy is the cost of a ``minimal cut'' 
that disconnects the initial-time and final-time nodes: see  Sec.~\ref{sec:classical} for a detailed definition.

Determining the scaling of this min-cut cost is a toy problem that provides intuition for the generic ``quantum'' problem.
The minimal cut becomes an exact description of the operator entanglement only in special limits, as described in
Sec.~\ref{sec:classical} 
(specifically, for projective measurements in the case where the local Hilbert space dimension $q$ goes to infinity,
and for a generic local Hilbert space dimension if we consider the somewhat unphysical zeroth R\'{e}nyi entropy, $S_0$).
The ``classical'' problem has its transition at a measurement rate $r_c^\text{cl}$ that is, for spin-1/2, strictly larger than the critical measurement rate $r_c$ for the true quantum transition, as diagnosed for example by all the $S_n$ with $n\geq 1$.

We first identify the critical point $r_c^\text{cl}$ associated with percolation on the graph, which illustrates the importance of local tree structure in all-to-all circuits. 
We then present an effective continuum field theory for percolation on this graph, which gives the relevant scaling forms near $r_c^\text{cl}$.  
We demonstrate this critical scaling using extensive numerical simulations for percolation observables and correlation functions.  This demonstration is possible despite significant finite time-corrections, which arise because the critical timescale scales as $N^{1/5}$ and is modest even for simulations with very large $N$.

We demonstrate the plateau in the cost of the minimal cut that was described above, ${S_0 \sim s_0(r) N}$ over a long timescale.  Close to criticality at ${r\lesssim r_c^\text{cl}}$ we find the entanglement density (min-cut tension)
\be
s_0(r) \sim  \lf r_c^\text{cl} - r \ri^{5/2},
\ee
which is an appropriate limit of a general scaling form $S_0=H(t/N^{1/5}, \delta r N^{2/5})$,
and the corresponding long timescale
\ba
    \tau & \sim \exp \lf a(r) \times N \ri, &
    a(r) & \sim (r_c^\text{cl} - r)^{5/2}. 
\end{align}
The scaling we identify applies not only for the all-to-all problem, but also for spatially local circuits with spatial dimension $d \geq 5$, as follows from a standard crossover scaling argument.

{
The all-to-all percolation model has also been analyzed independently in Ref.~\cite{Gullans_unpublished}, using a different method in which rate equations for the number of percolation clusters of a given size are solved. This analysis  also implies that the scaling variables are ${t/N^{1/5}}$ and ${\delta r N^{2/5}}$, in agreement with what we find.}

\subsection{Tree tensor networks: exact results}
\label{sec:overviewtree}

When the system size $N$ is large, the structure of the quantum circuit in the vicinity of a given unitary is tree-like 
(the smallest loops involve a parametrically large number of unitaries).
This means that it is trivial to locate the classical critical point mentioned above. 
But in some cases (forced measurements) it also allows exact results for the quantum problem.
This motivates us to study entanglement transitions on ``quantum trees'', i.e. tree tensor networks, in Sec.~\ref{sec:quantumtree}.

While our approach could be generalized, we focus on trees with bond dimension 2,
 where each node is a random tensor whose probability distribution is invariant under $\mathrm{U}(2)$ 
 rotations on its legs. 
 This includes trees that appear spontaneously in 
 the spin-1/2 FMPT circuit for unitaries drawn from the Haar measure, for example. 

Formally we can think of an (upside-down) tree like that shown in Fig.~\ref{fig:summary} as a tensor network wavefunction for a single spin at the apex (root) and many spins (leaves) at the base.
Our starting point is to characterize the entanglement between apex and base, which for a bond-dimension 2 tree is characterized by a single number, $Z$.
For an asymptotically large tree, $Z$ has a critical vanishing at a particular measurement rate $r_c$.
(In more general trees, $r$ can be thought of as a parameter in the node tensors' distribution.)

We write a random recursion relation for $Z$ as a function of the generation number $k$ of the tree.  
This recursion relation allows us to derive the location of the critical point $r_c$ analytically for the case with Haar-random unitaries (we also study a slightly broader class of distributions):
\be\label{eq:treeoverviewrc}
r_c  = \f{212+75\pi}{362+75\pi}. 
\ee
This critical point $r_c$ is detected by any R\'{e}nyi entropy $S_n$ with ${n>0}$; $S_0$ instead detects the classical transition, at the strictly larger value $r_c^\text{cl}$, discussed in the previous section.
This is the difference between the existence of a percolating path connecting the root of the tree to infinity (for ${r<r_c^\text{cl}}$) and the ability of the tree to broadcast a nonzero amount of quantum information from the root of the tree to infinity, rather than an amount that decays exponentially with the distance from the root. 

Assuming a plausible conjecture, Eq.~\ref{eq:treeoverviewrc} is also the value of $r_c$ for the FMPT in the all-to-all Haar circuit, and yields a bound on the critical scaling of the entanglement density $s_2(r)$ (Eq.~\ref{eq:snfirstdefn}).
While the treatment of the tree may hold lessons for the MPT in addition to the FMPT, we do not discuss the MPT from this perspective:  the measurement correlations encoded in Eq.~\ref{eq:overviewMPTprob} hamper our approach.

We also obtain the the critical scaling of $Z$ for  ${r\lesssim r_c}$.  
Since the full nonlinear recursion relation for $Z$ is complicated, this requires us to make a \textit{conjecture}, 
which is that the universal features of the scaling are faithfully retained in a simplified nonlinear recursion relation.
We can then write a continuum description that describes a Fisher-Kolmogorov-Petrovsky-Piskunov-like traveling wave \cite{derrida1988polymers}. 
This is a standard description for the partition function of a directed polymer on a tree \cite{derrida1988polymers}, with the addition of a  diffusion constant that varies with the (fictitious) spatial coordinate, reflecting the nonlinearity of the recursion. For the parameters of the trees we treat, there is a surprisingly rapid scaling close to the critical point: the entanglement between apex and base of an infinite tree scales as (${S_2^\text{tree} \simeq 2 Z}$)
\be\label{eq:overviewS2tree}
S_2^\text{tree} \sim \exp \lf - \f{\text{const.}}{\sqrt{r_c -r}} \ri. 
\ee
We also address the scaling of $S_n$ as a function of tree size exactly at $r_c$.

Using a nonrigorous argument, we extend these formulas to give the entanglement of a subset of the spins, 
in a tree tensor network wavefunction for a spin chain (whose spins are the leaves of the tree).
These results are not relevant to the all-to-circuit, but are interesting in the context of tree tensor network states, which are toy models for some features of scale invariance in 1+1D, and are also useful numerical tools \cite{Shi2006,Tagliacozzo2009,Murg2010,Silvi2010,Li2012,Nakatani2013,Murg2015, Vidal2007,Swingle2012,lopez2020mean}.
We obtain a ``modified minimal cut'' formula for the tree, 
in which the cost of cutting a bond in the tree is loosely speaking weighted by appropriate factors of the quantity $Z$, which is parametrically small close to $r_c$.
This gives a quantitative picture of how  the entanglement of $\ell$ consecutive spins in a tree tensor network state goes from the well-known logarithmic scaling, ${S_\ell \sim c(r) \ln \ell}$, suggested by its hierarchical structure \cite{Swingle2012,pfeifer2009entanglement, lopez2020mean}, to an area law state, ${S_\ell = \mathcal{O}(\ell^0)}$. We find that $c(r)$ vanishes exponentially as $r\rightarrow r_c$ and that the state is area-law even at~$r_c$.

Recently,  Ref.~\cite{lopez2020mean} studied the entanglement transition in a quantum tree state, with bond dimension 3, using a different approach. 
The authors conjectured that the scaling was the same as in a statistical mechanics model that shares some of the features of a replica formulation derived from the tree (the exact replica formulation was not tractable). 
Surprisingly, the findings in Ref.~\cite{lopez2020mean} are quite different from those we obtain 
(assuming the conjecture mentioned in the previous paragraph) 
in the trees  studied here. 
For example, Ref.~\cite{lopez2020mean} finds that the coefficient  $c$ in ${S\sim c \ln \ell}$ is a power law in the control variable close to the transition, and that entanglement is super-area-law at $r_c$.
The reason for the different results in these two models remains to be understood.

Our conjectured continuum description for scaling in the tree  has a parameter $\Delta$ that describes the degree of disorder in the tensor network, and which determines the critical exponents.   For the trees we study, whose node tensors have a  distribution with a $\mathrm{U}(2)$ invariance property on the legs, this parameter is fixed to  $\Delta = 1/4$ at $r_c$. This corresponds to a ``strong disorder'' regime \cite{derrida1988polymers}.   We raise the question of whether general distributions of tensors allow us to explore the phase transition at other values of $\Delta$. If so, it is possible to obtain a range of universality classes for the tree transition, analogous to a renormalization group fixed line. However, we have not determined whether this is possible.

\subsection{Direct simulations of quantum circuits}
\label{sec:overviewnumerics}

We perform direct simulations of the all-to-all measurement circuit and forced measurement circuit, and interpret the results in the light of the tree calculation and the replica approach described below.
These simulations are computationally demanding: we are limited to system sizes $N \leq 20$ for quantities involving states and to smaller sizes for the operator entanglement. 
Determining $r_c$ accurately (the value of which is expected to differ for measurements and forced measurements) is not possible, 
but we are able to confirm many of the key features of the entangled phase in Sec.~\ref{sec:overviewdetecting}. 

We give evidence for the plateau (\ref{eq:snfirstdefn}) in the operator entanglement, with a nonzero information transmission per spin $s(r)$ that is asymptotically time independent, and for a positive exponential growth coefficient ${a(r)>0}$ for the characteristic timescale within the entangled phase.

It is convenient to define this timescale $\tau$ via the late-time convergence of two distinct, initially orthogonal, states $\ket{\psi_1(t)}$ and $\ket{\psi_2(t)}$ that are postselected to undergo  the  \textit{same} sequence of measurement outcomes, so that they are evolved with the same $V_{\vec m}$.
These states remain approximately orthogonal for a long time in the entangled phase: a kind of effective unitarity of the nonlinear, nonunitary time evolution Eq.~\ref{eq:overviewevolution}  for a given $\vec{m}$. (This orthogonality is related to the error-correction property of the dynamics  \cite{choi2020quantum,gullans2019dynamical}.) The two states collapse at late times. We show that $a(r)$ is positive at small $r$  and vanishes at large $r$. 

For forced measurements our expectation is that $r_c$ is given by the result of the tree calculation.
Numerically, it in fact becomes unmeasurably small at a significantly smaller value of $r$. 
Our interpretation of this is that, because of exponential scaling in  Eq.~\ref{eq:overviewS2tree}, the quantities $s(r)$ and $a(r)$ vanish extremely fast as ${r\rightarrow r_c}$.  A more stringent test of the identity of the two transition points would be valuable.

We have also examined the observables discussed here in 1+1D circuits, motivated by the fact that, 
since they do not require us to introduce a spatial bipartition of the system, they avoid introducing a lengthscale that is smaller than the system size. 
We will report on this elsewhere.

\subsection{Replicas and field theories}
\label{sec:overviewreplicas}

A key question is whether useful continuum field theories 
can be written for the MPT and FMPT, and also for entanglement transitions in (reasonably generic\footnote{The term ``random tensor network'' allows for almost any structure, so infinite numbers of universality classes can in principle be accessed, most of them extremely fine-tuned.})  random tensor networks.
This question has not been resolved, despite  progress on mapping the quantum problems to effective ``classical'' lattice models \cite{hayden2016holographic,nahum2018operator,zhou2018emergent,vasseur2019entanglement,jian2020measurement,bao2020theory, zhou2019entanglement}.
A basic issue is the need to handle disorder. 
The most familiar approach to this is to use the replica trick  
\cite{vasseur2019entanglement,zhou2018emergent,jian2020measurement,bao2020theory}. 
(In this section we use $N$ to denote the number of replicas: this should not be confused with the number of physical  spins in the previous sections.)
However the complicated $N$-dependence of the interactions makes it unclear a priori how to coarse-grain these effective lattice models.

In Sec.~\ref{sec:landautheory} we start by reviewing the approach 
of mapping circuits and tensor networks to effective lattice models for permutations.
We discuss coarse-graining of such models in a heuristic way. 
We then suggest an alternative way of thinking about effective statistical mechanics models for circuits (motivated by a physical picture for the emergence of permutations, in terms of phase cancellation in sums over Feynman histories \cite{zhou2018emergent,zhou2019entanglement}).
This picture connects entanglement transitions to approaches familiar from disordered magnetism,  the  random field Ising model,  spin glasses, etc. \cite{fischer1993spin}. 

With this motivation, we construct  the simplest Lagrangians that capture the global symmetry associated with the replica formulation 
\cite{vasseur2019entanglement,zhou2018emergent, zhou2019entanglement}, which we denote
\be
G_N \equiv \lf S_N \times S_N \ri \rtimes \mathbb{Z}_2,
\ee
and which pass some basic consistency tests.

 The limiting number of replicas $N$ is distinct for the case of  \textit{(i)} the MPT and \textit{(ii)} both the FMPT and the RTN \cite{vasseur2019entanglement,jian2020measurement, bao2020theory}. For the FMPT  and RTN we need to take ${N\rightarrow 0}$, as in standard quenched disorder problems. For the MPT, realizations are weighted by the additional Born rule factor, which increases the number of replicas:  we need to take ${N\rightarrow 1}$ \cite{jian2020measurement,bao2020theory}.
Previously, properties in the vicinity of a fine-tuned point have been used to motivate the suggestion that all of these problems may have similar universal properties, despite the differing numbers of replicas \cite{jian2020measurement}. 
However, we find that  the simplest  field theory candidates
(which may of course be too simple)
 are strikingly different in the two different cases.

The Lagrangians we propose have the schematic forms 
\ba
\label{eq:overviewlagrangians}
\mathcal{L}_X&  = \sum_{ab} \, \left[ \,
 (\partial  X_{ab})^2 + \mu  X_{ab}^2 +  X_{ab}^3
\, \right],
\\
\mathcal{L}_Y & =
 \sum_{ab} \left[
(\partial Y_{ab})^2 
+   \nu Y_{ab} 
+  Y_{ab}^3
\right] +  \sum_{abcd} Y_{ab} F_{ab,cd} Y_{cd}.
 \notag
\end{align}
$F$ is the tensor $F_{ab,cd} = \delta_{bd} + \delta_{ac}$.
We have suppressed all coupling constants except the crucial one that drives the transition, denoted $\mu$ or $\nu$.
Both space and time derivatives are grouped together in the derivative term: in the case of the circuit there 
will in general be a nonuniversal speed $v$ appearing, so that the derivative terms have the form 
${(\partial_t X)^2 + v^2 (\nabla X)^2}$. The plus sign means that there is an emergent Euclidean, rather than Lorentzian, spacetime symmetry 
\cite{skinner2019measurement,li2019measurement, li2020conformal}.\footnote{For generic versions of the MPT and FMPT, the emergent spacetime symmetry is of course partly a conjecture. It is perhaps made more plausible by the existence of such symmetries in some simpler limiting models. The minimal cut problem \cite{skinner2019measurement}, and also some alternative $q\rightarrow\infty$ limits \cite{jian2020measurement, bao2020theory}, map to percolation problems which have this symmetry. Some  measurement induced critical points with a free fermion structure also map to conformally invariant models 
\cite{nahum2020entanglement, sang2020measurement,lavasani2020measurement,lang2020entanglement, alberton2020trajectory}. Conformal invariance in 1+1D Clifford measurement circuits has been demonstrated numerically \cite{li2019measurement, li2020conformal}. 
There is numerical evidence that the dynamical exponent is unity for the Haar-random  MPT \cite{skinner2019measurement}.
Finally, we may use dual-unitary circuits  to set up  measurement circuits that have $90^\circ$ rotational invariance in spacetime even microscopically (we will discuss this elsewhere).}

In the Lagrangian $\mathcal{L}_X$, the field $X_{ab}$ is a real ${N\times N}$ matrix satisfying $\sum_a X_{ab}=0$ and $\sum_b X_{ab}=0$. It may be thought of (modulo a constant shift) as a coarse-grained permutation matrix.
This Lagrangian is appropriate for the replica limit ${N\rightarrow 1}$. 
It has upper critical dimension ${D=6}$ 
(this is the spacetime dimension in the case of the circuit).
This is a candidate Lagrangian for describing the MPT.

At first we might assume that the same Lagrangian $\mathcal{L}_X$ for the measurement transition could be continued to the distinct limit ${N\rightarrow 0}$ in order to describe the random tensor network and the forced measurement transition. 
We argue in Sec.~\ref{sec:landautheory} that this is not the case.
Instead, the  simplest candidate for the FMPT and RTN is the Lagrangian $\mathcal{L}_Y$. 
Here, the field $Y$ is a real ${N\times N}$ matrix, with ${N\rightarrow 0}$, that does \textit{not} satisfy any constraints on its row and column sums.
The upper critical dimension for this theory is the unexpectedly large value ${D=10}$. See Sec.~\ref{sec:landautheory} for further discussion.

We caution that these theories are conjectures based on symmetry considerations and certain limited consistency checks. Further investigation is required to determine whether they are in fact sufficient to describe the problems of interest. It is possible that more elaborate continuum descriptions are required, 
either for  a particular microscopic model or in general.

Indeed, the trees described in Sec.~\ref{sec:overviewtree}, which have exponential order parameter scaling close to the critical point, appear to be one case that is not captured by $\mathcal{L}_Y$. 
(Contrary to the naive guess that the high-dimensional limit of the field theory and the tree would show similar ``mean field'' critical scaling.) We defer an examination of the reason for this to a future work.

In Sec.~\ref{sec:landautheory} we also present some results that are independent of the speculative field theories above. 
In particular we use effective domain wall pictures to obtain the scaling within the phases
(mentioned above in Sec.~\ref{sec:overviewdetecting}).

We also briefly discuss the use of Ising toy models for the properties of the second R\'enyi entropy in measurement dynamics, pointing out that the formalisms of \cite{zhou2018emergent} or  \cite{zhou2019entanglement} allow these to be justified in certain strongly entangled regimes, rather than being regarded simply as toy models as in previous work. However, quenched disorder must be taken into account in the resulting Ising model. Additionally, the Ising picture breaks down close to the critical point  (or in the disentangling phase) and also at long times.

Finally we discuss variants of the MPT, FMPT and RTN phase transitions. We point out that quite different scaling obtains for models of free fermions subjected to measurements, as a result of continuous rather than discrete replica symmetry.

\section{Minimal cut problem}
\label{sec:classical}

A natural starting point for understanding the MPT is to map the  quantum circuit to a classical graph
on which one can study a classical ``minimal cut'' optimization problem  \cite{skinner2019measurement}. 
In this mapping there is a phase transition at the point where the graph percolates.

We  think of this classical  min-cut problem  as a toy model for the generic quantum transition. 
In the circuits we study, the cost of the minimal cut  gives the exact value of the (somewhat unphysical) zeroth R\'enyi entropy,\footnote{$S_0$ counts the (logarithm of the) number of nonzero Schmidt values in the singular value decomposition of a state or, as we focus on here, an operator.} $S_0$. 
It also gives exact results for the other R\'enyi entropies in the limit of large local Hilbert space dimension (e.g., a large value of each spin), with Haar random gates.
But in general, the minimal cut is only an upper bound on the entanglement entropies $S_n$ with $n\geq 1$. (There can be no quantum information propagated from the initial to the final time if the associated classical graph is disconnected; in this regime, the ``cost'' of the minimal cut vanishes.) The true ``quantum'' transition in general occurs at a smaller value of $r$ than the classical transition discussed in this section  (and in general has distinct universal properties). 
Despite this, the classical problem conveys some useful lessons.

Viewed as a graph, the circuit is a bond percolation configuration, as described below. 
The frequency of projective measurements determines the fraction of broken bonds in this percolation configuration.
The minimal cut is the minimal number of additional bonds that must be severed in order that two parts of the boundary of the circuit, $A$ and $\bar A$ , no longer have any percolating path between them.
This minimal cut is a unifying heuristic \cite{Swingle2012, pastawski2015holographic, casini2016spread, hayden2016holographic, nahum_quantum_2017, jonay2018coarse} for the entanglement of various objects, depending on how we choose $A$ and $\bar A$.  
If these are taken to be two complementary subsets of the legs of the circuit at the final time, then the minimal cut gives the entanglement $S_0$ of a subset $A$ of the spins in the final state quantum state, assuming the initial state was a product state.
Here we are more interested in a  minimal cut separating the top boundary of the circuit from the bottom. That is, $A$ contains all the circuits ``legs'' at the final time, and $\bar A$ all those at the initial time.
This ``horizontal'' minimal cut is a measure of information transmitted from the initial to the final time, equal to the operator entanglement $S_0$ for the nonunitary time evolution operator $V$ (Sec.~\ref{sec:overview}).

In the percolating regime, this horizontal minimal cut must sever a number of bonds that is extensive in the number of spins $N$, so that $S_0 \simeq s N$. The coefficient $s$ is a ``surface tension''  \cite{jonay2018coarse}  for the minimal cut, which vanishes continuously at the percolation threshold.
In 1+1D this transition is conformally invariant. 
Many of the critical exponents, such as the correlation length exponent $\nu$, are standard percolation exponents, while others are less familiar, since the minimal cut is an additional optimization problem built on top of the percolation configuration~\cite{skinner2019measurement, chayes1986critical}.

In the circuit without fixed spatial structure, where any qubit can couple randomly to any other, the location of the critical point, and the basic critical exponents, can be determined exactly, as we show in this section.
These exponents also apply to the finite-dimensional minimal cut problem when the spatial dimensionality $d$ is greater than or equal to 5 (Sec.~\ref{sec:largefinited}). 
Interestingly, there is also reason to speculate that the exponents apply for some versions of the quantum measurement transition in high dimensions, even without the minimal cut approximation (see Sec.~\ref{sec:landautheory}, where we discuss Landau theory for the measurement transition and entanglement transitions).

\begin{figure}
    \centering
    \includegraphics[width=\linewidth]{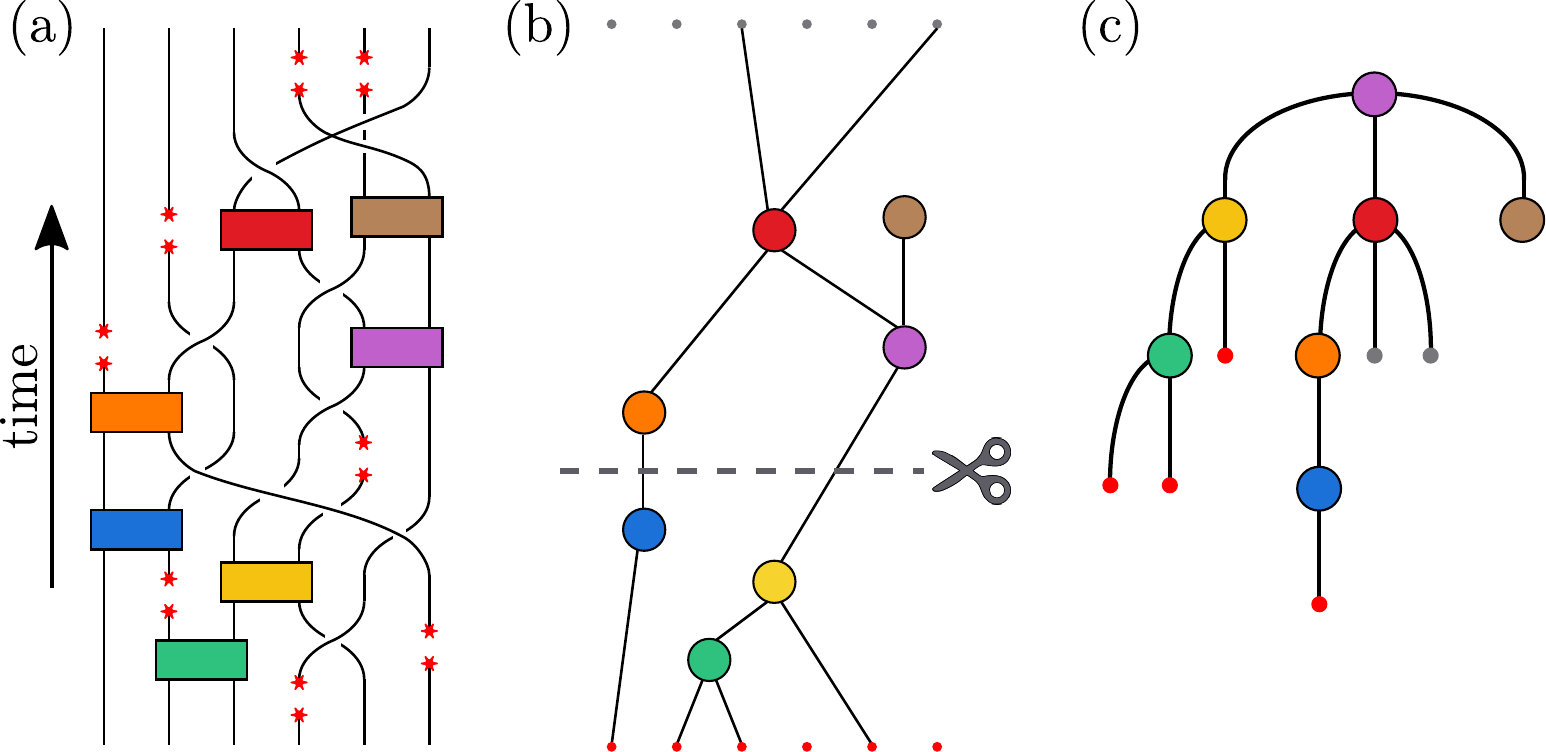}
    \caption{
    (a) Example of a small unitary circuit with $N=6$ spins. Black worldlines represent the evolution of a particular spin, with time proceeding vertically.  Colored blocks indicate two-spin unitaries, and broken lines (marked with red crosses) indicate single-spin measurements. (b) The equivalent graph, with nodes representing unitaries (node of a particular color corresponds to the unitary of the same color in (a)) and edges representing unbroken segments of worldline.
    Small red/gray circles denote the initial/final time for a given spin. A possible minimal cut for this graph is shown by the dashed line: removing the two indicated edges disconnects the initial and final times. (c) The classical graph arranged as a tree, with the purple node used as a seed and generation number $k$ proceeding downward. (This illustrative circuit forms a tree; in general the structure of a large circuit is only locally treelike.)
    }
    \label{fig:circuit-schematic}
\end{figure}

For all-to-all circuits, the classical percolation problem is defined as follows. The circuit defines a random graph, in which the nodes (vertices) correspond to unitaries and the edges are the sections of spin worldline that are not broken by measurements.
In other words, an edge connects two nodes whenever (i) the two nodes correspond to successive unitaries in the time evolution of a particular spin; and (ii) that spin is not measured during the time in between the two unitaries.  
Figure \ref{fig:circuit-schematic}(a) shows an example circuit, and Fig.\ \ref{fig:circuit-schematic}(b) shows the corresponding graph.
Each node has at most four edges connected to it, corresponding to the four legs of each unitary in Fig.~\ref{fig:circuit-schematic}(a).
The minimal cut in the figure indicates an operator entanglement $S_0=2$ for this small circuit.

We take the number of spins to be very large, ${N\gg 1}$, while by definition the degree (connectivity) of each node is only of order 1.
In this situation, standard considerations \cite{bollobas2001random} imply that the local structure of the graph is \textit{treelike} on both sides of the percolation transition. 
Above the percolation transition closed loops do exist, but their length is of order $\ln N$.

\subsection{Local tree structure and percolation}
\label{sec:classicaltreestructure}

To relate the classical graph to a tree,  imagine starting at an arbitrarily chosen ``seed'' unitary in the bulk of the circuit (far from the initial and final time boundaries) and tracing out its cluster:  finding the nodes connected to the seed by an edge, then those connected to the seed by a path of length 2 edges, etc.
In this way the cluster containing the seed may be arranged in a tree, with the seed at the top and subsequent generations of connected nodes below: see Fig.~\ref{fig:circuit-schematic}(c). We denote the generation number by $k$, with $k=0$ being the seed.

The probability $p$ that a given one of a unitary's four possible edges is \textit{absent} is equal to the probability that (as we travel along that segment of worldline) the spin undergoes a measurement before it is involved in another unitary. This probability is given by
\be\label{eq:pandr}
p = \f{r}{2-r}.
\ee

The small circuit shown in Fig.~\ref{fig:circuit-schematic} contains no loops. In general the circuit can contain loops.  However, a  sub-cluster of any finite size is guaranteed to be free of loops in the limit $N\rightarrow\infty$ (since the probability that two unitaries in generation $k$ both connect to the same unitary in generation $k+1$ is of order $1/N$).

To understand the location of the critical point, note that the average branching number of the tree (the average number of descendants of a given node with $k>0$)
is $3\times (1-p)$. The percolation transition in the graph occurs when the branching number is 1, i.e.\ at $p_c = 2/3$ (as also noted in Ref.\ \cite{gullans2019dynamical}), or 
\be
r_c = \f{4}{5}.
\ee
(In this section only, $r_c$ denotes the classical transition point, $r_c=r_c^\text{cl}$.)
When $r$ is greater than $r_c$, all trees are finite even in the limit $N, t\rightarrow \infty$, where the graph itself is infinite: starting at a seed node, the tree inevitably dies out after a finite number of generations. 
Therefore at $r > r_c$ all unitaries are in finite clusters; this is the non-percolating phase.
When $r<r_c$, however, there is a nonzero probability $f_\infty$ that a tree continues forever, or rather until it includes a number of nodes proportional to $N$.
In the percolation problem, $f_\infty$ is the order parameter --- the probability that the unitary lies in the infinite cluster. The critical exponent $\beta$ for this order parameter is 1, which is the mean field value for percolation.  A simple recursive treatment (App.~\ref{app:finfinity}) shows that, close to $r_c$,
\ba
\label{eq:finfty}
f_\infty & \simeq \frac{50}{3} (r_c - r)^\beta, 
&
\beta & = 1.
\end{align}
Note that the window of $r$ in which we can hope to observe critical scaling, corresponding to $0 < f_\infty \ll 1$, is rather narrow as a result of the large (nonuniversal) prefactor in Eq.~\ref{eq:finfty}.

\subsection{Effective 1D continuum theory}
\label{sec:percscalingforms}

We now show that near the critical point the basic scaling variables for the percolation and minimal cut problems are:
\ba
\label{eq:basicscalingvariablesclassical}
&\frac{t}{N^{1/5}}, 
&
\delta r N^{2/5},
\end{align}
where $\delta r = r - r_c$ and $t$ is, say, the temporal duration of the evolution.
For example, the characteristic timescale for a large system at its critical point scales as $N^{1/5}$.
In Secs.~\ref{sec:percolationnumerics},~\ref{sec:classicalmincut} and Apps.~\ref{app:percpotts}-\ref{app:correlationfunctions}
we will show how these variables appear in scaling forms for the minimal cut and other observables.
{  The critical exponents in Eq.~\ref{eq:basicscalingvariablesclassical} have also been  obtained in independent work \cite{Gullans_unpublished},
by an approach that is complementary to the one below (Sec.~\ref{sec:overviewmincut}).}

This problem is similar to one of crossover scaling, in which a system that is effectively very high dimensional on short timescales crosses over to one that is one-dimensional on long scales. This analogy can be used to obtain the above exponents, as we discuss in Sec.~\ref{sec:largefinited}. This approach also sheds light on the quantum problem (Sec.~\ref{sec:landautheory}). Here, however, we solve the classical problem directly. 

To simplify the discussion, let us consider a percolation problem with the same basic features as the circuit, but with a simpler connectivity rule inspired by the Erd\H{o}s-R\'enyi random graph \cite{bollobas2001random}. This simplification does not change the universality class, as we show numerically in App.\ \ref{app:layeredER}.
The random graph we consider has a layered structure, with one layer for each timestep. 
This graph may be contrasted with one studied in Ref.\ \cite{vijay2020measurement}, which maps a measurement transition in a class of ``instantaneous quantum polynomial time'' circuits to the percolation transition in an Erd\H{o}s-R\'enyi graph without a time dimension.

We discretize the time $t$ in integer steps. At each $t$ we have $N$ nodes, labelled $(i, t)$ with $i=1,\ldots, N$. We allow edges only between sites in adjacent $t$ layers,  each edge being present with probability $b'/2N$, independently of the others. This scaling with $N$ ensures that the average degree of a site, $b'$, is $\mathcal{O}(1)$, as in the circuit. It is easy to see by thinking about the local tree structure that the phase transition is at $b'_c=1$.
As in the circuit, connectivity is local in time, but there is no notion of spatial structure with a layer at a fixed time.

Classical percolation can be mapped to the $Q$-state Potts model in the limit $Q\rightarrow 1$ \cite{fortuin1972random,zia1975critical,amit1976renormalization,cardy1996scaling}. For our problem, the fact that each site couples to \textit{all} the sites on the adjacent layers means that the Potts partition function simplifies  after a Hubbard Stratonovich transformation with a field $\Phi(t)$ that depends only on time. This transformation is shown in detail in App.~\ref{app:percpotts}.  The field $\Phi$ may be taken to be a $Q\times Q$ traceless diagonal matrix, on which Potts symmetry acts by permuting the diagonal components.

It is possible to take the continuum limit in a controlled way, to give an effective one-dimensional field theory.
Close to the critical point, such that ${b' - 1 = \delta b' \ll 1}$, the partition function for this field theory is
\be\label{eq:Zpotts}
Z = 
\int \mathcal{D}\Phi
\exp\lf -\int \dd t \, \mathcal{L} \ri,
\ee
with 
\be \label{eq:Lpotts}
\mathcal{L}
=
\f{1}{4} \tr \, (\partial_t \Phi)^2
- 
\f{\delta b'}{2} \tr \Phi^2
-
\f{1}{6 \sqrt{N}}
\tr \Phi^3.
\ee
Modulo the values of the order 1 constants, we expect the same field theory to apply to the percolation model arising from the circuit.

The factor of $1/\sqrt{N}$ in Eq.~\ref{eq:Lpotts} allows a long timescale and nontrivial scaling forms to emerge at the critical point ${\delta b' = 0}$, despite the fact that the effective field theory is one-dimensional. One-dimensionality implies that for any fixed $N$, correlations decay exponentially at sufficiently large $t$, but the timescale diverges with $N$.

The critical exponents for the minimal cut  problem in the all-to-all circuit follow from the observation that the change of variables
\ba\label{eq:needtolabel}
\tilde{t} & = \frac{t}{N^{1/5}}, &
u & = \delta r \, N^{2/5}, &
\widetilde \Phi & = \f{\Phi}{ N^{1/10}},
\end{align}
eliminates $N$ from the action.  Scaling forms for correlation functions follow from this fact together with the corresponding scailings for operators. We discuss some examples in the following subsection and in App.\ \ref{app:correlationfunctions}.

We may also obtain these exponents from a crossover scaling argument if we assume that they are the same as those in a system which \textit{does} have spatial structure, but with a very high spatial dimensionality $d$. 
This crossover is described in Sec.~\ref{sec:largefinited}.  

In fact, the exponents in Eq.~\ref{eq:basicscalingvariablesclassical} apply  for any $d>5$ (in an appropriate regime of timescales) with logarithmic corrections in $d=5$. This is because $d=5$ gives a total spacetime dimension of 6, which is the upper critical dimension for percolation. This fact allows an even simpler mnemonic for the above exponents. Suppose for a moment that we are considering a graph with a regular lattice in spatial dimension $d=5$, with $N=L^d=L^5$, where $L$ is the system size. $d=5$ is the lowest dimension in which mean-field exponents apply (up to logarithms).
In this picture, the first scaling variable above is simply $t/L$, corresponding to the dynamical exponent $z=1$ in the 5-dimensional theory, and the second scaling variable is $u=\delta r L^{1/\nu}$, with the mean field correlation length exponent $\nu = 1/2$. 
In $d>5$ we must also consider the dangerous irrelevance of the interaction term in the field theory \cite{aharony1984scaling} (which means that the relevant timescale is no longer $t/L$),
but this term can be treated using a standard coarse-graining argument  (Sec.~\ref{sec:largefinited}).

\subsection{The percolation probability}
\label{sec:percolationnumerics}

Before describing the minimal cut itself (Sec.~\ref{sec:classicalmincut}), we first consider an observable that is simpler to study both analytically and numerically -- namely, the probability $\pperc$ of percolation between initial and final times in the classical graph.  The value of $1-\pperc$ is equivalent to the probability that the operator entanglement is exactly zero, since non-percolation of the classical graph implies that the initial and final times are causally disconnected.

$\pperc$  has scaling dimension zero, i.e.\ it has no power-law prefactor in $N$,  so it is useful for numerical tests of the scaling defined by Eq.~\ref{eq:basicscalingvariablesclassical}. In App.~\ref{app:correlationfunctions} we present numerical results for two  observables with nontrivial scaling dimension: namely, the probability of two nodes on either the same or opposite time boundaries being connected to the same cluster. We show that these observables are also described by the scaling variables in Eq.~\ref{eq:basicscalingvariablesclassical}.

In the Potts language, $\pperc$ is expressed in terms of the free energy cost of twisted boundary conditions \cite{cardy1992critical}. (In the 1D field theory this free energy involves boundary magnetic fields that are parametrically large in $N$; this is discussed in App.~\ref{app:percpotts}.) We obtain the scaling form:
\ba
\label{eq:scalingformpperc}
\pperc & = F\lf \f{t}{N^{1/5}} , \,  N^{2/5}  \delta r \ri.
\end{align}
(Here $t$ denotes the full temporal duration of the dynamics.) 
First consider the critical point $r = r_c$, for which 
\ba
\pperc & = F_\text{crit} \lf \f{t}{N^{1/5}} \ri.
\end{align}
In principle we should obtain a scaling collapse simply by plotting ${\pperc}$ as a function of the scaling argument. Practically speaking, however, the characteristic timescale $N^{1/5}$ is modest for the values of $N$ we can access numerically, and it appears to be necessary to include a subleading correction. This correction is of a type that is generically present for non-periodic boundary conditions, and corresponds to replacing the scaling variable with $(t-c_0)/N^{1/5}$, for a nonuniversal $\mathcal{O}(1)$ constant $c_0$.  

\begin{figure}
    \centering
    \includegraphics[width=\linewidth]{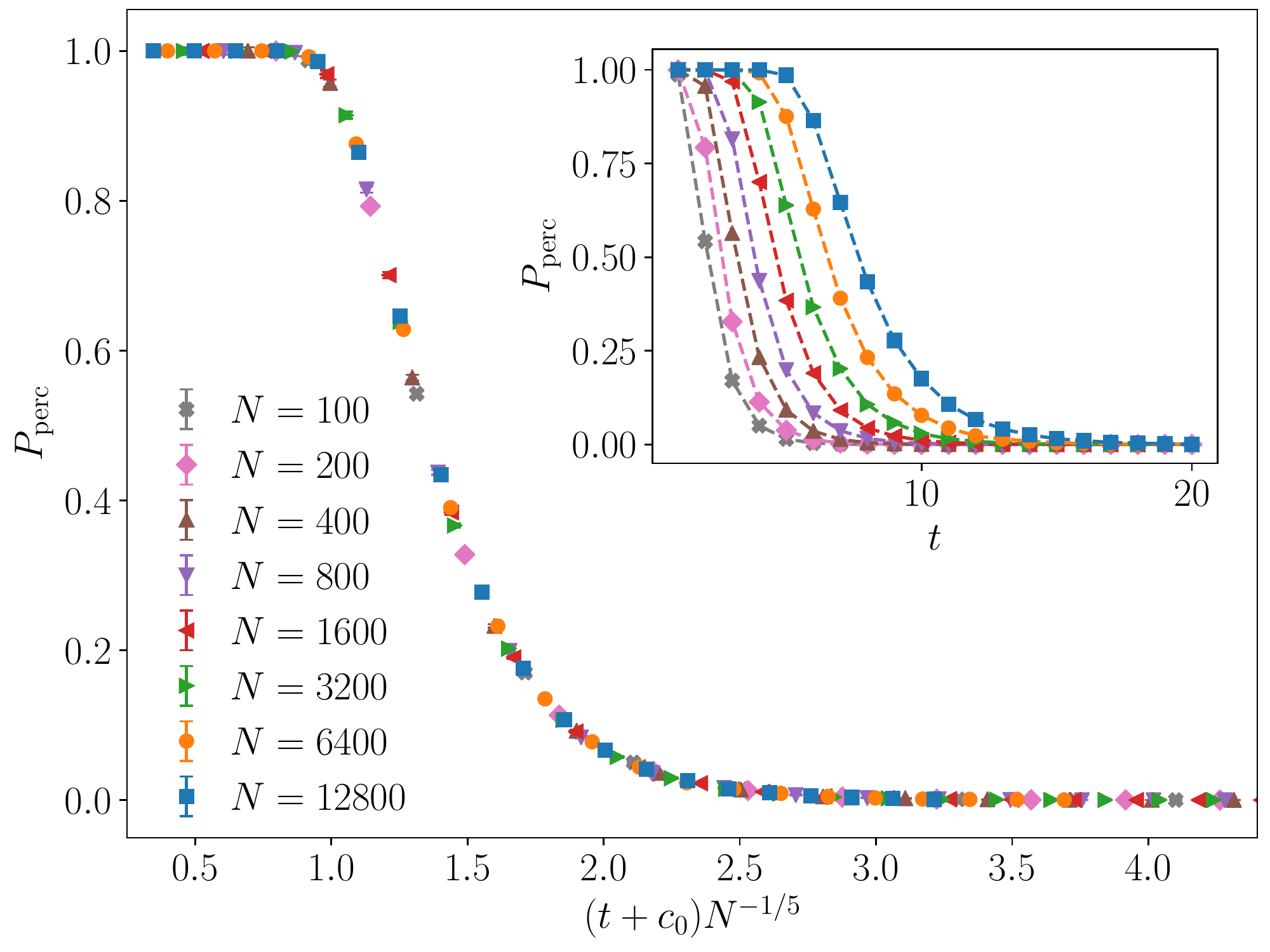}
    \caption{The probability of percolation, $\pperc$, as a function of time for the classical circuit  with the critical measurement rate $r = r_c = 4/5$. Different curves correspond to different system sizes $N$. The inset shows the raw data. In the main panel the time is rescaled by $N^{1/5}$ and a shift $c_0$ is introduced, with $c_0 \approx 1.3$. All data is averaged over 40,000 realizations.
    }
    \label{fig:Pperc}
\end{figure}

Figure \ref{fig:Pperc} (inset) shows raw data for the percolation probability $\pperc$ of the classical graph (for the all-to-all circuit) as a function of time and $N$. As can be seen in the main panel, this data collapses onto a single curve when $\pperc$ is plotted against $(t+c_0)/N^{1/5}$, where $c_0 \approx 1.3$.

\begin{figure}[!h]
    \centering
    \includegraphics[width=\linewidth]{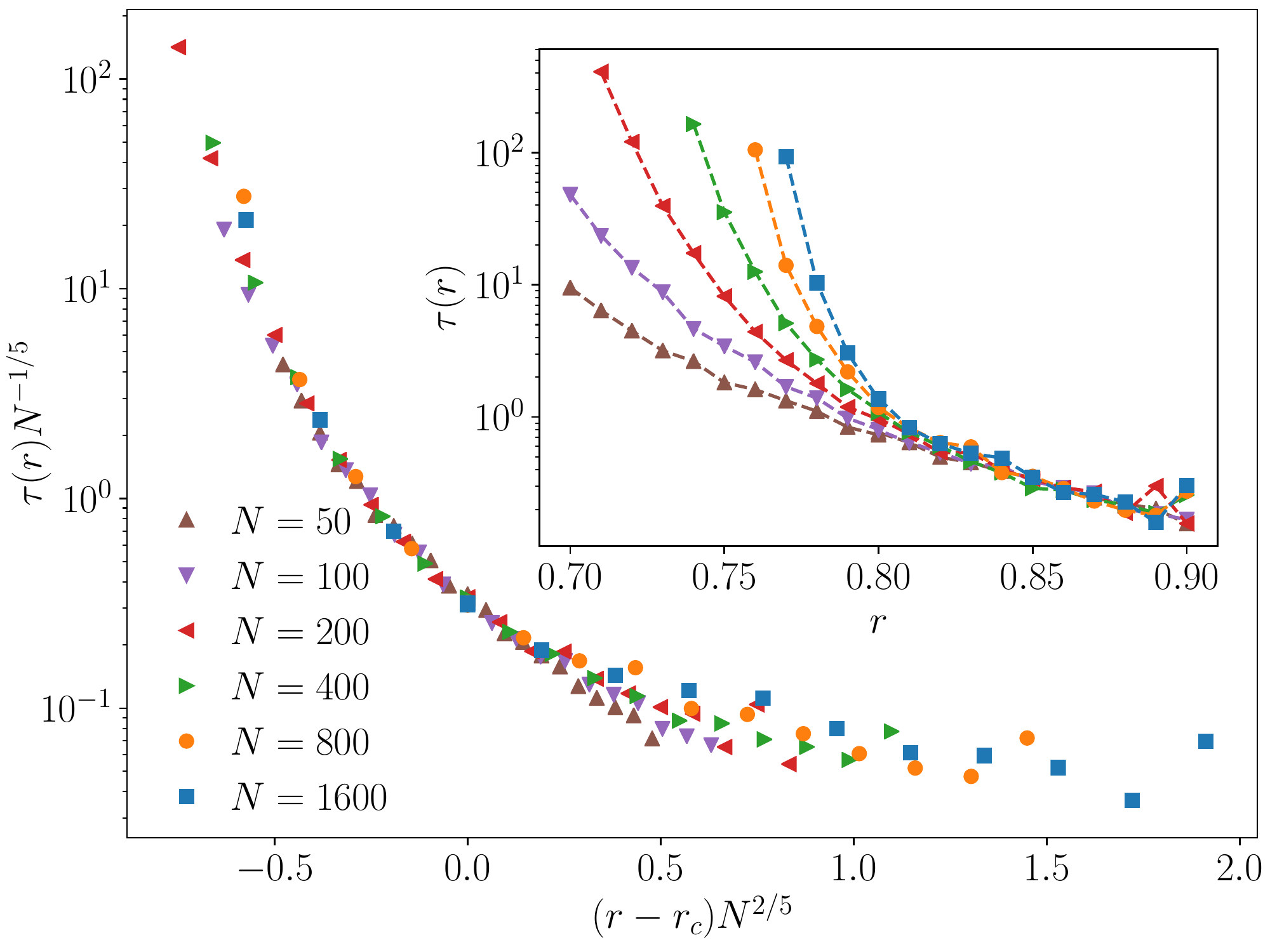}
    \caption{The characteristic decay time $\tau$ for the percolation probability.  The inset shows $\tau$ as a function of the measurement rate $r$ for different values of the system size $N$.  The main figure shows this same data plotted as a function of the scaling variables in Eq.~\ref{eq:basicscalingvariablesclassical}.
    }
    \label{fig:tau_scaled}
\end{figure}

At any fixed values of $r$ and $N$, the probability $\pperc$ decays exponentially with time $t$ at large enough values of $t$. One can extract the associated decay time $\tau(r,N)$, which according to Eq.~\ref{eq:scalingformpperc} has the scaling form 
\be\label{eq:tauscalingform}
\tau (r,N) = N^{1/5} \, W \left( N^{2/5}\, \delta r \right).
\ee
This scaling is confirmed in Fig.~\ref{fig:tau_scaled}. Figure \ref{fig:tau_scaled} comprises a check of off-critical scaling close to $r_c$ as well as the scaling at $r_c$ that is shown in Fig.~\ref{fig:Pperc}.

The decay time $\tau (r,N)$ of the percolation probability constitutes one way of defining a characteristic timescale over which information is able to propagate between the initial and final times. A key feature of the classical graph, which carries over to the quantum case, is that the timescale $\tau$ grows very rapidly with $N$ within the entangling phase.
At any fixed $r<r_c$, we can argue that as $N\rightarrow \infty$ the timescale $\tau$ grows as
\be\label{eq:classicaltimescaleisexp}
\tau_{r<r_c} \sim \exp \lf a(r) \times N \ri,
\ee
neglecting power-law prefactors.
In the present classical problem, this exponential growth can be understood in terms of rare events that disconnect the cluster. 
Close to the transition we must have
\be\label{eq:timescaleexponentialgrowthrate}
a(r) \sim (\delta r)^{5/2}
\ee
in order to match the scaling form.\footnote{A conclusive numerical check of the exponent in Eq.~\ref{eq:timescaleexponentialgrowthrate} would require larger system sizes since it requires the scaling function in Fig.~\ref{fig:tau_scaled} to have large negative argument.} 
We expect that $\pperc$ is close to 1 for $t\ll \tau$.
This exponentially long timescale can also be seen directly from the field theory in Eq.~\ref{eq:Lpotts},  in terms of ``instantons''  in the field theory (domain walls in time); see App.~\ref{app:percpotts}.

As mentioned in the previous subsection, one can model the minimal cut problem using the simpler setup of a sequence of layered Erd\H{o}s-R\'{e}nyi random graphs. Within this model one can calculate the percolation probability $\pperc$ and characteristic decay time $\tau$.  In App.~\ref{app:layeredER} we show that this layered Erd\H{o}s-R\'{e}nyi model gives the same scaling behavior as in Figs.~\ref{fig:Pperc} and \ref{fig:tau_scaled}.

Percolation two-point functions give further information on the connectivity of the circuit. These are analyzed in App.~\ref{app:correlationfunctions}.

\subsection{Scaling for the minimal cut}
\label{sec:classicalmincut}

Because of the lack of spatial structure in the all-to-all model, it is natural to focus on the transmission of information between the initial and final times. One measure of this transmission is the operator entanglement of the linear, but nonunitary, operator  $V$ that defines the time evolution for a particular sequence of measurement outcomes (Sec.~\ref{sec:overview}). 

In the minimal cut picture, the operator entanglement between initial and final times is the cost of the minimal cut through the circuit that separates the initial and final times [as illustrated in Fig.~\ref{fig:circuit-schematic}(b)].
We refer to this cost as $S_0$ (the Hartley entropy), although in some cases (including the limit of infinite local Hilbert space dimension, mentioned above) it is equal to the other R\'enyi entropies as well. 

The behavior of $S_0$ is most interesting within the entangled phase, so let us consider some fixed ${r<r_c}$. As illustrated in the previous subsection, in this phase there is an exponentially large (in $N$) timescale over which the percolation probability is close to $1$. 
Correspondingly, there is a parametrically large time range, corresponding to
\footnote{We write this formula for the case where ${r_c-r}$ is of order 1. Otherwise the lower limit on the range may involve a critical timescale that is larger than 1 but much smaller than $\tau$.}
${ 1\ll \ln t \ll  N}$,  over which $S_0/N$ is approximately constant. The crudest picture for the subleading corrections gives\footnote{For a naive picture of the scaling of $S_0$ with $N$, consider a minimal cut that has zero temporal width within the entangled phase. There are $\mathcal{O}(t)$ choices for the time at which to place the cut. Each choice has a random cost, which for the present illustration we assume to be Gaussian with variance $N$, arising from a sum of $\mathcal{O}(N)$ random contributions. Taking the minimum gives the formula above.
The second term is subleading so long as ${\ln t \ll N}$.}
\be\label{eq:S0plateausubleading}
{S_0/N = s  - \mathcal{O}\lf \sqrt{ N^{-1}\ln t } \ri.}
\ee
We refer to the range of times where subleading terms are negligible as the ``plateau'' in the entanglement.
Over this large time window, the horizontal minimal cut has a well-defined \text{cost per spin}, $s$.
This cost per spin $s$ is the infinite-dimensional version of the line  tension for the minimal cut in the 1+1D  case or the  surface tension in the  2+1D case  \cite{skinner2019measurement}. These quantities all vanish at the critical point.
In the plateau regime, the information \textit{per spin} transmitted by the circuit is nonzero, up to an exponentially large time.

\begin{figure}
    \centering
    \includegraphics[width=\columnwidth]{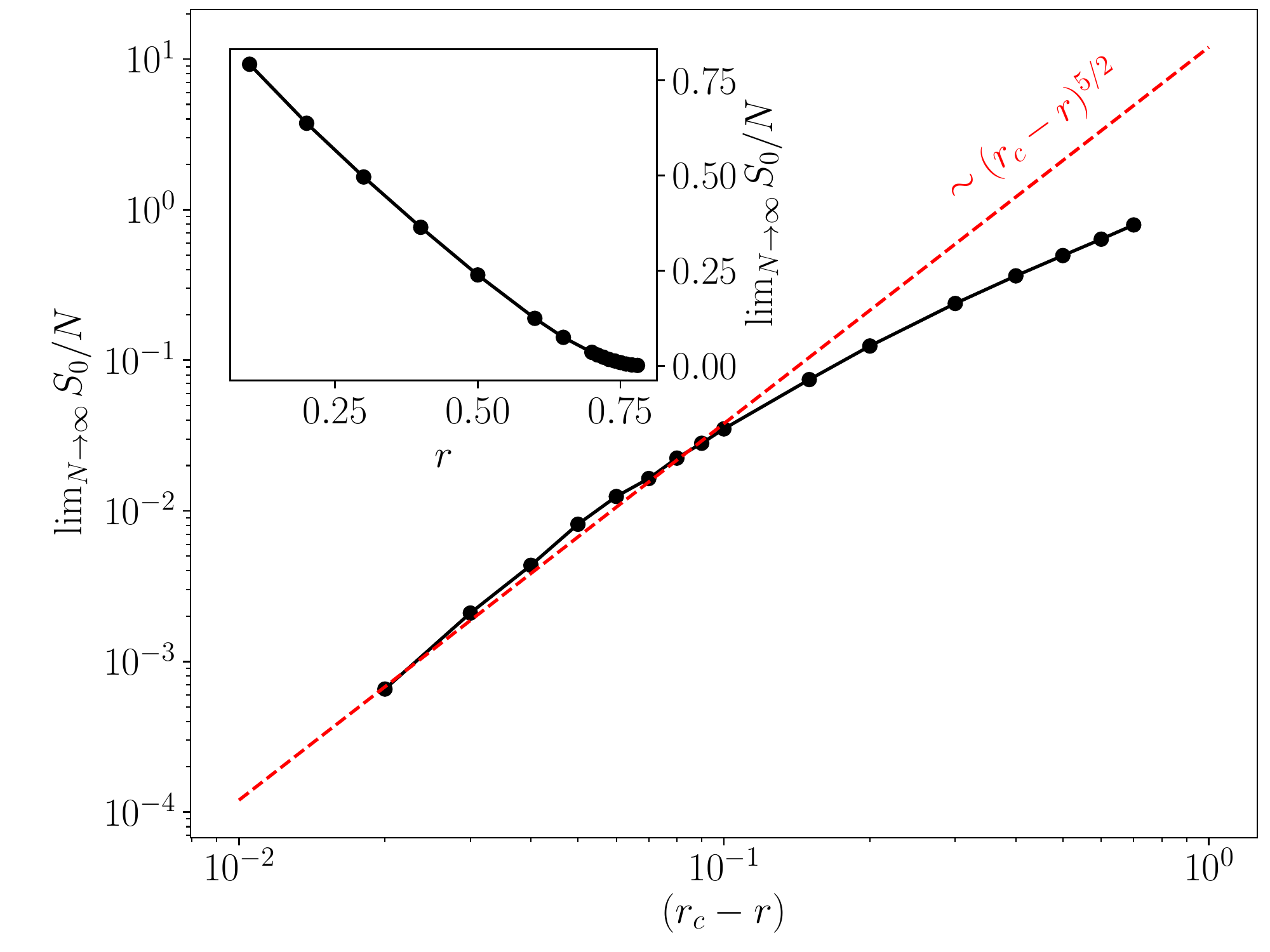}
    \caption{ The cost of the minimal cut, $S_0$, divided by the system size $N$, extrapolated to the limit of $N \rightarrow \infty$. The main figure shows $S_0/N$ as a function of $r$, while the inset shows $S_0/N$ as a function of $r_c - r$ in double-logarithmic scale.  The dashed red line shows the dependence $S_0/N \propto (r_c -r)^{5/2}$. Details of the extrapolation procedure are discussed in the text. Error bars are smaller than the symbol size. 
    }
    \label{fig:S0N-extrapolated}
\end{figure}

The inset of of Fig.~\ref{fig:S0N-extrapolated} shows this cost per spin in the entangled phase, as measured by a numerical simulation using the Ford-Fulkerson method \cite{cormen2009introduction}.  The details of the extrapolation to large $N$ are described below and in App.~\ref{app:extrapolation}.

Let us relate this minimal cut to the scaling theory close to the critical point.
We expect the scaling form
\be
\label{eq:mincutscaling}
S_0 =  H \lf \f{t}{N^{1/5}} , \,  N^{2/5} \delta r \ri.
\ee
(see Sec.\ \ref{sec:percscalingforms}).
If we assume that within the entangled phase there is a time regime during which $S_0$ is extensive in $N$ and time-independent (i.e. independent of the first scaling variable above), then we obtain in this regime 
\ba
S_0 & = s(r) \, N, 
\end{align}
with the entropy per spin $s(r)$ scaling as 
\ba\label{eq:costperspin}
s(r) & \sim (r_c - r)^{5/2} 
& 
(r & \lesssim r_c).
\end{align}
The main panel of Fig.~\ref{fig:S0N-extrapolated} shows $s(r)$ close to the critical point on a double logarithmic scale. Though we cannot extract a clear power law from the data, it seems roughly consistent with the  prediction (\ref{eq:costperspin}). 

In order to numerically obtain the value of $s(r)$ for the plots above,
we measure $S_0$  as a function of the time $t$ and the system size $N$ from simulations.  
For a fixed $t$, we find that $S_0(t, N) / N$ has a linear dependence on $1/\sqrt{N}$ at large $N$ (in line with the simple picture in Eq.~\ref{eq:S0plateausubleading}).  This dependence allows us to estimate a value of $S_0(r)/N$ in the limit of $N \rightarrow \infty$ by extrapolating the linear relationship to $1 / \sqrt{N} = 0$.  Further details of this extrapolation procedure are presented in Appendix~\ref{app:extrapolation}.

\subsection{Finite dimensions with $d\geq 5$}
\label{sec:largefinited}

The scaling exponents that we found in Secs.~\ref{sec:percscalingforms} and \ref{sec:percolationnumerics} also apply to the classical problem in a system with a regular spatial lattice (and unitaries applied only between nearest-neighbors) in a large enough number of spatial dimensions $d$, as we now discuss. 
The total spacetime dimension, $d+1$, should be greater than $6$, which is the upper critical dimension for percolation (in $d=5$ we will have the same exponents with additional logarithms).

We start with the standard Potts representation of percolation \cite{fortuin1972random,zia1975critical,amit1976renormalization,cardy1996scaling} in ${d+1}$ dimensions. Suppressing all $\mathcal{O}(1)$ constants, as well as a nonuniversal velocity scale, a continuum action is
\be\label{eq:finitedimpercaction}
S = \int \dd t \dd^d x \, \tr \lf 
(\partial_t \phi)^2 + 
 (\nabla \phi)^2 + \delta r \phi^2 +  \phi^3
 \ri.
\ee
Here $\phi$ is a traceless diagonal ${Q\times Q}$ matrix, as in Sec.~\ref{sec:percscalingforms}. Our system is of extent $L$ in each of the spatial dimensions, with 
\be
\label{eq:NL}
N = L^{d},
\ee
and extent $t\gg L$ in the time direction. We take the UV cutoff (``lattice spacing'') to be 1.

We coarse-grain the system by a factor of order $L$, so that the spatial system size becomes comparable with the UV cutoff, and we have an effective 1D theory as far as correlations on scales $\gg L$ are concerned. 
Since the cubic coupling is irrelevant, with RG eigenvalue ${y_3 = -(d-5)/2}$, it decreases during the flow, leading to (again we suppress order 1 constants):
\be
S_\text{eff}  = \int \dd s  \, \tr \lf 
 (\partial_s \phi')^2 + \delta r  L^2   \phi'^2 + L^{-(d-5)/2}  \phi'^3
 \ri.
\ee
Here $s$, the coarse-grained time coordinate, is equal to $t/L$, and $\phi'\sim L^{(d-1)/2} \phi$  from the scaling dimension of the field in ${d+1}$ dimensions. If we now write the action in terms of $t$ and $\Phi \equiv L^{1/2} \phi'$, we recover the form of the action in Eq.~\ref{eq:Lpotts} with $N=L^d$. 

Because of the dangerous irrelevance of $\phi^3$ \cite{aharony1984scaling}, 
a finite-dimensional model with $d>5$ has two distinct large timescales, 
\be
L=N^{1/d} \quad \text{and} \quad  L^{d/5}=N^{1/5}.
\ee
The shorter timescale (which is compressed to order 1 in the all-to-all model) marks the crossover between ${(d+1)}$-dimensional and $1$-dimensional scaling for correlation functions.
The longer timescale is the one of more interest to us, and indicates the time at which the percolation probability starts to vary away from unity. This longer time becomes the characteristic critical timescale in the all-to-all model.

The scaling forms that we have already discussed carry over to the present case ($5<d<\infty$) with $N\rightarrow L^d$.

\subsection{Lessons for the full quantum problem}
\label{sec:lessonsquantum}

So far we have discussed the (classical) minimal cut problem in all-to-all and high-dimensional circuits. 
A priori, one can expect the universal properties of the generic measurement transition to be different from those for the minimal cut transition: the minimal cut is only an exact representation of the entanglement in certain special cases (as described at the beginning of Sec.~\ref{sec:classical}). 
Nevertheless, as in 1+1D, the solution of the minimal cut problem provides more general lessons.

First, there are qualitative features that  carry over to the generic problem.
The most basic feature is the existence of a transition between a phase in which the operator entanglement $S(t)$ 
--- the information propagated from the initial to the final time --- 
decays quickly with time, and a phase in which an extensive value of entanglement, $S(t)\sim s N$, persists over a time that grows \textit{exponentially} with the number $N$ of spins. In Secs.~\ref{sec:simulationscircuits} and \ref{sec:replicatimescale} we demonstrate that these features carry over to the operator entanglement (as measured by the von Neumann or  R\'enyi entropies $S_{n\geq 1}$), and related observables, in spin-1/2 circuits with measurements or forced measurements.
Another generic feature is that close to the critical point, the scaling of the exponential timescale is tied to that of the plateau entanglement: $\ln \tau(r) \sim N s(r)$ (Sec.~\ref{sec:replicatimescale}).

The minimal cut model also illustrates a possible relationship between the all-to-all case and the case of a high-dimensional regular lattice. In the classical problem, the exponents of the all-to-all model are those of  finite but high dimensions, once we take account of  the dangerous irrelevance of interactions in high dimensions, which leads to a critical timescale $L^\text{const.}$ that is parametrically larger than the linear system size $L$ (a timescale $\tau\sim L$  is what one would naively expect from $z=1$ scaling). 
In Sec.~\ref{sec:landautheory} we discuss similar crossovers in field theories 
for generic quantum models. 
However, we caution that our results in Sec.~\ref{sec:quantumtree}  suggest more complex possibilities in the all-to-all systems.

Finally, we saw that in the classical problem, the percolation order parameter and the value of $r_c$ could be obtained exactly by studying a simpler problem on a tree.  In the next section, we propose that exact results for the full quantum version of the FMPT can also be obtained by studying trees: not only their classical connectivity, as here, but their ``quantum'' connectivity as defined by entanglement measures for tree tensor networks.

\section{Entanglement transitions in quantum trees}
\label{sec:quantumtree}

\subsection{Motivation for studying quantum trees}

\label{sec:treesmotivation}

Locally the all-to-all circuit has the structure of a tree  (Sec.~\ref{sec:classicaltreestructure}). Viewing the circuit as a graph whose nodes are unitaries and whose edges are segments of spin worldline, the size of the smallest loops diverges when ${N\rightarrow \infty}$.
This is true for all values of the measurement or projection rate, including deep in the entangled phase.
We propose that this allows some exact results for the phase transition in the circuit, in certain cases (the FMPT), by studying the entanglement transition in a tree tensor network. As a by-product, we give exact results for general tree tensor networks.

\begin{figure}
    \centering
    \includegraphics[width=\linewidth]{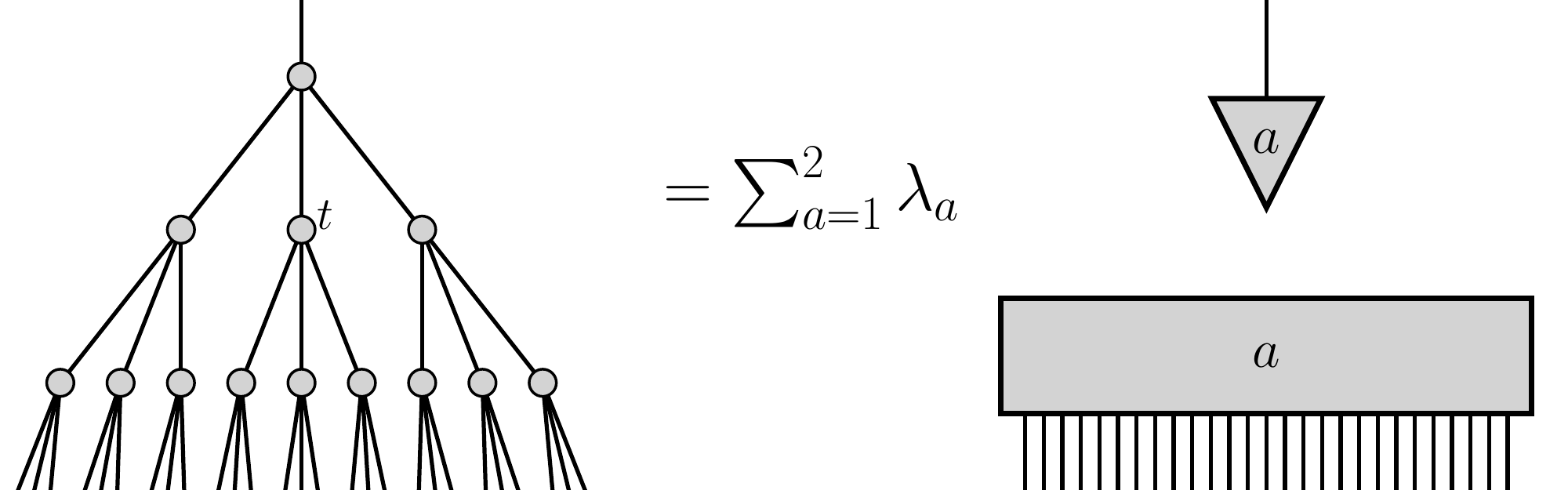}
    \caption{A tree tensor network with $k=3$ generations of nodes, and its singular value decomposition between apex and base. The individual node tensors are denoted by $t$ (see for example, Eq.~\ref{eq:tdef}).}
    \label{fig:recursive-tree}
\end{figure}

Fig.~\ref{fig:recursive-tree} (Left) is a schematic of the first $k=3$ generations of the tree that is connected to one end of a link somewhere in the bulk of the circuit.
For later convenience we have used a slightly different definition of the tree to that in  Sec.~\ref{sec:classicaltreestructure}. Previously we ``pruned off'' all the branches below a projection operator, while in Fig.~\ref{fig:recursive-tree} (Left) we leave them in place, so that the number of descendants after $k$ generations (the number of links at the base of the tree) is always $3^k$. 
Each four-coordinated node in this figure, such as the one denoted $t$, 
includes a unitary, together possibly with projectors on its legs --- we describe this below.  

This tree is a tensor network. It has  one free tensor index at the top and $3^k$ free indices at the bottom, and tensors $t$ in the interior (built from a unitary and projectors).
A basic way to characterize such a tensor network is via the amount of quantum information shared between apex and base. We can quantify this by the entanglement entropy between apex and base (Sec.~\ref{sec:entanglementapexbase} below). This language suggests analogous, but distinct, criteria for the classical and quantum transitions.

The \textit{classical} percolation transition at $r_c^\text{cl}$ (Sec.~\ref{sec:classical} above)  has a simple interpretation in terms  of the  tree tensor network.
For an asymptotically large tree,
$r_c^\text{cl}$ is the projection rate beyond which the apex and the base are guaranteed to be strictly disconnected by projectors. 
That is, once we go beyond the classical transition, the quantum information shared between apex and base vanishes for simple geometric reasons.\footnote{As usual, this geometrical disconnection is reflected in the vanishing of the zeroth R\'enyi entropy $S_0$ between apex and base.}

This suggests that we can also diagnose the \textit{quantum} transition in the circuit, occurring at a value $r_c^\text{qm}$ (we will see below that ${r_c^\text{qm} < r_c^\text{cl}}$) using the properties of the tree.
{ We will show that the tree has a transition at a critical value $r_c$,
which we conjecture is also the location of the critical point for the circuit ($r_c=r_c^\text{qm}$).}
For $r>r_c$ the amount of  information shared between apex and base decreases exponentially with the number of tree generations, even though the apex and the base may not be disconnected in the trivial geometrical sense.
For $r< r_c$, the von Neumann entanglement  entropy between apex and base instead remains positive: ${\lim_{k\rightarrow \infty} \< S_1\> > 0}$.

Motivated by this connection between the circuit and trees, in this Section we derive some universal results for  entanglement transitions of tree tensor networks.
We will argue that the tree structure allows us to find the exact location of the critical point for the simplest version of the all-to-all circuit model exactly. In the  language of Sec.~\ref{sec:overview}, this is the FMPT rather than the MPT. We explain in Sec.~\ref{sec:treenetworkstructure} immediately below why it is necessary for us to restrict to the FMPT in this section. 

Tree tensor networks are also interesting quite apart from the connection to the all-to-all circuit
\cite{Shi2006,Tagliacozzo2009,Murg2010,Silvi2010,Li2012,Nakatani2013,Murg2015, Vidal2007,Swingle2012,lopez2020mean}. They are instructive toy models for 1D wavefunctions with a scale-invariant entanglement structure \cite{Vidal2007,Swingle2012}, and they also allow efficient numerical tensor contraction algorithms \cite{Shi2006}. 
Many of the results of the following subsections apply to more general disordered tree tensor networks that are unrelated to the circuit (see the discussion in Sec.~\ref{sec:othertree}).

{
We obtain specific universal results for a broad class of trees that includes those arising in the FMPT circuit. 
These trees have bond dimension 2, and the probability distribution of the local tensors has a simple invariance property.
We also discuss, speculatively, what happens for trees with more general disorder distributions. 
Our conjectured continuum theory allows, a priori, for the the entanglement transition to be in distinct universality classes --- a phenomenon analogous to a line of fixed points (there is an overview in Sec.~\ref{sec:treeoverview}). 
Strikingly, for the class of trees that we study here, the transition is  constrained to lie on a specific point on this line. 
It remains to be seen whether other points on the line can be obtained by varying the model.

Heuristically, these different possibilities for the tree transition can be related to different possibilities for the disentangled phase close to the transition. 
In the disentangled phase the entanglement between apex and base is exponentially small in $k$. 
But we can distinguish, in principle, between a ``strong disorder'' regime where this small amount of entanglement is (loosely speaking) dominated by a single path from apex to base, and a ``weak disorder regime'' where exponentially many paths through the tree contribute. For the tree tensor networks we study here, we show that the former (strong disorder) case applies.}
The possibility of these two regimes is due to the existence of a glass transition in the classical problem of a directed polymer on a tree \cite{derrida1988polymers}. 
We will rely heavily on the methods developed in Ref.~\cite{derrida1988polymers} for the directed polymer problem, which relate a linear recursion relation for  the  polymer's partition function to a travelling wave equation.

\subsection{Structure of tree tensor network}
\label{sec:treenetworkstructure}

\subsubsection{Generalities}
\label{sec:treestructuregeneralities}

The trees we consider have  branching number  three and  bond dimension 2 for each bond (these are not essential restrictions). 
The four-index tensor $t^a_{bcd}$ at a given node has bond  index $a=1,2$  for the upper bond and ${b,c,d}$ for the lower bonds.
Below we describe the structure of $t$ for the circuits we consider. We note that they fall within a special class of tree tensor networks with a simplifying feature, for which we will be able to make strong statements.

Let us first consider trees like such as Fig~\ref{fig:recursive-tree} (Left) in general terms, without assuming that they arise from a circuit problem.

First, our analytical treatment will assume that the individual random tensors $t^{a}_{bcd}$ for the nodes are statistically uncorrelated. 
This is important as it allows a simple recursive equation for the entanglement between top and bottom. We will also take them to be identically distributed.

Second, for most of this section we will assume that the probability distribution of the local tensors $t^{a}_{bcd}$ has a simple invariance property. Namely, the distribution is invariant under multiplying an arbitrary  $\mathrm{U}(2)$ matrix $u$ on any index: for example under
\be
t^{a}_{bcd}\rightarrow t^{a}_{bcd'}u_{d'd}.
\label{eq:tdef}
\ee
This feature simplifies the recursive equation: it means we can write a recursion for singular values alone, without having to keep track of singular vectors.
In fact we only need a weaker condition:  
below the invariance property will hold for only the lower indices of $t^{a}_{bcd}$, which is sufficient.

The two assumptions above will be satisfied naturally for the circuit ensembles we consider, for example those built from Haar-random two-site gates.
They turn out to lead to surprisingly strong constraints on the structure of the recursion.

{
Towards the end of our discussion of trees  (Sec.~\ref{sec:othertree}) we speculate about what happens when we relax the second condition.
}

\subsubsection{Application to FMPT in circuit}

In applying our results on trees to the transition in a circuit, 
the first assumption above (on the statistical independence of the node tensors)
restricts us to the FMPT rather than the MPT.

Recall that for the MPT the measurement outcomes in the circuit are determined by Born's rule. That means they have nontrivial statistics that depend on the random unitaries, violating the first assumption above.
But in the FMPT the local projection operators are fixed \textit{independently} of the choice of unitaries, not with Born's rule. This means all the nodes of the tree are statistically independent, allowing a recursive statistical treatment.
The nodes are described explicitly below.
For the ensembles of two-site unitaries we study, it does not in fact matter \textit{how} the directions of the local projections are fixed, so long as this is done independently of the realization of unitaries. For definiteness we take all the local projectors to be onto the spin-up state.

To complete the specification of the circuit model, we just need to fix the distribution from which each two-site unitary $U$ is drawn. (As in Sec.~\ref{sec:classical}, the rate at which projection operators is applied is $r$.)

\subsubsection{Choice of ensemble of unitaries}
\label{sec:unitaryensemble}

The simplest choice is to take each $U$  independently Haar-random in $\mathrm{U}(4)$, i.e. drawn from the circular unitary ensemble.
\be\label{eq:uHaardistribution}
\text{Ensemble 1:} \qquad
U\sim \operatorname{Haar}.
\ee
{ For this ensemble we find a ``quantum'' transition point $r_c=0.749$ that is quite close to the classical one, ${r^\text{cl}_c=0.8}$. 
In order to increase the separation between these transitions, we also consider a second ensemble with  more weakly entangling gates. 
Unitaries in this second ensemble, referred to below as the ``$\Delta t$ ensemble'', are of the form
\be\label{eq:ufixeddistribution}
\text{Ensemble 2:} \qquad
U = (V_1 \otimes V_2)\,  U_\text{fixed}\, (W_1\otimes W_2),
\ee
where $V_{1}$, $V_2$, $W_1$ and $W_2$ are Haar-random one-site unitaries (ensuring the invariance property mentioned above), and $U_\text{fixed}$ is a non-random, \textit{fixed} unitary:
\ba\label{eq:Ufixed}
U_\text{fixed} = \exp \lf -i\, \Delta t\,H \ri.
\end{align}
The parameter $\Delta t$ controls the strength of the unitary, and consequently the position of the quantum transition (see Fig.~\ref{fig:deltatphasediag}). However, for concreteness we will mostly refer to the fixed value $\Delta t=0.3$. For $H$ we use \cite{skinner2019measurement}
\begin{equation}
\begin{split}
    H  = &~0.3 X_1 X_2+0.2(X_1+X_2)+\\&~0.4 Z_1 Z_2+0.5(Z_1+Z_2),
\end{split}
\end{equation}
where $X_i$, $Y_i$, $Z_i$ are the Pauli matrices on site $i$.
These parameter values are not fine-tuned, and we do not expect results to depend qualitatively on the precise values.}

The recursive treatment for $Z_k$ below applies to a more general family of distributions for the 2-site unitaries  for which the assumptions in Sec.~\ref{sec:treestructuregeneralities} are obeyed. 
First, the distribution of $U$ should be  invariant under left/right multiplication by \textit{single}-site unitaries.\footnote{This is a weaker condition than that satisfied by the Haar ensemble of two-site unitaries, which is invariant under left or right multiplication by any \textit{two}-site unitary.}
Second, it should have the property of being statistically invariant under exchange of the two spins acted on  by the unitary, and under transposition of the unitary. (This is less crucial, but simplifies the recursion.\footnote{This is because when we redraw the section of the circuit as a tree, we turn some of the unitaries upside down (which amounts to taking  a transpose) or reflect them left-right. The discrete invariances of the distributions mean that we do not need to keep track of this.})

\subsubsection{Node tensor in tree}

Recall from Sec.~\ref{sec:classical} that we can ``grow'' the tree by starting at some seed location in the circuit and following links (segments of spin worldline) to form a cluster of   unitaries at greater and greater distance from the seed. 
In fact, if we start on a link, we can think of it as a seed for   \textit{two} trees, one attached to each end of the link. It suffices to consider the properties of one of these trees separately. Truncating the tree at $k$ generations gives a  tensor network with a single bond at its apex and $3^k$ bonds at the base (we follow the convention in Sec.~\ref{sec:treesmotivation} where all branches are kept, even if they contain projections).

First consider a tree with no projections, where each node is a unitary. Now, when we include projections, each link of the tree has a probability $p$ to contain a projection.\footnote{The link may contain multiple projections, but this reduces to the case with a single projection. In the case where all the projection operators are identical, this is immediate. More generally, it holds because the distribution of unitaries is invariant under single site rotations (and because the overall normalization factor for the tree is not important).} If  a projection is present, we choose to incorporate it into the node below the link.
The node tensor is therefore: 
\be
\label{eq:4legtensordef}
\includegraphics[width = 0.6\linewidth]{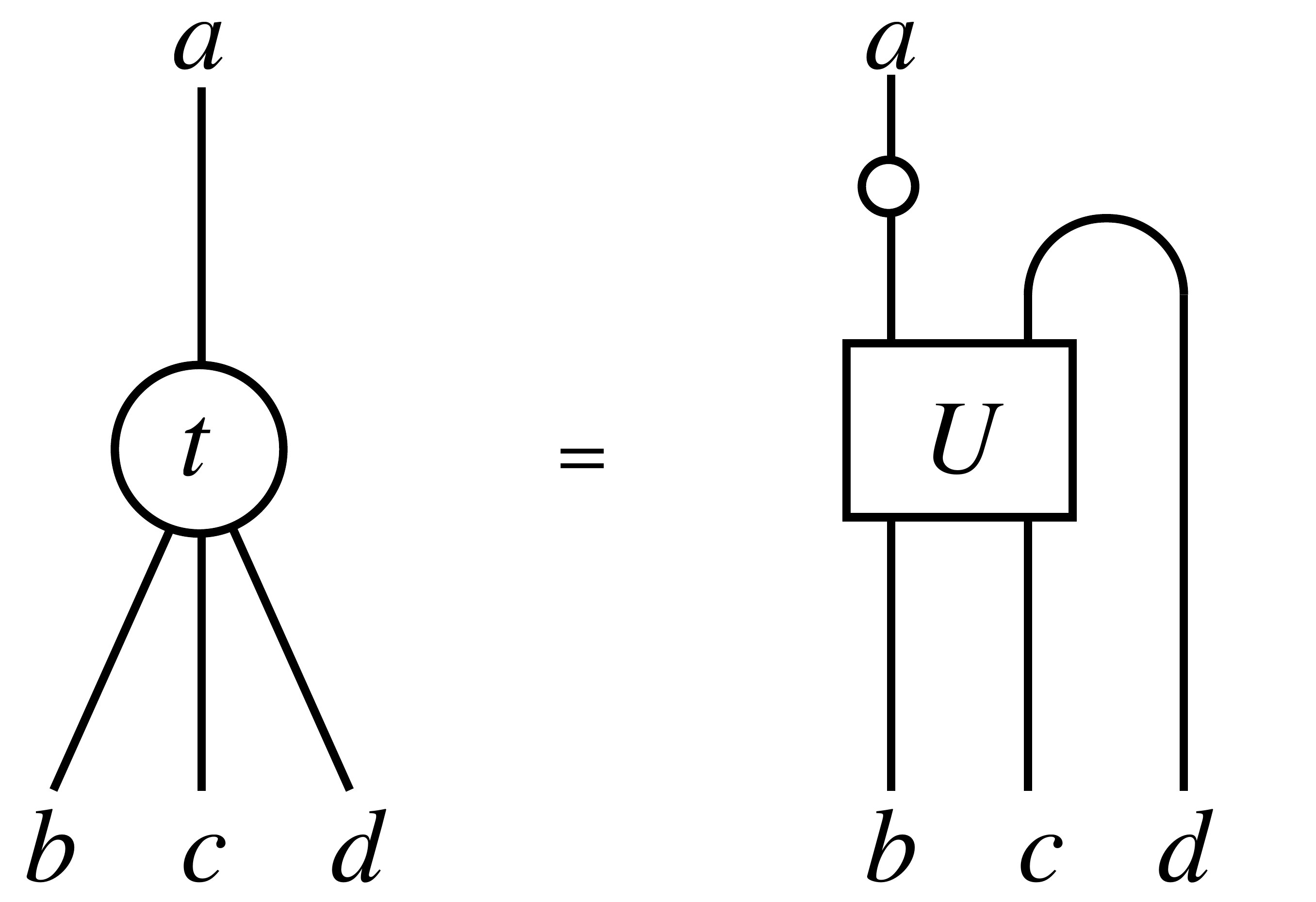},
\ee
or in components (we write the row index of $U$ as a superscript, and the column index as a subscript; both are multi-indices, since the unitary acts on two spins):
\be\label{eq:treenodetensorindices}
t^{a}_{bcd} = Q^{a}_{a'} U^{{a'}d}_{bc}.
\ee
The matrix $Q$, shown as a circle in the picture, is either the identity,  or the projector onto up, with probabilities $1-p$ and $p$ for each of the options. 
Recall that, in terms of the measurement rate (Eq.~\ref{eq:pandr}), 
\be\label{eq:prrelationrepeat}
p=\f{r}{2-r}.
\ee
In the case where the projector is present, we could simply prune off all the branches below it, but it is simpler to treat the geometry of the tree as fixed. Note that the distribution of Eq.~\ref{eq:treenodetensorindices} is invariant under multiplication of $\mathrm{U}(2)$ matrices on any of the lower indices, as required in Sec.~\ref{sec:treestructuregeneralities}.

\subsection{Entanglement between apex and base}
\label{sec:entanglementapexbase}

We will characterize the phase that the tree is in by the amount of quantum information shared between its apex and its base.
Depending on the phase, this can either be exponentially small in the number $k$ of generations of the tree, or it can be order 1 even for asymptotically large trees. 

We can always think of such a tree tensor network as a wavefunction for a single spin at the apex and multiple spins at the base. 
The information shared between apex and base is then quantified by the entanglement entropy between top and base, 
or more formally,
by the singular value decomposition when we partition the tensor network between the top and the base: see    Fig.~\ref{fig:recursive-tree} (Right).\footnote{In linear algebra terms, the tree represents a $2\times 2^\ell$ matrix: $2$ is the number of states associated with the bond at the apex, and $2^\ell$ is the number of states associated with the $\ell$ spins at the base. We are interested in the singular values in the  SVD for this matrix.}

Since the bond at the apex has a bond dimension of 2, there are only 2 singular values. After normalizing the tree, their squares sum to one, so we are in fact characterizing the tree by just a  \textit{single} number. We will take this to be the square of the smaller singular value, and will denote it by $Z_k$ for a tree with $k$ generations of nodes: 
\be
Z_k = \lambda_\text{min}^2.
\ee
The entanglement mentioned above is quantified by the R\'enyi entropies,
\be
S_n = \f{1}{1-n} \ln \lf Z_k^n + (1-Z_k)^n \ri,
\ee
which at small $Z_k$ are approximately
\ba 
S_{n>1} &\simeq \f{n}{n-1} Z_k, &
S_1 &\simeq Z_k \lf \ln \f{1}{Z_k} + 1 \ri.
\end{align}
In the random tensor network $Z_k$ is of course random.
Its distribution can be obtained recursively, using the fact that a larger tree can be built up by combining subtrees.

\subsection{Overview: classes of quantum tree}
\label{sec:treeoverview}

Let us summarize our basic conclusions for $Z_k$ before getting into calculations. 
Depending on the location in the phase diagram, the random variable $Z_k$ may have a broad distribution, and it will be vital to define its typical value using the average of $\ln Z_k$:
\be
\ln Z_k^\text{typ} = \< \ln Z_k \>_{\neq 0}.
\ee
We must condition on $Z_k$ not being strictly zero in order to define the typical value.\footnote{The trees we encounter in the circuit have a nonzero probability of terminating before they achieve $k$ generations, in which case $Z_k$ is identically zero. 
So long as we are below the \textit{classical} transition, 
an order 1 fraction of the probability distribution for $Z_k$ is supported on nonzero values even when $k$ is finite but large. For other tree tensor networks $Z_k$ may be nonzero with probability 1 (for any finite $k$), in which case we simply define $\ln Z_k^\text{typ} = \< \ln Z_k\>$.} 

For the quantum circuit with two-site unitaries and projections, the tree undergoes an entanglement transition at a critical value $r_c$. The value of $r_c$ depends on the ensemble of unitaries, but a basic point is that it is strictly below the classical transition point for any ensemble satisfying our assumptions: 
\be
r_c< r_c^\text{cl}.
\ee 
We compute the value of $r_c$ analytically for the Haar circuit:
\ba\label{eq:haartreerc}
r_c & = \f{212+75\pi}{362+75\pi}. 
\end{align}
This critical point at $r_c\simeq 0.749004$ lies not that far from the classical transition at $r_c^\text{cl}=0.8$.
For the ${\Delta t=0.3}$ ensemble the spacing is increased,
\begin{align}
\label{eq:deltatrc}
r_c & = 0.621(3).
\end{align}

In the disentangling phase the information shared between apex and base tends to zero exponentially with the size of the tree:
\ba 
Z_k^\text{typ} &\sim \exp ( - |c_r| \, k )
&
&\text{for $r>r_c$,}
\end{align}
with the ``speed'' $|c_r|$ vanishing linearly as $r\rightarrow {r_c}$.

In the entangling phase $Z_k^\text{typ}$ is instead nonzero as ${k\rightarrow \infty}$, so that information is shared between apex and base even in the limit of an infinitely large tree.
This information becomes small as we approach the transition from the entangled side. The scaling is very rapid:
\ba \label{eq:sqrtscalingfirst}
Z_\infty^\text{typ} &\sim \exp \lf - \f{C}{\sqrt{r_c -r }} \ri
&
&\text{for $r\lesssim r_c$,}
\end{align}
with $C$ a nonuniversal constant. The distribution of $Z$ is also very broad when $r_c-r$ is small. 
For example $\< Z_\infty\> \sim \sqrt{Z_\infty^\text{typ}}$, so that the mean is parametrically larger than the typical.

If we are right at the critical point, the value of $Z$ decays more slowly with $k$ than in the disentangled phase.  A somewhat heuristic argument in Sec.~\ref{sec:rightatrc} suggests
\be
\ln  Z_k^\text{typ} \sim  - k^{1/3}.
\ee

{ The results above rely on an exact treatment of the linearized form of the recursion relation for $Z_k$, together with the conjecture that the effect of nonlinearity is captured by a simplified, analytically tractable model.
Making this assumption (for which we provide numerical evidence),}
the results above hold for the trees derived from any forced measurement circuit ensemble with the structure described in Sec.~\ref{sec:unitaryensemble} (recall that we assumed invariance of the distribution of $U$ under rotations on each leg.)

In fact, they apply for the entanglement transition in \textit{any} tree tensor network that obeys the two assumptions described in Sec.~\ref{sec:treestructuregeneralities}, in particular the $\mathrm{U}(2)$ invariance property of the node tensor.
In this context $r$ is no longer interpreted as a measurement rate: instead it is any parameter characterizing $t^{a}_{bcd}$ that can be used to drive the entanglement transition.
{ However for the purposes of the discussion we will  use the notation appropriate to the forced measurement circuit.

An analysis of random tree tensor networks outside the above class is a task for the future (see also Ref.~\cite{lopez2020mean}). However, our conjectured effective description suggests the interesting \textit{possibility} that there may be multiple universality classes for the tree entanglement transition.
The effective description includes a parameter ${\Delta >0}$, which  controls the scaling of $Z_\infty^{\mathrm{typ}}$ near $r=r_c$.
At first glance $\Delta$ is a nonuniversal parameter that will depend on the model. But, surprisingly,  the $\mathrm{U}(2)$ invariance property fixes ${\Delta = 1/4}$ at the entanglement transition. Eq.~\ref{eq:sqrtscalingfirst} applies for ${\Delta<1}$ in the effective description, but for ${\Delta>1}$ there is instead power law scaling of the ``order parameter'' $Z$ close to the transition.
For completeness, we solve the effective model in this ${\Delta>1}$ regime also. 
We find a regime with a variable exponent, and a regime where this exponent is pinned to 1:
\ba 
\label{eq:sqrtscalingintermediateregimefirst} 
Z_\infty^\text{typ} 
&\sim (r_c - r)^{\f{1}{\Delta-1}}
&
&\text{for $1<\Delta<2$.}\\
 \label{eq:sqrtscalinglargedeltafirst}
Z_\infty^\text{typ} 
&\sim (r_c - r),
&
&\text{for $\Delta>2$.}
\end{align}
However, at present it is unclear whether these regimes of the effective model can be accessed by any tensor network, or whether they exist only in the effective model (Sec.~\ref{sec:othertree}).
}

\subsection{Recursion relation for singular values}
\label{sec:recursionrelationfortree}

Let us think of the tree as a quantum state for a spin at the top and $3^k$ spins at the base: this is just to fix notation for bras/kets.  We may write its Schmidt (singular value) decomposition:
\be\label{eq:Tsvd}
T = \sum_{i=1,2} \lambda_{i} \, \ket{i}_\text{top} \ket{i}_\text{bottom}.
\ee
The states are Schmidt states in the appropriate Hilbert spaces (the second ket lives in the $2^{(3^k)}$-dimensional Hilbert space associated with the base). 
In the problem we are studying, the overall normalization of the tree is not important, so we will always take the Schmidt/singular values to be normalised: ${\lambda_1^2 + \lambda_2^2 =1}$. 

Given three trees $T_k$, $T_k'$, and $T_k''$, each  of $k$ generations, we may form a tree $T_{k+1}$ of $k+1$ generations by attaching $T_k$, $T'_k$ and $T''_k$ to the base of the  $t$ node shown in  Eq.~\ref{eq:4legtensordef}.
The statistical invariance of $U$ under single-site rotations means that we are free to take the  the Schmidt states $\ket{i}_\text{top}$ (Eq.~\ref{eq:Tsvd})  for $T_k$, $T_k'$ and $T_k''$ to be simply the two basis states (up and down spin states), which we denote $\ket{1}_\text{top}$ and $\ket{2}_\text{top}$. Then
\be\label{eq:T_recursive_relation}
T_{k+1} = \sum_{a,b,c,d = 1,2} t^{a}_{bcd} \lambda_b \lambda'_c \lambda''_d \ket{a}_\text{top} \ket{bcd}_\text{bottom}.
\ee
Here $\lambda$, $\lambda'$ and $\lambda''$ are singular values for $T_k$, $T'_k$ and $T''_k$,
$\{\ket{a}\}_\text{top}$ are computational basis states, 
and $\{ \ket{bcd}_\text{bottom}\}$ is a set of 8 orthonormal states associated with the base of the full tree, formed from the Schmidt states of the three sub-trees. Equivalently, in this basis,
\be
\lf T_{k+1}\ri^{a}_{bcd} =  t^{a}_{bcd} \lambda_b \lambda'_c \lambda''_d.
\ee
It is straightforward to compute the normalised singular values of $T_{k+1}$.\footnote{Considering $T\equiv T_{k+1}$ as a state, then 
 ${\lambda_1^2 + \lambda_2^2 =1}$ and 
${\lambda_1^4 + \lambda_2^4 = \Tr \rho^2/ (\Tr \rho)^2}$, where $\rho$ is the un-normalized reduced density matrix for the spin at the apex: ${\rho_{a,a'} = \sum_{bcd} T^a_{bcd} T^{*a'}_{bcd}}$.} 
Let us denote the smaller singular value squared of $T_{k+1}$ by $Z_{k+1}$. 
If $t$ includes the projector, then trivially 
\be
\label{eq:zequalszero}
Z_{k+1}=0
\ee
We write the other case explicitly for completeness, though we will only need a simple limit of it:
\ba\notag
Z_{k+1}^2 & + \lf 1 - Z_{k+1} \ri^2 = 
\\
& \f{
\sum U^{ad}_{bc}(U^{a'd}_{bc})^* U^{a'g}_{ef} (U^{ag}_{ef})^* 
\,\, \lambda_b^2 \lambda_e^2 
\,\, \lambda'^2_c \lambda'^2_f 
\,\, \lambda''^2_d \lambda''^2_g 
}{
\lf 
\sum  |U^{ad}_{bc}|^2 \lambda_b^2\lambda_c'^2 \lambda_d''^2 \ri^2 
}.
\label{eq:fullrecursion}
\end{align}

We are interested in a transition between a phase where $Z_{k}$ vanishes as $k\rightarrow \infty$, and a phase where the typical value of $Z_k$ remains positive in this limit. 
Even in this phase, if we are close to the phase transition, this typical value of $Z_k$ is small.
Therefore to understand the critical properties we can study the recursion relation in the regime where the minimal singular values are close to $0$ for all the trees. We order the singular values of any tree such that $\lambda_2 \leq \lambda_1$, and define $Z=\lambda_\text{min}^2=\lambda_2^2$ for each tree. 

The first step is to examine the linearized recursion relation. Taking Eqs.~\ref{eq:zequalszero},~\ref{eq:fullrecursion} to order $Z$,
\be\label{eq:linearizedrecursion}
Z_{k+1}=
\left\{ 
\begin{array}{ll}
A_1 Z_{k} + A_2 Z_{k}' + A_3 Z_{k}''   & \text{ with probability $1-p$}\\
  0   &  \text{ with probability $p$} 
  . 
\end{array}\right.
\ee
Here $A_i$ are three positive constants that depend on the random 
unitary: 
\ba\notag
A_1  &= \frac{ \left|U^{11}_{11} U^{21}_{21} - U^{11}_{21} U^{21}_{11}\right|^2}{\lf |U^{11}_{11}|^2 + |U^{21}_{11}|^2 \ri^2}, \\ \notag
A_2 &= \frac{ \left|U^{11}_{11} U^{21}_{12} - U^{11}_{12} U^{21}_{11}\right|^2}{\lf |U^{11}_{11}|^2 + |U^{21}_{11}|^2 \ri^2}\\
A_3 &  = \frac{ \left|
U^{11}_{11} U^{22}_{11} - U^{12}_{11} U^{21}_{11}
\right|^2}{\lf |U^{11}_{11}|^2 + |U^{21}_{11}|^2 \ri^2}.
\label{eq:Aconstants}
\end{align}
Analogous formulas hold for more general choices of the node tensor $t$, see Sec.~\ref{sec:othertree}.

Let us consider the meaning of this equation. $Z_k$, $Z_k'$ and $Z_k''$ refer to trees of the same size, so they are drawn from the same probability distribution. The recursion relation then defines the probability distribution for a variable
$Z_{k+1}$ at the next level in the hierarchy. 
This defines a sequence of probability distributions $P_k(Z)$ for increasing $k$. 
The initial condition at the lowest level of the hierarchy is $Z_0 = 1/2$  (for a single bond, the two singular values are equal), i.e. $P_0(Z) = \delta(Z-1/2)$. 
(This initial condition is far outside the linear regime, but close to the transition, $Z_k$ becomes small at large $k$.  The specific choice of the initial condition in the linear tree is unimportant as long as it is non-zero and positive.)

The linearized recursion relation (\ref{eq:linearizedrecursion}) is crucial. It is sufficient to obtain the exact location of the entanglement transition, although we will need to add nonlinearity to understand what happens close to this transition in the entangled phase.

Let us collect here some properties of the $A_i$ that will be useful below. It turns out that the statistical invariance of $U$ under single-site rotations allows some exact statements, regardless of the precise choice of distribution for $U$. We demonstrate these in App.~\ref{app:unitaryaverages}. In particular we will need the following identities, which  hold for all $i=1,2,3$ (so long as $U$ is nontrivially entangling with probability 1):
\ba
\label{eq:Aiunity}
\<A_i \> & = 1,
&
\<A_i^{1/2}\ln A_i\> & = 0.
\end{align}
The first of these is at first sight surprising, since if the unitary $U$ is the identity (for example if $\Delta t \rightarrow 0$ for the distribution in Eqs.~\ref{eq:ufixeddistribution},~\ref{eq:Ufixed}) then
$A_2$ and $A_3$ are exactly equal to zero.\footnote{This can be seen from the fact that $Z_{k+1}$ is manifestly independent of $Z_k'$ and $Z_k''$ in this limit.}
However, this is a singular limit for $\<A_i\>$, see below.

In the case where $U$ is a Haar-random $\mathrm{U}(4)$ matrix, we can obtain  more general analytic results (App.~\ref{app:unitaryaverages}):
\ba\label{eq:HaarA1A2}
\< A_1^\lambda\> = \<A_2^\lambda\> &  = \f{12}{12+8 \lambda - 7 \lambda^2 - 2\lambda^3+ \lambda^4},\\
\< A_3^\lambda\> &= \f{\pi \lambda (1-\lambda)}{\sin(\pi \lambda)}.
\label{eq:HaarA3}
\end{align}
To determine the location of the phase transition we will need the special cases $\langle A_i^{1/2}\rangle$. These are given in Table~\ref{Table:As_averages} (App.~\ref{app:unitaryaverages}) for both Haar and ${\Delta t=0.3}$ ensembles.

The asymptotics of the probability distributions of the $A_i$ 
are also obtained in App.~\ref{app:unitaryaverages}. These three variables are correlated, but here we discuss only the marginal distribution of a given one.
Let us define 
\be\label{eq:Videfn}
V_i \equiv \ln A_i.
\ee
For any generic distribution of $U$, the tails of the $V_i$ distribution are exponential:
\be
P(V_i) \, \dd V_i \sim \left\{
\begin{array}{ccc}
  e^{-2 V_i}\,\dd V_i   & \qquad V_i \gg 0,  \\
  e^{- |V_i|}\,\dd V_i   & \qquad  V_i \ll 0. 
\end{array}\right.
\ee
If the distribution of unitaries is taken to be weakly entangling, for example if $\Delta t\ll 1$ in Eq.~\ref{eq:Ufixed}, then the right-hand tail of the distribution has an intermediate part, extending over the range ${\ln \Delta t^{2} \ll V_i \ll \ln \Delta t^{-2}}$, that decays with the smaller exponent $-1/2$ (App.~\ref{app:unitaryaverages}). This slowly decaying tail, cut off at a parametrically large $V_i$,  is responsible for the failure of the limit $\Delta t\rightarrow 0$ to commute with the average in Eq.~\ref{eq:Aiunity} that was mentioned above.

\subsection{Linearized recursion relation}
\label{sec:linearizedtree}

\begin{figure}
    \centering
    \includegraphics[width=\linewidth]{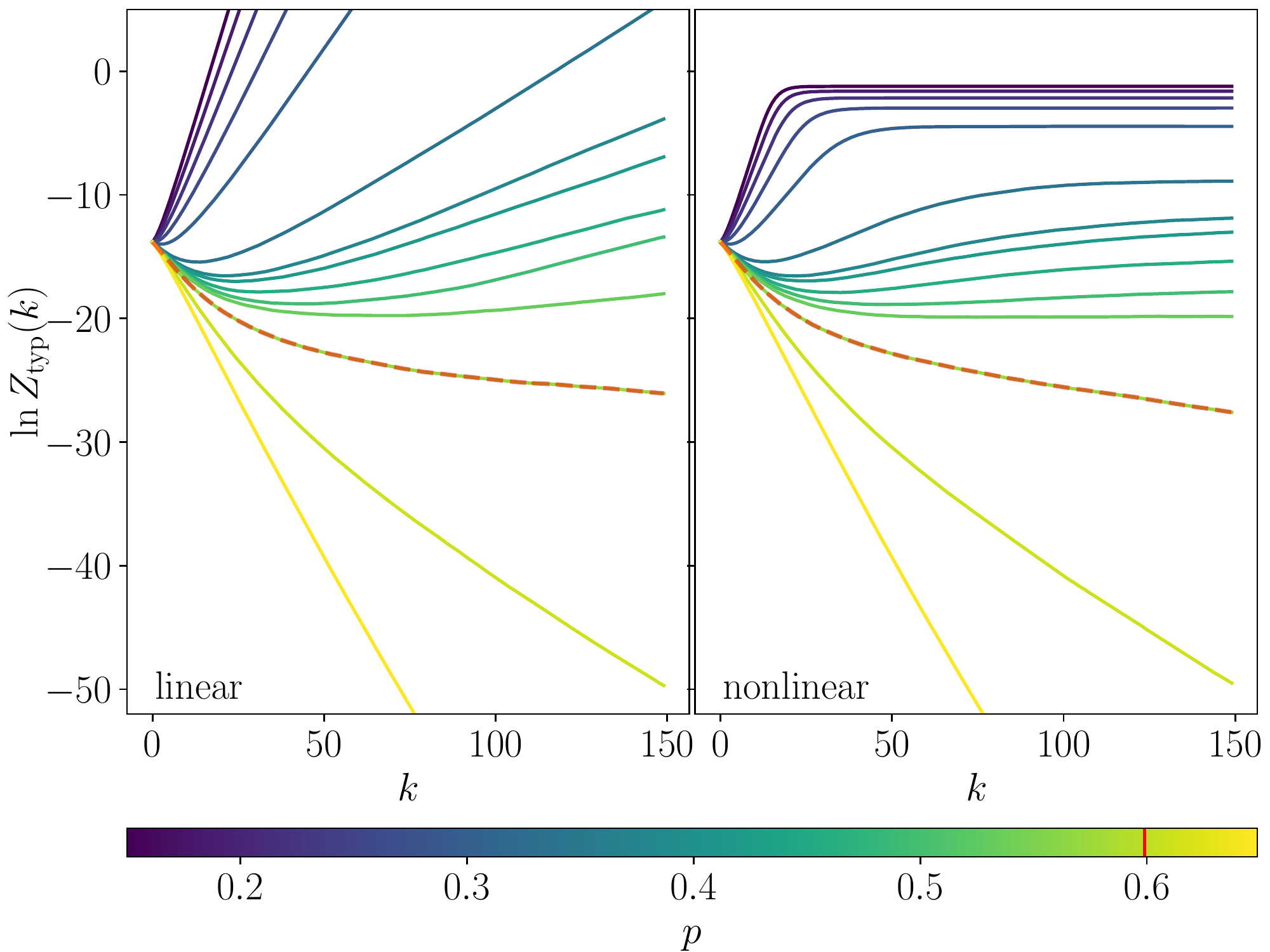}
    \caption{The evolution of the typical value of $Z$ with the depth $k$ of the tree, for the linear recursion (Left) and the non-linear recursion (Right), for various values of $p$. 
    The evolution is similar in the two cases for $p>p_c$, but for $p<p_c$, the nonlinearity causes $Z^\text{typ}$ to saturate. { The initial value ${Z_0=10^{-6}}$ has been used.}}
    \label{fig:Z_evolution}
\end{figure}

The most basic question about the linearized recursion (\ref{eq:linearizedrecursion}) is whether the typical value of $Z$ is exponentially growing or exponentially shrinking at large $k$ \cite{derrida1988polymers}. We may define an exponential growth speed $c_p$:
\be\label{eq:growthspeeddefn}
\ln Z_k^\text{typical} \sim  c_p \,  k.
\ee
In this section we will usually use $p$ as the parameter, rather than the equivalent $r$, Eq.~\ref{eq:prrelationrepeat}, since $p$ has a more direct interpretation in terms of the tree.

Define the point $p_c$ to separate a regime of exponential growth,  which we will see occurs for   $p<p_c$, from a regime of exponential decay at larger $p$:
\ba\label{eq:cpcdefn}
c_{p_c} & = 0
\qquad (\text{definition of $p_c$}).
\end{align}
We will see that $p_c$ is precisely the location of the entanglement phase transition for the tree.
When the linear recursion predicts that ${\Ztyp}\rightarrow 0$ at large $k$, this remains true when higher powers of $Z$ are included in the recursion. On the other hand, when the linear recursion predicts that ${\Ztyp}\rightarrow \infty$ at large $k$, then the nonlinear terms in the recursion replace ``$\infty$'' with a finite value, in a universal manner that we discuss in Sec.~\ref{eq:nonlineartree}.

This is illustrated using simulations for the case of Haar-random unitaries in Fig.~\ref{fig:Z_evolution}. This  compares $Z_k^\text{typ}$ for the linear and nonlinear recursion relations. In the linear case, evolution follows ${\ln Z_k^\text{typ}\propto k}$ at large $k$, for all $p$. In the nonlinear case, this is only true for ${p>p_c}$. Details of these simulations are described in App.~\ref{app:quantumTree}.

The speed $c_p$ can be extracted using the method of  Ref.~\cite{derrida1988polymers} which relates the linear recursion to a travelling wave problem.
Define the generating function
\be\label{eq:definegeneratingfn}
G_k(x) = \< \exp \lf - e^{-x} Z_k \ri \>,
\ee
where the average is over $Z_k$.
The recursion relation (\ref{eq:linearizedrecursion}) then becomes: 
\be\label{eq:recursionforG}
G_{k+1}(x) = p +(1-p)\< G_{k}(x-V_1)G_{k}(x-V_2)G_{k}(x-V_3) \>,
\ee
where the remaining average is only over the $V_i = \ln A_i$
defined in Eq.~\ref{eq:Videfn}.
(The fact that the $V_i$ appear additively in the arguments of the generating functions here is the reason why the generating function in  Eq.~\ref{eq:definegeneratingfn} is usually written with the double exponential.)

It may be helpful to think of $G_k(x)$, defined in Eq.~\ref{eq:definegeneratingfn}, as a smeared version of the cumulative probability distribution for $\ln Z$.
This definition shows that for ${x\gg \ln Z^\text{typ}}$, $G_k$ plateaus at the value 1, 
while for  ${x\ll \ln Z^\text{typ}}$,  $G_k$ plateaus at the probability of $Z_k$ being exactly zero.\footnote{The latter  converges at large $k$ to a constant below 1, so long as we are in the classically percolating phase for the tree. The value of this plateau can be set to zero by a linear transformation of  $G$. See Sec.~\ref{eq:nonlineartree}.} 
$G_k$ has a ``front'' at 
\be\label{eq:fronttypreln}
x_\text{front}(k) = \ln Z_k^\text{typ} + o(k) 
\ee
that interpolates between  these two plateaus.
It is useful to think of $x$ as a fictitious spatial coordinate, and of $k$ as fictitious time coordinate \cite{derrida1988polymers}.
Then, at late time, this front propagates  as a traveling wave with speed $c_p$, and obeys the traveling wave ansatz:
\be
\label{eq:travelingwaveansatz}
G_k(x) = G^{(\lambda)}\hspace{-0.7mm}\lf x - v_p(\lambda)\, k\ri.
\ee
The wave speed $v_p(\lambda)$ depends on a parameter $\lambda$ of the solution  $G^{(\lambda)}$. This parameter, which must be determined, is the exponential decay constant of $G^{(\lambda)}$ at large argument \cite{derrida1988polymers}:
\be\label{eq:gtail}
G^{(\lambda)}\hspace{-0.2mm}(u) \sim 1 - \alpha\, e^{-\lambda u}.
\ee
Substituting this form into (\ref{eq:recursionforG}) gives an explicit formula for the speed $v(\lambda)$ of the traveling wave solution with a given $\lambda$:
\be
\label{eq:travelingwavevelocityfmla}
v_p(\lambda) = \f{1}{\lambda} \ln \left[
(1-p) \left(
\< A_1^\lambda \> +
\< A_2^\lambda \> +
\< A_3^\lambda \> 
\right)
\right].
\ee

We must then determine the correct value of $\lambda$, i.e. \textit{which} traveling wave solution the initial condition converges to.
This is done by standard considerations of velocity selection for travelling waves \cite{fisher1937wave,derrida1988polymers}.

In outline, there is a privileged  \textit{minimal speed} traveling wave defined by the parameter value ${\lambda=\lambda_*}$ where ${v_p(\lambda)}$ is minimal:
\ba
v_p^\text{min} & \equiv v_p(\lambda_*),
&
v_p' (\lambda_*) & = 0.
\end{align}
$G_k$ will converge to this minimal speed solution \textit{if} $\lambda_*$ is less than 1,\footnote{The value $\lambda=1$ appears here because it is the exponential decay constant for the initial condition at $k=0$: ${G_0(x)\sim 1-\<Z_0\> e^{-x}}$ at large $x$. At late times the solution converges to a traveling wave in which the decay constant is ${\min\{1, \lambda_*\}}$. That is, the decay constant is either that of the initial condition, or that of the minimal-speed wave, whichever  decays more slowly as $x\rightarrow\infty$.} while it will converge to the solution with $\lambda=1$ if $\lambda_*>1$. In the latter case the speed is $v_p(1)$, which we refer to as the ``annealed'' value of the speed (for reasons described in  Sec.~\ref{sec:glassaside}):
\be\label{eq:vpanndefn}
v_p^\text{ann} \equiv v_p(1).
\ee
Therefore the desired exponential growth rate is given for any $p$ by
\be\label{eq:cptwocases}
c_p = 
\left\{
\begin{array}{cc}
    v_p^\text{min} &   \quad \text{if $\lambda_* < 1$,} \\
   v_p^\text{ann} &   \quad \text{if $\lambda_* > 1$.}
\end{array}
\right. 
\ee
Recall that $\lambda_*$ is determined using Eq.~\ref{eq:travelingwavevelocityfmla}, via ${v_p'(\lambda_*)=0}$, so it depends on $p$.

The above equation (\ref{eq:cptwocases}) can lead to a nonanalyticity in $c_p$ as $p$ is varied. This has a meaning in terms of the statistical mechanics of the linearized recursion relation \cite{derrida1988polymers}, which we review in Sec.~\ref{sec:glassaside}.
For now we simply note that, for the present class of circuits\footnote{Recall that we assumed various invariances of the distribution of unitaries to simplify the treatment (Sec.~\ref{sec:treenetworkstructure}).}
the first line in Eq.~\ref{eq:cptwocases} is always the one that applies for $p$ close to $p_c$. This is shown in Sec.~\ref{sec:glassaside}. 
Given this, $p_c$ is determined by solving 
\ba\label{eq:vvprimezero}
v_{p_c} (\lambda_*) & = 0, & 
v_{p_c}'(\lambda_*) & = 0.
\end{align}
for $\lambda_*$ and $p_c$.

\begin{figure}
    \centering
    \includegraphics[width=\linewidth]{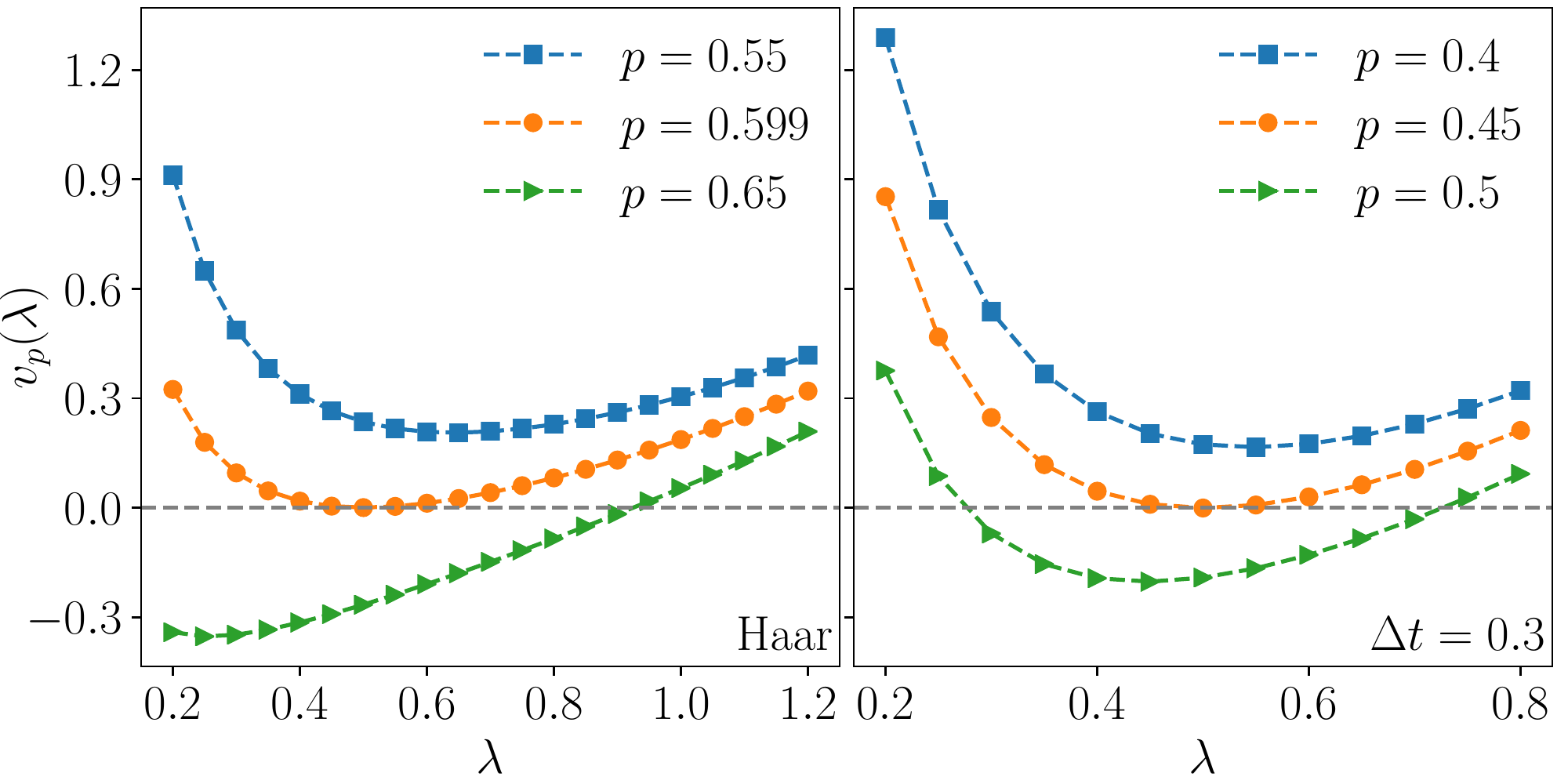}
    \caption{The velocity $v_p(\lambda)$ vs. $\lambda$ for three values of $p$ (above, at and below the critical point). Left panel is for Haar evolution and right is for $\Delta t = 0.3$. }
    \label{fig:velocity}
\end{figure}

Fig.~\ref{fig:velocity} shows $v_p(\lambda)$, defined in Eq.~\ref{eq:travelingwavevelocityfmla}, for the Haar tree and the ${\Delta t=0.3}$ tree, in the vicinity of their  respective $p_c$ values. 
(Numerically, these are obtained by simple averages using a single tensor. In the Haar case, Eqs.~\ref{eq:HaarA1A2},~\ref{eq:HaarA3} also give the exact form.) 
$c_p$ is given by the minimal value of the curve, ${c_p = v_p(\lambda_*)}$, which passes through zero at ${p=p_c}$.

\begin{figure}
    \centering
    \includegraphics[width=\linewidth]{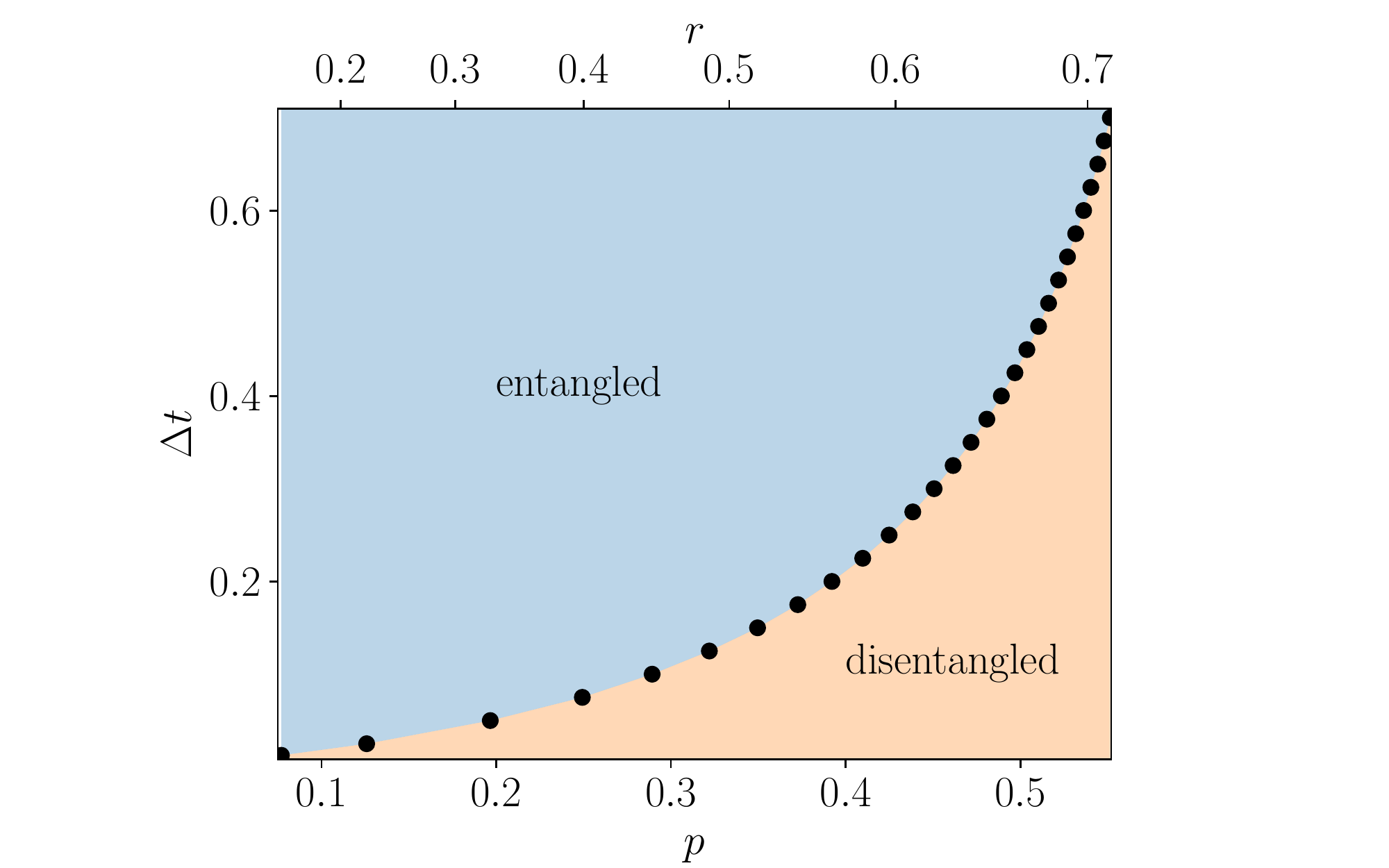}
    \caption{Phase diagram of the quantum tree generated by the unitary gate in Eq.~\ref{eq:Ufixed}, in the space of $\Delta t$ and $r$.}
    \label{fig:deltatphasediag}
\end{figure}

This can be used to determine $p_c$ numerically, but in fact further analytical progress is possible.
Using the definition of $v_p(\lambda)$ in Eq.~\ref{eq:travelingwavevelocityfmla},
the equations (\ref{eq:vvprimezero})  reduce to 
\ba
\sum_{i=1}^3 \< A_i^{\lambda_*}\> & = \f{1}{1-p_c}, 
& 
\sum_{i=1}^3 \< A_i^{\lambda_*} \ln A_i \> & = 0.
\end{align}
Remarkably, the second identity in Eq.~\ref{eq:Aiunity} shows that the solution is always at ${\lambda_*=1/2}$, for \textit{any} ensemble of unitaries satisfying our assumptions. This fact gives an explicit expression for $p_c$ as a simple average for the local node tensor, 
\be\label{eq:treecriticalpointgeneral}
p_c= 1 - \f{1}{\sum_{i=1}^3 \langle A_i^{1/2}\rangle}.
\ee
This may be evaluated analytically for the Haar case (Eqs.~\ref{eq:HaarA1A2},~\ref{eq:HaarA3}), giving 
$p_c = \f{212+75\pi}{512+75\pi}$
(equivalent to the $r_c$ value quoted in Sec.~\ref{sec:treeoverview}) and  numerically for the $\Delta t$ ensemble. The location of the critical point in the $\Delta t$ ensemble is shown for various values of $\Delta t$ in Fig.~\ref{fig:deltatphasediag}.

\subsection{Aside: glass transition in linear recursion}
\label{sec:glassaside}

\begin{figure}
    \includegraphics[width=0.6\linewidth]{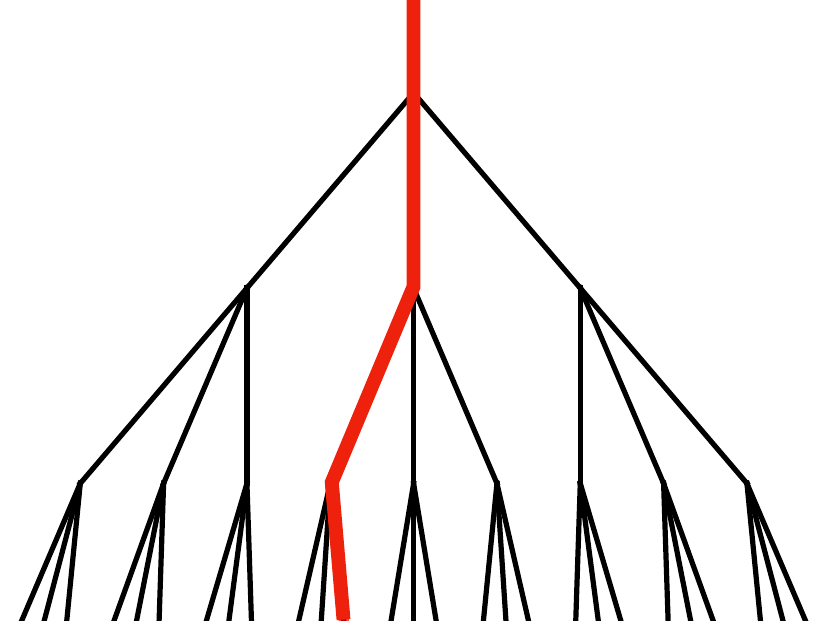}
    \caption{Schematic of a directed polymer on a tree.}
    \label{fig:tree-red-branch}
\end{figure}

The canonical example of linear recursion relations like Eq.~\ref{eq:linearizedrecursion}  
 is the problem of the directed polymer on a tree 
 \cite{derrida1988polymers}: see Fig.~\ref{fig:tree-red-branch}.
In the disentangled phase, where the linear treatment is valid at large $k$, this gives another interpretation of the singular-value-squared $Z_k$ as a sum over paths through the tensor network.
Here we briefly review this mapping and use it to clarify which of the regimes in Eq.~\ref{eq:cptwocases} is relevant.
This subsection is not essential to the subsequent development.

Within the linear approximation Eq.~\ref{eq:linearizedrecursion},  $Z_k$ is exactly equal to the partition function of a polymer that lies along a path from the top to the bottom of a tree of depth $k$, as in Fig.~\ref{fig:tree-red-branch}.
We view $-V_1$, $-V_2$, $-V_3$  in Eq.~\ref{eq:Videfn} as random potentials on the three bonds below a given node. The energy of the polymer is the sum of the potentials for the bonds it visits:
\be\label{eq:polymerZ}
Z_k = \sum_{\text{paths}} e^{ \sum_{
\substack{\text{bonds} \\ \text{on path} }
} 
V_\text{bond}}.
\ee
This is easily seen to satisfy the recursive Eq.~\ref{eq:linearizedrecursion}.
There are minor differences from the standard polymer model.
First, if $p>0$, there are some bonds that the polymer cannot visit, where $A=0$ or $V=-\infty$ (these bonds and the subtrees below them can simply be removed). Second, the $V$s have a nontrivial distribution, with links that share the same parent node having correlated potentials.

{
The polymer can be in either a glass phase or a paramagnetic phase  \cite{derrida1988polymers}.
These are distinct thermodynamic phases in the polymer problem, but to avoid confusion we will refer to them as ``regimes'' , because they do not correspond to distinct phases of the entanglement problem. 
(The distinction between the glass and paramagnet is a feature of the linearized problem only, and is unrelated to the distinction  between entangled and disentangled phases.)

The glass obtains when the pinning effect of disorder on the polymer defeats the depinning effect of entropy.
Usually the glass would be entered by decreasing the temperature (increasing the scale of $V$). 
Here we increase the strength of disorder by increasing $p$.
In the paramagnetic regime the polymer has extensive entropy (propotional to $k$) while in the glass the entropy per unit length vanishes. 

The glass and paramagnet regimes have a simple translation to the language of the traveling wave (Sec.~\ref{sec:linearizedtree}), which we only state \cite{derrida1988polymers}.
The polymer is in the glass regime if  ${\lambda_*<1}$, and in the paramagnetic regime if  ${\lambda_*>1}$ \cite{derrida1988polymers}.
These correspond to the two lines in  Eq.~\ref{eq:cptwocases} for the growth rate $c_p$, which is 
simply (minus) the free energy per unit length of the polymer. 

In our problem, the entanglement transition necessarily takes place in the glass regime of the linear recursion, essentially because of the fact that ${\<A_i\>=1}$. Let us give an intuitive picture.

To begin with, imagine that the polymer is in the paramagnetic regime. 
In this regime (but not in the glass\footnote{Eq.~\ref{eq:vpann} is the exact growth rate of $\<Z_k\>$ for any $p$ in the linear problem, but it is only for ${p<p_\text{glass}}$ that $Z_k^\text{typ}$ has the same growth rate as  $\<Z_k\>$. It is $c_p$ as defined by $Z_k^\text{typ}$ that will be relevant when we include nonlinearity. The failure  of the annealed approximation when $\lambda_*<1$ is because, in this regime the distribution of $Z$ becomes broad in the sense that  $\lim_{k\rightarrow \infty}\< Z/Z^\text{typ}\>=\infty$. (In this regime the  tail in the  probability distribution for $\ln (Z/Z^\text{typ})$  decays as $e^{-\lambda_*\ln (Z/Z^\text{typ})}$.)}) the ``annealed'' expression for the free energy/growth rate $c_p$ applies (the second line of Eq.~\ref{eq:cptwocases}). 
This expression is in fact just the annealed approximation to the free energy, in which we average the partition function of the polymer, $\<Z_k\>$, instead of averaging its logarithm. 
In the present  linearized problem this gives:
\be\label{eq:vpann}
 v_p^\text{ann} = \ln \lf 3 [1-p] \ri,
\ee
using ${\<A_i\>=1}$.

Recall that the entanglement transition is at the value of $p$ where $c_p=0$. 
We see from Eq.~\ref{eq:vpann} that \textit{if} the polymer was in the paramagnetic regime in the vinicity of $p_c$, then the entanglement transition would coincide with the classical percolation transition at $p_c^\text{classical}=2/3$!

We can see that this is inconsistent as follows. 
Consider  the structure of large trees when we approach the classical percolation transition at $p_c^\text{classical}=2/3$ from below. 
After deleting subtrees that terminate before reaching the base,\footnote{These have $Z=0$, so do not contribute to the recursion.} 
a large tree with ${Z\neq 0}$ is made up of one-dimensional chains connected by branching events.
Close to the classical transition, the typical length of one of these 1D chains grows like ${(2/3-p)^{-1}}$.\footnote{This is because, close to the classical transition, only a small fraction, of order ${(2/3-p)}$, of subtrees survive. 
As a result most nodes in a ``pruned'' tree have a single descendant, with a fraction of order ${(2/3-p)^{m-1}}$ having $m$ descendants ($m\leq 3$).}
Treating them as renormalized bonds in the polymer problem, one may check that the effective disorder strength on these renormalized bonds grows without bound as they get longer. 
This increasing disorder strength implies that we must enter the glass regime before we get to the classical transition.
That is, either the linear recursion relation is in the glass regime for all $p$, or it is in the glass regime for all $p>p_\text{glass}$ for some  $p_\text{glass}< 2/3$.

When the polymer is in the glass phase, $c_p$ is strictly smaller than the annealed approximation above (Eq.~\ref{eq:cptwocases}). Therefore $c_p$ in fact hits zero at a smaller value of $p$ than  $v_p^\text{ann}$ does. In other words, $p_c$ is strictly smaller than $p_c^\text{classical}$.

The value of $p_\text{glass}$ is determined by the equation $v_p'(1) = 0$.
For the $\Delta t=0.3$ ensemble the value of $p_\text{glass}$ is evaluated numerically and found to be negative, indicating that this ensemble is always in the glassy phase.  
For the Haar ensemble,  $p_\text{glass} =  (3-e^{7/9})/3\approx 0.274$ (from Eqs.~\ref{eq:HaarA1A2},~\ref{eq:HaarA3},~\ref{eq:travelingwavevelocityfmla}). 
But since this value lies inside the entangled phase, where the linearized recursion is not valid, we do not expect that the glass transition is physically significant for the tensor network.

The arguments here, showing that the entanglement transition must take place within the glass regime of the linear recursion, extend to the class of tree tensor networks described in Sec.~\ref{sec:treestructuregeneralities}.
The possibility of  other universality classes of entanglement phase transition for other kinds of  quantum trees is discussed in Sec.~\ref{sec:othertree}.

}

\subsection{Including the nonlinearity}
\label{eq:nonlineartree}

Having understood the linear approximation to the recursion relation for the singular value squared, Eq.~\ref{eq:linearizedrecursion}, we must now consider the effect of nonlinearity. The nonlinearity is necessary to make sense of the entangled phase, where $Z_k$ is of order 1, rather than being exponentially large in $k$ as the linear equation would predict. 
Our aim in this section is to determine the scaling of $Z$ close to the transition, on the entangled side. Our basic conclusions have already been summarized in Sec.~\ref{sec:treeoverview}.

\subsubsection{Numerical results}
\label{sec:sqrtsigmanumerics}

\begin{figure}
    \centering
    \includegraphics[width=1\linewidth]{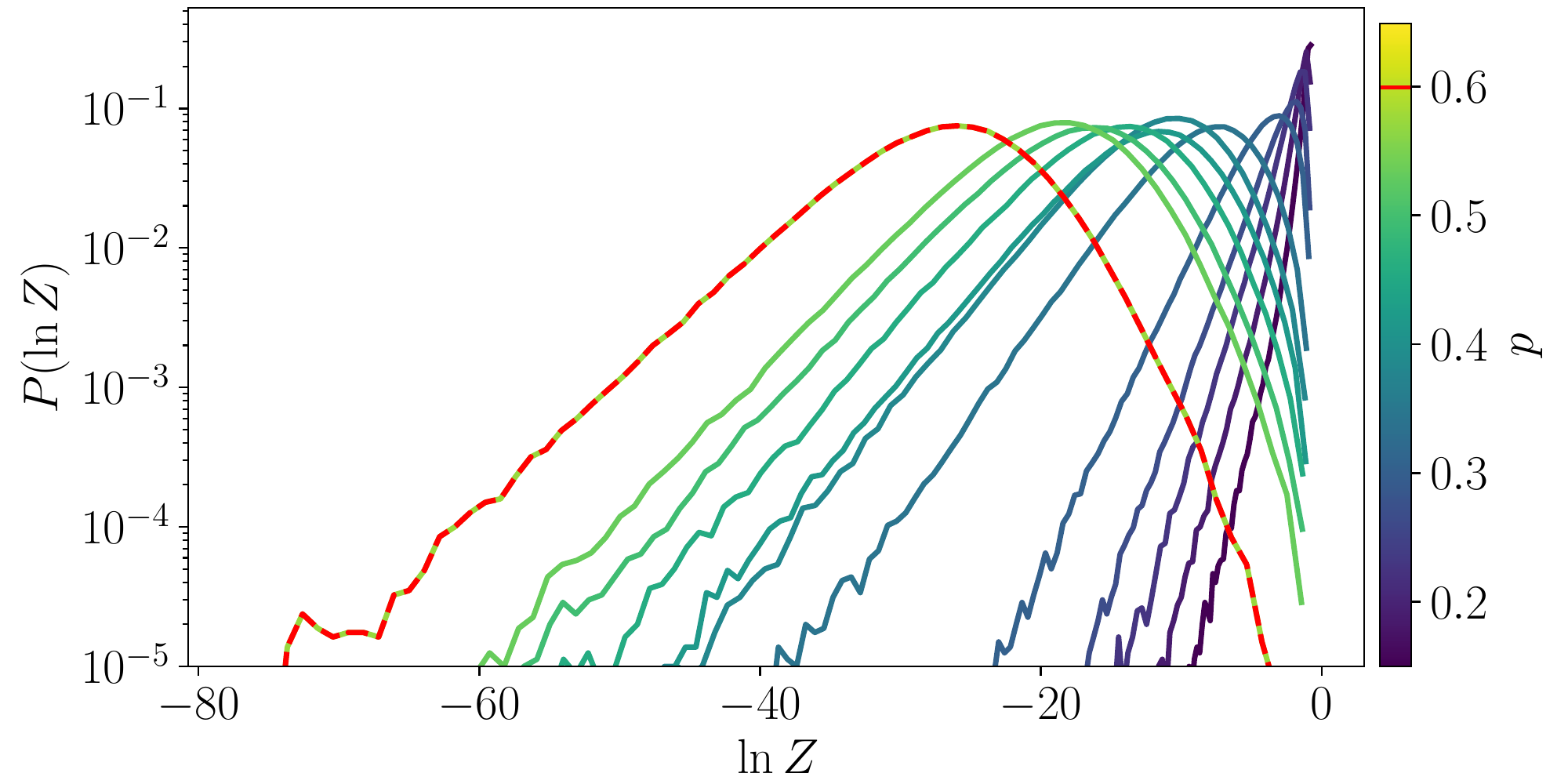}
    \caption{The distribution of $\ln Z$  in a tree of $k=150$ generations (for the full nonlinear problem). Here we have removed instances where $Z$ is exactly zero (which have a finite support due to the forced measurements).}
    \label{fig:Z_dists}
\end{figure}

Let us show numerical results before turning to an analytical treatment.

First, Fig.~\ref{fig:Z_dists} shows the probability distribution of ${\ln Z_k}$ for the Haar ensemble (Sec.~\ref{sec:unitaryensemble}) in a tree of $k=150$ generations, where we have removed instances where $Z$ is exactly zero.\footnote{We note that due to the forced measurements there is a finite probability for $Z$ to be exactly zero, i.e. the distribution function has a delta function with a finite weight. These are instances where the tree is classically disconnected. When we compute $Z^{\text{typ}}$ and present distribution functions we do not include these trivial instances.} 
Various values of $p$ less  than or equal to $p_c$ are shown.
The maximal possible value of $Z_k$ is 1/2:
deep in the entangled phase the distribution is concentrated near this upper limit, but as we approach the critical point ${\ln Z^\text{typ}}$ moves to the left. 
The shape of the distribution also stabilizes.  (In fact 
it approaches the shape for the linear problem, except on the right where $Z$ is of order 1.)

\begin{figure}
    \centering
    \includegraphics[width=\linewidth]{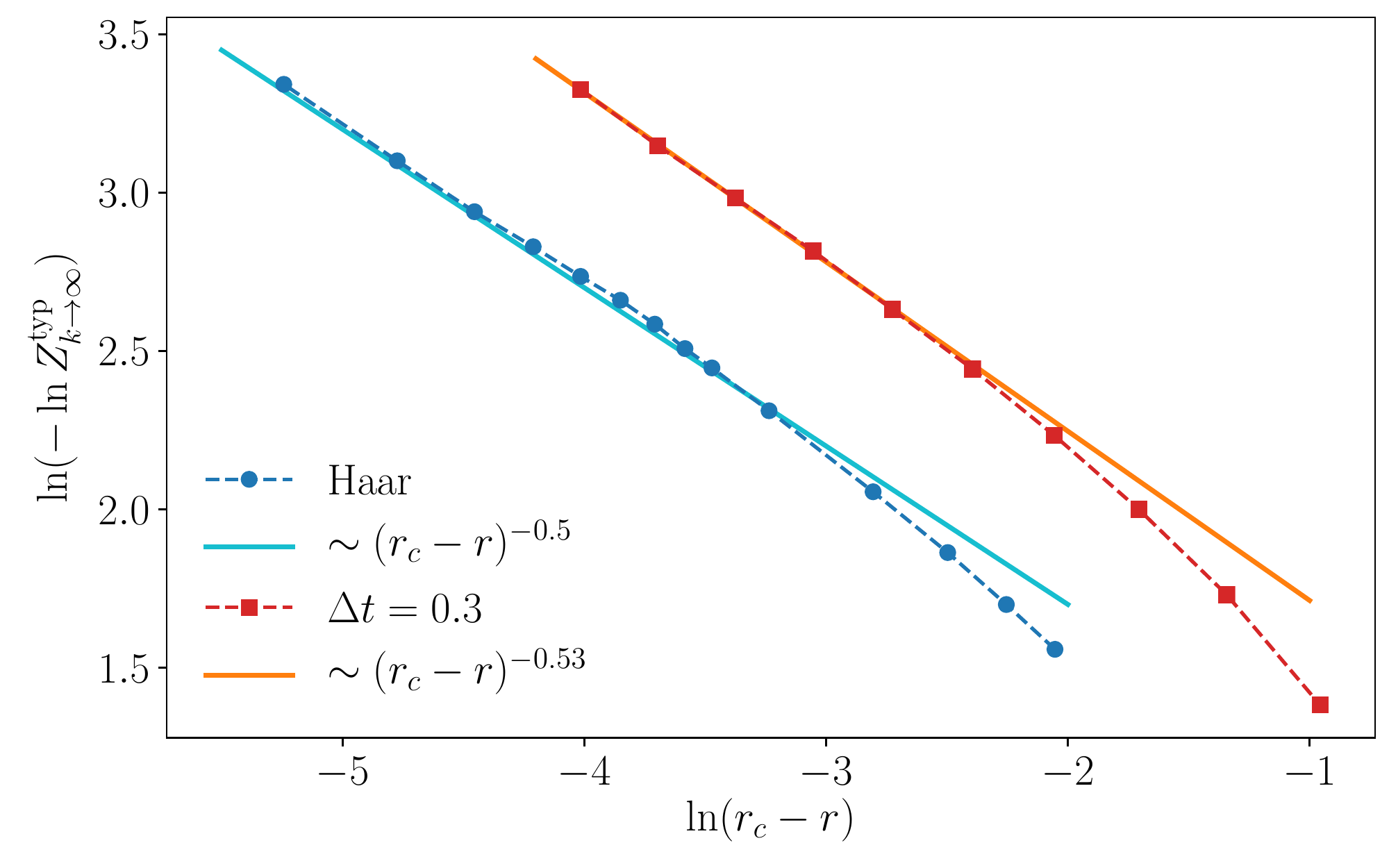}
    \caption{Critical behaviour for the Haar and $\Delta t=0.3$ trees as obtained from $Z^\mathrm{typ}_{600}$ and $Z^\mathrm{typ}_{700}$, for Haar and $\Delta t = 0.3$, respectively, for a pool size $N = 3\times 10^4$. The data suggest $Z^\mathrm{typ}_{k\to\infty}\sim e^{-D(r_c-r)^b}$ with the best fits for $b$ being $-0.5$ and $-0.53$ for the Haar and $\Delta t=0.3$ trees respectively; the corresponding values of $D$ are 2.01 and 3.24, respectively. 
    }
    \label{fig:Z_critical}
\end{figure}

Next, in Fig.~\ref{fig:Z_critical} we show the scaling of $Z^\text{typ}$ for both choices of the ensemble of unitaries (Sec.~\ref{sec:unitaryensemble}), close to the critical point.
The analytic treatment below gives ${\ln Z^\text{typ} \simeq -D / \sqrt{r_c - r}}$, which corresponds to a straight line with slope $-1/2$ in the plot. This slope is indicated by the trend line.
The data is consistent with this value of the exponent. 

However, the value of the non-universal constant $D$ that we extract from fitting this data is $D = 2.01$ for Haar and $D = 3.24$ for $\Delta t = 0.3$, which is far from that predicted below, for both ensembles.
Experimenting with simpler toy models suggests that this may just be because of finite ${r_c-r}$ effects, i.e. not being close enough to $r_c$.
The numerical method we use is afflicted by severe finite size effects (see Refs. \cite{miller1994weak,monthus2008anderson,garcia-mata2017scaling} and Appendix \ref{app:quantumTree}), 
associated with correctly sampling the right hand tail of the distribution in Fig.~\ref{fig:Z_evolution},
which mean we cannot approach too close to the critical point.
Details of the numerical method are in  Appendix \ref{app:quantumTree}.

\subsubsection{Nonlinear toy model}
\label{sec:nonlineartoymodel}
\begin{figure}
    \centering
    \includegraphics[width=\linewidth]{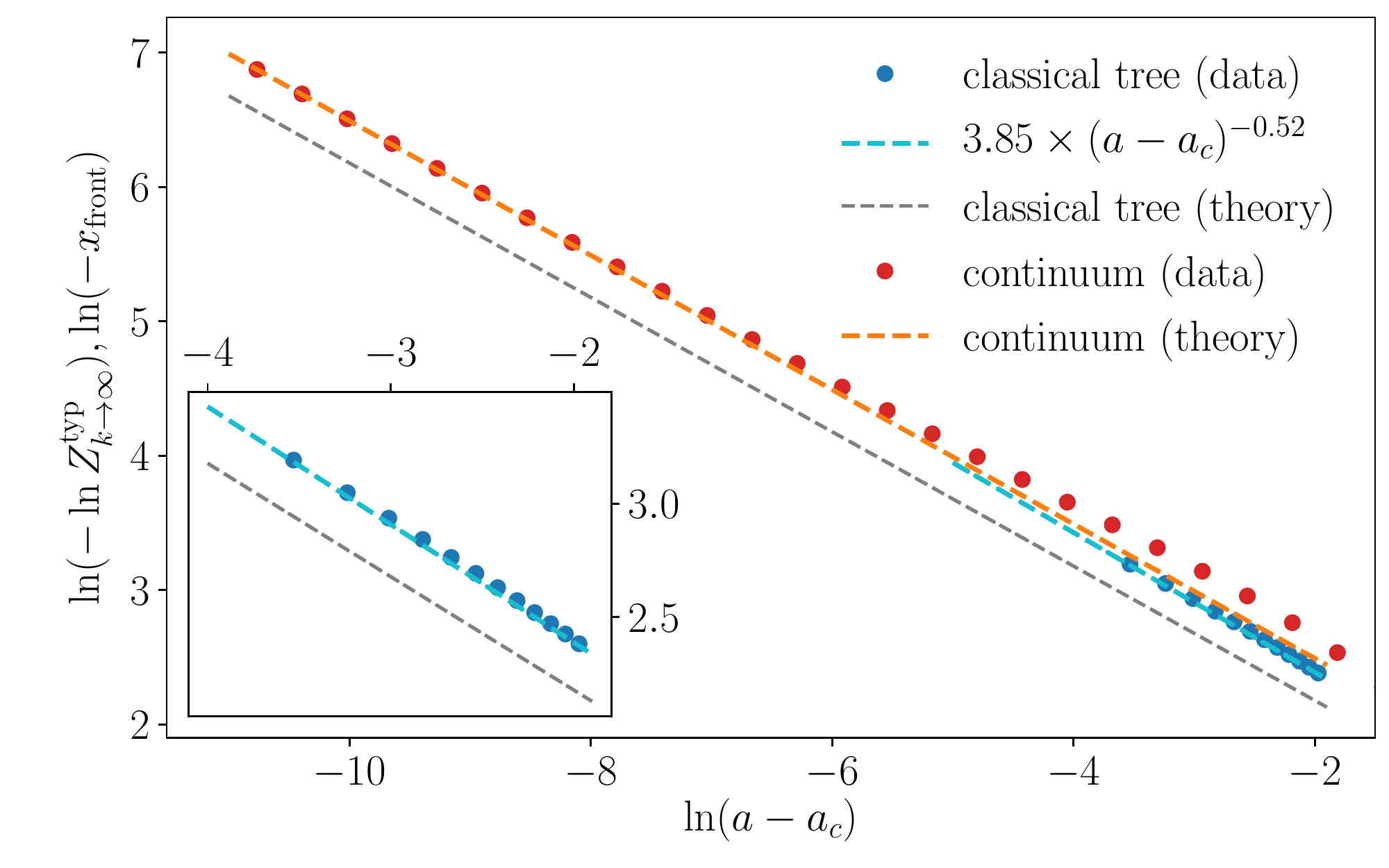}
    \caption{Critical behaviour of the classical non-linear tree and the continuum equation in the glass phase. Similar to the quantum trees, the data for the classical tree (blue dots) suggests a scaling $\sim e^{-c(a_c-a)^b}$ for $Z^\mathrm{typ}_{k\to\infty}$. The best fits are $b=-0.52$ and $c=3.86$. For the parameters for the tree used, $p_0=0.2$, $p_1=p_2=0.4$, $\epsilon=1.411$, and $\gamma=0.5$, the theoretical result shown in gray corresponds to $c=3.247\cdots$ and $b=0.5$. Data obtained by numerically solving the continuum equation (red dots) is in excellent agreement with the theoretical prediction close to the critical point, although some deviations are observed further away from the critical point. Inset: zoom of the data for the classical tree.}
    \label{fig:glass-toy-tree}
\end{figure}

The nonlinear recursion relation in Eqs.~\ref{eq:zequalszero},~\ref{eq:fullrecursion} is not very approachable, even if expanded only to quadratic order. 
To make progress, we conjecture that the universal properties can be understood in a simpler model that retains a few basic features. 
We study a  recursion relation satisfying two requirements. 
First, it contains both linear terms and nonlinear terms of  order $Z^2$ which tend to suppress $Z$ 
(naively, the terms of higher order than $Z^2$ should be negligible when we are parametrically close to the transition and $Z\ll 1$).
Second, its linearized form  is in the glass regime, as for the circuit. (Though in fact we will study both this case and the paramagnetic case for completeness.)

We first write down a toy model for a tree with a discrete generation number $k$, but as in Ref.~\cite{derrida1988polymers} it will be convenient to take a continuum limit in $k$. We assume that this continuum limit preserves the universal properties, as is the case for the linear problem.

For the toy model,
define the random variable $Z_{k+1}$ at level ${k+1}$ in terms of a sum of $\ell$ random variables $Z_k^{(1)},\ldots, Z_k^{(\ell)}$ at level $k$. Here $\ell$ is the branching number of a node,  and is taken to be random with a distribution $p_\ell$ for $\ell\geq 0$. The precise range allowed for $\ell$ is not important, so for simplicity we allow ${\ell = 0,1,2}$.
We also include nonlinearity of strength $\gamma$, and a multiplicative random variable written as $e^V$, with Gaussian $V$:
\be\label{eq:toynonlinearproblemdefn}
Z_{k+1} = \bigg( e^V \sum_{i=1}^\ell Z_k^{(i)} \bigg) \exp\left[  - \gamma \bigg( e^V \sum_{i=1}^\ell Z_k^{(i)} \bigg) \right].
\ee
This can be viewed as the composition of a linear transformation analogous to Eq.~\ref{eq:linearizedrecursion},
\be
Z_{k+1} =  e^V \sum_{i=1}^\ell Z_k^{(i)}
\ee
(but slightly simpler because we avoid having correlated random variables) and a nonlinear one,
\be
Z_{k+1} \longrightarrow Z_{k+1} e^{- \gamma Z_{k+1} }.
\ee
The exponential form is arbitrary: for the continuum limit below it will anyway be sufficient to expand only to order $\gamma$, giving a quadratic recursion relation for $Z$.
However the above form guarantees that $Z_{k+1}$ is  positive for any input values, which was important for our numerical explorations.

We conjecture that by solving this simple nonlinear system we also capture universal scaling for the problem of interest (Sec.~\ref{sec:recursionrelationfortree}). 

\subsubsection{Continuum traveling wave equation}
\label{sec:toymodelcontinuum}

The equation for the generating function (cf. Eq.~\ref{eq:definegeneratingfn}) that follows from expanding Eq.~\ref{eq:toynonlinearproblemdefn} to order $\gamma$ is:
\ba\label{eq:toymodelgenerationfunction}
G_{k+1}(x) = 
\exp\Big( 
\gamma \, \partial_x e^x \partial_x \Big)
\sum_\ell  p_\ell \, \Big\langle G_k(x-V)^\ell\Big\rangle_V,
\end{align}
where $p_0$ and $p_2$ are the probabilities of a termination and a branching, respectively.
Now we take the continuum limit in the ``time'' $k$.
When $\gamma=0$, this gives the Fisher-KPP traveling wave equation \cite{derrida1988polymers}.
We introduce a ``time'' step $\delta \tau$ which will be sent to zero and define ${\tau = k \, \delta\tau}$.
The probabilities $p_0$ and $p_2$ are taken to be of order $\delta \tau$ (i.e. ${p_1=1-p_0-p_2}$ is close to 1) so that in the limit the tree becomes a continuous time branching process. 
The parameter $\gamma$ is taken of order $\delta \tau$ (note that the prefactor does not matter: it can be absorbed into the normalization of $Z$)  
and the strength of the random potential is also taken to vanish with $\delta \tau$.
It is convenient to parameterize its mean and second moment as:
\ba\label{eq:Vvariancemean}
 \<V\> & =  \f{a}{2}  \langle V^2\rangle
 &
 \langle V^2\rangle & = 2 b \times \delta \tau.
\end{align}
Finally we absorb some constants into the generating function  by defining:\footnote{The multiplicative factor appearing here is the probability that a tree never terminates. In the limit of large $k$ this is also the probability that $Z$ is nonzero.}
\be
G_k(x) = 1 - \f{p_2 - p_0}{p_2} H_k(x).
\ee
The asymptotics of $H$ may be taken to be 
\ba
H( -\infty)&=1,
&
H( \infty)&=0. 
\end{align}
After absorbing a constant into the definition of  $\tau$, and shifting $x$ by a constant, $H$ satisfies:
\be\label{eq:htravelingwave}
\partial_\tau H =
\partial_x  \Big( D(x) \partial_x  - a \Big) H 
+ \Delta H \big( 1-H \big)
\ee
where the growth rate is ${\Delta = 2 ( p_2- p_0)/\<V^2\>}$ (which is finite in the $\delta\tau\rightarrow 0$ limit),
the drift coefficient is $a$ in Eq.~\ref{eq:Vvariancemean}, 
and there is a spatially varying diffusion coefficient
\be
\label{eq:diffusioncoefficient}
D(x) = 1 + e^x.
\ee
The exponential term in Eq.~\ref{eq:diffusioncoefficient} is the effect of the nonlinearity $\gamma$ in the tree problem.

Note that nonlinearity in the tree is unrelated to   nonlinearity in the Fisher-KPP field $H(x,\tau)$ (which instead reflects branching of the tree).

$H$ forms a traveling wave, whose speed $c$ sets the exponential growth rate of $Z$ (cf. Eq.~\ref{eq:growthspeeddefn} and Eq.~\ref{eq:fronttypreln}).
On their own, the combination of ordinary diffusion and logistic growth (the $\Delta$ term) in Eq.~\ref{eq:htravelingwave} would give a traveling wave propagating to the \textit{right} ($c>0$) which corresponds to exponential growth of $Z$. 
Here, in one phase, this wave instead propagates backwards ($c<0$): this is possible because of the drift term in Eq.~\ref{eq:htravelingwave}.
In the other phase, the wave attempts to propagate to the right but is stopped by the exponential growth of the diffusion constant at positive $x$, which prevents the buildup of $H$ at large $x$. This results in $c=0$. 

There is therefore a transition between a phase where $Z$ is exponentially small at large generation number and a phase where $Z$ remains order 1. This is the toy model's version of the entanglement transition.
 
In the absence of the $e^x$ term in $D(x)$, the velocity of a traveling wave with tail ${H\sim e^{-\lambda x}}$ is \cite{fisher1937wave}
\be
\label{eq:wavespeed}
v(\lambda) = \lambda + a + \Delta \lambda^{-1}
\ee
(as we see by keeping only the order $H$ terms in Eq.~\ref{eq:htravelingwave}), with a minimum at ${\lambda_* = \sqrt{\Delta}}$. 
Therefore the linearized tree is in the glass regime \cite{derrida1988polymers}
(Sec.~\ref{sec:glassaside}),
where the traveling wave travels at speed ${v_\text{min} = v(\lambda_*)}$, so long as ${\Delta < 1}$.
This is the case we are interested in for the current circuit models, where $\lambda_*=1/2$ at the entanglement transition.

The wavespeed is then $c=2\sqrt{\Delta}+a$, so in this toy model the analogue of the entanglement transition is at  ${a_c = - 2\sqrt{\Delta}}$.
Let us therefore write
\be
a=-2\sqrt{\Delta} + \sigma.
\ee
We are interested in small positive $\sigma$, just inside the entangled phase.
In principle we would like to solve for the stationary solution at late times, 
\be\label{eq:hstationary}
\partial_x  \Big( D(x) \partial_x  + 2\sqrt{\Delta} - \sigma \Big) H 
+ \Delta H \big( 1-H \big)=0,
\ee
which we expect to exist when ${\sigma>0}$.
In the absence of a full solution, we consider the equation piecewise \cite{brunet1997shift}.

Let the position of the front, whose scaling with $\sigma$ we wish to determine, be denoted $x_\text{front}(\sigma)$. We assume (and confirm below) that $x_\text{front}(\sigma)$ is large and negative at small $\sigma$.

First, at large positive $x$, the leading term in the equation is simply ${\partial_x e^x \partial_x H = 0}$, so the only solutions satisfying $H\rightarrow 0$ at large $x$ have $H\sim e^{-x}$.

Second, consider $-|x_\text{front}(\sigma)| \ll x \ll 0$. In this regime we neglect both the variation of the diffusion coefficient and the $\mathcal{O}(H^2)$ term. From Eq.~\ref{eq:wavespeed}, we can find a stationary solution for positive $\sigma$ only by making $\lambda$ complex \cite{brunet1997shift}. Keeping only the leading $\sigma$ dependence,
\be
\label{eq:Hsine}
H \sim e^{- \sqrt{\Delta} \,  (x-x_0)} \sin \lf 
\phi + 
\Delta^{1/4} \sqrt{\sigma} \, x
\ri.
\ee
We would like to use the as-yet-undetermined constants $x_0$ and $\phi$ in order to allow this solution to match onto the solutions at large positive and negative $x$. 
Note that the slope of this solution on a logarithmic plot is 
\be\label{eq:hslope}
\partial_x \ln H = - \sqrt{\Delta} +  \f{\sqrt{\sigma} \, \Delta^{1/4}}{\tan \lf \phi + 
\Delta^{1/4} \sqrt{\sigma} \, x \ri}
\ee
For generic $x$, this slope is close to $\sqrt{\Delta}$, because of the small factor $\sqrt{\sigma}$ in the second term. However, close to the zeroes of the tangent this is not true.  
This allows us to match on the right hand side of the range,\footnote{By examining the first order equation satisfied by ${R=\partial_x \ln H}$ after dropping the $\mathcal{O}(H^2)$ and $\mathcal{O}(\sigma)$ terms, we see that the value of this slope at some arbitrarily chosen value $x$ in the vicinity of the origin must be tuned to the correct $\mathcal{O}(1)$ value, $R_x$, in order to match onto the correct solution at large positive $x$, where the slope is $-1$. Further this value $R_x$  is necessarily less than $-\sqrt{\Delta}$, which is the approximate slope of Eq.~\ref{eq:Hsine} in the region where the sine is of order 1. To achieve this, we must take advantage of the negative divergence of the the cotangent in Eq.~\ref{eq:hslope} at argument $\pi$. This means that to leading order in $\sigma$ we must have ${\phi=\pi}$.} where the slope is steeper, so long as we take ${\phi=\pi}$ to leading order in $\sigma$.

Similar considerations on the left show the argument of the tangent must approach $0$ as the vicinity of the front is approached. Therefore, to leading order in $\sigma$, the position of the front is
\be
\label{eq:frontpositionnonlinear}
x_\text{front}(\sigma) = - \f{\pi}{\Delta^{1/4} \sqrt{\sigma}}.
\ee
The constant $x_0$ in Eq.~\ref{eq:Hsine} then has the same leading term, to to ensure that $H$ is of order 1 in the front region.

Since $x_\text{front}(\sigma)$ also sets average value of $\ln Z$,
\be
\label{eq:toyproblemtypical}
{\Ztyp} \sim \exp \lf  - \f{\pi}{\Delta^{1/4} \sqrt{\sigma}} \ri.
\ee
By considering the tail of the distribution, we see that in the regime we are discussing, where ${\Delta < 1}$, the mean scales as
\be
\< Z \> \sim (\Ztyp)^{\sqrt{\Delta}} \sim 
\exp \lf  - \f{\pi \Delta^{1/4}}{\sqrt{\sigma}} \ri.
\ee

Notice that Eq.~\ref{eq:toyproblemtypical} is 
\ba
Z^\text{typ}  \sim
\exp\lf -\f{\pi}{| \operatorname{Im} \lambda_\sigma |}\ri
\end{align}
(also ${\< Z\>  \sim (Z^\text{typ})^{\lambda_0}}$)
where $\lambda_\sigma$ solves  ${v_\sigma(\lambda) = 0}$.
That is, it depends only on the function $v(\lambda)$ for the linear problem!
Indeed the  strength of the nonlinearity $\gamma$ in Eq.~\ref{eq:toynonlinearproblemdefn} cannot appear, since it can be absorbed into a rescaling of $Z$ (which does not affect $\ln Z$ at leading order).

This suggests that we can  apply the result to the quantum tree of Sec.~\ref{sec:recursionrelationfortree}, using Eq.~\ref{eq:travelingwavevelocityfmla} for $v$. This gives
\ba
\label{eq:treescalingentangledphase}
{\Ztyp} & \simeq \exp \lf - \f{C}{\sqrt{p_c-p}} \ri,
\end{align}
with 
\be
C = 
\f{\pi (1-p_c)}{\sqrt{2}} \sqrt{ \< \sum_i A_i^{1/2} (\ln A_i)^2 \>}.
\ee
For the Haar-random case $C$ is given exactly by Eqs.~\ref{eq:HaarA1A2},~\ref{eq:HaarA3}. In terms of $r$ (Eq.~\ref{eq:prrelationrepeat}),
\ba
\label{eq:treescalingentangledphaser}
{\Ztyp} & \simeq \exp \lf - \f{1.482...}{\sqrt{r_c-r}} \ri.
\end{align}

\subsubsection{Tree entanglement at critical point}
\label{sec:rightatrc}

\begin{figure}
    \centering
    \includegraphics[width=\linewidth]{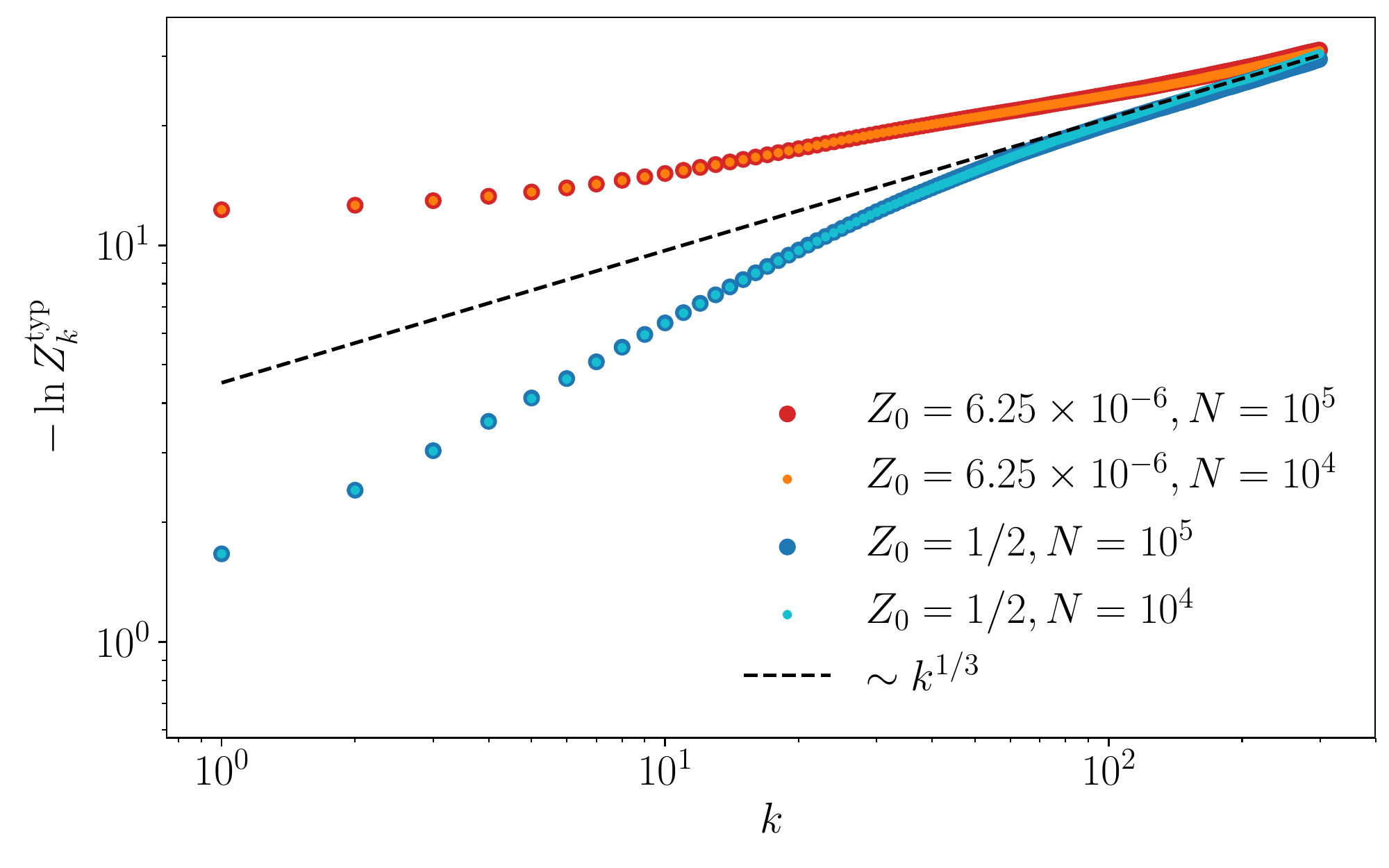}
    \caption{The behaviour of $Z^\mathrm{typ}_k$ with $k$ at the critical point for two initial conditions, $Z_0$, and two pool sizes, $N$. The data is consistent with a $k^{1/3}$ scaling of $\ln Z^\mathrm{typ}_k$. Results are shown for the Haar ensemble and errorbars are smaller than the data points. }
    \label{fig:Z-at-pc-haar}
\end{figure}

So far we have discussed scaling in the two phases.
Exactly at the transition, we might expect that $Z_k$ tends to zero with $k$, but more slowly than in the distentangled phase. 

Figure~\ref{fig:Z-at-pc-haar} shows data for this for the Haar ensemble.
The data is compatible with, though it does not clearly establish, the scaling ${\ln Z_k^\text{typ}\sim k^{1/3}}$ which is suggested by the following argument for the continuum model. 

We expect that for the time-dependent equation $x_\text{front}$ drifts sub-ballistically to the left. 
Let us conjecture that at a given time $t$, and in the range ${x_\text{front}\ll x \ll 0}$,
the instantaneous solution of the nonlinear equation  approximates sufficiently closely the traveling wave solution $G^{(\lambda)}$ of the linear equation with the same instantaneous speed, ${v=\dot x_\text{front}}$.
At the critical point ($\sigma=0$), $v(\lambda)$ has a double zero at ${\lambda=\sqrt{\Delta}}$, so this means that ${\lambda(t) \simeq \sqrt{\Delta}\pm i \Delta^{1/4} \sqrt{|\dot x_\text{front}|}}$. This gives a solution like Eq.~\ref{eq:Hsine}, but with $\sqrt{|\dot x_\text{front}|}$ in place of $\sqrt{\sigma}$. Eq.~\ref{eq:frontpositionnonlinear} then becomes 
\be
x_\text{front} \sim - \f{\pi}{\Delta^{1/4} |\dot x_\text{front}|^{1/2}},
\ee
which gives ${x_\text{front}\sim
(3 \pi^2 t / \sqrt{\Delta})^{1/3}}$. These values for the exponent and the prefactor are in good agreement with a numerical solution of Eq.~\ref{eq:htravelingwave} at ${a=-2\sqrt{\Delta}}$ (we checked the case ${\Delta = 1/4}$).

If $\sigma$ is small but positive there must be a crossover at a large time ${t_\text{sat}}$ from  ${x_\text{front}\sim
-t^{1/3}}$ to  ${x_\text{front}\sim-1/\sqrt{\sigma}}$. 
This suggests ${t_\text{sat}\sim \sigma^{-3/2}}$, which also agrees well with numerical solutions.

\subsection{Quantum trees: other universality classes?}
\label{sec:othertree}

{

Above we noted that a priori there were two possibilities according to whether the entanglement transition takes place within the glass or the paramagnetic regime of the linearized recursion relation: $\lambda_*<1$ and $\lambda_*>1$ respectively. 

However, our approach required the statistical invariance of the node tensor $t^{a}_{b_1, b_2, b_3,\ldots, b_\ell}$ under $\mathrm{U}(2)$ rotations on a leg. 
(We are free to allow for  an arbitrary branching number $\ell$.)
This invariance was necessary so that we could write a recursion relation for singular values only: otherwise we need a combined recursion relation for singular values and singular vectors.
For any such tree, the argument of Sec.~\ref{sec:linearizedtree} and App.~\ref{app:unitaryaverages}
shows that ${\lambda_*=1/2}$ at the transition.
That is, the recursion relation is of the form\footnote{Defining $\vec t_{b_1\ldots b_\ell}$ as the vector with components $t^a_{b_1\ldots b_\ell}$, then
$A_1 =_\text{law} 
|\vec t_{11\ldots 1}|^{-4}
\lf
|\vec t_{21\ldots 1}|^2 |\vec t_{11\ldots 1}|^2 - |\vec t_{21\ldots 1}^\dag \vec t_{11\ldots 1}|^2
\ri
=
|\vec t_{11\ldots 1}|^{-4} 
\left| 
t^{1}_{21\ldots 1} t^{2}_{11\ldots 1} - t^{1}_{11\ldots 1} t^{2}_{21\ldots 1}
\right|^2$ 
(for $A_m$ the ``2'' is the $m$th subscript). The argument in App.~\ref{app:unitaryaverages} making use of the invariance of $t$ under single-leg rotations also applies here.} 
\be
Z_{k+1} = \sum_{i = 1}^\ell A_i Z_k^{(i)} + \mathcal{O}(Z^2),
\ee
with ${\langle A_i\rangle=1}$ and ${\langle A_i^{1/2}\ln A_i\rangle=0}$, which is sufficient to ensure  $\lambda_*=1/2$ at the critical point  (Sec.~\ref{sec:linearizedtree}).

In this class of trees the weak correlations between top and base in the disentangled phase are dominated by only a subgraph of the tensor network that contains a few paths from top to bottom.
For this broad class of trees we expect the universal scaling described above.
Therefore within the class of trees that our formalism applies to there is no freedom to vary $\lambda_*$. 

However it is interesting to ask what happens in trees where the  unitary invariance property is broken.
Breaking this invariance introduces correlations between singular values and singular vectors.
A plausible guess, 
at least if these correlations are not too strong, 
is that in this setting the same toy model nevertheless captures the universal scaling. If this is the case (which we will not determine here) then the next question is whether in these more general models it is possible  to vary the  critical value of $\lambda_*$ away from $1/2$.

With this somewhat speculative motivation (and for completeness), below we extend the analysis of critical scaling in the toy model to the regime $\lambda_*>1$.

Our analysis has also been restricted to trees with bond dimension two. A recursion relation (for the subleading squared singular values squared) may be formulated for trees with larger bond dimension, but has a more complicated structure, even at lowest order. It would be interesting to study this further.

}

\subsubsection{Scaling of nonlinear recursion: $\Delta>1$}
\label{sec:toymodellargedelta}

We return to the toy model of Sec.~\ref{sec:nonlineartoymodel} in the continuum limit, now with $\Delta>1$.
The near-critical regime inside the entangled phase is now 
\be
{a = -(\Delta + 1) + \sigma}
\ee
with ${0<\sigma\ll 1}$ representing the control parameter that drives the entanglement transition.

The difference from the case studied above (cf. Eq.~\ref{eq:Hsine}) is that the solutions $\lambda$ of $v_\sigma(\lambda)=0$ are no longer complex: instead there is a real solution at ${\lambda = 1+\mathcal{O}(\sigma)}$, and a larger real solution at ${\lambda_\text{+}= \Delta +\mathcal{O}(\sigma)}$.
That is, if we neglect both the nonlinearity in $H$ and the $x$-dependence of the diffusion constant, the stationary solution is a sum of two exponentials, in contrast to Eq.~\ref{eq:Hsine}.

In App.~\ref{app:FKPPcontinuum} we study this regime via the equation for ${R=\partial_x \ln H}$, which interpolates between $0$ for ${x\ll x_\text{front}}$ and $-1$ for ${x\gg 0}$. We conclude that 
\ba
Z^\text{typ} &\sim \sigma^\kappa, &
\kappa & \equiv \max \left\{ 
\f{1}{\lambda_+-1}, 1
\right\},
\label{eq:kappafmla}
\end{align}
where the denominator appearing in $\kappa$ is the difference of the two solutions to $v(\lambda)=0$ at the critical point ${\sigma=0}$. In the present model, $\kappa=\max\{ 1/(\Delta-1), 1\}$, 
but we conjecture that the form in Eq.~\ref{eq:kappafmla}, which requires only knowledge of the speed function $v(\lambda)$ of the linearized problem, applies to a wider set of models.

\begin{figure}
\includegraphics[width=\linewidth]{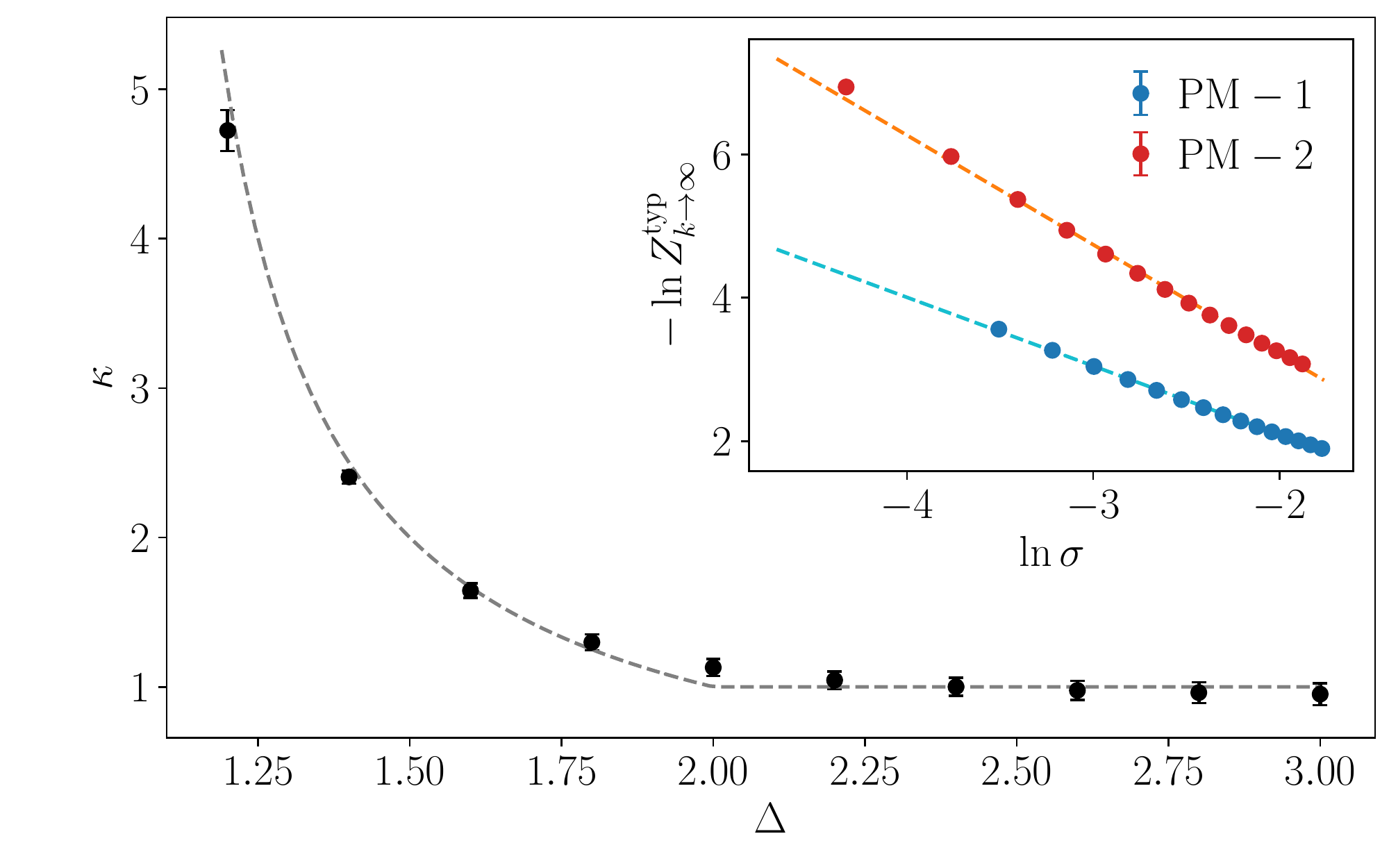}
\caption{Main panel: The exponent $\kappa$ that defines the critical divergence of $Z^\textrm{typ}$, (see Eq.~\ref{eq:kappafmla}) as a function of the parameter $\Delta$ in the power-law regime ${\Delta>1}$. Black points show results from a numeric solution of Eq.~(\ref{eq:htravelingwave}) whereas the gray dashed line shows $\kappa = \max\{ 1/(\Delta - 1), 1 \}$. Details of the numeric solution and definition of error bars are given in App.~\ref{app:FKPPcontinuum}. Inset: The critical divergence of $Z^\mathrm{typ}$ from the discrete toy tree for two sets of parameters in the paramagnetic phase, $\mathrm{PM-1}$ and $\mathrm{PM-2}$, see text for details. For $\mathrm{PM-1}$, the expected exponent is $-1$ and the best fit exponent is $-0.96$ whereas for $\mathrm{PM-2}$, the expected and the fitted exponents are $-1.97$ and $-1.53$ respectively.
}
\label{fig:FKPPslopes}
\end{figure}

Our argument in App.~\ref{app:FKPPcontinuum} is not rigorous, so we have compared the formula ${\kappa=\max\{ 1/(\Delta-1), 1\}}$ with a numerical solution of the continuum equation. Results are shown in Fig.~\ref{fig:FKPPslopes} and are in fairly good agreement with the prediction. 

Numerical solution of the continuum equation suggests that the above exponent $\kappa$ also determines the decay of $Z$ right at $r_c$,
\ba
Z & \sim t^{-\kappa}.
\end{align}

We have also studied  the discrete toy tree model in Eq.~\ref{eq:toynonlinearproblemdefn} numerically in the regime with ${\lambda_*>1}$.
We find polynomial scaling of $Z^\mathrm{typ}$ near the critical point as expected. The numerical estimates 
of the exponents differ somewhat from the predicted ones, which we attribute to finite size limitations.
See Fig.~\ref{fig:FKPPslopes} (inset) for examples. 
The parameters corresponding to $\mathrm{PM-1}$ are $p_1 = 0.15$, $p_2=0.85$, $\epsilon=0.95$, and $\gamma=0.5$, as such $\Delta = 3.97391$ and we expect $Z_{k\to\infty}^\mathrm{typ}\sim \sigma^{-1}$. Indeed the best fit exponent from our simulation is $\kappa=-0.96$. On the other hand, the parameters corresponding to $\mathrm{PM-2}$ are $p_2=1$, $\epsilon=1.5$, and $\gamma=0.5$ such that $\Delta=1.5066$ as such the expected exponent is $\kappa=-1.97$. We however find a best fit exponent of $\kappa=-1.53$ and attribute the discrepancy to finite-size of the pool and distance from the critical point.

\subsection{Trees, entanglement and min-cut}
\label{sec:treemincut}

So far we have characterized the entanglement between the top of the tree and the base. We now  apply this to more  general entanglement quantities in the tree.

\begin{figure}[t]
    \includegraphics[width=0.9\linewidth]{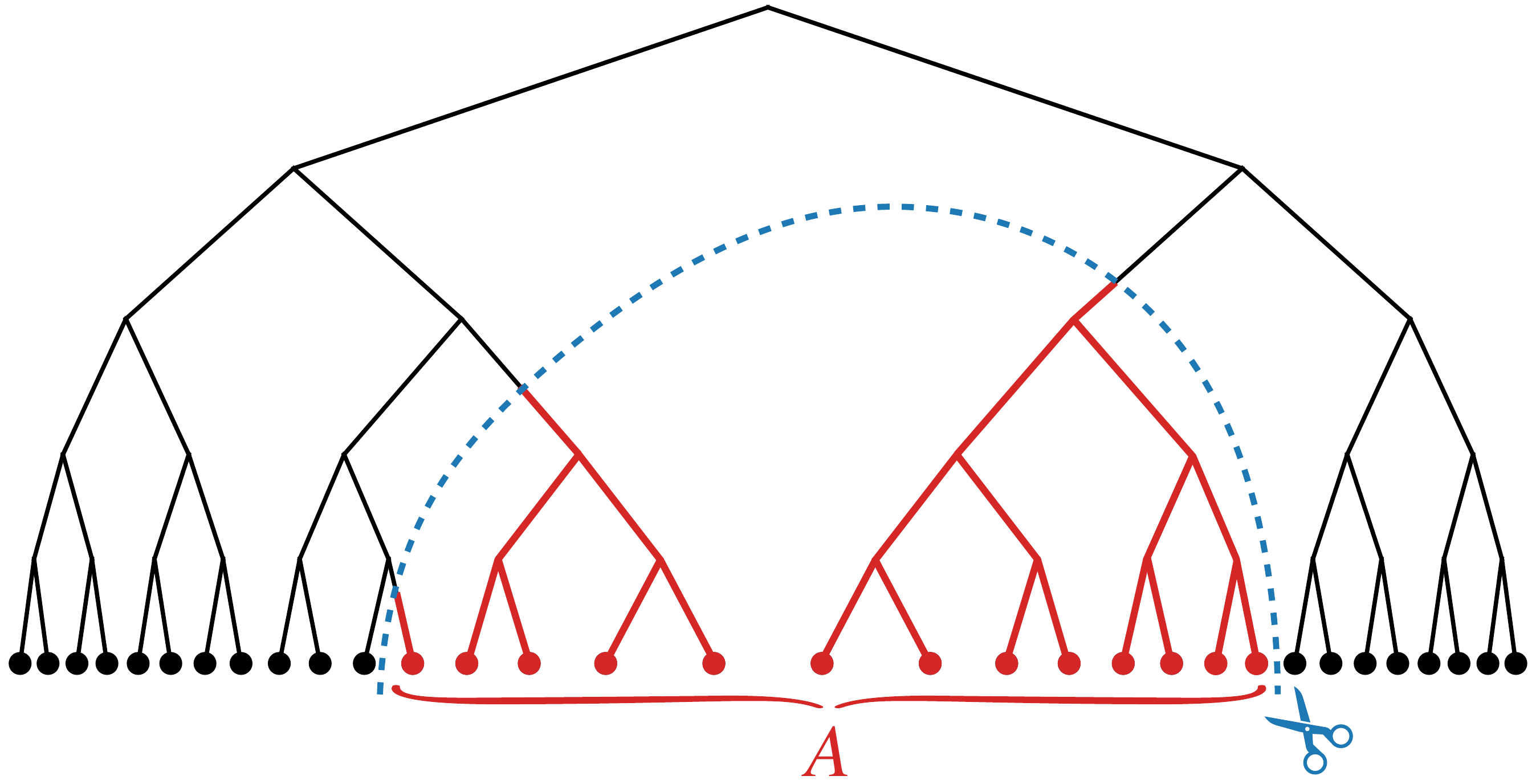}
    \caption{Schematic: tree tensor network wavefunction for a chain of spins, with a subset $A$ of spins and the corresponding minimal cut indicated. We propose a modified minimal cut formula for the entanglement.}
    \label{fig:min-cut-tree}
\end{figure}

Fig.~\ref{fig:min-cut-tree} is a schematic of a wavefunction for a chain of spins that is given by a tree tensor network (note that there is no longer a free bond at the top). 
Here we will consider the entanglement $S(R)$ of a set $A$ of ${R\gg 1}$ contiguous spins in a much larger chain.
This problem has also been tackled recently in Ref.~\cite{lopez2020mean} using a different method: see  Sec.~\ref{sec:overviewtree}.

As is well known, in such a geometry the minimal cut cartoon suggests the scaling $S\sim \ln R$ 
\cite{Swingle2012,pfeifer2009entanglement, lopez2020mean}
, which is the number of bonds cut for ``typical'' choices of the placement of the region $A$ (the tree strongly breaks translational invariance).
Figure~\ref{fig:min-cut-tree} shows an example of a minimal cut in a small tree. 
The logarithmic scaling is presumably correct in the entangled phase, but what happens close to the transition? For simplicity we consider the second R\'enyi entropy.

Note that the minimal cut in Fig.~\ref{fig:min-cut-tree} lops off a disjoint set of smaller subtrees, marked in red/thick. 
We will assume that the region is placed so that this is the case.
In this setting, a natural conjecture for the tree is that the universal scaling forms for the entanglement close to the transition and in the disentangled phase are given by a   ``modified minimal cut'' formula:
 we first find the geometrical minimal cut, but then \textit{weight} the contribution to the entanglement of a bond at height $k$ ($k$ generations above the base) by an amount that depends on $Z_k$.
 Since we are interested in the region close to the critical point and large $k$, we assume $Z_k\ll 1$.

What should this weight be? 
The simplest case is where the minimal cut only breaks one bond, i.e. where region $A$ corresponds to a single connected sub-tree.
In such cases the minimal cut breaks the full tree into two subtrees, and each one is characterized by a $Z$ value.
(In general  one of them has an irregular structure,  with different numbers of generations for different branches, 
but we can still use the recursion relation to compute its $Z$ value.)
A $2\times 2$ matrix calculation shows that the second R\'enyi entropy $S_2$ is proportional to the product of these $Z$ values, ${S_2\propto Z Z'}$, to leading order.\footnote{The prefactor depends on the singular vectors at the top of the sub-trees, but is of order one.}
Using the fact that $Z_k^\text{typ}$ is asymptotically non-increasing in $k$,
the $Z$ values of the subtrees are both of typical size $Z_k^\text{typ}$  for $k$ equal to the height of the cut bond.

This suggests the conjecture
\be\label{eq:sumtree}
S_2(R) \sim \sum_{k=1}^{c \ln R} Z_k Z_k'.
\ee
In this schematic formula, $c \ln R$ is the maximum height  reached by the minimal cut, and 
$Z_k$, $Z_k'$ are random variables.
All order one constants have been neglected, since we only aim to capture the asymptotic scaling with $R$ and with the distance from the critical point.
We can confirm Eq.~\ref{eq:sumtree} explicitly in an artificial limit in which the scale of the $Z$s tends to zero, with $R$ arbitrary but fixed: this is described in App.~\ref{app:treemincut}. However in the physical problem we wish to take $R$ to infinity, so this does not prove the conjecture.

Consider the first class of trees (including those with the statistical invariance property of the node tensors, for example those appearing in the Haar circuit). 
Let $r$ be an arbitrary parameter that drives the tree's entanglement transition.
The results in the previous sections and Eq.~\ref{eq:sumtree} yield:
\ba
\overline{S_2(R)}  & \sim \exp\lf - \f{\text{const.}}{\sqrt{r_c-r}}\ri 
 \ln R
 & &(r\lesssim r_c),
\\ 
\overline{S_2(R)}  &\sim \mathcal{O}(1) & & (r=r_c).
\end{align}
{ Surprisingly, the entanglement is order 1 at large $R$ at the critical point, because of the rapid decay of $Z$ with $k$  (Sec.~\ref{sec:rightatrc}). This is also true in the disentangled phase. 

In the previous section we speculated about the existence of trees with an effective value of ${\Delta>1}$. If such trees exist, then the same reasoning as above gives
${\overline{S_2(R)} \sim (r_c-r)^{2 \kappa}}$ for ${r\lesssim r_c}$, with a variable exponent 
${\kappa = \max\{[\Delta-1]^{-1},1 \}}$. The entanglement right at $r_c$ is again $\mathcal{O}(1)$.}

\subsection{Connecting back to the quantum circuit}

{

Our original motivation for studying the tree was the conjecture that,
for the forced measurement circuit models described in detail in Sec.~\ref{sec:unitaryensemble},
the critical point $r_c$ of the appropriate tree ensemble was also the critical point for the circuit.

Here we give an argument  
which  bounds the operator entanglement in the circuit in terms of the entanglement in the tree.
This argument is very heuristic: a task for the future is to make the connection between the circuit and the tree more precise.

The basic idea is to imagine breaking a bond in the interior of the FMPT circuit, and to ask how much effect this can have on properties of the nonunitary time evolution operator $V$.
Let $b$ be a bond inside the circuit at time coordinate $\sim t/2$.
Then $V_b$ will be a modification of $V$ in which bond $b$ is broken. 

Starting from $b$ we imagine marking the two trees $T$ and $T'$ attached to either end it, using the convention in Sec.~\ref{fig:circuit-schematic}, where bonds with projectors on them are removed.
We stop after $k$ generations, choosing the largest possible $k$ such that these are indeed two disjoint trees (no loops). Therefore $k$ should be of order $\ln N$.
We assume that the number of spins $N$ is very large, so that the typical size of $Z$ and $Z'$ for these trees (the minimal singular value squared) is given by the asymptotic large $k$ result. Close to $r_c$, this typical value is small.

Together $T$ and $T'$, connected by $b$, form a tensor network $\widetilde T$. This can be seen as  a state in a tensor product Hilbert space ${\mathcal{H}\otimes \mathcal{H}'}$ associated with the bonds on the boundary of $T$ and $T'$ respectively.
We may form the corresponding singular value decomposition of $\widetilde T$:
the smaller of its two singular values
is of order $\sqrt{Z Z'}$, in terms of the $Z$ values or $T$ and $T'$ (assumed small, since we are in the critical regime). 

Breaking the bond is defined to mean dropping this minimal singular value.
\textit{Formally}, this induces an error that is of order $\sqrt{Z Z'}$
in $Z$ and $Z'$.
After averaging over the local unitaries, the error in any physical quantity is --- again,  formally --- of order $Z Z'$.\footnote{Here we used the $\mathrm{U}(2)$ invariance property of the unitaries to show that lower terms vanished by phase cancellation.}
This suggests that the average change in $S_2$ when we break a single bond is \textit{at most} of order $\<Z_\infty\>^2$.
This is of course far from being a proof, because in principle the  small term $ZZ'$ in the formal expansion could be  systematically compensated by a large prefactor. 

Assuming this bound, we may straightforwardly bound the plateau value $s(r) N$ of the operator entanglement close to $r_c$ (Sec.~\ref{sec:overview},~Sec.~\ref{sec:replicatimescale}). 
Since breaking  all the $N$ bonds in a timeslice reduces the entanglement to zero,  
we must have
 ($c_1$ and $c_2$ are constants):
\be\label{eq:s(r)qmbound}
s(r) \leq c_1 \exp \lf - \f{c_2}{\sqrt{r_c-r}} \ri
\ee
for small $r_c-r$.
Assuming also our conjecture that $r_c$ is the same for the tree and the circuit, 
this indicates that $s(r)$ vanishes extremely rapidly as the critical point is approached. In turn, $s(r)$ is related to the scaling of the exponential timescale in the entangled phase  (Sec.~\ref{sec:replicatimescale}), just as it was in the classical problem (Sec.~\ref{sec:classical}).

We note that the bound (\ref{eq:s(r)qmbound}) on the scaling need not be tight. 
This can be understood by considering the analogous argument for the classical minimal cut problem. 

Above, our bound used $\<Z\>$.
The analogous quantity in the classical problem is the the probability that a given tree is infinite.
This is essentially the order parameter in the classical problem, scaling like  ${f_\infty \sim (r_c^\text{cl}-r)}$. 
We can bound the cost of the classical minimal cut $S_0$ as follows.
Consider all the bonds of the percolation configuration that traverse some timeslice, say at time $t/2$.
Each bond has a probability $\sim f_\infty^2$ that the two trees attached to either end of it are \textit{both} infinite. 
These are the only bonds we need to cut (the others lie in disconnected clusters or dangling ends).
This shows that  $s_\text{cl}(r)$  goes to zero \textit{at least} as fast as ${(r_c^\text{cl}-r)^2}$ close to the transition.

This bound is consistent with, but weaker than, what we have argued  is the true scaling in the classical problem, ${s_\text{cl}(r)\sim (r_c^\text{cl}-r)^{5/2}}$ (Sec.~\ref{sec:classicalmincut}).

}

\section{Simulations of quantum circuits}
\label{sec:simulationscircuits}

Having made a connection between trees and all-to-all circuits, we now turn to the numerical simulation of the latter.
Exponentially large in system size Hilbert-space dimensions restrict us to  systems with $N\le 20$ spins-1/2. 
We simulate both measurement circuits 
and forced measurement circuits, keeping in mind that the results obtained from the tree
apply only to the latter.
Unless specified, the results shown here are for the Haar ensemble Eq.~\ref{eq:uHaardistribution} (we also comment briefly on the $\Delta t=0.3$ ensemble, Eq.~\ref{eq:ufixeddistribution}).

All-to-all circuits have no spatial structure. Consequently, the entanglement transition does not entail a  volume-to-area law transition in the entanglement associated with a spatial bipartition of a state. Instead, we consider two observables 
which quantify the amount of quantum information transmitted from the initial to the final time: 
(i) the time-evolution of the \textit{operator entanglement entropies} (opEE) of the non-unitary evolution operator $V$,
and (ii) the overlap of two initially orthogonal states that are both evolved using $V$.

We will show that the  entanglement transition separates an entangled phase at ${r<r_c}$, wherein an extensive amount of quantum information is retained for an exponentially long time, from a disentangled phase at $r>r_c$ wherein memory of the initial state is rapidly lost. 
This is in agreement with analytical results for the quantum problem in  Sec.~\ref{sec:replicatimescale}, and is qualitatively similar to what we found in the  classical toy model in Sec.~\ref{sec:classical}.

First, we will give evidence for a plateau in the operator entanglement for $r$ below a critical value.
We will then turn to observable (ii) above: since this does not require exact diagonalization of $V$, it allows larger $N$ to be accessed. 
We use this observable to define a timescale $\tau(r,N)$, and show that this timescale scales exponentially with $N$ inside the entangled phase.
Details of numerical calculations are relegated to App.~\ref{app:circuit-numerics}.

\subsection{Operator entanglement}
\label{sec:numericsopEE}

The amount of information carried from the bottom of the circuit to the top can be quantified via the opEE of $V$. 
In the case of measurements, where we must choose an initial state in order to define the Born rule probabilities, we take this state to be a product state (with spins aligned in the positive $x$ direction, $\ket{\rightarrow\rightarrow\cdots\rightarrow}$).

The opEE is obtained from the singular value decomposition
    \be 
    V = \sum_{j=1}^{\mathcal{D}_H} \mu_j |j\rangle_t \langle j |_0 \,,
    \ee
    where $\mathcal{D}_H=2^N$ is the Hilbert-space dimension,
    and $\{|j\rangle_0\}$ and $\{|j\rangle_t\}$ are bases corresponding to the initial and final time.
    (We leave the $t$--dependence of $V$ implicit.)
    The opEE is:
    \be
    S_n ={1\over 1-n}\,\ln\, \sum_{j=1}^{\mathcal{D}_H}\, \lambda_j^{ 2n}\,,
    \label{eq:opee-def}
    \ee 
    where $\lambda_j \equiv \mu_j/\sqrt{\sum_j \mu_j^2}$. For a unitary $V$, $S_n $ takes on its maximal value of $N \ln 2$. Any reduction compared to this value reflects loss of information between initial and final time due to worldlines of the spins broken by measurements.
   $S_n$ is bounded from above by the  minimal cut separating the initial and final times. 
Close to the FMPT transition, we also have the conjectural bound Eq.~\ref{eq:s(r)qmbound}
on the scaling form for $s_2(r)=S_2/N$ in the plateau region.

\begin{figure}
\includegraphics[width=\linewidth]{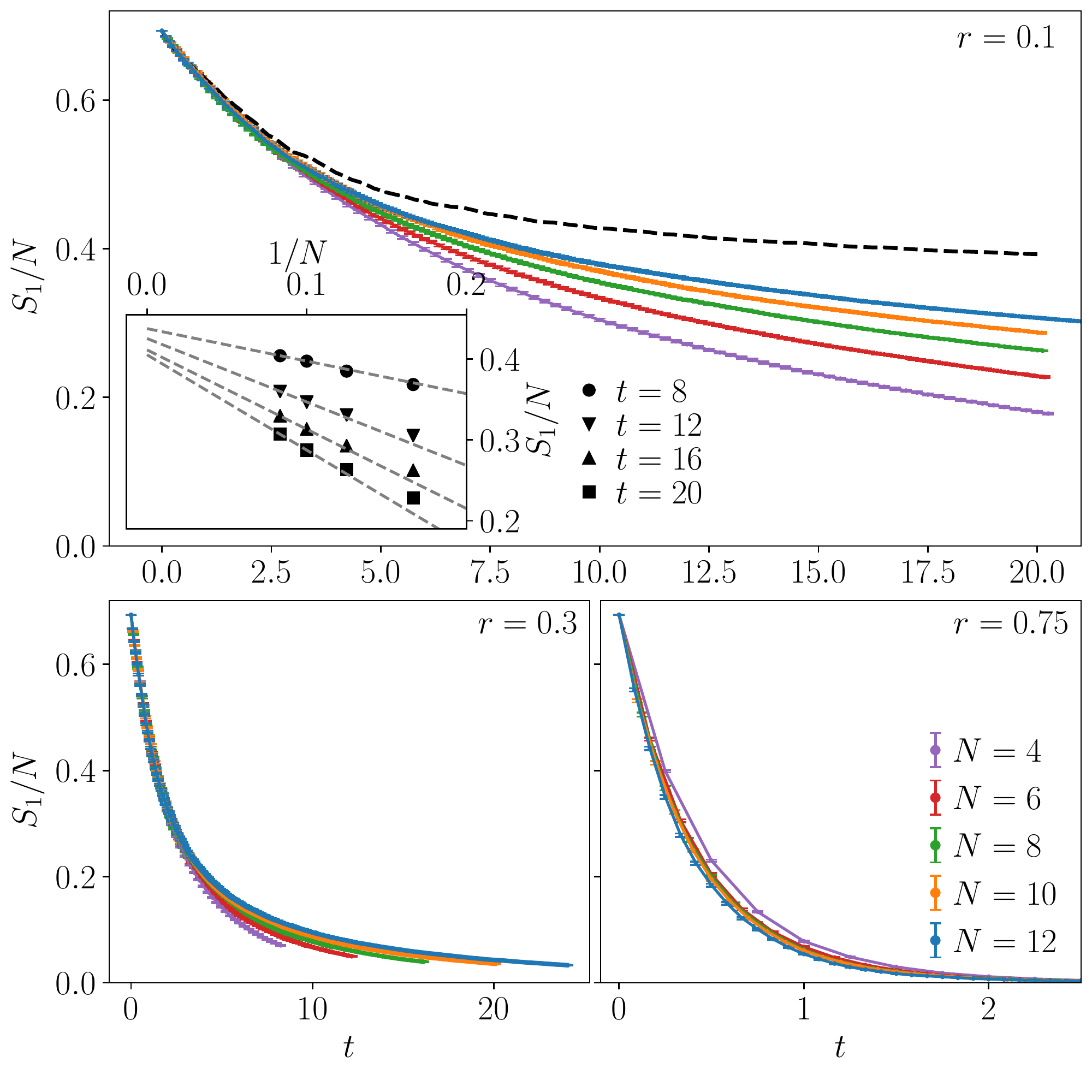}
    \caption{The opEE, $S_1$, of the non-unitary time-evolution operator (see Eq.~\ref{eq:opee-def}) for the Haar ensemble with forced measurements. The top panel is for $r=0.1$, deep in the entangled phase where there is a plateau in $S_1/N$ in the $N\rightarrow \infty$ limit. The extrapolated $N\rightarrow \infty$ value is shown as the black dashed line; the extrapolation in $N$ is shown for a few exemplary time points in the inset. 
    The bottom panels correspond to $r=0.3$ and $r=0.75$, the latter being the putative $r_c$ obtained from the Haar tree. 
    Note that the value of $S_1/N$ is already quite small in the entangled phase at $r=0.3$. 
    All data is averaged over 5000 realisations.}
    \label{fig:S1-haar}
\end{figure}
    
 Results for $S_1$ for the Haar ensemble with forced measurements are shown in Fig.~\ref{fig:S1-haar}. In the top panel, we plot the entanglement density $S_1/N$ {\it vs.} $t$ for $r = 0.1$ and various systems sizes. 
 After an initial linear decrease with $t$ associated with the first measurements, there is a time regime where $S_1/N$  increases with $N$. This suggests the emergence of a plateau in $S_1/N$ at large $N$ (Sec.~\ref{sec:overview}).
 
 Recall that in the entangled phase we expect
 a nonzero value for 
 \be
 s_n(r) \equiv \lim_{t\rightarrow \infty} \lim_{N\rightarrow \infty} \f{S_n(N, t, r)}{N},
 \ee
 and that when $N$ is finite but large, $S_1/N$ remains close to $s_1(r)$ over a range of times that grows exponentially with~$N$.

To give evidence for the nonzero value of $s_1(r)$ at ${r=0.1}$, we extrapolate the data to $N=\infty$ for each value of the time.\footnote{We use a naive linear extrapolation: the detailed functional form of the subleading corrections to $S/N$ may depend on the precise regime of $N$ and $t$   (Sec.~\ref{sec:replicatimescale}).}
This ${N\rightarrow \infty}$ extrapolation is shown as a dashed line in the figure.
It is consistent with a plateau, extending to $t=\infty$, with  $s_1(0.1)>0$. 
(We defer an analysis of timescales to the following subsection.)
A similar plateau was observed in Ref.\ \cite{gullans2019dynamical} in  Clifford circuits.

It is clear that the plateau value $s(r)$
decreases very rapidly with increasing $r$.
In the lower left panel we plot $S_1/N$ for the same set of $N$ and $r = 0.3$.
This $r$ value is still far from the conjectured location of the critical point obtained from the tree ($r_c\simeq 0.749$).
An increase of $S_1/N$ with $N$ is still observed, but it is clear that $s_1$ (assuming it is nonzero) is small.
It is tempting to associate this with the exponential scaling in Eq.~\ref{eq:s(r)qmbound}, 
which suggests that $s(r)$ goes to zero very fast as the critical point is approached, so that a plot of $s(r)$ against $r$ would be very flat for $r\lesssim r_c$.

On the other hand, at $r = 0.75$ 
(lower right panel),
$S_1/N$ decays exponentially to zero, with a very weak $N$-dependence and no indication of saturation at large $t$.
In fact, the trend with increasing $N$ is in the opposite direction to the cases  $r=0.1$ or $0.3$.

Thus, the opEE for these system sizes is consistent with an entanglement transition, occurring below the classical critical point, and with the expected plateau for $S/N$ in the entangled phase.
However, the rapid decay of the plateau value $s(r)$, and the weak $N$-dependence,
make it hard to pin down the position of the transition. We have checked (but do not show) that the data for $S_2$ is qualitatively similar to that for $S_1$.

\begin{figure}
    \includegraphics[width=\linewidth]{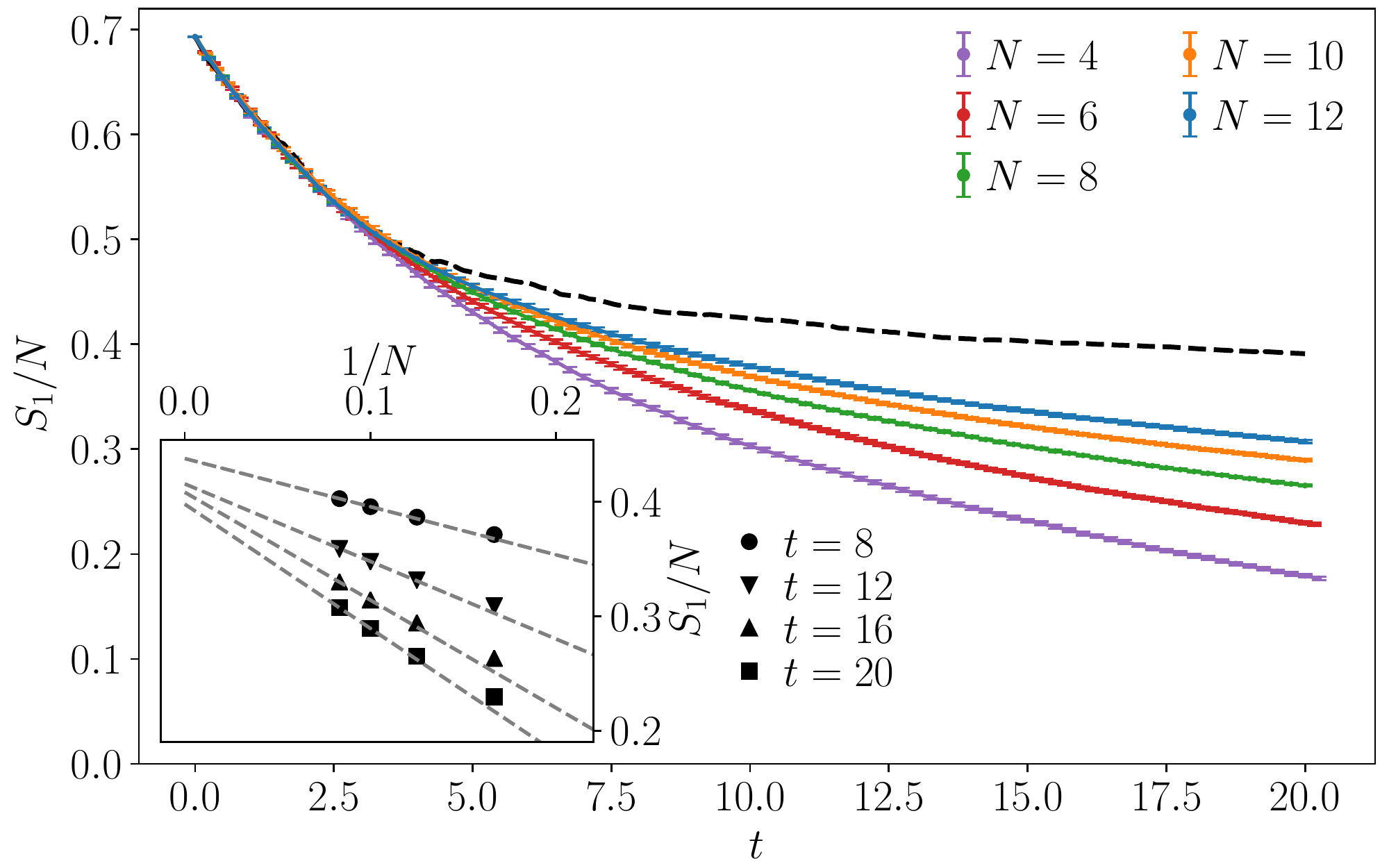}
    \caption{As in Fig.~\ref{fig:S1-haar} (top), the opEE density, $S_1/N$, of the non-unitary time-evolution operator for the Haar ensemble but with measurements for $r=0.1$. The data is consistent with there being a plateau in $S_1/N$ in the $N\to\infty$ and $t\to\infty$ limit. All data is averaged over 5000 realisations.}
    \label{fig:S1-haar-measurement}
\end{figure}

We find the same qualitative features for the Haar circuit  with true measurements.
Figure~\ref{fig:S1-haar-measurement} shows the case $r=0.1$.
In fact at this relatively small value of $r$, the data for forced measurements and measurements is almost indistinguishable.

\begin{figure}[t]
    \includegraphics[width=\linewidth]{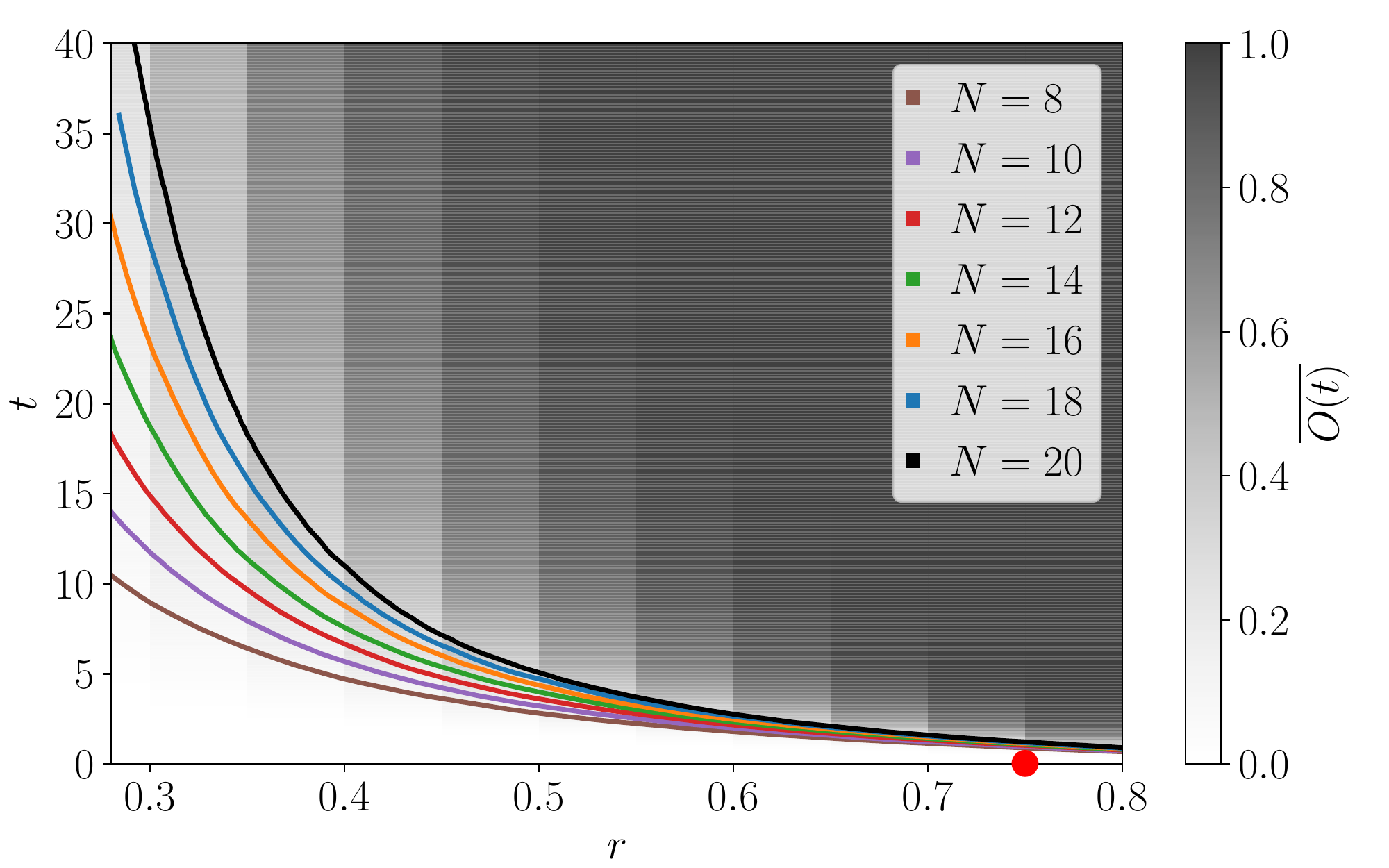}
    \caption{Heatmap: the squared overlap $\overline{O(t)}$ between the states $\vert{\psi^{(1)}(t)}\rangle$ and $\vert{\psi^{(2)}(t)}\rangle$ (see Eq.~\ref{eq:Ot-def}), as function of time $t$ and measurement rate $r$ for $N=20$. The different lines show the contours for $\overline{O(t)}=1/2$ for different $N$. The data corresponds to the Haar circuit with forced measurements. The red mark at $r=0.75$ denotes the critical point from the Haar tree. }
    \label{fig:density-overview}
\end{figure}

\subsection{State overlap and timescales}
\label{sec:overlap}

We now turn to the overlap of two initially orthogonal states undergoing time-evolution with the non-unitary operator $V$:
\be 
O(t) \equiv |\langle\psi^{(2)}(t) | \psi^{(1)}(t) \rangle |^2\,.
\label{eq:Ot-def}
\ee
We also define the ``distance'' $\mathcal{D}(t)$ as
\be
\mathcal{D}(t) =  1 - O(t).
\ee    
The states are initiated as product states in the $\sigma^x$-basis, $|\psi^{(1)}(0)\rangle =\ket{\rightarrow\rightarrow\cdots\rightarrow}$ and $\vert\psi^{(2)}(0)\rangle =\ket{\leftarrow\leftarrow\cdots\leftarrow}$ and are evolved using the non-unitary operator 
    \be 
    | \psi^{(j)}(t) \rangle = {V| \psi^{(j)}(0) \rangle \over  \sqrt{\langle \psi^{(j)}(0) |V^\dag V| \psi^{(j)}(0) \rangle}} \,.
    \ee 
In the case of forced measurements the spins are always projected along the positive $\sigma^z$-direction, and there is a symmetry between $\ket{\psi^{(1)}(t)}$ and $\ket{\psi^{(2)}(t)}$. In the case of measurements we use $\ket{\psi^{(1)}(t)}$ to determine the Born rule probabilities, so this symmetry is absent.

$O(t)$ is an another way to quantify the amount of information retained from the initial state.
In the limit $r=0$, where $V$ is unitary, the two states remain orthogonal for all time, $O(t)=0$. {In the opposite limit $r=1$, where no unitaries are applied, $O(t)$ will be exactly one as soon as all spins have been measured.}
For any fixed $r>0$, and for a fixed value of $N$, the states will inevitably converge to 1 as $t\rightarrow\infty$, because they are being subjected to the same projections. 
However  we expect the timescale for this to grow exponentially with $N$ in the entangled phase.

For a broadbrush view, 
we first show $O(t)$  as a heatmap in the space of $r$ and $t$ for the Haar forced measurement circuit with $N=20$ spins: Fig.~\ref{fig:density-overview}.
$O(t)$ grows towards unity with an $r$-dependent timescale. 
One way to define a timescale is using the 
contour $\overline{O(t)}=1/2$.
This is shown not only for $N=20$, 
but also for smaller values of $N$, in the figure.
The conjectured $r$ value of the phase transition is marked by the red dot.
The timescale grows rapidly as $r$ is decreased. It also shows a clear $N$-dependence in the entangling phase, which becomes much weaker on approaching the transition point.

\begin{figure}[t]
    \includegraphics[width=\linewidth]{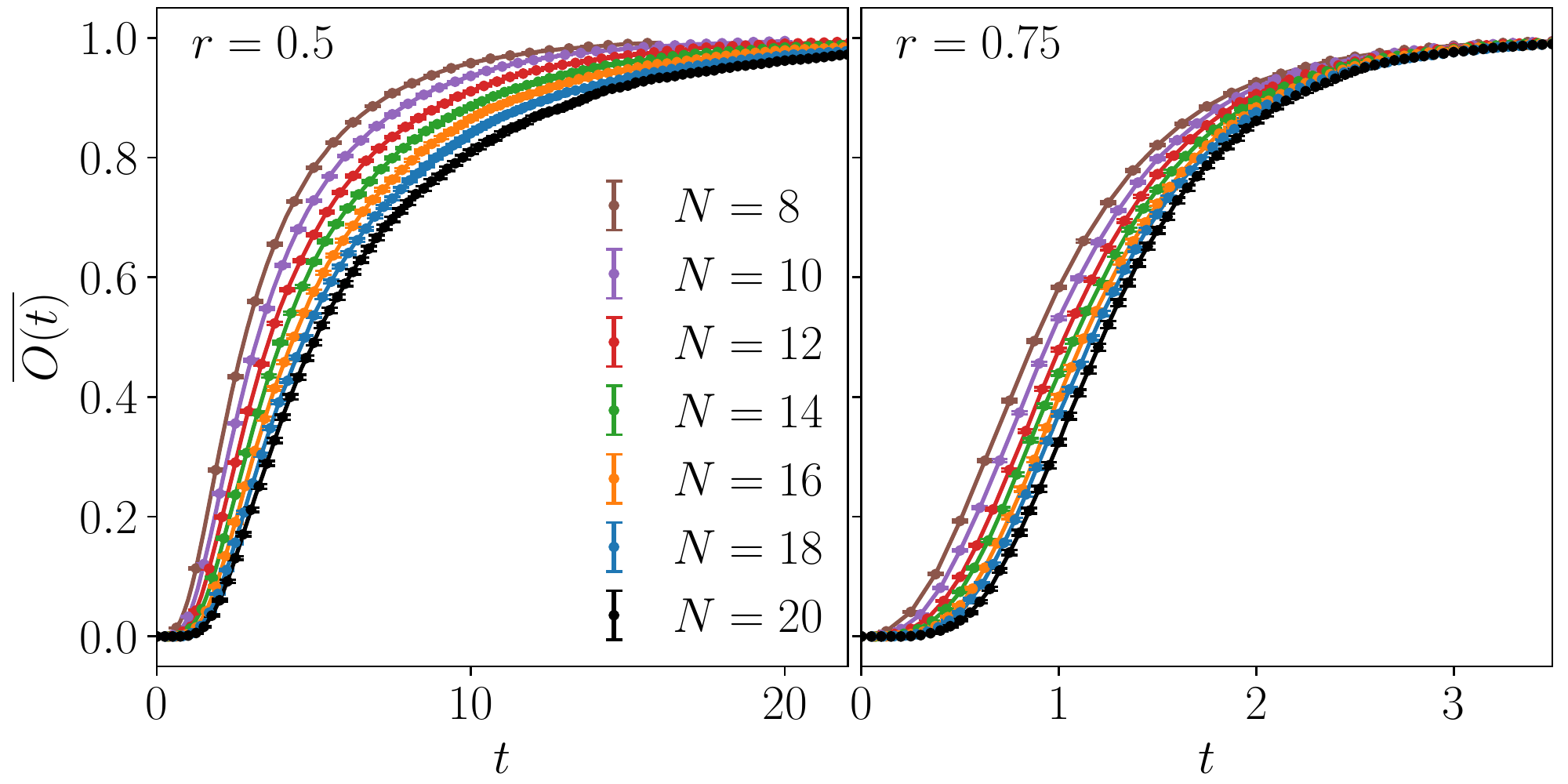}
    \caption{The squared overlap between the states $\overline{O(t)}$ as function of time $t$ for two exemplary values of $r$ for different $N$. The data corresponds to the Haar circuit with forced measurements. Note the difference in the range of times shown in the two panels. It is also possible to see the emergence of a plateau at ${\overline{O(t)}=0}$ at early times.}
    \label{fig:overlap-linear}
\end{figure}
\begin{figure}[t]
    \includegraphics[width=\linewidth]{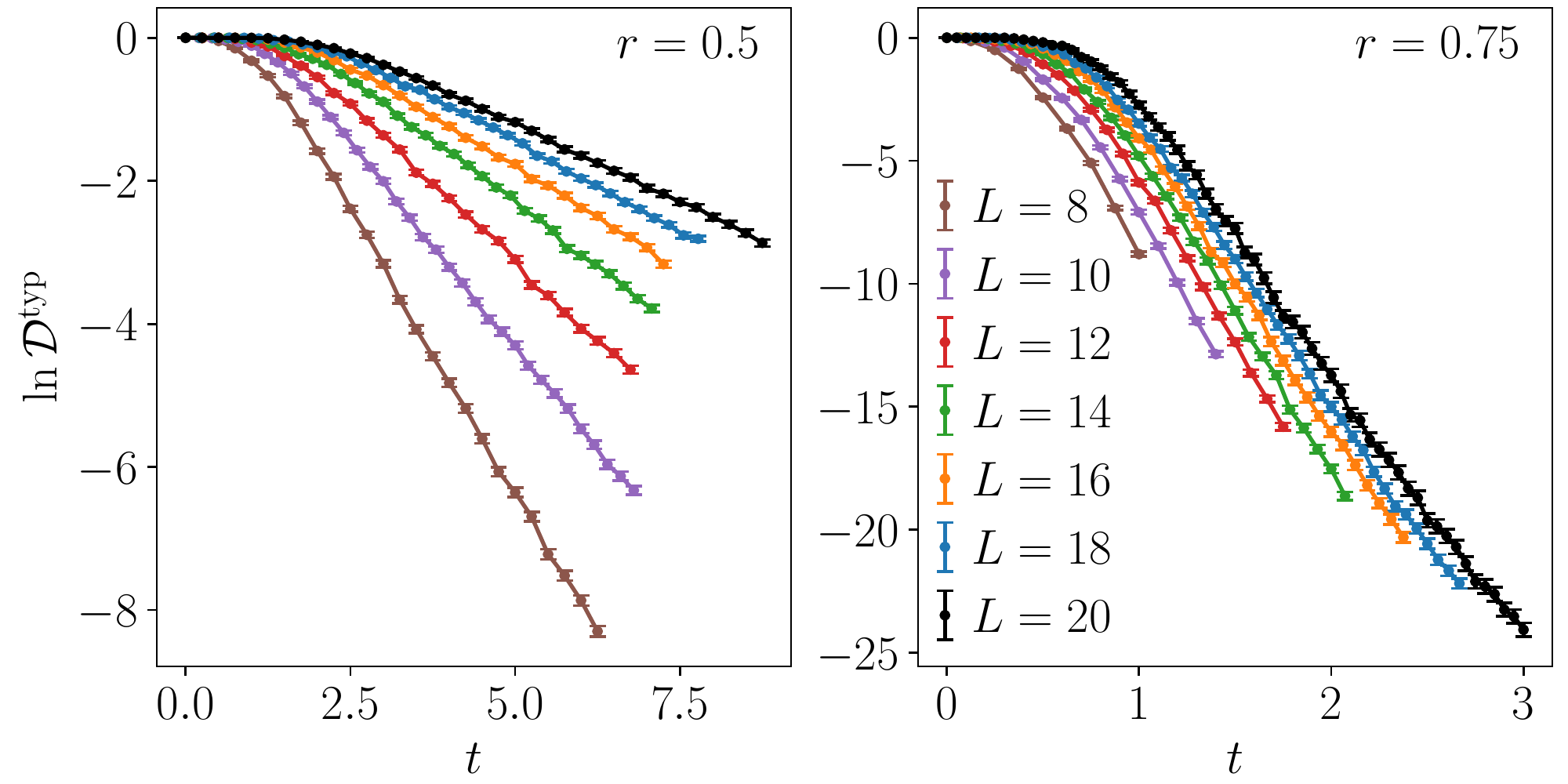}
    \caption{The typical distance $\mathcal{D}(t)=1-O(t)$ between the states $\vert{\psi^{(1)}(t)}\rangle$ and $\vert{\psi^{(2)}(t)}\rangle$  as function of time $t$ for the same parameters as in Fig.~\ref{fig:overlap-linear}, showing exponential convergence of the two states at late time. Note the difference in the scales shown in the two panels.
    }
    \label{fig:overlap-typ}
\end{figure}

Fig.~\ref{fig:overlap-linear}   shows the time-dependence of the overlap in more detail for $r = 0.5$ and $r = 0.75$.\footnote{ Note that the latter value of $r$ is right at the putative critical point. This value was chosen to avoid being above, or to close to, the classical critical point $r = 0.8$ where the network is trivially disconnected.}.
In both cases the time required to achieve a given value of $O$ increases with $N$, but in the latter case this is mostly a shift of the curve, whereas in the former case there is the clear sign of an increasing time constant for the exponential approach of $O$ to 1.

We use the exponential approach of the overlap to unity to define a timescale $\tau(r,N)$. 
Since at late times ${\mathcal{D}=1-O}$ is exponentially small, and may have a broad distribution, we choose to look at its typical value. We define this by $\ln\mathcal D^{\text{typ}}(t) \equiv \overline{\ln \mathcal{D}(t)}$,
where instances in which  $\mathcal{D}(t)$ is exactly zero are excluded from the average (similar to the treatment of the  singular value $Z$ in the quantum tree, Sec.~\ref{sec:treeoverview}).
At late times this  shows an exponential decay,
\be \label{eq:D_asymptot}
\ln \mathcal{D}^\mathrm{typ}
\sim -\f{t}{\tau(r,N)}\, .
\ee

\begin{figure}[t]
    \includegraphics[width=\linewidth]{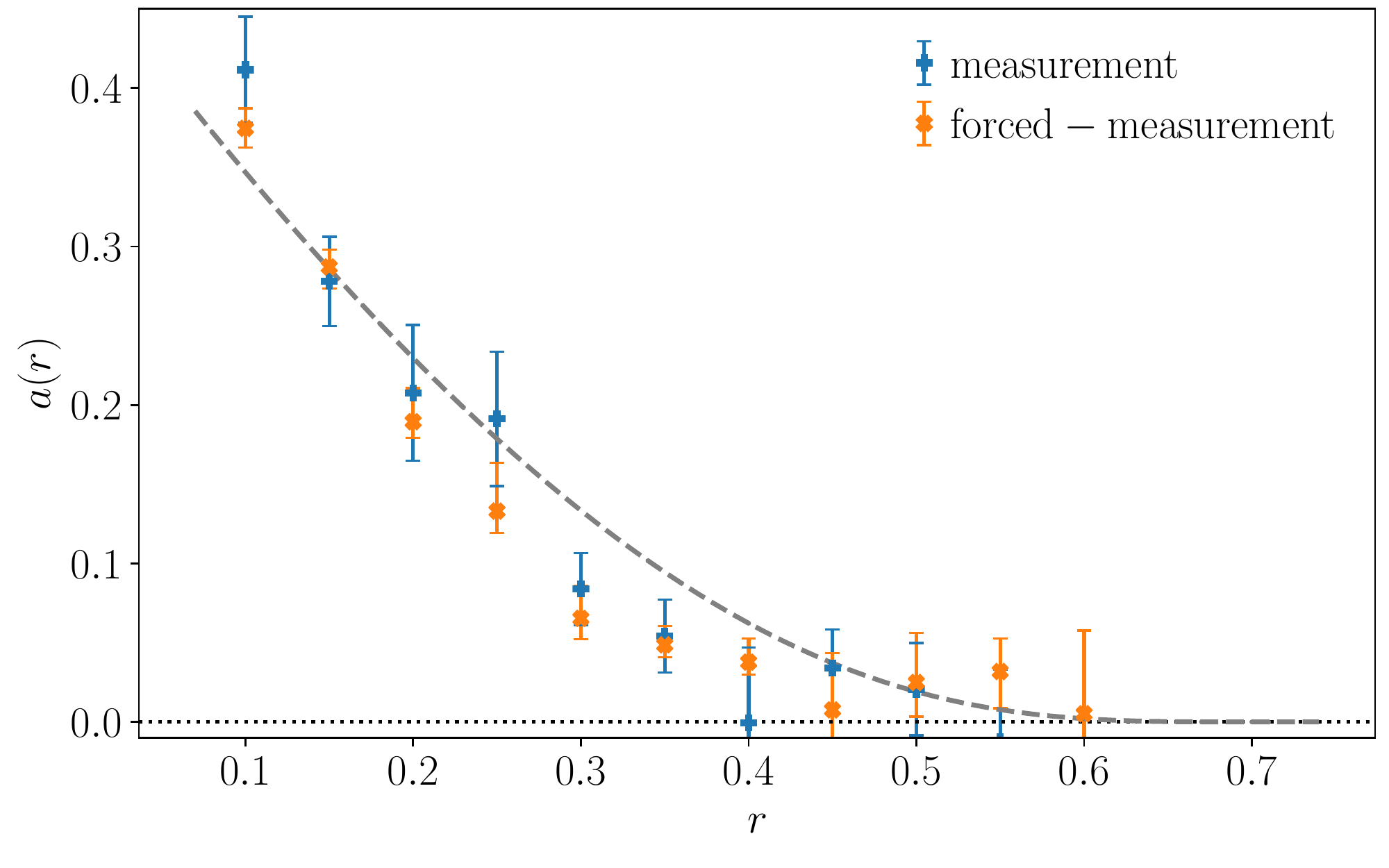}
    \caption{The coefficient $a(r)$ in the exponential dependence of the timescale ${\tau\sim e^{a(r) N}}$ on $N$,  as a function of $r$ for the Haar circuit with measurements and that with  forced measurements. The dashed line is a guide to eye for a { function $c_1 e^{-c_2/\sqrt{r_c-r}}$ with $r_c=0.75$, $c_1=39.4$} and $c_2=3.8$. See Figs.~\ref{fig:haar-fit-tau} for the fits used to extract $a(r)$.}
    \label{fig:ar-haar}
\end{figure}

Data for $\ln \mathcal D^{\text{typ}}(t)$ {\it vs.} $t$ are shown in  Fig.~\ref{fig:overlap-typ}, for the same values of $N$ and $r$ as in Fig.~\ref{fig:overlap-linear}. 
We see clear exponential decay.
In fact, Fig.~\ref{fig:overlap-typ} vividly shows the qualitative difference between the cases of $r=0.5$ and 0.75: while $\tau$ grows with $N$ for $r=0.5$, it appears essentially $N$-independent for $r = 0.75$.
Data for the circuit with measurements (not shown) is qualitatively similar.

We now analyze $\tau(r,N)$  in the entangled phase. 
This is the asymptotic slope of plots like Fig.~\ref{fig:overlap-typ}.
We extract this from a plot of 
${\tau_\mathrm{eff}(t)= -\left(\dd\ln\mathcal{D}^\mathrm{typ}/\dd t \right)^{-1}}$,
the time-dependent slope: at late times, $\tau_\text{eff}(t)$ should stabilize at the value $\tau$.
Representative data for $\tau_\mathrm{eff}(t)$ and the plateaux therein are shown in App.~\ref{app:circuit-numerics}, see  Figs.~\ref{fig:haar-log-derivs} and~\ref{fig:haar-mes-log-derivs}. 

It turns out that finite-time effects become significant at larger values of $r$, 
but for $r$ not too large we are able to obtain an estimate of $\tau$.
The data (shown in App.~\ref{app:circuit-numerics}) is consistent at small $r$ with exponential-in-$N$ growth of the timescale:
\be
\ln \tau(r,N) \sim a(r) N.
\ee
The coefficient $a(r)$ is plotted against $r$ in Fig.~\ref{fig:ar-haar}.
This figure also shows data for the case of true measurements.

We expect $a(r)$ to vanish at the critical point with ${a(r)\sim s(r)}$ 
(see Sec.~\ref{sec:replicatimescale}).
Unfortunately, hamstrung by severe finite-size effects, we are not able to estimate the critical point accurately.\footnote{These finite size effects, together with the fact that extracting $a(r)$ requires two separate fits (to go from $\mathcal{D}_\text{typ}$ to $\tau(r,N)$  and then to $a(r)$) also make it hard to estimate error bars on $a(r)$ accurately.} 
The data is certainly consistent with a critical point for the FMPT which is below the conjectured value $\simeq 0.749$.
However, we speculate that this is instead  a symptom of $a(r)$ vanishing very rapidly as $r_c$ is approached, as is suggested by the essential singularity in Eq.~\ref{eq:s(r)qmbound}.
The dashed line in the figure shows this exponential form with $c_1 = 39.4$ and $c_2 = 3.8$. These values have no theoretical significance: this line is simply to indicate the possibility of $a(r)$ remaining very small even for $r$ considerably below $r_c$.

\begin{figure}
    \includegraphics[width=\linewidth]{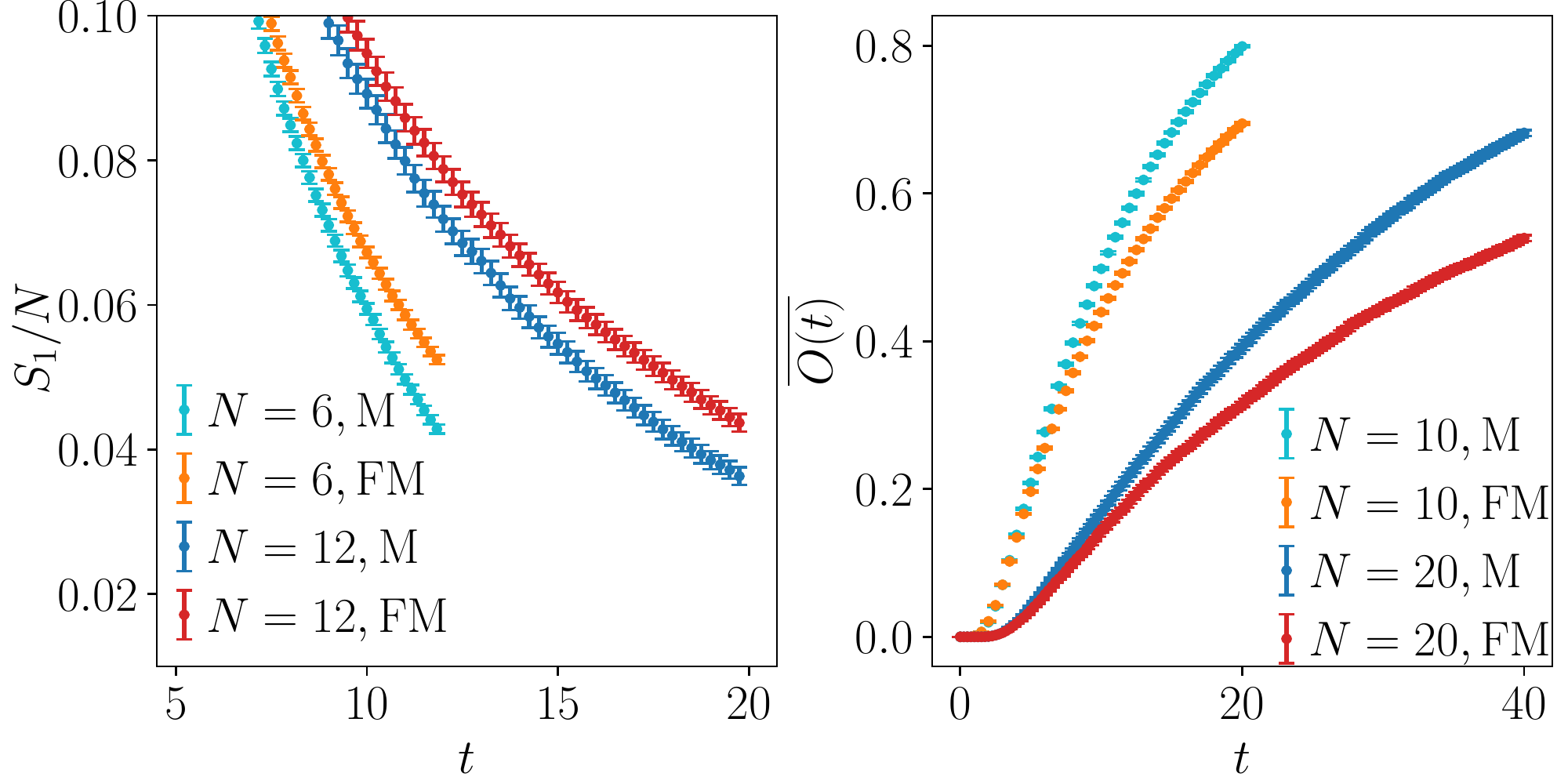}
    \caption{Evidence that the forced measurement (FM) protocol leads to higher entanglement than the measurement (M) protocol at the same value of $r$. The data is for $r=0.3$. The left panel shows $S_1/N$ whereas the right panel shows the average overlap $\overline{O}(t)$.}
    \label{fig:compare-M-FM}
\end{figure}

The data for $S_1$ at small values of $r$ is very close for measurements and forced measurements, as noted above. We do see differences between the two cases at intermediate $r$, with the forced  measurement circuit having slightly larger entanglement at a given $r$. This is shown for  $S_1/N$ in Fig.~\ref{fig:compare-M-FM}, Left. 
The comparison between the overlap data for the two cases at the same value of $r$ (Fig.~\ref{fig:compare-M-FM}, Right), is also consistent with the above, with $\overline{O(t)}$ growing more slowly in the forced measurement case. This hints that $r_c$ for the measurement case may be lower than that for the forced measurement case, but our data does not allow us to determine this.

While the data above was shown for the Haar circuit, we performed the same set of numerical calculations for the $\Delta t=0.3$ circuit as well. The results were qualitatively similar. Consistent with the results from the quantum tree, $r_c$ appeared to be smaller for the $\Delta t=0.3$ circuit compared to the Haar.

\section{Field theories for measurement and entanglement transitions}
\label{sec:landautheory}

A key question about the measurement phase transition (MPT),   not previously  resolved, is whether there is a simple Landau-Ginsburg-Wilson-like field theory that captures its universal properties. 
This question is also unresolved for entanglement transitions in random tensor networks (RTNs), and for the closely related FMPT.
In this section we propose candidates for these field theories.
(In this section the spacetime dimensionality ${D=d+1}$ is allowed to be arbitrary.)
We obtain two Lagrangians, one for the MPT, and one for both FMPT and RTN.
Surprisingly, these two Lagrangians are quite different in their structure, having for example different values for the upper critical dimension.

Microscopically,  random circuits and random tensor networks can be mapped to lattice statistical mechanics models \cite{hayden2016holographic, nahum2018operator,zhou2018emergent,vasseur2019entanglement,jian2020measurement,bao2020theory,hunter2019unitary,liu2020entanglement}.
These are effective spin models where the ``spin'' is a group element in the permutation group $S_N$ for $N$ objects (we review this below; here $N$ is a replica number and not the number of qubits as in previous sections). 
However, using these lattice models to guess  appropriate continuum field theories is nontrivial for various reasons, one of them being a replica limit that is necessary to handle randomness.
``Replica'' lattice models were described for a random  tensor network in Ref.~\cite{vasseur2019entanglement}, for Haar circuits in \cite{zhou2018emergent}, and for circuits with  measurement in Refs.~\cite{jian2020measurement,bao2020theory}.

Previous work pointed out that in certain limits (either by artificially deforming the weights in the effective spin model \cite{vasseur2019entanglement}, or by taking a $q\rightarrow \infty$ limit in the measurement problem \cite{jian2020measurement,bao2020theory})
 one could access a fine-tuned point where the effective spin model had a simple continuum theory, namely that of percolation.
While this was a useful step,  this fine-tuned  point has an infinite number of relevant perturbations \cite{vasseur2019entanglement} so unfortunately this does not provide  a definite Lagrangian for the physical phase transitions of interest. 
Another approach has been to study Ising models that are obtained by simply omitting the replica limit, 
roughly in the spirit of an annealed average in conventional disordered systems \cite{hayden2016holographic,fan2020self,li2020statistical}.
These are useful toy models for various phenomena in the entangled phase \cite{fan2020self,li2020statistical} 
(we will give an explanation for why this is, building on \cite{zhou2019entanglement}) but they cannot capture the correct critical properties. Therefore we attempt here to formulate explicit continuum replica field theories.

We emphasize that these theories are speculative conjectures, based on writing down the simplest Lagrangians compatible with the basic symmetries of the problem. It is certainly possible that in fact something more complicated happens in the continuum.
Indeed, the exponential scaling we found in the tree seems to mean that it is not described by the high-dimensional limit of the  field theory for RTNs proposed below  (see Sec.~\ref{sec:FTconsequences}). How to resolve this tension is a question for the future.

We will first review the replica approach, the inevitable global symmetry of the field theories we are looking for,.
 and the emergence of permutations in the simplest Haar-random models
(Secs.~\ref{sec:multilayer},~\ref{sec:replicaexample} are largely review). 
We then discuss coarse-graining of these degrees of freedom (Sec.~\ref{sec:permutationscoarsegraining}). 
Next we note that these degrees of freedom have a more general meaning in terms of Feynman trajectories in the circuit \cite{zhou2019entanglement,garratt2020many}.  
This picture motivates an alternative derivation of a lattice field theory which in turn suggests a simpler continuum formulation (Sec.~\ref{sec:isingmapping}).
Our discussion also suggests an alternative way of thinking about the effective statistical mechanics of random tensor networks, in a way that is closer to traditional replica formulations of random magnets.

Then we discuss the issue of ``replica group theory'' for the MPT on one hand, and the RTN and FMPT on the other: 
that is, constraints on the field theories associated with the replica symmetry \cite{cardy2013logarithmic}.
We propose the simplest candidate Lagrangians in each case (Secs.~\ref{sec:MPTFT} ---\ref{sec:FTRTNFMPT}).  
We discuss some of the basic consequences of the simpler of these Lagrangians, that for the MPT (Sec.~\ref{sec:FTconsequences}). 
Our discussion of these field theories is relatively schematic: further details will be given in Ref.~\cite{replicasforthcoming}.

Sec.~\ref{sec:replicatimescale}, which is independent of the field theories proposed here, addresses scaling within the two phases, not necessarily near the critical point. Finally Sec.~\ref{sec:commentscaveatsetc} describes variations of the measurement problem that are in distinct universality classes, for example models with free fermion structure or with additional symmetries.

\subsection{Multi-layer circuits and replica symmetry}
\label{sec:multilayer}

\begin{figure}[t]
    \includegraphics[width=0.6\linewidth]{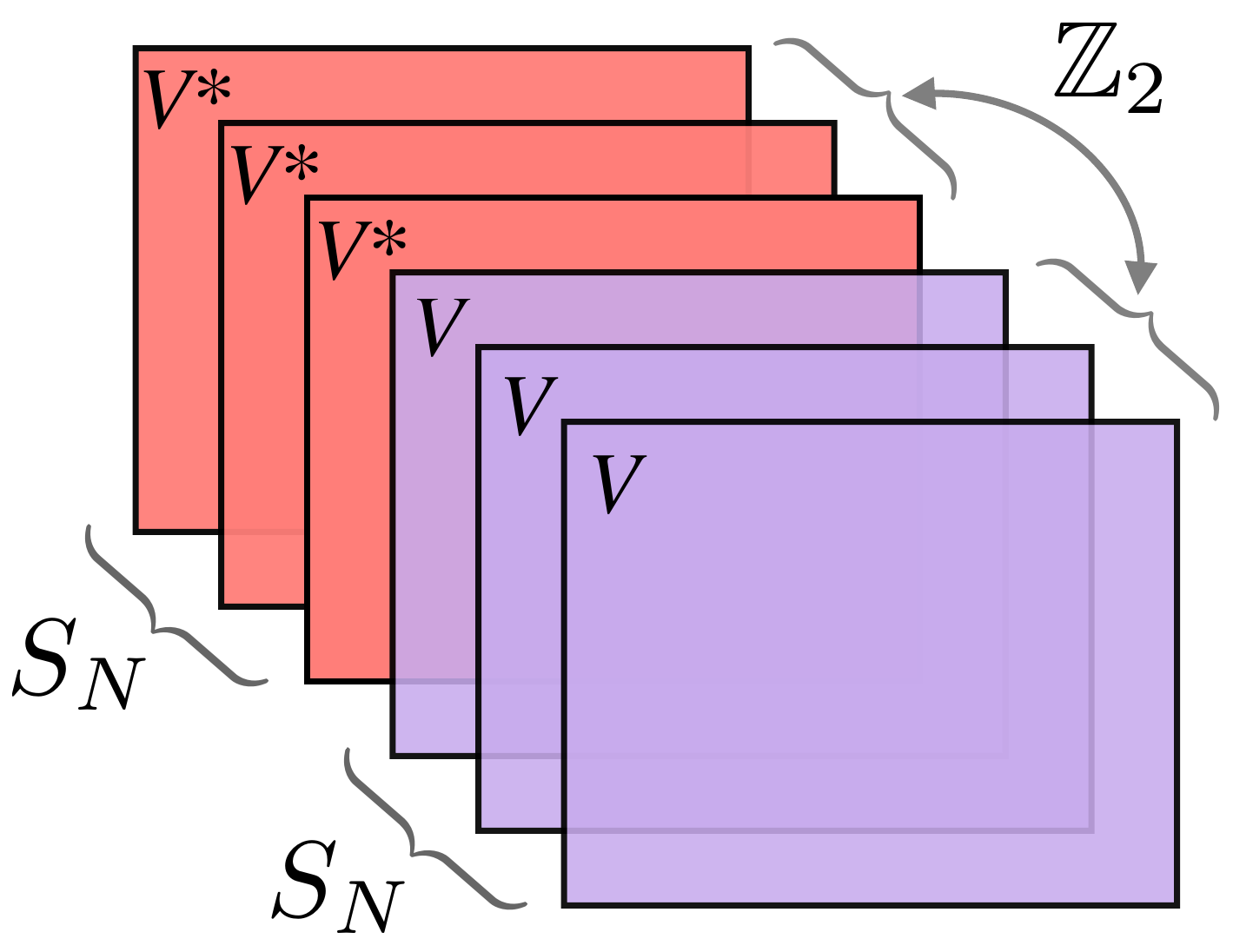}
    \caption{Schematic:  multi-layer circuit (or tensor network) with $N$ copies of the circuit $V$ and $N$ copies of its complex conjugate $V^*$. Each layer of the circuit has free bond indices at the bottom and top (initial and final time)  which are not shown.
    The actions of the left and right permutation symmetries and the $\mathbb{Z}_2$ exchange that make up the internal symmetry group ${G_N = \lf S_N \times S_N \ri \rtimes \mathbb{Z}_2}$ of the replica field theories are illustrated.
    }
    \label{fig:stack}
\end{figure}

The crucial symmetries of the problem arise when dynamical quantities are written in terms of a multi-layer circuit, illustrated schematically in Fig.~\ref{fig:stack}. 
(We will use the language of a circuit, with $d$ spatial dimensions and one time dimension, but
but analogous considerations apply to a ${D=d+1}$ dimensional RTN.)
This multi-layer circuit is a discrete analogue of a path integral with multiple forward and backward paths, and it arises when we write powers of the reduced density matrix, say for the final state, in terms of the circuit. Let us briefly review this.

The layers are $N$ identical copies of the original circuit $V(t)$ and $N$ copies of its complex conjugate $V(t)^*$. 
We will call these ``forward'' and ``backward'' layers respectively.
Formally, the multi-layer circuit with a given $N$ may be written 
\be\label{eq:multicopyV}
V^{(N)}  \equiv 
\underbrace{V \otimes  \ldots \otimes V}_{N} \otimes \underbrace{V^* \otimes \ldots \otimes V^*}_N.
\ee
The physical quantity of interest will dictate the boundary conditions at the top and bottom: 
for example contractions of indices between layers, or contraction of the bond indices at the bottom of a layer with an initial wavefunction.
We review this in a simple setting in Sec.~\ref{sec:multilayer}.
An important feature is the replica trick: 
 $N$ must be left free at intermediate stages of any calculation and then sent to a limiting value at the end \cite{vasseur2019entanglement,zhou2018emergent,jian2020measurement,bao2020theory}.
 (A special case where replicas can be omitted is mentioned in Sec.~\ref{sec:replicaexample} below.)
Replicas allow us to handle denominators that arise because of the normalization of states \cite{vasseur2019entanglement}
(non-unitarity means these normalization factors are nontrivial, see Eq.~\ref{eq:overviewevolution}) 
and/or to deal with logarithms in the definition of the entanglement entropies. See Refs.~\cite{vasseur2019entanglement,jian2020measurement,bao2020theory} for detailed discussions of this for the RTN and MPT.

The global symmetry of the effective models arises ultimately from a simple invariance of $V^{(N)}$ under various operations.
$V^{(N)}$ is clearly invariant under (i)~permutations of the forward layers among themselves; (ii)~permutations of the backward layers among themselves; and (iii)~complex conjugation accompanied by exchange of all the forward layers with all the backward layers \cite{zhou2019entanglement}. Together these make up the symmetry group:
\be
G_N \equiv \lf S_N \times S_N \ri \rtimes \mathbb{Z}_2.
\ee
Here the $\mathbb{Z}_2$ is generated by (iii) above.
$G_N$ is a symmetry of the bulk structure of the tensor network; it will in general be broken by boundary conditions, e.g. by a choice of index contractions at the boundary  of $V^{(N)}$.

\begin{figure}[t]
    \includegraphics[width=0.9\linewidth]{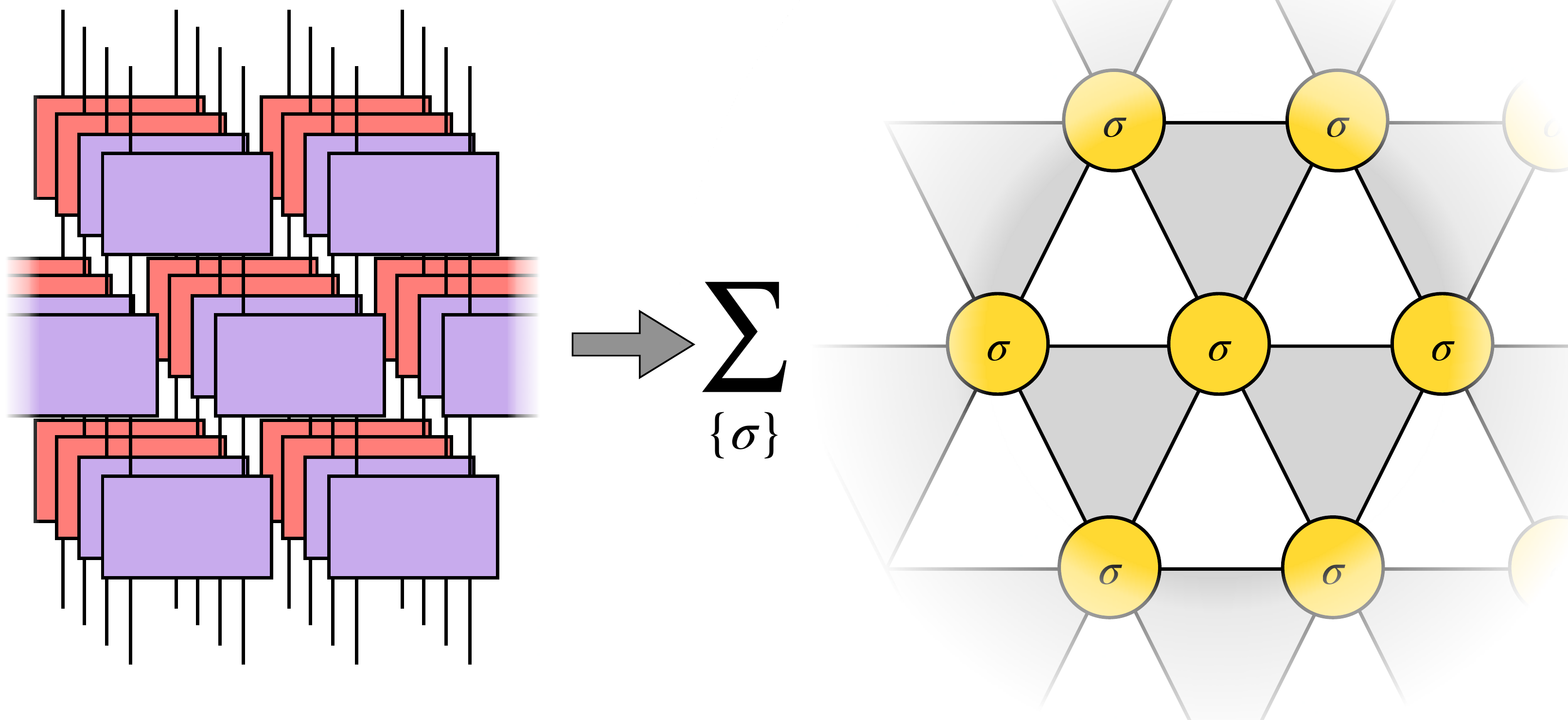}
    \caption{Schematic: mapping a circuit with unitaries and possibly measurements to an effective spin model. 
    Each physical unitary gives rise to a separate spin degree of freedom ${\sigma\in S_N}$ (yellow circles). These spins have interactions on downward pointing triangles (shaded). 
    We refer to $\sigma$ and its continuum versions as the ``pairing field''.}
    \label{fig:triangles}
\end{figure}

{
A formal way to see the importance of this symmetry is via explicit mappings of random circuits or random tensor networks onto effective lattice spin models.  We review this next. We will give an alternative picture below in Sec.~\ref{sec:isingmapping}, by introducing an Edwards-Anderson-like 
field in a multilayer tensor network (this alternative picture may be more intuitive for those familiar with random magnets).
}

In simple  models, averaging over the random  tensors or unitaries leads to effective lattice magnets in which the ``spins'' $\sigma$ (not to be confused with the physical spins that the circuit acts on)
are valued in the permutation group \cite{hayden2016holographic, nahum2018operator,zhou2018emergent,vasseur2019entanglement,jian2020measurement,bao2020theory,hunter2019unitary,liu2020entanglement}:
\be
\sigma\in S_N.
\ee
We will not need details of the lattice construction, but
Fig.~\ref{fig:triangles} shows an example for a 1+1D circuit geometry.
For each unitary in the original circuit, we obtain a spin degree of freedom $\sigma$ in the effective statistical mechanical model. 
We may write the partition function for these spins schematically as 
\be \label{eq:latticeZ}
\mathcal{Z}_{N} = \sum_{\{\sigma\}} W(\{ \sigma \}).
\ee
The boundary conditions on the $\sigma$ depend on the observable (Sec.~\ref{sec:replicaexample}).
The Boltzmann weight $W(\{ \sigma \})$ is a product of local weights on each of the shaded triangles in 
Fig.~\ref{fig:triangles}:
the form of the weight ${J(\sigma_a, \sigma_b, \sigma_c)}$  for the three spins $\sigma_a$, $\sigma_b$, $\sigma_c$ on a given triangle interact via  interactions whose form may be found in Refs.~\cite{jian2020measurement,bao2020theory} for a circuit  with measurements and in Ref.~\cite{nahum2018operator,zhou2018emergent} for the purely unitary case.
Constructions for the random tensor network  with random Gaussian tensors were discussed earlier in Ref.~\cite{hayden2016holographic} and extended to take into account the replica trick in Ref.~\cite{vasseur2019entanglement}. 
In all these cases the interaction terms are, loosely speaking, ferromagnetic, in that the Boltzmann weight is maximized when the $\sigma$ configuration is uniform.

Physically, the spin $\sigma$ should be thought of as a way to label a choice of  \textit{pairing} of the forward layers with the backward layers. 
Let the permutation ${\sigma\in S_N}$ map a given element ${i\in \{1,\ldots, N\}}$ to ${\sigma(i)}$.
Then $\sigma$ stands for the pairing in which forward layer $i$ is paired with backward layer $\overline{\sigma(i)}$ and so on. 
For example the identity permutation, $\sigma=\mathbb{I}$, denotes the pairing of $1$ with $\bar 1$, of $2$ with $\bar 2$, and so on, i.e. in the pattern:
\ba
\sigma=\mathbb{I}&: & 
&\includegraphics[width = 0.28\linewidth]{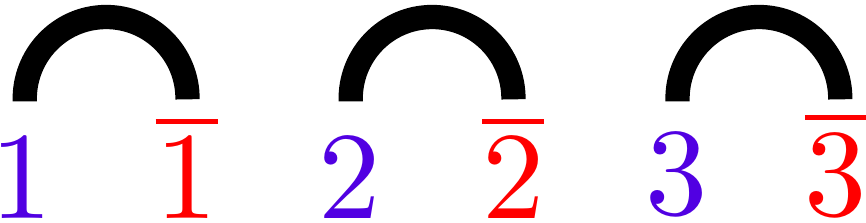}.
\end{align}
We have taken $N=3$ for this example, and we have reordered the layers in comparison with Fig.~\ref{fig:ar-haar} so the pairing can be drawn without crossings.
For the transposition, $\sigma=(12)$, layer $1$ is paired with $\bar 2$ and layer $2$ with $\bar 1$:
\ba
\sigma=(12)&: & 
&\includegraphics[width = 0.28\linewidth]{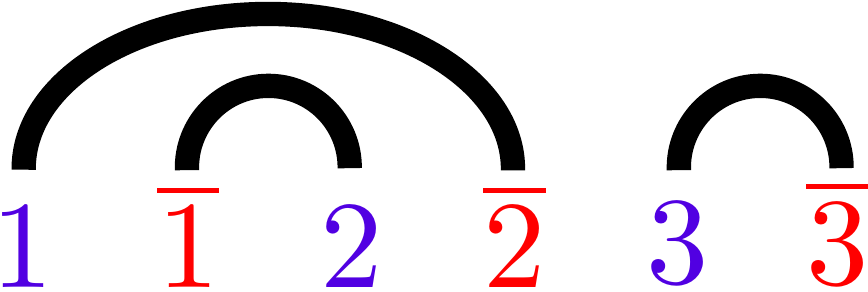}.
\end{align}

Since $\sigma$ specifies a pairing of layers, we will sometimes refer to it (and the continuum versions in the subsequent sections) as the ``pairing field''. 
The physical interpretation of these pairings of layers is discussed in Sec.~\ref{sec:isingmapping} below.
Heuristically, pairing Feynman histories in the discrete time evolution
 allows phase cancellation to be avoided \cite{zhou2018emergent,zhou2019entanglement}, in the spirit of the diagonal approximation in periodic orbit theory \cite{garratt2020many}.
 
 {
That is, we may think of the multi-layer circuit as a discrete path integral for $N$ forward and $N$ backward copies of the system. A Feynman trajectory is specified by a sequence of spin states in each of the copies.  
In a given layer, the corresponding product of matrix elements of local gates is the discrete analogue of the exponentiated action for a continuum  Feynman trajectory: $e^{iS}$ or $e^{-iS}$ depending on whether it is a forward or a backward layer.
After averaging 
(or, in some cases, even without averaging \cite{zhou2019entanglement}) this multi-layer path integral may be dominated by configurations in which  forward and backward layers form ``pairs'' with similar spin configurations, contributing opposite phases to the total weight.
Such a pairing allows  the effect of phase cancellation to be reduced.
(See also Sec.~\ref{sec:isingmapping}.)
The pattern of pairing will in general differ at different locations in spacetime, corresponding to spacetime dependence of the pairing field $\sigma$.
If the boundary conditions  ---  say at the final time ---  involve pairwise index contractions of layers, as arise in the expressions for R\'enyi entropies (Sec.~\ref{sec:replicaexample}),  this will act as a boundary ``magnetic field'' which selects out a particular value for the pairing field $\sigma$ at the boundary.
}

$G_N$ acts on $\sigma$ via both left and right multiplications, for the two $S_N$ factors respectively, and via inversion for the  $\mathbb{Z}_2$ generator, so that we have the symmetry transformations:
\ba\label{eq:GNaction}
\sigma& \rightarrow g_L^{\phantom{1}} \sigma g_R^{-1},
 &
\sigma & \rightarrow \sigma^{-1},
\end{align}
for permutations $g_L$ and $g_R$, together with combinations of the above.

The effective spin interactions in Eq.~\ref{eq:latticeZ}
are local for simple choices of the random tensors or gates, 
but in general depend nontrivially on $N$, and may even be negative.\footnote{Simplifications arise in the fully unitary case. Even there, for general $N$ it is possible to have negative Boltzmann weights $W_\sigma$. Simplifications also arise at large local Hilbert space dimension \cite{zhou2018emergent, vasseur2019entanglement, jian2020measurement, bao2020theory}.} 
However there is a relatively simple picture of the entangling phase as a phase where $\sigma$ is ferromagnetically ordered, so that $G_N$ is spontaneously broken, and of the disentangling phase as a disordered phase.
Entanglement entropies may be expressed as free energy costs for non-uniform boundary conditions \cite{hayden2016holographic,vasseur2019entanglement,zhou2018emergent}  (see Sec.~\ref{sec:replicaexample}).
We will appeal only to these facts and the symmetry structure above.
Note that the simplest nontrivial case is $N=2$: then there are only two possible pairings, $\mathbb{I}$ and $(12)$. Denoting these $+$ and $-$ leads to an effective Ising model \cite{hayden2016holographic,nahum2018operator}. In this case $G_N$ reduces to a simple $\mathbb{Z}_2$ symmetry relating the two states.

Finally, we must specify the replica limits of interest. 
Loosely speaking, the required value of $N$  \cite{vasseur2019entanglement,jian2020measurement}  can be seen by counting powers of $V$. It is
\ba
N&\rightarrow 1 & &\text{for the MPT} \\ 
N&\rightarrow 0 & & \text{for the RTN and FMPT}.
\end{align}
$N\rightarrow 0$ is what we typically have for systems with quenched randomness (Sec.~\ref{sec:isingmapping}).
The additional power of $V$ and $V^*$ for the MPT comes from the Born's rule factor\footnote{Recall that we label ${V=V_{\vec m}}$ by the sequence ${\vec m}$ of measurement outcomes obtained in a given realization of the dynamics.}
\be
P(\vec m) = \bra{\psi}V_{\vec m}^\dag V_{\vec m}^{\phantom{\dag}} \ket{\psi}
\ee
which must be included in every average for the MPT.  We review this more carefully in Sec.~\ref{sec:replicaexample}.

\subsection{Boundary conditions in replica formalism}
\label{sec:replicaexample}

\begin{figure}
    \includegraphics[width=0.9\linewidth]{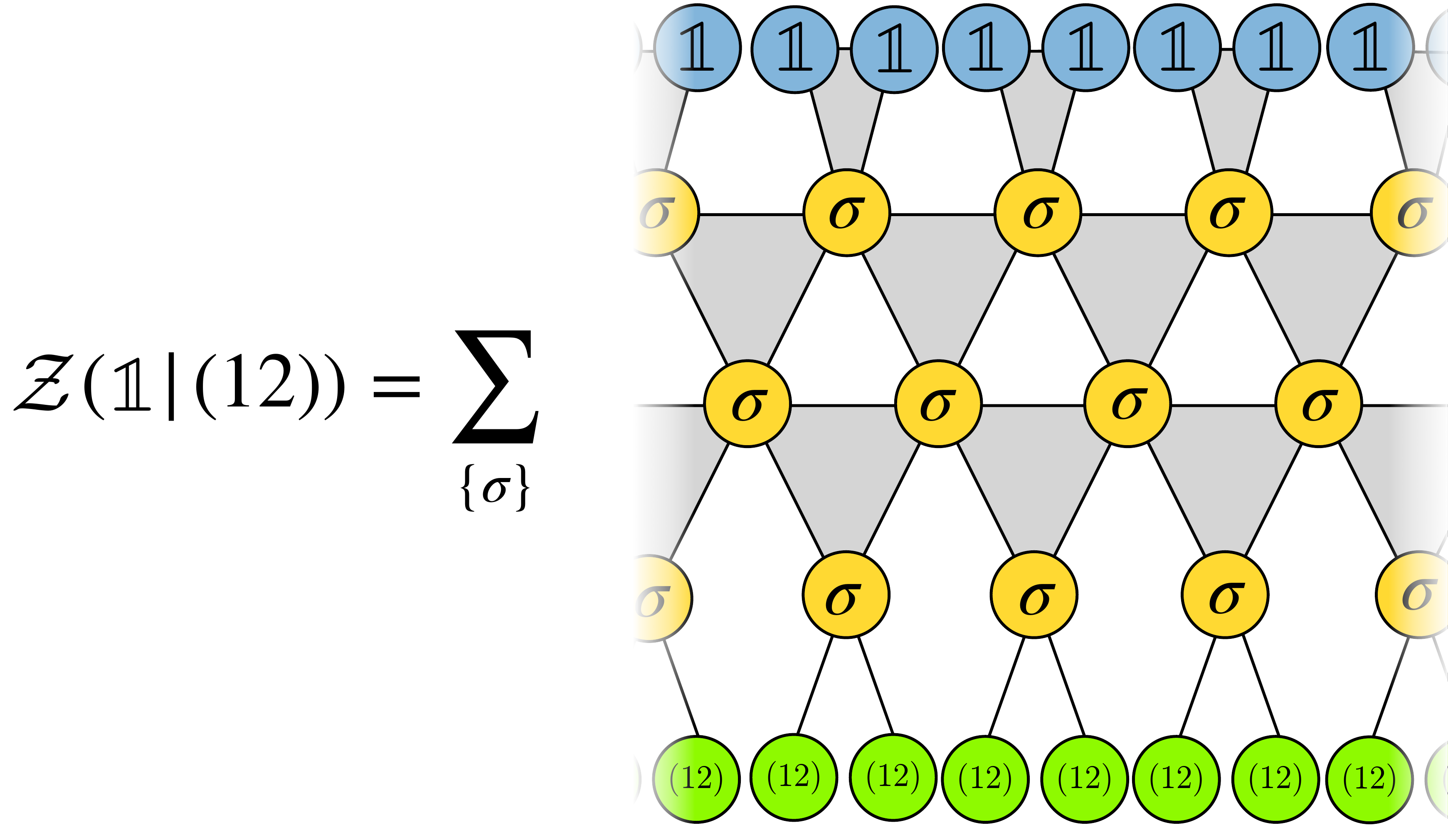}
    \caption{Schematic: boundary conditions in  model for effective spins $\sigma$
    obtained for a 1D unitary circuit with or without measurements. Lines/triangles indicate interactions. The values of the boundary spins are fixed, and determined by the choice of  contractions at the initial and final time in the multi-layer circuit.}
    \label{fig:replicabcs}
\end{figure}

In order to review the replica formalism \cite{vasseur2019entanglement,jian2020measurement,bao2020theory}, let us express the operator entanglement $S_2$ of the nonunitary time evolution operator $V$ in a measurement or forced measurement circuit. The latter case is precisely analogous to a random tensor network, except that for the case of time evolution there is a natural division of the external legs of the tensor network into those associated with the initial time and those associated with the final time.
We focus in this subsection only on reviewing how the boundary conditions in the  effective partition function arise formally (see Sec.~\ref{sec:isingmapping} for more on how the ``pairing field'' arises in the bulk).

We defined the operator entanglement in Sec.~\ref{sec:overview}.  Recall that, if we view $V$ formally as a tensor network \textit{wavefunction} for $2N$ spins, then  $\rho_t$ is the unnormalized reduced density matrix associated with the final-time legs. 
Let us start with the case of the FMPT, where expectation values (denoted by $\mathbb{E}[\ldots]$ or $\overline{[\ldots]}$), are simple averages over the unitaries and projections in $V$. The expectation value of the second R\'enyi entropy is 
\ba\label{eq:S2noreplica}
\overline S_2 & = - \mathbb{E} \ln \f{\tr \rho_t^2}{(\tr \rho_t)^2}  = - \mathbb{E} \ln \f{
\includegraphics[height = 0.1\linewidth]{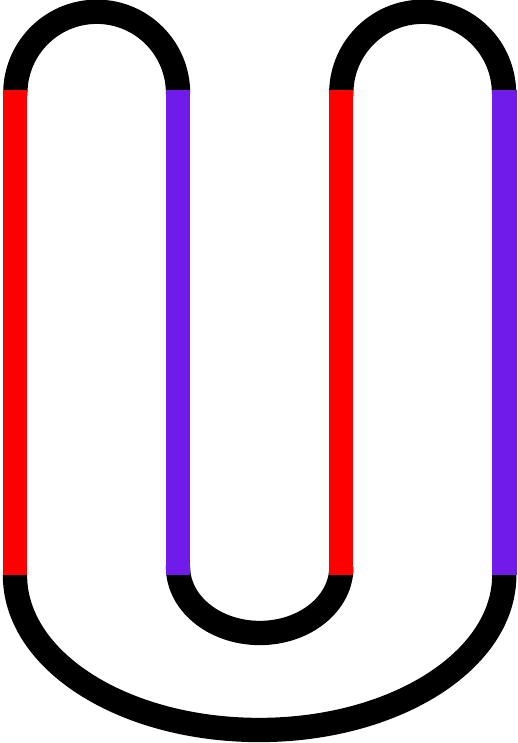}
}{
\includegraphics[height = 0.1\linewidth]{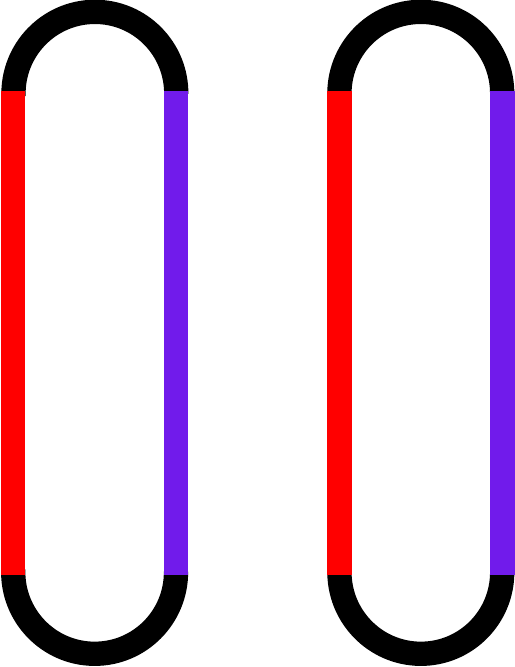}
}
\end{align}
On the right we have indicated the pattern of index contraction graphically. The vertical lines represent a stack of copies of $V$ and $V^*$, like that in Fig.~\ref{fig:ar-haar}, but viewed from the side. For convenience, we have ordered the four layers in the stack as follows: ${\red V^*}$, ${\color{violet} V}$, ${\red V^*}$, ${\color{violet} V}$ (instead of grouping all of the $V^*$s together as we did in Fig.~\ref{fig:ar-haar}). The arcs at the top and bottom indicate the pattern of index contractions between layers. Index contractions are done separately for each of the physical sites. 

Next let us define ``partition functions'' that are averages of the multi-layer circuit with particular choices of boundary conditions. 
We use the notation $\mathcal{Z}_N(\sigma|\tau)$ for the average of the circuit with $N$ layers of ${\color{violet} V}$ and $N$ layers of ${\red V^*}$, and with index contractions in the pairing pattern $\sigma$ at the top and $\tau$ at the bottom. For example, 
\ba
\mathcal{Z}_2 (\mathbb{I}|\mathbb{I}) & =  \mathbb{E} \,\,
\begin{gathered}
 \includegraphics[height = 0.1\linewidth]{contractionid2}
 \end{gathered}
&
\mathcal{Z}_2 (\mathbb{I}|(12)) & =  \mathbb{E} \,\, 
\begin{gathered}
\includegraphics[height = 0.1\linewidth]{contractiontransposition}
 \end{gathered}.
\end{align}
$\mathcal{Z}_N(\sigma|\tau)$  maps to a partition function for the pairing field with an effective ``magnetic field'' favouring pairing state $\sigma$ at the final time (top) and $\tau$ at the initial time (bottom).

Eq.~\ref{eq:S2noreplica} is not immediately written in terms of such partition functions, because of the logarithm and the fraction, but this can be dealt with using the replica trick \cite{vasseur2019entanglement}. Eq.~\ref{eq:S2noreplica} is  trivially equivalent to 
\ba
\overline S_2 & 
= - \f{1}{m} \, \mathbb{E} \, \ln \,
\f{
\includegraphics[height = 0.1\linewidth]{contractiontransposition}^m
}{
\includegraphics[height = 0.1\linewidth]{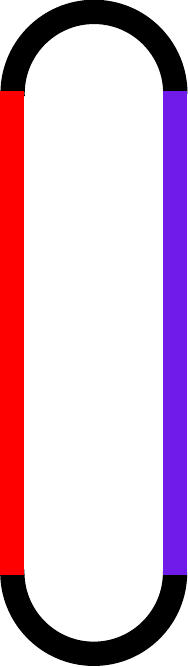}^{2m}
}
\end{align}
for \textit{any} $m>0$, since the factors of $m$ cancel. But one may check (by expanding in $m$ in the numerator and denominator below) that in the limit $m\rightarrow 0$ the expectation value may be taken for the numerator and denominator separately:
\ba
\overline S_2 & 
= - \lim_{m\rightarrow 0} \, \f{1}{m} \, \ln \,
\f{ \mathbb{E}  \, 
\begin{gathered}
\includegraphics[height = 0.1\linewidth]{contractiontransposition}^m
\end{gathered}
}{ \mathbb{E} \,
\begin{gathered}
\includegraphics[height = 0.1\linewidth]{contractionid1}^{2m}
\end{gathered}
}
\end{align}
As usual, we treat $m$ as a positive integer at intermediate stages of the calculation. The above then becomes 
\ba\label{eq:S2replicaratio}
\overline S_2 & 
= - \lim_{m\rightarrow 0} \, \f{1}{m} \, \ln \,
\f{
\mathcal{Z}_{2m}(\mathbb{I}|\,\tau_{2,m})
}{ 
\mathcal{Z}_{2m}(\mathbb{I}|\mathbb{I})
}
\end{align}
Here $\tau_{2,m}$ denotes a permutation in $S_{2m}$ that is a product of $m$ commuting 2-cycles \cite{zhou2018emergent}
\be\label{eq:tau2mdef}
\tau_{2,m} = (12)(34)\ldots (2m-1,2m).
\ee
Eq.~\ref{eq:S2replicaratio} may now be interpreted as the free energy cost of imposing distinct boundary conditions for the pairing field $\sigma$ (represented by the continuum field $X$ in the sections below) at the initial and final times. If the free energy cost for given boundary conditions $\sigma$ and $\tau$ is\footnote{The dependence is only on the cycle structure of $\sigma^{-1}\tau$ thanks to $G_N$ symmetry.} $\mathcal{F}_N(\sigma^{-1}\tau)$ then
\be
\overline{S_2} = \lim_{m\rightarrow 0} \, \f{1}{m} \, \mathcal{F}_{2m} ( \tau_{2,m}). 
\ee
This generalizes directly to higher R\'enyi entropies. (The von Neumann entropy can either be obtained using an additional limit $n\rightarrow 1$, or by a slightly different construction with a single replica limit \cite{vasseur2019entanglement}.) Note that the total number $N$ of replicas (denoted $2m$ above) tends to zero as stated above for the FMPT and the RTN.

The simplest situation, discussed in the next subsection, is where the pairing field is well-ordered across the entire sample. Then the free energy cost $\mathcal{F}$ is essentially  the free energy cost of inserting a single domain wall in this order \cite{hayden2016holographic, vasseur2019entanglement, zhou2018emergent}. 
See for example Fig.~\ref{fig:quasi1ddomainwall} in Sec.~\ref{sec:replicatimescale}.

In fact, in this situation  (the strongly-ordered regime ) results from the unitary case suggest that in 
the replica limit can be dispensed with:
we can map the entanglement to the free energy cost of a single domain wall \textit{in an effective classical disordered system} \cite{zhou2018emergent}.
The most direct way to understand this is to avoid the replica trick entirely \cite{zhou2019entanglement}. 
It is possible to make a formal mapping of the multilayer circuit in Eq.~\ref{eq:S2noreplica} (with $N=2$) to  an ``Ising model'' without any averaging. In general this model has complicated long range interactions, so that it is is not useful for discussing the critical point. But in the strongly ordered regime we expect (assuming the considerations for the unitary case in  \cite{zhou2019entanglement} carry over)  that the interactions are effectively local after sufficient coarse-graining. $S_2$ in a given realization can then be understood as a domain wall cost in a \textit{disordered} Ising model. This is a route for justifying the the use of an Ising model to discuss for example  subleading corrections to the volume law in the entangled phase \cite{fan2020self, li2020statistical}.
When the critical point is approached we must however return to the replica description above. 

The application to the MPT is similar to the case of the FMPT. Graphically, the Born probability ${P(\vec m) = \bra{\psi}V_{\vec m}^\dag V_{\vec m}^{\phantom{\dag}} \ket{\psi}}$ for a sequence of measurement outcomes may be denoted by 
\be
P(\vec m)  = \begin{gathered} \includegraphics[height = 0.1\linewidth]{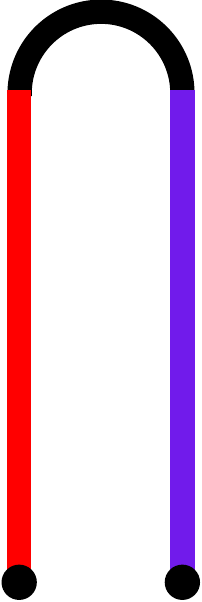} \end{gathered}\, ,
\ee
where the dots represent contraction with $\psi$ or $\psi^*$ as appropriate. Let us absorb a trivial constant into ``$\mathbb{E}$'' so that it denotes the average  over the structure of the circuit together with the unweighted sum over $\vec m$:
\be
\overline{S_2} = - \mathbb{E}  \,
\begin{gathered} \includegraphics[height = 0.1\linewidth]{contractionborn} \end{gathered} \,
\ln \f{
\includegraphics[height = 0.1\linewidth]{contractiontransposition}
}{
\includegraphics[height = 0.1\linewidth]{contractionid2}
}.
\ee
We may  simplify the formulas  slightly by averaging over the initial state, which yields
\be
\overline{S_2} = - \mathbb{E}  \, 
\begin{gathered} \includegraphics[height = 0.1\linewidth]{contractionid1} \end{gathered} \,
\ln \f{
\includegraphics[height = 0.1\linewidth]{contractiontransposition}
}{
\includegraphics[height = 0.1\linewidth]{contractionid2}
}.
\ee
The replica trick then allows us to write
\be\label{eq:S2replicabcsMPT}
\overline{S_2} = -\lim_{m\rightarrow 0} \, \f{1}{m} \, \ln \, \f{\mathcal{Z}_{2m+1} (\mathbb{I} | \tau_{2,m} )}{\mathcal{Z}_{2m+1} (\mathbb{I} | \mathbb{I} )}.
\ee
Formally this is similar to (\ref{eq:S2replicaratio}), but  the total number of replicas ${N=2m+1}$ is taken to 1 rather than 0~\cite{jian2020measurement, bao2020theory}.

\subsection{Permutations and coarse-graining}
\label{sec:permutationscoarsegraining}

We will now focus on the critical properties. 
Let us first make a brief detour to consider coarse-graining a lattice model of permutations, such as that shown schematically in Eq.~\ref{eq:latticeZ}, in an abstract sense, in order to understand one of the basic challenges. (This section is not an essential prerequisite for the following developments --- the reader who wants to get to the concrete results may wish to skip it.)

We work throughout with a system in some finite number of dimensions ${D=d+1}$ (the spacetime dimension in the case of a circuit). Naively we might expect the limit of large $d$ to match the all-to-all circuit (as  in Sec.~\ref{sec:largefinited}) but this is unclear (Sec.~\ref{sec:FTconsequences}).

Let us first imagine attempting a block-spin RG procedure in a naive way, by simply ``averaging'' the spins $i$ within each $D$-dimensional local block:
\be\label{eq:mublockspin}
\mu_\text{block} \propto \sum_{i\in \text{block}} \sigma_i.
\ee
What does this expression mean? 
At this point, each $\sigma_i$ on the RHS is a formal group element in $S_N$. 
Their linear combination, $\mu_\text{block}$, is no longer in $S_N$, since addition is not a group operation (only multiplication). Instead it is an element of the group algebra of $S_N$ \cite{stone2009mathematics}.
A general element of the group algebra is a linear combination of the elements $g$ of the group with numerical coefficients $M_g$,
\be
\mu_\text{block} = \sum_{g\in S_N} M_g \, g, 
\ee
where in the present case $M_g \in \mathbb{R}$. In other words, we can think of the coefficients $M_g$ as forming a vector $\vec M$ of length $N!$, which is the order of $S_N$.
The coarse-grained spin above is equivalent to this vector.

However, $\mu_\text{block}$, or equivalently the vector $\vec M$, is not a natural coarse-grained field in general.
The reason for this is that $\vec M$ does not form a single representation of the global symmetry $G_N$. 
Instead, the $N!$--dimensional vector space splits into many distinct representations, in fact a number of representations that grows exponentially as $N$ grows. 
Standard results for the group algebra 
imply that the representations of $G_N$ that appear when we decompose $\mu_\text{block}$  are in one-to-one correspondence with the irreducible representations of $S_N$ \cite{stone2009mathematics}.
To extract a particular representation of $G_N$, we simply replace the formal group elements in Eq.~\ref{eq:mublockspin} with their \textit{matrix} representatives in the corresponding representation of $S_N$.

This means that our initial attempt to form a block spin has led us not to a single coarse-grained field, but to an indeterminate number (because $N$ must be left free) of different coarse-grained fields, each in a different representation of the global symmetry group $G_N$.

In principle, we could try to write down a Lagrangian including all of these fields. However, since the number of these fields, and therefore the number of couplings, depends on $N$, this does not seem promising. 
Instead, it is natural to hope that only one or a small number of the fields become massless at the critical point, and the other fields do not need to be included in a continuum Lagrangian.
This is the assumption we will make, motivated by the more explicit picture in the following section.

This picture of splitting $\mu_\text{block}$ into separate fields gives an alternative view on the discussion of the percolation fixed point in Ref.~\cite{vasseur2019entanglement}. 
The authors imagined starting with a lattice model with a much enlarged symmetry, $S_{N!}$ (not $S_N$ or ${S_N\times S_N}$ or $G_N$). 
This much larger symmetry group is allowed to arbitrarily permute all the $N!$ values $\sigma\in S_N$ that the spin can take.
Such a lattice model is simply a Potts model with $Q=N!$ states, for which the continuum theory is well known (becoming percolation when ${Q\rightarrow 1}$). 
The authors then considered deforming model in the direction of the physical model of interest (cf.~\ref{eq:latticeZ}) which does \textit{not} have $S_{N!}$ symmetry. They found that the lowest order perturbation that could be added was quadratic in the Potts field, and so relevant. However there was considerable freedom in the index structure of this perturbation, which could be formed from any class function of $S_N$. 

From the present point of view, this perturbation is a sum of mass terms, with one independent mass for each of the infinite number of fields  that appear when we decompose $\mu_\text{block}$ above into representations of $S_N$ for arbitrary~$N$.

\subsection{Motivating a simple Landau theory}
\label{sec:isingmapping}

A familiar way to represent a permutation in $S_N$ is as an $N\times N$ matrix $X_{a,b}$ of ones and zeros, with a single 1 in each row and in each column,
\ba
\sum_{a}X_{ab}& = 1,
&
\sum_{b}X_{ab}& = 1.
\end{align}
Under the global symmetries in Eq.~\ref{eq:GNaction}, this matrix transforms as:
\ba\label{eq:Xsymm}
X & \rightarrow L X R^{-1}, &
X & \rightarrow X^T,
\end{align}
where $L$ and $R$ are permutation matrices representing $g_L$ and $g_R$. 

We might hope that we can  build a Landau theory from such a matrix. In terms of the discussion in the previous section, this will correspond to the simplest choice of  representations of $G_N$ to include in the continuum theory (discussed below).
In fact we can motivate such a Landau theory in a more direct way, without the need to go through the mappings discussed above involving permutations. 

\begin{figure}
    \includegraphics[width=0.9\linewidth]{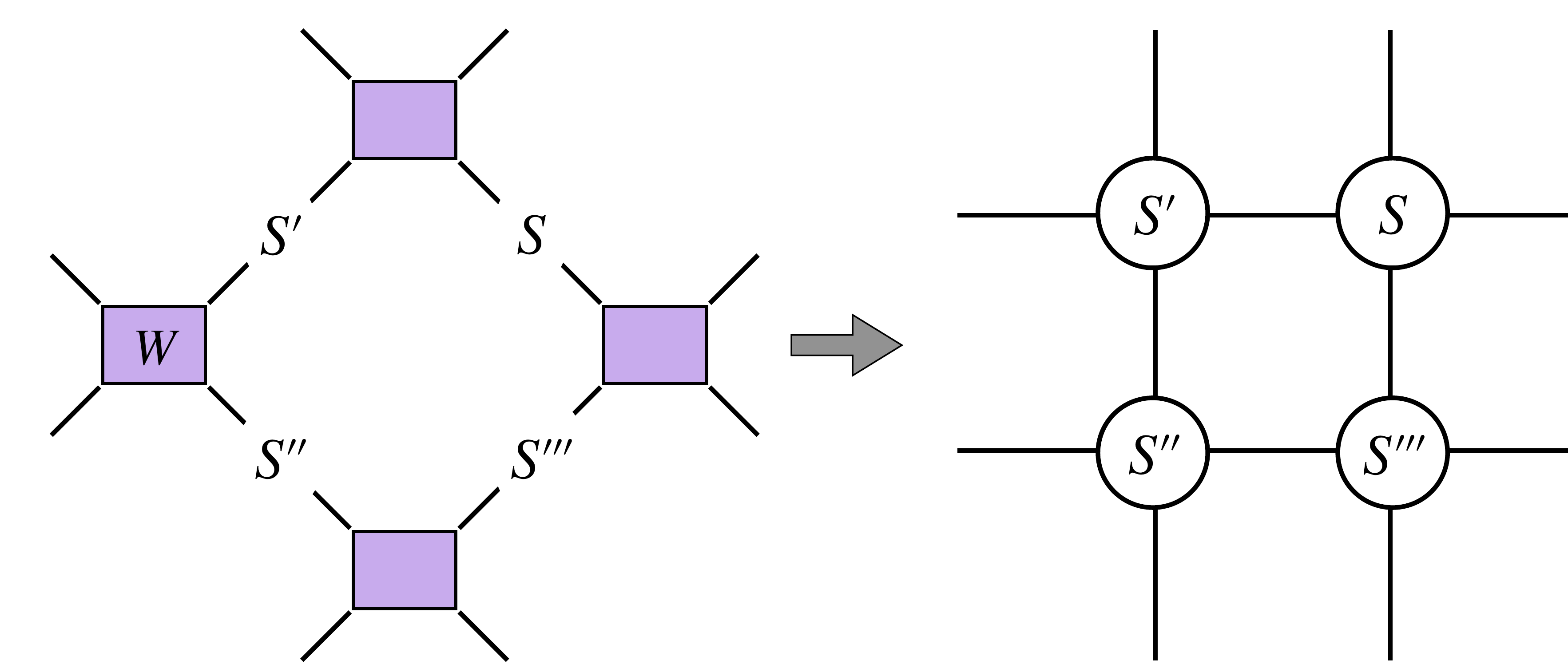}
    \caption{Left: The simple random tensor network described around Eq.~\eqref{eq:isinggatew}, with bond dimension 2, which reduces to a square-lattice Ising model with complex interaction constants (Right). The Ising spin values on the right represent values of bond indices on the left.}
    \label{fig:isingtensornetwork}
\end{figure}

For this we appeal to the basic physical picture for why the pairings of layers arise in the multi-layer circuit, which is to avoid phase cancellation. 
To make this explicit, let's consider a particularly simple example of a tensor network $V$ (which we can interpret formally as a nonunitary time evolution for qubits) with the geometry in Fig.~\ref{fig:isingtensornetwork}, Left.
Label the bond index values by  $S=\pm 1$
(these are the spins' $\sigma_z$ values if $V$ is interpreted as a time evolution). 
Take the local gates $w$, with bond indices   $S_1$, $S_2$, $S_3$, $S_4$, to have the simple form
\be\label{eq:isinggatew}
w_{S_1, S_2, S_3, S_4} = \exp\lf{\sum_{1\leq i \leq 4} h_i S_i + \sum_{1\leq i < j \leq 4} J_{ij} S_i S_j}\ri,
\ee
where each $h$ is an independent, identically distributed \textit{complex} Gaussian variable with mean zero, and equal variance $\Delta_h^2/4$ for its real and imaginary parts, and similarly for the $J$s, with variance $\Delta_J^2/2$. (These couplings are taken complex since tensors in a generic tensor network are complex.) 

The tensor contraction defining $V$ involves a sum over all the indices carried by the internal bonds in Fig.~\ref{fig:isingtensornetwork}, i.e. over all Feynman trajectories, if we think of the vertical direction as time. This tensor contraction is an Ising partition function for the indices $S_i$ on the bonds $i$. With the choices above, this Ising model lives on a rotated square lattice. We may write its partition function as
\be
Z \equiv \sum_{\{S_i \}} \exp \lf \, i \mathcal{S}[\{S\}] \, \ri,
\ee
where the exponentiated lattice ``action'' $e^{i\mathcal{S}}$ is just a product of terms of the form (\ref{eq:isinggatew}), so that $i \mathcal{S}[\{S\}]$ is an Ising Hamiltonian with random complex magnetic fields and random complex nearest-neighbour couplings. 
This is schematic as we have left the boundary conditions unspecified. 
(Fixed boundary conditions on the spins give a matrix element of $V$, for example; in practise we are interested in taking several layers of $Z$ which are coupled at their boundaries.)

Quantities of interest involve the replicated partition function
(cf.~Fig.~\ref{fig:ar-haar}). Up to boundary conditions, this is given by averaging $Z^N\times Z^{*N}$ over all of the random $h$ and $J$ parameters, as in the standard application of the replica trick to the Ising model with random bonds or random fields \cite{cardy1996scaling}.
Introducing $N$ replicas of the Ising spin for the forward layers, denoted $S^a$ for $a=1,\ldots, N$, and $N$ replicas for the backward layers denoted $\overline{S}^b$, the
replicated partition function $\mathcal{Z}_N$ has the form
\be
\mathcal{Z}_N
= 
\sum_{
\{S_i^a\}, \{ \overline{S}_i^b\} 
}
\exp\lf - \mathcal{S}_\text{eff} \ri
\ee
(we do not include an $i$ in the definition) with 
\be\label{eq:SeffX}
\mathcal{S}_\text{eff} = 
- \Delta_J^2 \sum_{\<ij\>} \sum_{ab} X_{ab}(i) X_{ab}(j)
- \Delta_h^2 \sum_{i} \sum_{ab} X_{ab}(i),
\ee
where we have defined the ``pairing field''
\be
\label{eq:isingreplicaXdef}
X_{ab}(i) = S^a_i \overline{S}^b_i.
\ee
This is similar to an Edwards-Anderson order parameter in an Ising spin glass. However the usual Edwards-Anderson order parameter would be of the form $S^a S^b$ (as there would be no distinction between forward and backward layers) and the replica permutation symmetry would act on both $a$ and $b$ together. In the present case we have separate permutation symmetries for the the row and column indices of $X$. 

We defer an explicit discussion of coarse-graining for this and other microscopic models to a separate publication \cite{replicasforthcoming}.
Here we note only that the form of Eq.~\ref{eq:SeffX}, with ferromagnetic interactions between the pairing field $X$ for different sites, motivates writing a continuum Lagrangian for an $N\times N$ matrix, as discussed above. 

In the present microscopic formulation, 
$X$ is not a permutation matrix, but the action of symmetry is the same (Eq.~\ref{eq:Xsymm}). This is what we will use, together with the assumption that the pattern of symmetry breaking in the entangled phase is the simplest one corresponding to pairing, i.e. to a choice of permutation.

Without loss of generality, let this permutation be the identity permutation (other cases are related by symmetry). Then the pattern of symmetry breaking is { captured by an expectation value of the form}
\be
X_{ab} = f \delta_{ab} + c,
\ee
where $f$ is the order parameter. { Here $c$ is a constant which is generically nonzero even in the disordered phase, since (unlike $f$) it  does not break any symmetry.} This order breaks ${G_N}$ down to ${S_N\times \mathbb{Z}_2}$, where the remaining permutation group is the subgroup of diagonal ${S_N\times S_N}$ transformations with ${g_L=g_R}$. (We will briefly discuss more complex possiblities for symmetry breaking in Sec.~\ref{sec:commentscaveatsetc}.)

The physical interpretation of $X$ is simple: if in some region the spin configuration in the forward layer $a$ is close to that  in backward layer $b$, then the coarse-grained $X_{ab}$ in this region will be large. Heuristically, we expect repulsive interactions between $X_{ab}$ and $X_{ac}$ for $b\neq c$: if the configuration in $a$ is close to that in $b$, the phases from the $a$ layer are already (partially) cancelled, so there is less gained by also pairing with $c$.

Let us briefly mention a caveat to the above discussion. A ``random tensor network'' is by definition a statistical mechanics problem with very little required structure. Similarly the complex Ising model discussed above (which is an example of a random tensor network) is close to being the most general Ising model that one could write down for this lattice geometry.\footnote{A given realization has no symmetry. Because of the distribution we chose for the disorder there are statistical symmetries, i.e. symmetries of the disorder distribution. For example the distribution is invariant under $S\rightarrow -S$ on a given site. However this symmetry does not act nontrivially on the effective field $X$, so we expect it could be broken without changing the universality class of the  transition under discussion.}
On the other hand, the true measurement dynamics in the MPT does have some structure (for example, structure associated with causality) which is not present in a generic tensor network.
In writing down the field theory in the next section we are assuming that the only aspect of the structure of the MPT that is important for the critical theory is the shift in the number of replicas from $N=0$ to $N=1$ that is induced by the Born probability. This assumption should certainly be examined further.

We note that the unitary limit, $r=0$, is a case where additional structure due to unitarity certainly \textit{is} important. There the  appropriate effective ``spin model''  has hard constraints on the allowed spin configurations, which for example enforce causality \cite{nahum2018operator, zhou2018emergent, hunter2019unitary} (these are relaxed when projection operators are included \cite{bao2020theory,jian2020measurement}).
As a result, the unitary models do not possess invariance under $\mathrm{O}(d+1)$ rotations in spacetime, even in the scaling limit, and are not described by the field theories below, which do possess this symmetry. However the unitary models do share some features with the ordered phases of these theories, such as a positive domain wall tension.

\subsection{A field theory for the measurement transition}
\label{sec:MPTFT}

With this motivation, let us write the simplest Lagrangian for $X_{ab}$, which can represent a coarse-graining either of a permutation matrix or of the composite field above.
We will see that this simplest Lagrangian passes a basic consistency check for the MPT. (In the next section we will see that we need to extend it for the  RTN and the FMPT.) 

Let us make subtractions so that the row and column sums of the matrix give zero:
\ba\label{eq:rowcolsums0}
\sum_a \hat X_{ab} &= 0, 
&
\sum_b \hat X_{ab} &= 0.
\end{align}
In the case where $X$ is microscopically a permutation, this simply requires us to subtract a constant:
\ba\label{eq:subtractionpermmat}
\hat X_{ab} & = X_{ab} - \f{1}{N}.
\end{align}
As a result of these linear constraints, which are preserved under coarse-graining, $\hat X$ has $(N-1)^2$ independent components, and forms an irreducible   representation of $G_N$.  Below we will omit the caret on $\hat X$.

Including terms in the potential only up to cubic order in $X$, and imposing $G_N$ symmetry gives a relatively simple Lagrangian. The theory we propose for the MPT is:
\be
\label{eq:MPTlagrangian}
\mathcal{L} = \sum_{ab} \, \left[ \,
\f{1}{2} (\partial  X_{ab})^2 + \f{\mu}{2}  X_{ab}^2 + g  X_{ab}^3
\, \right].
\ee
We have included both time and space derivatives in the first term with the same coefficient, i.e. we have set a nonuniversal speed to 1. This field theory has emergent Euclidean rotational invariance (not Lorentz invariance) in spacetime if this is not broken by boundary conditions.
The components of the matrix $X$ are not independent, because of the constraints in Eq.~\ref{eq:rowcolsums0}.
Note that as a result, in contrast to the theory discussed in the next section, the only linear term $\sum_{ab} X_{ab}$ that would be allowed by symmetry is in fact zero.
The replica limit $N\rightarrow 1$ is also implied.
The renormalized  squared mass  vanishes at the critical point, ${\mu^2 \propto (r-r_c)}$.

Alternately, we may write $X$ in terms of an unconstrained ${(N-1)\times(N-1)}$  matrix field $\phi_{\alpha\beta}$,\footnote{We use the set of ${N}$   vectors $\vec{e}^{1}, \ldots, \vec{e}^{N}$,
each of $N-1$ components,
that are familiar in the context of the Potts field theory \cite{zia1975critical, amit1976renormalization} (see App.~\ref{app:percpotts}) and satisfy ${\sum_{a} e^a_\alpha e^a_\beta = \delta_{\alpha\beta}}$:
${X_{ab} = \sum_{\alpha,\beta = 1}^{N-1} ( e^a_\alpha e^b_\beta ) \, \phi_{\alpha\beta}}$.}
\be
\mathcal{L} =
\f{1}{2} \sum_{\alpha\beta} \lf  (\partial \phi_{\alpha\beta})^2 + \mu  \phi_{\alpha\beta}^2 \ri + 
g \sum
D_{\alpha\beta\gamma}^{\mu \nu\lambda} 
\phi_{\alpha\mu}
\phi_{\beta\nu}
\phi_{\gamma\lambda}.
\ee
The tensor $D$ is a tensor product of that appearing in the cubic term of the Potts model  \cite{zia1975critical, amit1976renormalization}:
\ba
D_{\alpha\beta\gamma}^{\mu \nu\lambda} & = d_{\alpha\beta\gamma} d_{\mu \nu\lambda}, & 
d_{\alpha\beta\gamma} &  = \sum_{a=1}^N e^a_\alpha e^a_{\beta} e^a_\gamma.
\end{align}
The theory with the cubic term can only make sense for the replica limit
--- for $N>2$ we have an unstable potential and for $N=2$ the cubic term vanishes. This is also the case for the Landau-Ginsburg-Wilson-like  theory for percolation, which we have already discussed in Sec.~\ref{sec:percscalingforms}.
{
Like that theory, the upper critical spacetime dimension of (\ref{eq:MPTlagrangian}) is $D=6$.}

A basic consistency check on our picture is that this theory indeed sustains a stable ordered phase, with the simple pattern of symmetry breaking described in previous sections, when ${\mu^2<0}$.
That is, the masses of fluctuations about the ordered state should remain positive in the replica limit ${N\rightarrow 1}$: 
otherwise some more complex pattern of  symmetry breaking might be required \cite{pytte1979scaling,de1978stability,parisi1980sequence}. 
To check this we put ${X_{ab} = f (\delta_{ab} - 1/N)+W_{ab}}$, where $f$ is the magnitude of the order parameter, and $W$ represents fluctuations (with ${\sum_a W_{ab}=0}$, etc.). The saddle-point equation requires ${f=\f{-\mu^2}{3g}\f{N}{N-2}}$.
The mass terms in the Lagrangian for $W$ are then  ${\mathcal{L} = {\f{-\mu^2 N}{2 (2-N)} \big( \sum_{a b} W_{ab}^2 -2  \sum_a W_{aa}^2 \big)}}$. We may check that the eigenvalues of the mass matrix appearing here are indeed positive when ${\mu^2<0}$ and ${N\rightarrow 1}$ (App.~\ref{app:nto1masses}), so this consistency check is satisfied.

Now we consider another important consistency check.

\subsection{Counting fields}

Above we started with an $N\times N$ matrix $X_{ab}$ transforming under $G_N$ symmetry.
For integer ${N>1}$ we may split a general such matrix into four distinct fields, transforming under distinct representations of ${S_N\times S_N}$:
\ba 
\notag
S & \equiv \sum_{ab}X_{ab}, \\ \notag
R_a & \equiv \sum_{b}X_{ab} - \f{S}{N}  \\\notag
C_b & \equiv \sum_{a} X_{ab} - \f{S}{N}  \\
\hat X_{ab} & \equiv X_{ab} - \f{C_a + R_b}{N} 
- \f{S}{N^2}.
\label{eq:fourfields}
\end{align}
The last of these, $\hat X$, is in the fundamental (standard) representation for both $S_N$ factors. It lives in an irreducible representation of $G_N$ of dimension $(N-1)^2$.
$R$ and $C$ each transform under only one of the $S_N$ factors. Since they are exchanged by the $\mathbb{Z}_2$ generator, together they form a single representation of $G_N$ of dimension $2(N-1)$. $S$ is a singlet.

For the MPT we constructed a Landau theory that contained only the field $\hat X$.
This was the obvious thing to do for various reasons (for example, if we think of $\hat X$ microscopically as a permutation matrix, then $R$, $C$ and $S$ are trivial constants). We conjectured that for the MPT $\hat X$ is the only field that becomes massless at the critical point.

However the group theory  at $N\rightarrow 0$ \cite{cardy1999logarithmic} gives additional constraints which strongly suggest that  all of the representations in Eq.~\ref{eq:fourfields} become simultaneously massless at the critical point, so that we cannot throw away the representations $R$, $C$ and $S$. Therefore we have to work with a general matrix $X$ in which the row and column sums are not fixed to zero. The first indication of this is that the subtractions  in Eq.~\ref{eq:subtractionpermmat} and Eq.~\ref{eq:fourfields} diverge when $N\rightarrow 0$.

As with many other replica field theories, the partition functions that we are interested in become trivial --- exactly equal to 1 --- in the replica limit, for certain choices of boundary conditions.
An unusual feature of the circuit models with measurements or forced measurements is that this occurs at two values of $N$. 
When $N\rightarrow 0$  (FMPT) it occurs for the usual reason --- because the partition function is the average of something raised to the power zero. 
When $N\rightarrow 1$ (MPT) the partition function is the sum of the probabilities of all the measurement outcomes --- again giving 1 but for a different reason.

The fact that the microscopic partition function is equal to 1 implies constraints on the spectrum of operators in the continuum theory  \cite{cardy1999logarithmic,cardy2013logarithmic, vasseur2012logarithmic,vasseur2014operator}.
Here a minimal heuristic point will be sufficient: there should not be any massless fields  left when $N$ is set equal to $N_*$, the desired number of replicas, otherwise we will have a nontrivial free energy, contradicting $\mathcal{Z}=1$.

The Lagrangian (\ref{eq:MPTlagrangian}) for the MPT satisfies this condition, since the field is in a representation of dimension $(N-1)^2$, which tends to zero when $N\rightarrow 1$. Therefore it passes this basic consistency check.

At first we might have assumed that the same field theory could also be continued to $N=0$ in order to describe the RTN and FMPT. 
However this is not the case.
Since $(N-1)^2$ is equal to one in this limit, rather than zero,  this is not consistent.

However, the total multiplicity of all the representations in   Eq.~\ref{eq:fourfields} is just ${N^2}$ (the number of components of the matrix), which \textit{does} tend to zero in the replica limit $N\rightarrow 0$. 
This suggests that we should write a Lagrangian for a matrix  $X$ without imposing any condition on its row or column sums.\footnote{At first sight the interpretation of these additional fields may seem obscure, given that for a permutation matrix they are trivial constants. This may be more transparent in the approach of Sec.~\ref{sec:isingmapping}.} This is what we do next.

\subsection{Field theory for random tensor network/FMPT}
\label{sec:FTRTNFMPT}

Let us denote the unconstrained real $N\times N$ matrix by $Y$, to distinguish it from the  matrix $X$ above which obeyed linear constraints.
Assuming only $G_N$ symmetry and no constraints on $Y$, we argue below that the most relevant terms as ${N\rightarrow 0}$~are contained in
\be\label{eq:RTNactionYlanguage}
\mathcal{L} =
 \sum_{ab} \left[
 \f{1}{2}(\partial Y_{ab})^2 
+ r  Y_{ab} 
+ g Y_{ab}^3
\right]
+ \f{m_F^2}{2} \sum_{abcd} Y_{ab} F_{ab,cd} Y_{cd}
\ee
where $F$ is the tensor
\be
F_{ab,cd} = \delta_{bd} + \delta_{ac}.
\ee
The parameter that drives this theory off criticality is $r$, the coefficient of the linear term (not to be confused with the measurement rate in previous sections, also denoted~$r$).  Since no constraint is imposed on $Y$, this linear term does not vanish 
(contrast Sec.~\ref{sec:MPTFT}).
The term $\sum_{ab}Y_{ab}^2$ is absent because its coupling can be set to zero by a shift $Y_{ab}\rightarrow Y_{ab} + C$ with a constant $C$, i.e. it is redundant \cite{fisher1978yang}.
Surprisingly, we will find below that for this theory the  upper critical dimensionality of spacetime is $D=10$.

A peculiar feature of the ${N\rightarrow 0}$ limit of Eq.~\ref{eq:RTNactionYlanguage}, which is shared with some other replica field theories such as the Landau-Ginsburg formulation of the random field Ising model \cite{cardy1996scaling}, is the presence of a quadratic coupling 
which is \textit{not} zero at the critical point and which cannot be removed. This is the term  $m_F^2 Y.F.Y$.

If we instead study the above theory for a larger value of $N$, for example in the ${N\rightarrow 1}$ limit, then the effect of $m_F^2$ is simply to give a mass to certain representations in the decomposition of $Y$.
The corresponding fields can therefore be eliminated at large scales/low momenta.
Doing so returns us to the critical theory proposed in Sec.~\ref{sec:MPTFT} for the MPT, with ${\mu^2 \sim -r}$.  
This is shown explicitly in App.~\ref{app:mFreduction}.
However, writing the propagator explicitly shows that the  limit ${N\rightarrow 0}$ that is of interest to us in this section does not commute with the limit of small momentum \cite{cardy1996scaling}. Therefore we have to retain the $YFY$ term explicitly. 

Note that this term, which can be written
\be\notag
\sum_{abcd} Y_{ab} F_{ab,cd} Y_{cd} = \sum_a \big( \sum_b Y_{ab}\big)^2 + \sum_b \big( \sum_a Y_{ab}\big)^2,
\ee
includes contributions such as $Y_{12}Y_{13}$: this is consistent with the ``repulsion'' that was discussed heuristically towards the end of Sec.~\ref{sec:isingmapping}, between pairing patterns  involving a given layer.

$G_N$ symmetry allows many other  terms at order $Y^3$ but we argue that in the ${N\rightarrow 0}$ limit they contribute only less relevant couplings.
The dimensional analysis may be simplified using an  approach \cite{cardy1985nonperturbative,cardy1985field,cardy2001exact,kaviraj2020random} introduced by Cardy  for the field theories of the  random field Ising model \cite{parisi1979random,bray1985scaling} and the branched polymer \cite{lubensky1978field,parisi1981critical,brydges2003branched}.  Since decomposition into representations of $S_N$ fails in the ${N\rightarrow 0}$ limit, the next best thing is to exploit a decomposition into representations of an $S_{N-1}$ subgroup acting on indices $2,\ldots, N$. Here we must do this for both the row and column indices of $Y$.

We make a linear transformation to rewrite the field $Y_{ab}$ as a field $y_{\alpha\beta}$ whose indices $\alpha$ and $\beta$ take values in the set $\{ +, - , 2, \ldots, N\}$:
\be
y_{\alpha\beta} = \f{1}{2} \vec{v}^\alpha . Y . \vec{v}^\beta.
\ee
The index values $+$ and $-$ denote two distinct linear combinations that are invariant under $S_{N-1}$, while the values $2,\ldots, N$ are permuted by $S_{N-1}$.
The vectors $\vec{v}^\alpha$ are 
\ba \label{eq:vvectors}
\vec{v}^+ & = \f{1}{2} \lf   1,\f{1}{N-1} ,\ldots, \f{1}{N-1}  \ri 
\\
\vec{v}^- & = \f{1}{2} \lf   1,\f{-1}{N-1} ,\ldots, \f{-1}{N-1}  \ri \notag
\\
\vec{v}^i & =  \lf   0,\ldots,0, 1,0, \ldots, 0 \ri - \f{1}{N-1} \lf   0,1,\ldots,1  \ri ,\notag
\end{align}
where in the last line the extra ``$1$'' is in the $i$th place. The ${N-1}$ vectors $\vec{v}^2, \ldots \vec{v}^N$ add up to zero, so span only an ${N-2}$ dimensional space. Below, indices $i, j, k$ always run over ${2, \ldots, N}$. Technical details are in App.~\ref{app:ntozerolimit}. 

After this rewriting, the terms in Eq.~\ref{eq:RTNactionYlanguage} up to quadratic order in $Y$ become 
\be\label{eq:RTNquadraticL}
\mathcal{L}_\text{quadratic} = 
\mathcal{L}^{(1)} +\sum_{j} \mathcal{L}^{(2)}_j + \sum_{k} \mathcal{L}^{(3)}_k + \sum_{jk} \mathcal{L}^{(4)}_{jk},
\ee
with
\ba \notag
\mathcal{L}^{(1)} 
 = &   
 \lf \partial y_{++} \ri \lf \partial y_{--}\ri+ \lf \partial y_{+-}\ri \lf \partial y_{-+} \ri + 2 r y_{--} 
\\ 
& \qquad\qquad \qquad\,\,\,+ 2 m_F^2 y_{--} \lf y_{+-} + y_{-+} \ri 
\\
\mathcal{L}^{(2)}_j
 = &    \f{1}{2}  \lf   \partial y_{j+} \ri\lf  \partial y_{j-} \ri+ \f{m_F^2}{2} y_{j-}^2  
 \\
\mathcal{L}^{(3)}_k
 = &    \f{1}{2}    \lf  \partial y_{+k}\ri \lf  \partial y_{-k}\ri + \f{m_F^2}{2} y_{-k}^2 
\\
\mathcal{L}^{(3)}  = & \f{1}{4}  \lf  \partial y_{jk} \ri^2.
\end{align}
Because of the linear constraints $\sum_j y_{j+}=0$ etc., sectors $(2)$ and $(3)$ each contain ${N-2}$ copies of the same theory, and sector $(4)$ contains ${(N-2)^2}$ copies of the same theory.\footnote{At the quadratic level, these various sectors can be replaced by a theory with fermions but without a replica limit \cite{parisi1979random,cardy1985nonperturbative} (for example ${N-2}\rightarrow -2$ copies of a bosonic theory can be replaced by a fermionic version of the theory).}
In the above rewriting, terms with couplings that vanish as $N\rightarrow 0$ were dropped \cite{cardy1985nonperturbative}.

Before writing the interaction terms, we use the quadratic terms to assign engineering dimensions to the various fields (see App.~\ref{app:ntozerolimit} for details). 
We assign dimensions $x_{\alpha\beta}$ such that all the quadratic terms in the Lagrangian Eq.~\ref{eq:RTNquadraticL}  are marginal at the critical point $r=0$. This gives
\be
{x_{\alpha\beta}= w_\alpha+w_\beta}
\ee
with (recall that in the case of a circuit ${D=d+1}$ is the spacetime dimension)
\ba
w_+ & = \f{D-6}{4}, &
w_{i} & = \f{D-2}{4}, &
w_- & = \f{D+2}{4}.
\end{align}
The RG eigenvalue of a cubic interaction term $y_{\alpha\beta} y_{\alpha'\beta'} y_{\alpha''\beta''}$ is then 
determined by the difference in the number of  $+$ indices it contains and the number of $-$ indices it contains among ${\alpha, \ldots, \beta''}$ (App.~\ref{app:ntozerolimit}).

However, the terms that can appear are constrained by the $G_N$ symmetry 
(whose effects are less obvious in the new representation). 
We confirm in App.~\ref{app:ntozerolimit} that the cubic term $g\sum_{ab} Y_{ab}^3$ shown in Eq.~\ref{eq:RTNactionYlanguage} is strictly more relevant than the other symmetry-allowed cubic terms (at least for large enough $D$) and is of the form
\ba \notag
g {\sum}\, Y_{ab}^3  =  \f{g}{2} \big[ &6 y_{++}  \lf  
  y_{++}  y_{--} + 2 y_{+-} y_{-+}  \ri \\   \notag
  & + 6 y_{++}  \big(   y_{-k} y_{+k} + y_{j-} y_{j+} + y_{jk} y_{jk} /4   \big)
  \\   \notag
  & +
  3 \lf 
y_{+-} y_{j+} y_{j+}+  y_{-+} y_{+k} y_{+k}
  \ri \\  
  & + 
  3 y_{j+} y_{+k} y_{jk} \big]
   \notag
  \\
  & + \text{less relevant terms}.
  \label{eq:appcubictermexplicit}
  \end{align}
The RG eigenvalue of $g$ is $(10-D)/2$, so the upper critical dimension for this theory is ${D=10}$.

\subsection{Consequences of the MPT field theory}
\label{sec:FTconsequences}

We discuss some simple consequences of the putative field theory for the MPT, deferring  a detailed analysis, and a discussion of the more complicated theory in the previous section, to another time. 
However, first we note an important caveat to the discussion. 

Our initial hope was that  the large-$D$ limits of these field theories 
would give exact results both for the all-to-all circuits and for tree tensor networks.
For example, this is what we found for the classical minimal cut toy model (because  all-to-all percolation could be understood using the field theory for percolation in high dimensions, Sec.~\ref{sec:largefinited}.)
But the class of tree tensor networks that we understand best, including those derived from the all-to-all  FMPT circuit with Haar-random gates, seems \textit{not} to be described by the field theory of Sec.~\ref{sec:FTRTNFMPT}, simply because it is hard to imagine the exponential scaling of the order parameter in Eq.~\ref{eq:sqrtscalingfirst} being reproduced by a  mean-field treatment of Eq.~\ref{eq:RTNactionYlanguage}.
Therefore it seems unlikely that the all-to-all circuits studied in this paper are described by the $d\rightarrow\infty$ limit of the above field theories. We do not yet understand the reason for this difference.

It is not ruled out that our Lagrangians overlook some crucial structure, and that as a result they do not capture \textit{any} models of measurement circuits or random tensor networks, even in finite dimensions. For present purposes we will assume this pessimistic scenario does not hold, and that the two field theories in Secs.~\ref{sec:MPTFT} and \ref{sec:FTRTNFMPT} do capture at least some class of models for the MPT and for the FMPT/RTN. We will explore these issues further elsewhere.

The simpler of the two field theories is that in Sec.~\ref{sec:MPTFT} for the MPT, involving a field $X_{ab}$ with vanishing row and column sums.
As a result of the cubic term, this theory has upper critical spacetime dimension ${d+1= 6}$.
Interestingly, the logic of Sec.~\ref{sec:largefinited} for the percolation problem above 5 spatial dimensions applies in this case too, since it relied only on the engineering dimensions of the fields. 
We can therefore carry over the exact exponent values so that (neglecting physics on timescales shorter than $L=N^{1/d}$, see Sec.~\ref{sec:largefinited})  the natural scaling variables in high dimensions are again
\ba
& t/N^{1/5}, &  & N^{2/5} \delta r
\end{align}
where ${\delta r = r - r_c}$ is the parameter driving the transition (and the number $N$ of spins should not be confused with the replica number in the preceding sections).

Let us consider the operator entanglement in the ordered phase, still above the upper critical dimension.
The plateau value of the operator entanglement, $S_2 \sim s N$, is proportional to the energy cost of a domain wall in $X$ that spans the system in the spatial directions, as discussed in the following section. 
In high dimensions the scaling of $s$ follows from dimensional analysis, giving $s \propto \mu^5 g^{-2}$ (\ref{eq:MPTlagrangian}), or in terms of the deviation $\delta r$ from criticality,
\be
s \sim |\delta r|^{5/2},
\ee
which is the same exponent as for the classical problem in high dimensions. 
A similar scaling form will again apply, $\overline{S_2} = H [ t/N^{1/5}, N^{2/5} \delta r ]$, but with a different scaling function $H$.
The size of the order parameter $X$ itself, which may be measured using appropriate correlation functions, grows linearly with the distance from the critical point, $X\sim |\delta r|$.

Again we  have a characteristic timescale  ${\tau= N^{1/5} W(N^{2/5} \delta r)}$, for an appropriate scaling function $W$. 
In the entangled phase this timescale grows exponentially in $N$ (Sec.~\ref{sec:replicatimescale}), with
\be
{\ln \tau \sim s N \sim |\delta r|^{5/2} N}
\ee
close to the transition.

Below 5+1 dimensions the scaling is different, because the cubic term is no longer dangerously irrelevant. The appropriate scaling variables are as usual
\ba
& t/L,
&
& L^{1/\nu } \delta r,
\end{align} 
where $\nu$ is the correlation length exponent for the field theory (\ref{eq:MPTlagrangian}). 
Exponents could be computed  in a ${6-\epsilon}$ expansion and will differ from percolation exponents (since the structure of the field theory is different, despite sharing the same upper critical dimension).
In the ordered phase there is still an exponentially long timescale, with 
\be
\ln \tau \sim s N \sim |\delta r|^{\nu d} N
\ee
close to the transition (Sec.~\ref{sec:replicatimescale}).

\subsection{Long timescale in the entangled phase}
\label{sec:replicatimescale}

\begin{figure}
    \includegraphics[width=0.9\linewidth]{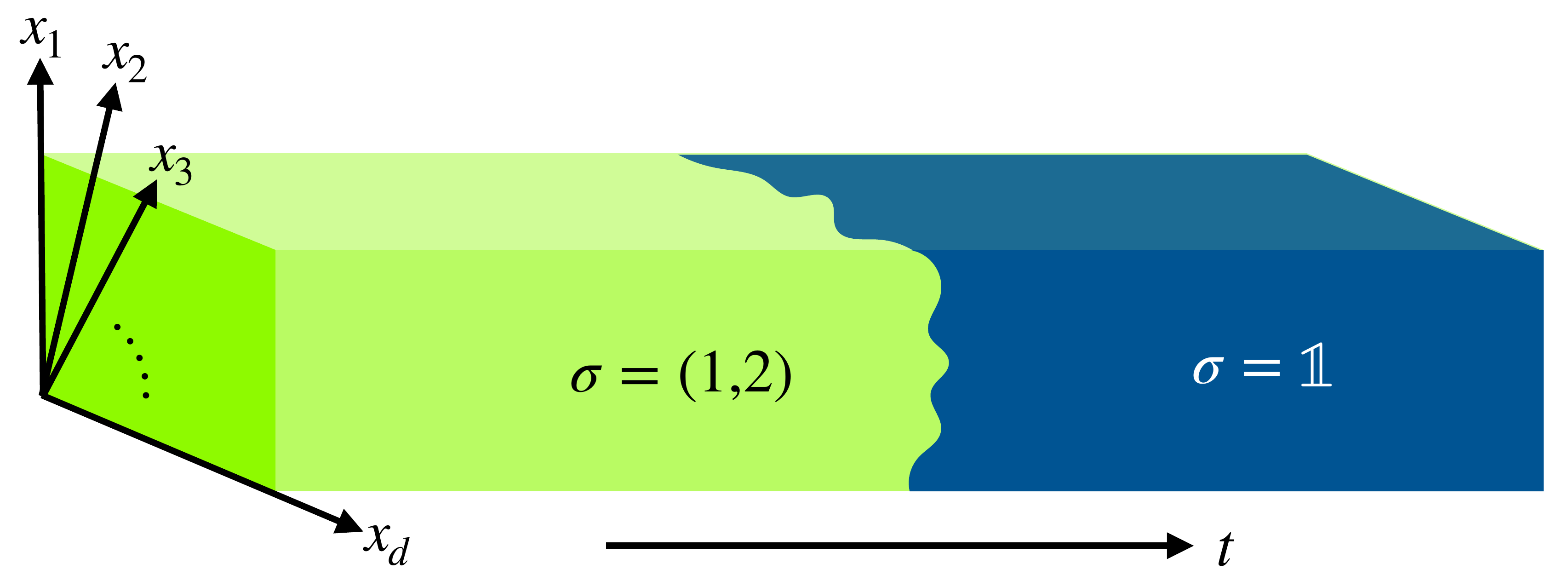}
    \caption{At large times, a system in $d+1$ spacetime dimensions is quasi-one-dimensional. The long timescale $\tau$ in the entangled phase, with ${\ln \tau\sim s N}$,  is due to the free energy cost ${\sim s N}$ of a domain wall in the effective spin model (Sec.~\ref{sec:replicatimescale}). At early times, the replica trick can be avoided, giving a domain wall in a disordered Ising model. More generally we must use the replica spin model.}
    \label{fig:quasi1ddomainwall}
\end{figure}

So far in this section we have focussed on the continuum description close to the transition. Here we discuss something simpler, namely the emergence of a  timescale that (in the entangled phase) is exponentially large in the number of spins, and the contrasting short timescale in the disentangled phase.  We may consider either a model in $d$ spatial dimensions with $N=L^d$ spins, or the all-to-all model. The results in this section are independent of the conjectural field theories above, as they rely only on more basic features of the effective spin model (pairing field) descriptions.

The appearance of a long timescale may be understood in analogy to standard 1D or quasi-1D classical models. 
Here the 1D coordinate is time: see Fig.~\ref{fig:quasi1ddomainwall}. 

In the ordered phase the pairing field (either $\sigma$ on the lattice or $X$ in the field theory) has long range order across a temporal slice and, after coarse-graining sufficiently, we may think of it as a function only of time. 
There is then a competition between the free energy cost of imposing a domain wall at a particular time, which scales as $s N$ with $s>0$, and the entropy $\ln t$ associated with translating the domain wall in the time direction. At a timescale $\tau$ with
\be\label{eq:tauproptosN}
\ln \tau \propto s \, N
\ee 
the translational entropy wins, and domain walls proliferate. Long-range order then no longer extends from the initial to the final time. By the identification of the entanglement with a free energy, this also means that  the entanglement begins to decay exponentially with time.

Recently the  exponentially long timescale in the entangled phase has been discussed from several points of view. Refs.~\cite{gullans2019dynamical} and \cite{fidkowski2020dynamical} consider a limit where the unitary evolution during a unit time can be treated as a $2^N\times 2^N$ Haar random unitary (see also App.~\ref{app:multiplyingmatrices} here for related considerations). 
Ref.~\cite{li2020statistical} has also given an analysis in terms of Ising domain walls that is similar to our considerations below.

The proportionality in Eq.~\ref{eq:tauproptosN} allows for an order 1 constant: however we expect that $N^{-1}\ln \tau$ vanishes in the same manner as $s$ when the critical point is approached from the entangled side (for example with the same power of the tuning parameter when this dependence is a power law).

At  times sufficiently shorter than $\tau$  the operator entanglement entropy has a plateau at an extensive value. 
The plateau value is corrected by a negative subleading term whose magnitude grows logarithmically with time.
In terms of the pairing field, the plateau regime is that where the number of domain walls is the minimal number allowed by the boundary conditions. 

For $S_2$, in the plateau regime, it is in fact sufficient to think about an Ising domain wall in a system with (Ising symmetric) disorder, for reason discussed  towards the end of  Sec.~\ref{sec:replicaexample}. 
That is, we expect that the replica trick can be avoided in the strongly entangled regime.
It is also possible to argue for the Ising picture using the replica treatment, by arguing that in this regime the replica theory is equivalent to the replica representation of a disordered Ising model \cite{zhou2018emergent}.\footnote{In the replica treatment we have $m\rightarrow 0$ ``elementary'' domain walls, each associated with one of the transpositions in $\tau_{2,m}$ (Eq.~\ref{eq:tau2mdef}). These may either bind together forming a composite domain wall or may separate for entropic reasons. See \cite{zhou2018emergent} for a discussion in the unitary case. However this system of $m\rightarrow 0$ domain walls can really just be thought of as the replica description of a \textit{single} Ising-like domain wall in a disordered environment. See \cite{zhou2018emergent, zhou2019entanglement} for details in the unitary case.}

If we neglect quenched disorder, then we obtain 
\be\label{eqS2logcorrection}
\overline{S_2}\sim s N - \ln t,
\ee 
in the plateau region.
The second term is the contribution from translational entropy, arising because the centre of mass temporal coordinate $t_\text{dw}$ of the domain wall can be located anywhere in $(0,t)$. 
The form in Eq.~\ref{eqS2logcorrection} was obtained in Ref.~\cite{gullans2019dynamical} in a limit of very dilute measurements, where the system can be viewed as completely scrambled  by a random unitary between each measurement.  
Ref.~\cite{li2020statistical} gave a picture in terms of Ising domain walls  equivalent to the one presented above. 
Here we have also suggested how the effective Ising model can be justified (in an appropriate regime and at the level of universal properties) rather than being only a heuristic model. Our consideration also implies that we should take into account quenched disorder, as discussed below.
(For another application of domain wall entropy in an effective 1D model to quantum chaos, see Refs.~\cite{chan2018spectral,garratt2020many}.)

As a check on the replica picture, we  have also considered a toy model for the entangled phase that involves multiplying large random matrices.  A crude treatment in App.~\ref{app:multiplyingmatrices}  (which neglects spatial structure, random fluctuations, and also the $n$-dependence of the R\'enyi entropies) reduces to computing the singular values of a sub-block of a large Haar-random unitary. This treatment also yields Eq.~\ref{eqS2logcorrection}, and shows that the plateau value $s N$ determines the timescale for exponential decay of $S_n$ in the regime of much later times, as expected from the above.
An analysis of related random matrix models has recently been presented in Ref.~\cite{fidkowski2020dynamical}.

Eq.~\ref{eqS2logcorrection} is the simplest picture, neglecting quenched disorder.
In reality there will be more complex crossovers. 
For example, in the all-to-all model there may be a regime of timescales where the subleading correction is not $\ln t$ but instead proportional to $\sqrt{N \ln t}$ as in the classical minimal cut problem (Sec.~\ref{sec:classicalmincut}).

This is because the conditional free energy $F(t_\text{dw})$, given by fixing $t_\text{dw}$,  will vary with $t_\text{dw}$ due to randomness: $F(t_\text{dw})= s N + \eta(t_\text{dw})$.
In high enough dimensions, and therefore presumably also in the all-to-all model, the typical fluctuations $\eta(t_\text{dw})$ will be Gaussian with a scale $\sqrt N$. 
Although these fluctuations are much smaller than $N$, they are in principle much larger than 1. 
Therefore at early enough times the free energy will be dominated by the optimal (most negative) value of $\eta(t_\text{dw})$, rather than by translational entropy.\footnote{This is similar to what happens for the classical minimal cut. If $\eta$ may be treated as Gaussian, the correction to the entanglement is of order $\sqrt{N\ln t}$ in this regime.}
But at larger times, there may be a regime where $\ln t$ entropy again dominates, giving the functional form in  Eq.~\ref{eqS2logcorrection}. At still larger times  multiple domain walls will proliferate (and the full replica treatment is required) and eventually $S_2$ decays exponentially in time.

The fact that only a single domain wall plays a role in the plateau regime means that there is an approximate factorization property for $S_2$ in a given realization of the circuit.
 If we divide $V$ into two parts, $V^{(1)}$ corresponding to evolution from $0$ to $t'$ and $V^{(2)}$  from $t'$ to $t$, then ${e^{-S_2}\simeq e^{-S_2^{(1)}} + e^{-S_2^{(2)}}}$.  The first term includes configurations with ${t_\text{dw}\in (0,t')}$ and the second those with $t_\text{dw} \in (t', t)$. (This is  approximate not only because it neglects configurations with multiple domain walls but also because it does not correctly treat domain walls with $t_\text{dw}$ close to $t'$.)

We now contrast the properties of the disentangled phase with those of the entangled phase.
Let us take the limit $N\rightarrow\infty$ first, so that as usual we can define the operator entanglement per spin at a given time:
\be
s_2(r,t) = \lim_{N\rightarrow\infty} \f{\overline{S_2}(r,t,N)}{N}
\ee
(we have written this equation for $S_2$, but the choice of R\'enyi index $n\geq 1$ should not be crucial). 
In contrast to the quasi-1D limit discussed above, this is the free energy cost, in an infinite slab of finite thickness, of imposing the domain wall boundary conditions described in Sec.~\ref{sec:replicaexample}.
In the disentangled phase  the free energy cost per unit transverse area decays exponentially with the thickness of the slab, so that $s_2(r,t)$ decays exponentially to zero with time.

\subsection{Variants and comments}
\label{sec:commentscaveatsetc}

In this subsection we discuss a few extensions of the field theory approach we have presented, as well as some open questions.

The measurement problems and random tensor networks that we have discussed so far have no internal global symmetries. 
One could also consider, say, measurement dynamics with an Ising symmetry \cite{sang2020measurement, lang2020entanglement}.
The definition of the pairing field in Sec.~\ref{sec:isingmapping} allows such symmetries to be incorporated, and suggests that in many cases they will change the universality class of the entanglement transition. 

For example, if the tensor network in Sec.~\ref{sec:isingmapping} has a  $\mathbb{Z}_2$ Ising symmetry that changes the sign of $S^a$ (and if we assume that the field whose mass vanishes at the transition is still ${X_{ab}\sim S^a \overline{S}^b}$) then odd powers of $X$ are forbidden by symmetry in the continuum Lagrangian,  which completely changes its structure in the limits of both ${N\rightarrow 0}$ and ${N\rightarrow 1}$.

This symmetry consideration highlights a feature of the discussion in Sec.~\ref{sec:isingmapping}, which is that the definition of $X_{ab}$ involves choosing a local basis.
In many cases this choice may not seem natural: 
for example, in many random models, the statistical invariance property emphasized in Sec.~\ref{sec:treestructuregeneralities}
ensures that any choice of local basis is equivalent to any other. 
(The exact mappings to models of permutations avoid having to choose a basis, but on the other hand it is less obvious how to coarse-grain them.) 
An open question is whether this necessity of choosing a basis is just an aesthetic issue, or a fundamental one.
Is it possible, for example, that the statistical invariance property imposes constraints on the continuum theory that we have neglected to take into account?

Other restrictions on the unitaries, not related to conventional symmetries,
can also  change the symmetries of the replica theory. For example, if all the unitaries are real-valued \cite{hunter2018operator} then there is no distinction in the bulk between forward and backward layers.
The symmetry  group $G_{N}$ is then enlarged to $S_{2N}$. In this case we can introduce a pairing field in a similar manner to Sec.~\ref{sec:isingmapping}, now with a replica symmetry action like that in standard disordered magnets and spin glasses.
(The restriction to Clifford unitaries \cite{FisherZeno,li2020conformal,gullans2020scalable,gullans2019dynamical,turkeshi2020measurement} is a more drastic change, which may require a different theoretical approach.)

The picture in Sec.~\ref{sec:isingmapping} relates random tensor networks (for which the limit ${N\rightarrow 0}$ is the appropriate one) to the language typically used to discuss spin glasses. 
This relation raises the question of whether other types of replica symmetry breaking, or other types of glass transition  \cite{parisi1980sequence}, are relevant to natural choices of circuit or tensor network.
{ (Of course we could always engineer, say, a glassy phase if we specifically design a tensor network with this in mind.)}
For example, one could imagine a second transition taking place inside the entangled phase for some choices of tensor network. At the entanglement transition, the $2N$ layers form a collection of $N$ pairs, breaking $G_N$ symmetry down to $S_N\times \mathbb{Z}_2$. Can the residual $S_N$ symmetry be broken in a subsequent transition? What are the entanglement properties of the resulting (presumably glassy) phase?

A statistical mechanics problem that provides a possible analogy for some of these phenomena is the directed polymer with random complex (or random sign) weights \cite{nguen1985tunnel,zhang1989directed,medina1989interference,
cook1990lyapunov,goldschmidt1992directed,derrida1993mean}.
The replica formulation of this problem involves $N$ copies of the polymer's partition function and $N$ copies of its complex conjugate. 
Averaging over random phases forces the copies to form pairs in order to avoid phase cancellation \cite{medina1989interference}, in analogy to the pairing phenomenon in the circuits.
Further, the paired object  --- a bound state of polymers from different copies --- may itself undergo phase transitions due to disorder. Perhaps this simpler problem can provide lessons for the circuit.

\subsection{Free fermion measurement dynamics}
\label{sec:freefermions}

Models of free fermions subjected to stochastic dynamics
\cite{cao2018entanglement,bernard2019open,bernard2020solution,frassek2020duality,nahum2020entanglement,sang2020measurement,chen2020emergent,alberton2020trajectory,thiel2018first,tobyforthcoming} can also show a transition in $d>1$ between two phases with differing amounts of entanglement \cite{nahum2020entanglement}. 
However, instead of an area law and a volume law phase (for states in finite dimensions), we instead have an area law phase and a phase with a  logarithmic violation of the area law \cite{nahum2020entanglement,chen2020emergent,alberton2020trajectory}. 

We may also characterize the two phases by transmission of information between initial and final time, which gives a distinction that makes sense in any dimension or for the all-to-all setup. For concreteness we may consider the latter case.
The model of Ref.~\cite{nahum2020entanglement}, which used the language of Majorana fermions, has a simple field theory description 
that is related to a model of classical loops (random walks) representing Majorana worldlines.
The quasi-one-dimensional regime which is relevant here has been studied in depth in  Ref.~\cite{amosjohnforthcoming}, which also characterizes the statistical properties of random samples. Here we consider only some more basic average quantities.

The characteristic timescale for the operator entanglement to decay is of order $N$, where $N$ is the number of lattice sites, 
rather being than exponentially large in $N$ as we found in the interacting case. This is a generic feature of free fermion models, as discussed below.
Within the ``more entangled'' of the two phases, the scaling of the operator entanglement is 
\be\label{eq:Snfree}
S_n \propto \left\{
\begin{array}{ll}
  {K(r) N}/{t}   &   \,\, t\ll K(r) N \\
  \exp\lf - c\, t/ [K(r) N] \ri  &  \,\, t\gg K(r) N.
\end{array}
\right.
\ee
Here $K$ (the sigma model stiffness) is an order-one constant deep in the phase, and vanishes as ${K(r) \sim (\delta r)^2}$ upon approaching the transition at $r = r_c$ to the disentangled phase ($c$ is a fixed order-1 constant). Note that the scaling in Eq.~\ref{eq:Snfree} is identical to the conductivity of a disordered $N$-channel wire, showing the crossover from Ohm's law to localization on a timesale of order ${K(r) N}$~\cite{amosjohnforthcoming,beenakker1997random}. 

\begin{figure}
    \centering
    \includegraphics[width=\linewidth]{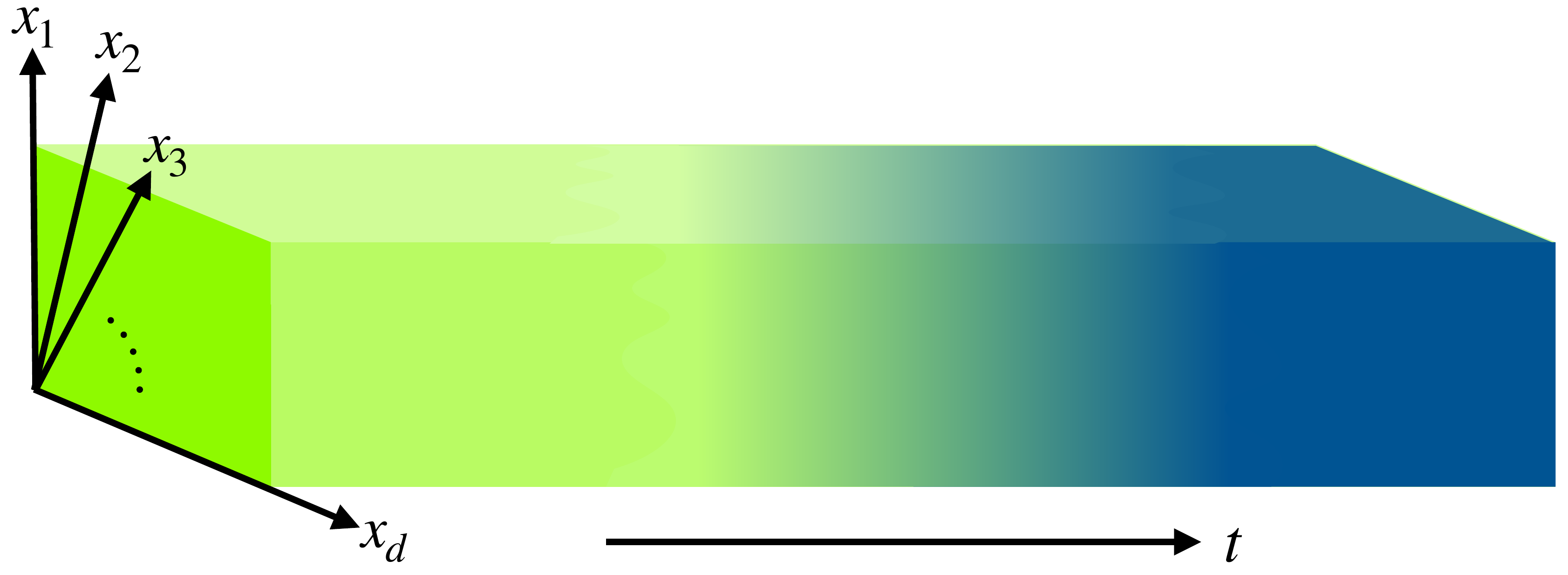}
    \caption{The field theory description of the  Majorana measurement model of Ref.~\cite{nahum2020entanglement} has a continuous replicalike symmetry, allowing smooth domain walls that give a more rapid decay of $S_n$ than in the interacting case where replica symmetry is discrete (App.~\ref{eq:appmajoranatimescale}). This exhibits a more general feature of free fermion models.}
    \label{fig:freefermiondw}
\end{figure}

The reason for the reduced timescale  in the entangled phase (of order $N$ compared to the exponential timescale in interacting models) is that the appropriate replica field theory has continuous, rather than discrete, replica symmetry. 
In the ordered regime, a nonlinear sigma model description may be used. 
Domain walls are smooth objects whose free energy cost decreases with their thickness, which in the case of interest is the temporal duration $t$ of the evolution: see Fig.~\ref{fig:freefermiondw}.

For this reason, we anticipate that the scaling in Eq.~\ref{eq:Snfree} applies to more general free fermion models with measurement. 
(The scaling of $K(r)$ close to $r_c$  will depend on symmetries and dimensionality. The constant $c$ may also depend on $n$ in general.)
General free fermion models can be formulated using the replica trick, in close analogy to replica sigma models for Anderson localization \cite{evers2008anderson}, leading to continuous replica symmetries. 
However in addition to the ${N\rightarrow 0}$ limit familiar from localization, the ${N\rightarrow 1}$ limit is now also of interest.
We will discuss this elsewhere \cite{replicasforthcoming}.

The timescale of order $N$ for free fermions agrees with the recent results of Ref.~\cite{fidkowski2020dynamical},
which  studied a model in which measurements of a single fermionic mode were alternated with Gaussian unitaries acting on the entire system. This model has even less locality structure than the all-to-all circuit.
In this limit also, the authors found that $\mathcal{O}(N^2)$ measurements were required to forget the initial state: this corresponds to $t=\mathcal{O}(N)$ in our conventions.

\section{Outlook}

It remains an open question to what extent the properties of the MPT, in various settings, will turn out to be tractable (either analytically or numerically). 
In this paper, however, we have shown that exact results are possible in certain regimes. We close by summarizing the  regimes we have studied, and some of the outstanding questions.

We began our analysis by considering the ``classical limit'' of the MPT in the all-to-all setting. We showed that a fairly complete picture is possible, including an analytical derivation of the critical point, critical exponents, and scaling forms for the entanglement.

Our results for quantum trees, including those obtained from a { spin-$1/2$} all-to-all circuit, show that exact results are also possible even far from this classical limit.
In this setting it was possible to demonstrate that an entanglement transition occurs at a definite nonzero measurement rate that is distinct from the classical value.
(It may even be possible to obtain rigorous results  on the phase diagram using the recursion relation approach.) 
The critical scaling on the tree is qualitatively different from a simple percolation picture.  

We argued that the critical point on the tree is the same as the critical point of the FMPT in the all-to-all circuit (which is locally treelike).
Since the location of the critical point in the circuit is difficult to check numerically, this equivalence has not yet been demonstrated clearly by our numerics. In the future we would like to have a clearer demonstration (or disproof) of this relationship between the tree and the all-to-all circuit.
Our results based on the tree were also restricted to the FMPT; it would be interesting to understand to what extent they are relevant to the MPT.

The scaling on the tree raised several questions that we hope to return to elsewhere. 
First, it will be worthwhile to examine the relationship between the random recursion relation studied here and approaches to tree tensor networks based on replicas \cite{lopez2020mean}.
Second, we raised the question of whether there are multiple universality classes on the tree.
This question remains to be settled, and could perhaps be addressed by generalizing our approach to a broader class of trees (with more general distributions of tensors or with larger bond dimension). 
Finally, it remains to be understood how to reconcile the scaling that we found on the tree with field theory.

In our numerical study of the MPT we have proposed observables that have benefits over the state entanglement, in that they do not require one to specify a spatial subregion. 
(Constructing such observables is crucial in the all-to-all setting, for which there is no meaningful distinction between area law and volume law phases, but they are also useful in 1+1D, where significant finite size effects make it important to avoid introducing lengthscales that are smaller than the system size.)
We demonstrated numerically that there is a long timescale in the entangling phase over which some aspects of unitarity are retained; for example, two initially orthogonal states remain approximately orthogonal.

The optimal numerical protocol for studying critical properties in the all-to-all  circuit remains to be settled.  
One complication is the lack of a priori knowledge of how the characteristic timescale scales with $N$ when $r=r_c$.
In the 1+1D problem, establishing that the dynamical exponent is equal to unity \cite{skinner2019measurement, li2020conformal}  allows one to reduce the number of independent variables in scaling collapses by fixing $t/L$ to a constant. 
Our candidate field theory for the MPT suggests that in high dimensions the appropriate scaling variable is $t/N^{1/5}$, but it is unclear whether this theory applies to the all-to-all circuit.

The proximity of the classical critical point ($r_c^\text{cl} = 0.8$) to the quantum one (e.g. $r_c = 0.749$ for the FMPT with Haar-random gates) in the ensembles we studied may also complicate the numerical analysis. 
For this reason it might be useful to study an all-to-all model (for example, involving weak measurements) in which the classical transition is eliminated entirely. 
It will also be interesting to relax the unitary invariance property of the gate distribution: the strong constraints imposed by this invariance  are a surprising feature of our analysis of the quantum tree.

Finally, we  addressed the replica approach to the MPT and to random tensor networks, both in the two phases and near the critical point, and we have made concrete proposals for field theories for these problems. { Determining the  domain of applicability of these theories will be the subject of further work. A basic ingredient in Sec.~\ref{sec:landautheory} was the construction of an ``overlap'' order parameter for the MPT and  random tensor networks that brings these problems closer to the language we use for disordered magnets and spin glasses (in comparison with  the more abstract language of permutation group elements used so far): this point of view may shed light on new possibilities for ordering.}

\acknowledgements
We thank Patrick Draper, John Chalker, Andrea De Luca,  Tianci Zhou, Toby Swann, Michael Gullans, David Huse, Ehud Altman,  Andreas Ludwig, Romain Vasseur, Yi-Zhuang You, and David Kesler for useful discussions. AN was supported by a Royal Society University Research Fellowship. SR is supported by EPSRC Grant No. EP/S020527/1. JR acknowledges the funding of the Israeli Science Foundation under grant No. 994/19 and the support of the Alon fellowship by the Israeli Council of Higher Education.

\appendix

\section{More on classical problem}

\subsection{Density of infinite cluster}
\label{app:finfinity}

Here we briefly derive Eq.~\ref{eq:finfty}, which describes the probability $f_\infty$ that a given node in the interior of the classical graph is connected to an infinite number of other nodes in the limit of infinite $N$ and $T$.  In other words, $f_\infty$ describes the density of the infinite cluster.

Consider the process of building a tree starting with an arbitrarily chosen node, as depicted in Fig.~\ref{fig:circuit-schematic}(c).  The starting node has four possible edges, each of which may be severed by a measurement.  If we denote by $e_\infty$ the probability that following a given edge will lead to a subtree with an infinite number of nodes, then
\be 
f_\infty = 1 - (1 - e_\infty)^4.
\label{eq:finfty_selfconsistent}
\ee 
The quantity $(1 - e_\infty)^4$ denotes the probability that \emph{none} of the four edges connected to the starting node leads to an infinite number of other nodes.

Following a particular edge, one may next encounter either a measurement (with probability $p = r/(2-r)$) or a node (with probability $1-p$).  The probability that this node is connected to an infinite number of other nodes at later generations is given by $1 - (1-e_\infty)^3$.  Thus we can write a self-consistency relation for $e_\infty$, given by
\be 
e_\infty = (1-p)[1 - (1-e_\infty)^3].
\label{eq:einfty}
\ee 
Near the critical point, $p = 2/3 + \delta p$, where $\delta p = (25/18) \delta r$ and $\delta r = r - r_c \ll 1$.  On the disconnected side of the transition, $e_\infty = 0$, while just on the connected side (small negative $\delta r$) $0 < e_\infty \ll 1$.  Expanding Eq.~\ref{eq:einfty} for small $\delta r$ gives $e_\infty \simeq - (25/6) \delta r$.  A similar expansion of Eq.~\ref{eq:finfty_selfconsistent} gives Eq.~\ref{eq:finfty} of the main text.

\subsection{Effective 1D field theory}
\label{app:percpotts}

We derive the mapping between the ``layered Erd\H{o}s-R\'enyi'' percolation model and a one-dimensional field theory that was described in Sec.~\ref{sec:percscalingforms}.

This is a bond percolation model with sites labelled $(i,t)$ with ${i=1,\ldots, N}$ and ${t=1,\ldots T}$. Generalizing slightly from the case in the text let a bond between sites $(i,t)$ and $(j,t)$ on the same time-slice be present with probability $b/N$, and a bond between sites $(i,t)$ and $(j,t+1)$ on the next slice be present with probability $b'/2N$. The average degree of a bulk node is ${z=b+b'}$, and from  considerations like those in Sec.~\ref{sec:classicaltreestructure} the critical case is $z=1$.

Bond percolation can be simply mapped to the Potts model with $Q\rightarrow 1$ states (see Ref.~\cite{cardy1996scaling} for a review). We introduce a Potts spin ${\sigma(i,t) = 1, \ldots, Q}$ on each site $(i,t)$, and couplings for pairs of sites that are allowed to be connected by a bond. For each pair of spins that is allowed to be connected there is a term 
\be
\lf (1-p) + p \, \delta_{{\sigma(i,t), \sigma(j,t)}} \ri
\ee
in the Boltzmann weight, where $p$ is the bond probability. The two terms correspond, in a diagrammatic expansion, to the presence and absence of the bond, respectively. Sites in the same percolation cluster have the same Potts spin state because of the Kronecker deltas on the bonds. 
Summing over spin states gives a factor of $Q^{\#\text{clusters}}$ which becomes $1$ in the replica limit. Spin correlation functions can be used to diagnose connectivity. The probability that two sites $(i,t)$ and $(j,t')$ are in the same cluster is \cite{cardy1996scaling}
\be\label{eq:connectednesscorrelatorapp}
p_\text{conn}(i,t;j,t')  = \lim_{Q\rightarrow 1} \f{\< \delta_{\sigma(i,t),\sigma(j,t')} - 1/Q \> }{1-1/Q}.
\ee
Below, the limit $Q\rightarrow 1$ will be left implicit.

Using the fact that the bond probabilities are of order $1/N\ll 1$, the partition function may be written
\ba\notag
Z  = \sum_{\{\sigma\}} \exp 
\bigg(
\f{b}{2N}& \sum_{t=1}^T \sum_{i,j}  \delta_{{\sigma(i,t), \sigma(j,t)}} \\
 & + 
\f{b'}{2N}  \sum_{t=1}^{T-1} \sum_{i,j} \delta_{{\sigma(i,t), \sigma(j,t+1)}}
\bigg).
\end{align}
As is standard in the field theory formulation of the Potts model \cite{zia1975critical,amit1976renormalization}, it is convenient to use a set of $(Q-1)$-component vectors $\vec e_\sigma$, for ${\sigma=1,\ldots,Q}$
to represent the spin states, with the vectors satisfying
\be\label{eq:pottsvectors}
\vec{e}^\sigma. \vec e^{\sigma'} = \delta_{\sigma,\sigma'} - Q^{-1}.
\ee
For $Q=2$ we can take $e^{1,2} = \pm 1/\sqrt{2}$. For $Q=3$ the three vectors point to the three corners of an equilateral triangle. For $Q=4$ they point to the vertices of a regular tetrahedron, etc. Note that
\ba
\label{eq:evectoridentities}
\sum_\sigma \vec{e}^\sigma & = 0,
&
\sum_\sigma e^\sigma_\mu e^\sigma_\nu & = \delta_{\mu, \nu},
\end{align}
as we see by considering $(\sum_\sigma e^\sigma_\mu e^\sigma_\nu ) e^\tau_\nu$ and applying (\ref{eq:pottsvectors}). 
Writing $\vec{e}(i,t)=\vec{e}^{\sigma(i,t)}$, and denoting the sum of the spins in a layer by 
\be
\vec{E}_t = \sum_i \vec{e}(i,t),
\ee
the partition function is (we drop an unimportant multiplicative constant)
\ba
Z & = \sum_{\{\sigma\} } \exp 
\left(
\f{b}{2N} \sum_{t=1}^T \vec{E}_{t}^2
+ 
\f{b'}{2N}\sum_{t=1}^{T-1} \vec{E}_{t}.\vec{E}_{t+1}
\right) 
\\ \notag
&= \sum_{\{\sigma\} }  \exp 
\bigg(
\f{b-b'}{2N} \sum_{t=1}^T \vec{E}_{t}^2
+ 
\f{b'}{4N}\sum_{t=1}^{T-1}
(\vec{E}_{t}+\vec{E}_{t+1})^2
\\ \notag
& \qquad \qquad \quad \qquad \qquad \qquad  +
\f{b'}{4N}
\left(
 \vec{E}_{1}^2 + \vec{E}_{T}^2
\right)
\bigg).
\end{align}
We can use two sets of Hubbard-Stratonovich fields, one set located at half-integer times, denoted $\vec{f}_{t+1/2}$, to decouple the $b'$ term, and one set located at integer times, denoted  $\vec{g}_t$, to decouple the $b-b'$ term. Each has ${Q-1}$ components. 
Once the $\vec{E}$ appear linearly in the exponent we can sum over the spins in a given timeslice $t$ (the prime indicates that the sum is only over these spins) via
\ba
{\sum_{\{\sigma\}}}' \exp\left[{\vec{E}_t.\vec y}\right] & = 
\bigg( 
\sum_{\sigma=1}^Q  e^{\vec{e}^\sigma.\vec y} \bigg)^N 
= 
Q^N \exp \lf {N \, V(\vec{y})} \ri,
\end{align}
which defines $V(\vec{y})$. Expanding in $\vec{y}$ for small $\vec{y}$ and using the identities mentioned above for the set of vectors $\{\vec{e}^\sigma\}$,
\ba
V(\vec{y})
=
\f{\vec{y}^2}{2Q}  + \f{d_{\mu\nu\lambda} y_\mu y_\nu y_\lambda}{6Q}  
+ \mathcal{O}(y^4).
\end{align}
The tensor $d$ is \cite{zia1975critical,amit1976renormalization}
\be
d_{\mu\nu\lambda} = \sum_{\sigma=1}^Q e^\sigma_\mu
e^\sigma_\nu e^\sigma_\lambda.
\ee
After integrating out the spins,
\ba\label{eq:ZHubbardStrat}
Z =
\int 
\mathcal{D} (f,g)
e^{
-\f{1}{2}\lf \sum f^2 + \sum g^2\ri
+ 
\sum_{t=1}^{T}
N V \lf \vec{y}_t
\ri},
\end{align}
where the final sum is over integer $t$. For $2\leq t\leq T-1$,
\be
\vec{y}_t = \sqrt{\f{b'}{2N}} (f_{t-1/2} + f_{t+1/2})
+ 
\sqrt{\f{b-b'}{N}} g_t.
\ee
At the boundaries we have e.g. 
\be
\vec{y}_1 = 
\sqrt{\f{b'}{2N}} f_{1+1/2}
+ 
\sqrt{\f{b-b'/2}{N}} g_1.
\ee
For the present we will neglect the boundary terms. The boundary condition on the field theory is important but we will fix it on physical grounds. 

The negative power of $N$ in $\vec{y}$ will allow us to truncate the action at cubic order. Let us combine $\vec{f}$ and $\vec{g}$ into a field $\vec{h}$ labelled by both integer and half-integer values, $\vec{h}_t = \vec{g}_t$, $\vec{h}_{t+1/2} = \vec{f}_{t+1/2}$. The lattice field theory is then (with $\tau,\tau'\in \mathbb{Z}/2$)
\be
Z = \int \mathcal{D} \vec{h} \exp \lf -\f{1}{2} \sum_{\tau,\tau'} h_\tau T_{\tau,\tau'} h_{\tau'}
- \f{1}{\sqrt N } A_3(\vec{h}) \ri,
\ee
where $A_3$ contains the cubic terms. To avoid clutter, let us immediately set $Q=1$ in the dispersion relation. The matrix $T$ is then (the first row/column shown correspond to a half-odd-integer index value):
\be
T = \left(\begin{array}{ccccccc}
   \cdots &  & & & &&   \\
& 1-b' &  -\alpha/2 &- b'/2 & & &  \\
&-\alpha/2  &  1-b+b' &-\alpha/2  & 0 & & \\
& - b'/2 & -\alpha/2    & 1-b' &-\alpha/2 &- b'/2&\\
&&0&-\alpha/2     & 1-b+b' &-\alpha/2 & \\
                &      &         &       & &  &\cdots
\end{array}\right)
\ee
We have defined
\be
\alpha = \sqrt{2b'(b-b')}.
\ee
This becomes imaginary when ${b'>b}$ --- which includes the line $b=0$ on which we do simulations of this model --- but this does not present a problem in the formal derivation below.\footnote{Formally, when ${b'>b}$, we use as our modes the coefficients of the right eigenvectors of $T$, which is equivalent to analytically continuing the formula for the ${b'<b}$ case. The square root nonanalyticity does not appear in the physical quantities below, only in the intermediate formulas.} 
Let us write
\ba
b&=\f{z+\Delta}{2}, 
&
b'& =\f{z-\Delta}{2},
\end{align}
where $z$ is the mean degree of a site, and the location of the critical point is $z=1$ for any value of $\Delta$. 
The dispersion relation has one ``massive'' mode, and one mode that becomes massless at the critical point $z=1$, at frequency $\omega=0$, with the eigenvalue of $T$ being  ${(1-z) + \f{z-\Delta}{4} \omega^2 + \mathcal{O}(\omega^4)}$.
At $z=1$ and $\omega=0$ the eigenvector of this mode is ${(g,f)\propto (\sqrt{\Delta}, \sqrt{1-\Delta})}$.
For the low-frequency theory we make the coefficient of this mode a slowly-varying field, $\vec{\phi}(t)$. Let us write 
\be
\delta z = z-1
\ee
for the parameter that vanishes at the phase transition. Let us drop the small parameter $\delta z$ except in the mass term, where it is the leading factor:
\be
Z = \int \mathcal{D} \vec{\phi} 
\exp \lf- \int \dd t \left[
\f{1-\Delta}{8} (\partial_t \vec{\phi})^2
- 
\f{\delta z}{2}\vec{\phi}^2
\right]
- \f{A_3}{\sqrt N }  \ri.
\ee
Thanks to the small prefactor $1/\sqrt{N}$ of the cubic term, we may take the continuum limit in a controlled manner. The cubic term is negligible for frequencies $\omega$ of order 1 due to the small prefactor, but important at parametrically small frequencies (since it is RG relevant). Since only small frequencies are important we can simply insert the form of the low-lying mode at $k=0$ into the cubic term without any need to explicitly integrate out high-frequency modes. The final result is
\be
Z = \int \mathcal{D} \vec{h} \exp \lf - \int \dd t \mathcal{L} \ri 
\ee
with the ``Lagrangian'' (again a factor of $Q$ has been set to 1 in the denominator of the final term)
\be\label{eq:lagrangianappendix}
\mathcal{L}= 
\f{1-\Delta}{8} (\partial_t \vec{\phi})^2
- 
\f{\delta z}{2}\vec{\phi}^2
-
\f{d_{\mu\nu \lambda}}{6 \sqrt{N}}
\phi_\mu \phi_\nu \phi_\lambda.
\ee
Above, all the $(Q-1)$ components of the field $\vec{\phi}$ are independent. We can write a more explicit form at the cost of using $Q$ fields that obey a linear constraint (summing to zero). For notational convenience we write them as the components of a $Q\times Q$ diagonal matrix with components
\be
\Phi_{\sigma, \sigma} = \vec{\phi}.\vec{e}^\sigma. 
\ee
The constraint is tracelessness
\be
\tr \Phi = 0.
\ee
Eq.~\ref{eq:lagrangianappendix} becomes
\be\label{eq:lagrangianappendix2}
\mathcal{L}= 
\f{1-\Delta}{8} \tr (\partial_t \Phi)^2
- 
\f{\delta z}{2} \tr \Phi^2
-
\f{1}{6 \sqrt{N}}
\tr \Phi^3.
\ee
This is the result given in the text, in special case $b=0$ (i.e. $\Delta = -z\simeq -1$ close to the critical point).

Close to the critical point, the connectedness correlation function for sites at distinct times is 
\be
p_\text{conn}(i,t;j,t') = 
\f{\< \vec\phi(t).\vec\phi(t') \>}
{N(1-1/Q)} 
\rightarrow
\f{\< \phi_1(t) \phi_1(t')\>}{N},
\ee
where $\phi_1$ is an arbitrarily chosen component. We can also write this as
\be
p_\text{conn}(i,t;j,t') = \f{\< \tr \Phi(t) \Phi(t') \>}{N(Q-1)}.
\ee

In App.~\ref{app:correlationfunctions} we present results for the connectedness correlation functions of boundary points. 
Since the boundary conditions on the Potts spins are free, this corresponds to the ``ordinary'' surface transition (discussed for percolation in Refs.~\cite{theumann1979bond,carton1980surface,de1981mean})
where the boundary spin operator is $\propto \partial_t \Phi$ in the continuum theory rather than $\Phi$ as in the previous equation \cite{cardy1996scaling}. This gives the scaling forms in App.~\ref{app:correlationfunctions}.

Finally let us consider the percolation probability $P_\text{perc}$. This can be used to  define a characteristic timescale $t_*(r,N)$ for the classical problem, and it is much simpler to formulate in field theory than the minimal cut cost.
As discussed in Sec.~\ref{sec:largefinited}, all this carries over to finite spatial dimensions $d>5$ by setting $N=L^d$.
$P_\text{perc}$ is equal to ${1-e^{-\Delta F}}$, where $\Delta F$ is the free energy cost of imposing twisted boundary conditions\footnote{Translating these BCs into the continuum field theory  Eq.~\ref{eq:PottsAppSeff} gives boundary magnetic field terms in Eq.~\ref{eq:lagrangianappendix} of the form ${h \, \delta(t) \, e_{\sigma}.\phi(t)}$ (and similarly at the final time boundary) with $h \propto N^{1/2}$. At first sight one worries that the $N$-dependence of $h$ introduces another exponent that could appear in scaling forms. However we believe that the basic point is just  that $h$ diverges with $N$, so that the asymptotic scaling forms are those of the $h\rightarrow\infty$ limit and the detailed $N$ dependence of $h$ determines only subleading corrections.} on the Potts spins \cite{cardy1992critical}.

The scaling form is
\be\label{eq:percscalingfinited}
P_\text{perc} = F\lf t/N^{1/5}, \delta r \, t^2 \ri
\ee
where $t$ is now the total time, and we have used the notation $r$ for the parameter driving the transition to match the circuit. Let us consider a few different regimes.

By a rescaling of the field and the time coordinate we can  choose to write the action in the form (suppressing order-one constants)
\be\label{eq:PottsAppSeff}
S_\text{eff} = \f{N}{t^5} \tr  \int_0^1 \dd u \lf  (\partial_u \widetilde \Phi)^2 + ( \delta r \, t^2)\widetilde \Phi^2 + \widetilde\Phi^3 \ri.
\ee
This rewriting suggests that if we take the limit of large $N$ and $t$  (and small $\delta r$) in such a way that the scaling variable ${\delta r\,t^2}$ is fixed while ${N/t^5}$ becomes large, $\Delta F$ is given by a saddle point action (we will not try to make this precise in this replica theory),
\ba
\ln\lf 1-P_\text{perc} \ri \sim  - \f{N}{t^5} A \lf  \delta r \, t^2  \ri, 
\end{align}
In particular, at the critical point
\ba
\ln\lf 1-P_\text{perc} \ri & \propto  - \f{N}{t^5} && \text{for $r=r_c$ and $t\ll N^{1/5}$}.
\end{align}
Note that for a system with finite ${d>5}$, unlike the case ${d<5}$, $P_\text{perc}$ is parametrically close to 1 at the critical point of a system with $t\sim L$, i.e. with ${t \sim N^{1/d}}$.
This is because, for percolation above the upper critical dimension,
there are many percolating clusters in a large hypercubic sample at $p_c$  \cite{conigliofinelydivided, aizenman1997number, de1987multiplicity, fortunato2004percolation,fortunato2004number, kenna2017universal}.

Next, let us take $N$ large with $\delta r<0$ small but fixed, in order to examine the exponential growth of the timescale with $N$ inside the percolating phase. By an alternative rescaling of the field,
\be\label{eq:PottsAppSeff2}
S_\text{eff} = N \tr  \int_0^t \dd t' \lf  (\partial_{t'} \hat \Phi)^2 + ( \delta r \, t^2) \hat \Phi^2 +  \hat\Phi^3 \ri.
\ee
Assuming again that we can make an analogy with saddle-point solutions in more conventional theories with discrete symmetry, we anticipate a localized domain wall or ``instanton'' solution interpolating between the two boundary condition values of the spin, with a classical action $N\times c(r)$. The scaling form will require ${c(r)\sim (\delta r)^{5/2}}$ close to the critical point.
At sufficiently early times there is at most one such instanton, which can be placed at any time in between $0$ and $t$:
\be
1- P_\text{perc} \sim t  |\delta r|^{1/2}\exp \lf - \text{const.} \, (\delta r)^{5/2} \, N  \ri.
\ee
The $\delta r$ dependence of the prefactor has been fixed by requiring consistency with the scaling form.\footnote{In the saddle-point language this factor would come from a fluctuation integral (c.f. e.g. \cite{cardy1978electron,hayn1991instanton}): it is consistent with  ${(Q-2)\rightarrow -1}$ modes with eigenvalue ${\propto |\delta r|}$.} 
Therefore the plateau at $P_\text{perc}\simeq 1$ lasts for an exponentially long time
\be\label{eq:Appplateautimescale}
t_* \sim 
\f{1}{\sqrt{|\delta r|}}
\exp \lf  \text{const.} \, (\delta r)^{5/2} \, N  \ri.
\ee
The interpretation is just that the probability of having a disconnection event at a given time is ${p_\text{break}\sim|\delta r|^{1/2}e^{-\text{const.}(\delta r)^{5/2}}}$. Since the probabilities of such events are independent, at long times we have exponential decay of $ P_\text{perc}$, with a timescale also given by Eq.~\ref{eq:Appplateautimescale}.

\subsection{Criticality in layered \er  graphs}
\label{app:layeredER}

As introduced in Sec.~\ref{sec:percscalingforms}, the layered \er model is a simplification of the classical random graph depicted in Fig.~\ref{fig:circuit-schematic}(b), in which a large number $N$ of nodes are arranged in discrete layers with time index $t$.  Each node may be connected only to nodes in adjacent time layers; there are no connections between nodes within the same time layer.  Edges between time layers $t$ and $t + 1$ are randomly-chosen, such that a total number of edges $cN$ are created between adjacent layers.  The connectivity $c$ is the major parameter of the model ($c$ is equal to $b'/2$, in the notation of Sec.~\ref{sec:percscalingforms}), and plays a similar role as the complement of the measurement rate, $1 - r$.  The critical value of $c$ is $c_\text{crit} = 1/2$, since a given node at time $t$ has connections to both $t-1$ and $t+1$ and its expected number of total connections is $2c$.  

Since the layered \er description is the basis for the theoretical derivation of the scaling forms in  in Sec.\ \ref{sec:percscalingforms}, we numerically simulate the layered \er graph to verify these scaling forms and ensure that they yield the same behavior as the data presented in {Figs.~\ref{fig:Pperc} and \ref{fig:tau_scaled}} for the full classical graph.  

\begin{figure}
\includegraphics[width=\linewidth]{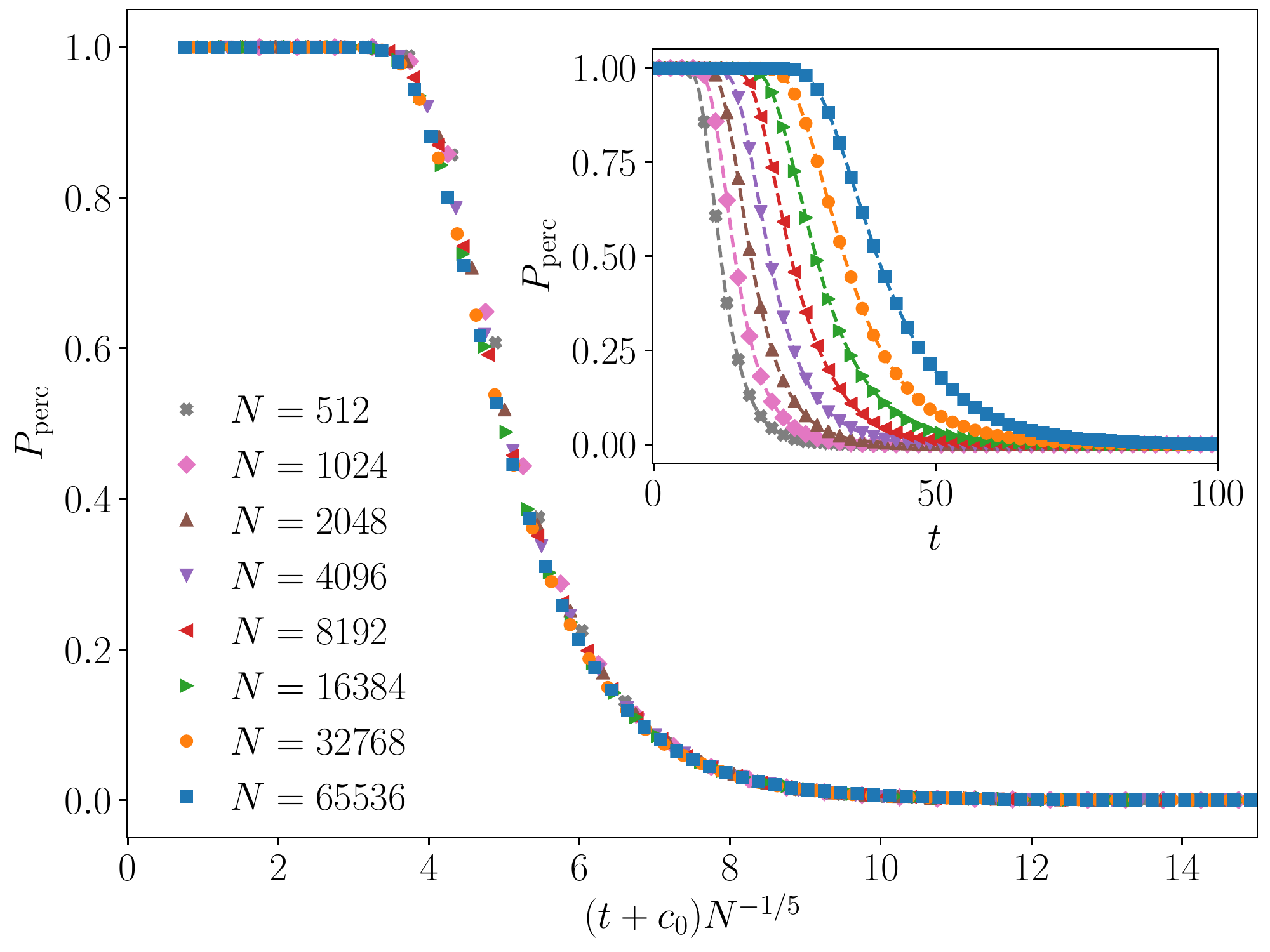}
\caption{The percolation probability $\pperc$ for the layered \er graph at the critical connectivity $b^\prime_\text{crit} = 1$ is plotted as a function of time.  The inset shows the raw data for different system sizes $N$, and the main figure shows the same data plotted as a function of $(t+c_0)/N^{1/5}$, where $c_0 = 6$ is a constant.
Compare Fig.~\ref{fig:Pperc} of the main text.}
\label{fig:Pperc-critical-ER}
\end{figure}

\begin{figure}
\includegraphics[width=\linewidth]{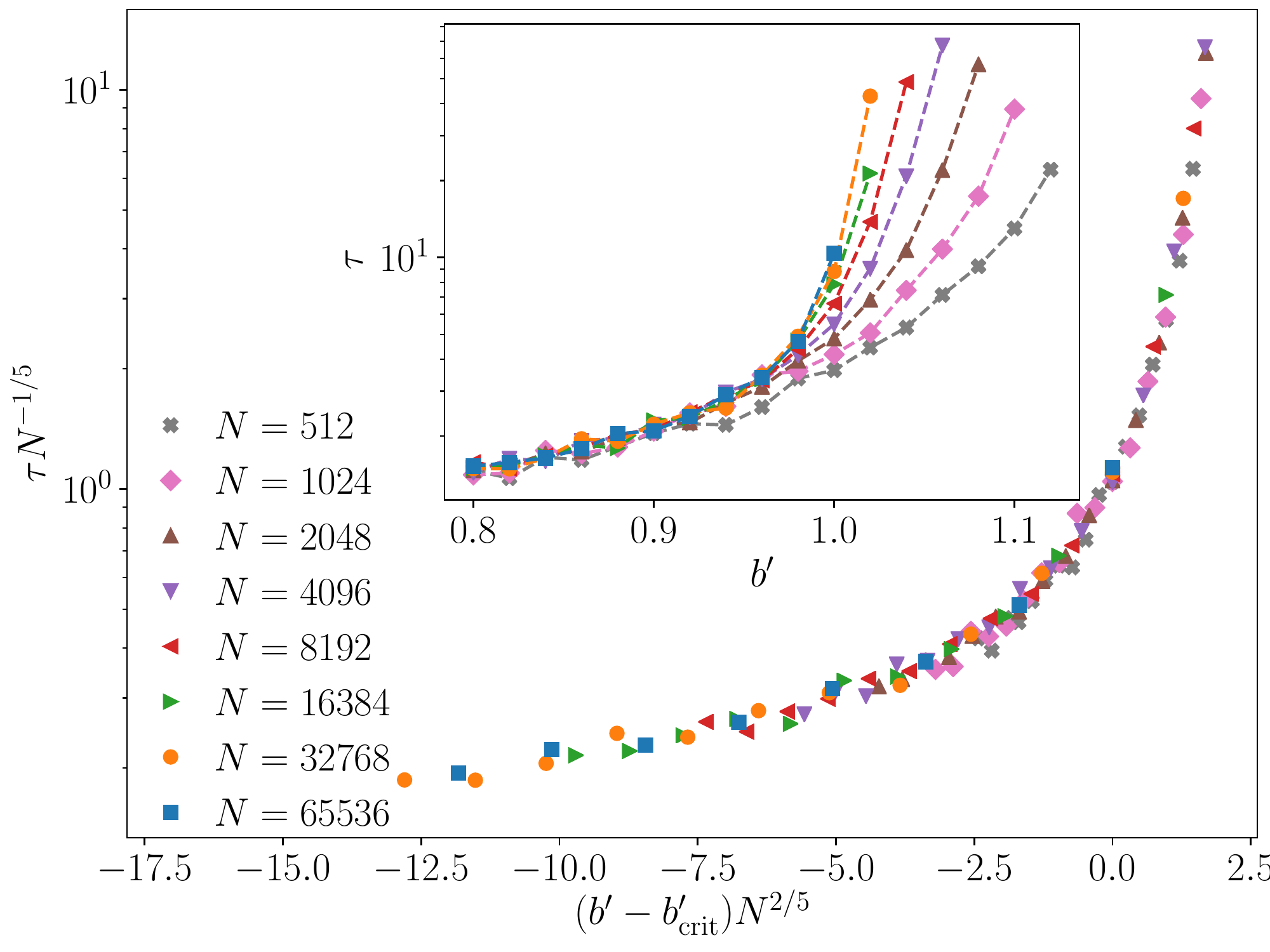}
\caption{The time scale $\tau$ for exponential decay of the percolation probability is plotted as a function of the connectivity $b^\prime$ and system size $N$ for the layered \er graph (inset).  This time scale exhibits critical scaling when plotted as a function of the scaled variables $\tau/N^{1/5}$ and $(b^\prime - b^\prime_\text{crit})N^{2/5}$, as suggested by Eq.~\ref{eq:basicscalingvariablesclassical} with critical connectivity $b^\prime_\text{crit} = 1$.
Compare Fig.~\ref{fig:tau_scaled} of the main text.}
\label{fig:tau-ER}
\end{figure}

Figure \ref{fig:Pperc-critical-ER} shows the percolation probability at the critical point, plotted as a function of time for different system sizes $N$.  A good scaling is observed as a function of the variable $t/N^{1/5}$, as suggested by Eq.~\ref{eq:basicscalingvariablesclassical}.  At a fixed value of $c$ and $N$, the percolation probability $\pperc$ is observed to decay exponentially with time.  As shown in Fig.~\ref{fig:tau-ER}, near the critical point the scaled exponential decay time $\tau(c,N)/N^{1/5}$ is a function only of the variable $(c - c_\text{crit})N^{2/5}$.  This is consistent with the scaling forms in Eq.~\ref{eq:basicscalingvariablesclassical}.

\subsection{Two-point correlation functions}
\label{app:correlationfunctions}

In Sec.~\ref{sec:percolationnumerics} we showed that the probability of percolation in the classical graph can be described in terms of the scaling variables $t/N^{1/5}$ and $(r - r_c) N^{2/5}$ (Eq.~\ref{eq:basicscalingvariablesclassical}).
Various correlation functions can also be understood in terms of these same scaling variables. The simplest correlation function, which we denote by $C$, is the probability that two distinct nodes in the graph belong to the same cluster.  The nodes may be on a temporal boundary (either the same boundary or different ones) or in the bulk of the graph. 

In the Potts language, the correlator is the spin two-point function. The bulk and boundary operators have different scaling dimensions, with the former scaling like $N^{-2/5}$ (or equivalently like $t^{-2}$) and the latter like $N^{-3/5}$ (or $t^{-3}$); see Sec.~\ref{sec:largefinited} and App.~\ref{app:percpotts}.

\begin{figure}
\includegraphics[width=\linewidth]{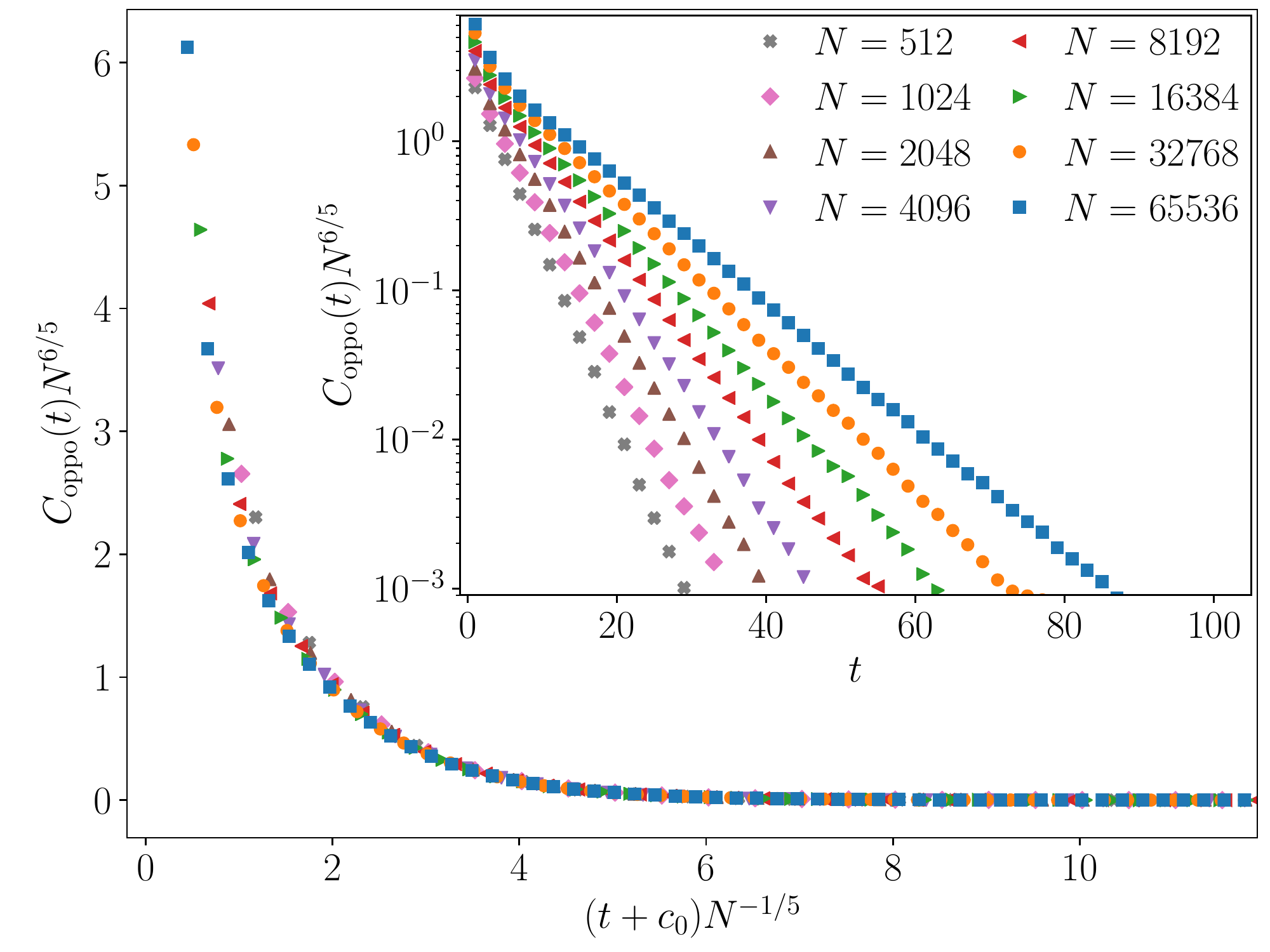}
\caption{The correlation function $C_\text{oppo}(t)$, which denotes the probability of two randomly-chosen nodes on opposite temporal boundaries being connected to the same cluster after a total time evolution $t$. Data here corresponds to the layered Erd\H{o}s-R\'{e}nyi model at the critical point, $c = 1/2$. The inset shows $C_\text{oppo}(t) N^{6/5}$ as a function of $t$ for different system sizes $N$, while the main figure shows that this same data scales onto a single curve when plotted as a function of $(t + c_0)N^{-1/5}$. Here $c_0 \approx 3.1$ is a constant.  }
\label{fig:Coppo-ER}
\end{figure}

As a result of this scaling, the probability $C_\text{oppo}$ that two nodes on opposite temporal boundaries are connected by a cluster has the form
\be
\label{eq:COppoScalingForm}
C_\text{oppo} (T, N) = \f{1}{N^{6/5}} F_\text{oppo}\lf T/N^{1/5}, \delta r N^{2/5}\ri.
\ee
In Fig.~\ref{fig:Coppo-ER} we test this scaling for case $r=r_c$, in order to confirm the theoretical value for the operator's scaling dimension. This data is for the simplified multi-layer Erd\H{o}s--R\'enyi model described in Sec.~\ref{sec:percscalingforms}.

\begin{figure}
\includegraphics[width=\linewidth]{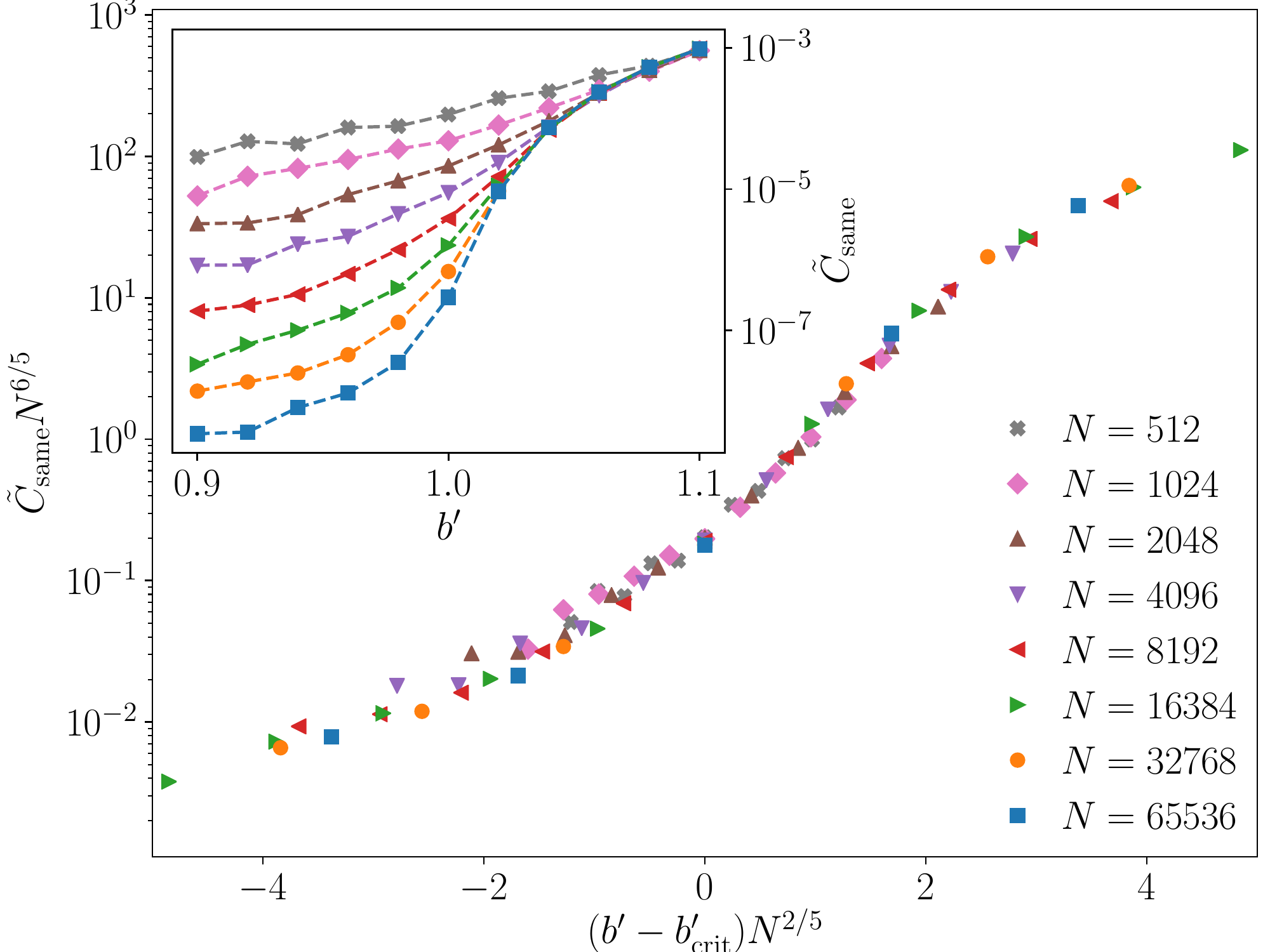}
\caption{The critical contribution $\widetilde C_\text{same}$ to the probability for two sites on the same temporal boundary to be connected, in the limit of long simulation time $T$. Data corresponds to the layered Erd\H{o}s-R\'{e}nyi model, and is plotted against $b^\prime$, twice the expected number of connections between a node at time $t$ and all other nodes at time $t+1$. The inset shows the raw data for different system size $N$, while the main figure shows the data plotted in terms of scaled variables. 
} 
\label{fig:Csame-ER}
\end{figure}

The probability $C_\text{same}$ for two sites on the \textit{same} temporal boundary to be connected has, in addition to the scaling term, a non-critical contribution of order $1/N$ which is in fact dominant at $r_c$.  This $1/N$ factor is on the order of the probability for the two sites to be connected by a ``microscopic'' path (for example by a single bond).  For simplicity, consider the limit $T\rightarrow \infty$, when only one of the arguments of the scaling function remains:
\be\label{eq:csamescaling}
C_\text{same}(N) = \f{1}{N^{6/5}}  F_\text{same}\lf \delta r N^{2/5}\ri + \f{A(r)}{N}. 
\ee
The noncritical term may be eliminated by a subtraction: ${\widetilde C_\text{same}(N)  =C_\text{same}(N)  - 2^{-1} C_\text{same}(N/2)}$. 
Fig.~\ref{fig:Csame-ER} demonstrates a reasonable scaling collapse for this quantity. This plot constitutes a second check that $\delta r N^{2/5}$ is the appropriate off-critical scaling variable.

If $r\lesssim r_c$ is fixed and $N\rightarrow\infty$, Eq.~\ref{eq:csamescaling} shows that $C_\text{same}(N) $ scales like $(r_c-r)^3$; this is the square of the \emph{surface} order parameter (the probability that a boundary site lies in the infinite cluster), which is parametrically smaller than the bulk order parameter (which scales as in Eq.~\ref{eq:finfty}) when $r_c$ is small. In the language of surface critical phenomena, this is the ``ordinary'' transition \cite{theumann1979bond,carton1980surface,de1981mean,cardy1996scaling}.

As an aside, let us make a distinction between the correlation function $C_\text{same}$ above and the mutual information between spins in the final state.
$C_\text{same}$ indicates whether in the final state two spins lie in the same connected tensor network. 
However, being in the same connected component does not imply that spins' mutual information, which can be detected with appropriate physical two-point functions, is large \cite{skinner2019measurement}.
The zeroth R\'enyi mutual information $I_0$
is given by a different classical correlation function to the one above, which in the finite-dimensional problem  behaves as a power law at $r_c$ \cite{skinner2019measurement}. 
A  related observable is the distribution of entanglement entropy $S_0$ for a single spin \cite{gullans2020scalable}.
These observables again map to boundary correlation functions of the Potts spins,\footnote{$S_0$ for a single spin is either 0 or 1 bits, with the former holding if the spin is connected to no other spins on the final time boundary. This may be written as a one-point function of the spin operator in a system where other boundary spins are fixed.
$I_0$ for two spins is either $0$, $1$ or $2$ bits, and is given by a minimal cut formula. If we assume that the probabilities for both $I_0=1$ and $I_0=2$ have the same scaling form (this can be demonstrated in 1+1D \cite{skinner2019measurement}) then we can focus on the simpler case $I_0=2$, which occurs only if the two spins are connected to each other, but not to any other spins on the final time boundary. This is the two-point function of the same Potts operator.} but  in a Potts system with a magnetized boundary condition, rather than with the free boundary conditions used above for $C_\text{same}$. 
(In the classification of surface criticality  this is the ``extraordinary''  transition \cite{lubensky1975critical,bray1977critical}.) The critical contribution to these quantities at $r_c$ is smaller than a trivial analytic contribution similar to that mentioned above, and we have not been able to see it numerically.

\subsection{Extrapolating min-cut tension to ${N \rightarrow \infty}$}
\label{app:extrapolation}

\begin{figure}
    \centering
    \includegraphics[width=\columnwidth]{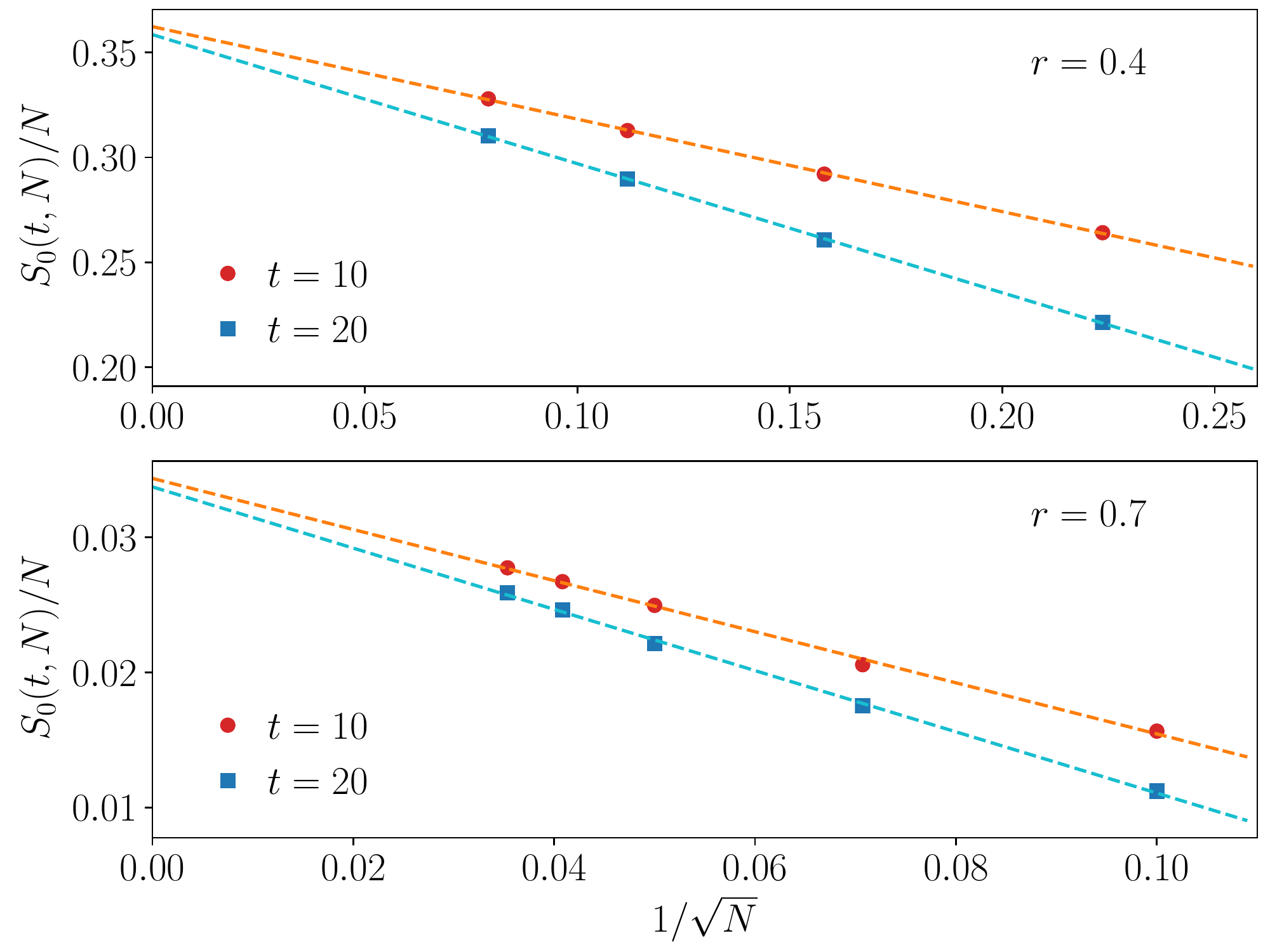}
    \caption{ 
    An example of the extrapolation of $S(t,N)/N$ to the limit of $N \rightarrow \infty$ in order to estimate the classical minimal cut per spin, $s(r)$. Points correspond to $S_0/N$ for different simulated system sizes at two different times $t$, and the dashed lines are linear best fit lines to $S_0/N$ as a function of $1/\sqrt{N}$.  The $y$-intercept of these lines gives the estimated value of $s(r)$, plotted in Fig.~\ref{fig:S0N-extrapolated}. The extrapolated values of $s(r)$ for two values of $t$ agree to within $3\%$. The top plot corresponds to $r = 0.4$, and the bottom plot corresponds to $r = 0.7$.
    }
    \label{fig:S0extrap}
\end{figure}

As mentioned in Sec.\ \ref{sec:classicalmincut}, the behavior of the classical minimum cut value $S_0(t, N)$ within the percolating phase, $r < r_c$, has the functional form
\be 
\frac{S_0(t, N, r)}{N} \simeq s(r) -  \frac{d(t,r)}{\sqrt{N}},
\ee 
at large $N$,
and 
for all times $t$
that are larger than an initial transient but short enough to satisfy $\ln t \ll \ln \tau$. Here, $d(t,r)$ is a constant for fixed $t$ and $r$ and $\tau$ denotes the  decay time of the percolation probability; $\tau$ grows exponentially with $N$ (see Fig.~\ref{fig:tau_scaled}).
An example of this scaling is shown in Fig.~\ref{fig:S0extrap}. One can estimate the value of the minimum cut per spin, $s(r)$, by extrapolating this relation to ${1/\sqrt{N} = 0}$. As illustrated in Fig.~\ref{fig:S0extrap}, this extrapolated value is relatively insensitive to the time $t$, so long as $t$ is larger than a short-time transient ($t \gtrsim 5$ is sufficient for the $r$-values plotted here and in Fig.~\ref{fig:S0N-extrapolated})  and shorter than $\tau$.

The data in Fig.~\ref{fig:S0N-extrapolated} corresponds to $t = 10$, and comes from an extrapolation using system sizes $N = 100, 200, 400, 600$, and $800$. As shown in Fig.~\ref{fig:S0extrap},  performing an extrapolation at $t = 20$ yields essentially identical results.  The extrapolation procedure becomes numerically difficult at very small $r_c - r$, since the decay time $\tau$ becomes short for all but very large system sizes.  The extrapolated data in Fig.~\ref{fig:S0N-extrapolated} is therefore limited to $r_c - r \geq 0.02$.

\section{Majoranas with pairwise measurement}
\label{app:majoranas}

Reference \cite{nahum2020entanglement} describes measurement-only dynamics for an even number of Majorana modes $\gamma_i$ for ${i=1,\ldots, N\gg 1}$.
Each measurement is of a fermion parity $i\gamma_i\gamma_j$ for some $i$ and some $j$. (If desired we can impose a bipartite structure, so that measurements are allowed only for $i\in A$ and $j\in B$ where $A$ and $B$ are say two sublattices of a bipartite lattice.) In finite dimensions, such a model allows a phase transition between an area law phase and a phase with a logarithmic violation of the area law. 
We briefly note the fate of this transition in the all-to-all setting.

We first select a single preferred grouping of the Majoranas: for $i$ odd, $i$ is grouped with ${i+1}$.
We can think of the two Majoranas within a group as forming a single complex fermion operator.
With probability $r$, a measurement is an  \textit{intra}group measurement,
and with probability $1-r$ it involves two Majoranas chosen uniformly at random, so that it is an \textit{inter}group measurement with probability 1 in the limit $N\rightarrow \infty$. 

The trivial phase arises at large $r$, when intragroup measurements predominate, and the nontrivial phase at small $r$ when intergroup measurements predominate. 
The dynamics can be thought of as the dynamics of an evolving \textit{pairing} between Majorana modes \cite{nahum2020entanglement}.
In spacetime it maps to a loop model \cite{candu2010universality,nahum2013phase}. The loops represent Majorana worldlines, and the operator entanglement is proportional to the number of worldlines connecting the initial to the final time.
The relevant case for the all-to-all model is a quasi-1D version of such a loop model~\cite{amosjohnforthcoming}.

The phase transition is described by the $RP^{n-1}$ sigma model in the limit $n\rightarrow 1$ (or $CP^{n-1}$ if we impose a bipartite structure). The relation between the all-to-all case and the finite-dimensional case is similar to that in Sec.~\ref{sec:largefinited}.
For the critical point it is natural to use a soft-spin formulation of the $RP^{n-1}$ or $CP^{n-1}$ model, with a cubic term and upper critical dimension 6 \cite{nahum2013phase,nahum2012universal}, so that by the logic of Sec.~\ref{sec:largefinited}  we expect the scaling variables in 
Eq.~\ref{eq:basicscalingvariablesclassical} to apply.
However, the associated scaling functions, and some of the other exponents, will be very different. The basic difference from the field theory discussed in Sec.~\ref{sec:largefinited} is that the $RP^{n-1}$ and $CP^{n-1}$ models have continuous replica-like symmetries [respectively $SU(n)$ and $SO(n)$], unlike the discrete  symmetries in both the field theory for percolation (Sec.~\ref{sec:largefinited}) and the replica descriptions of the generic quantum problem (Sec.~\ref{sec:landautheory}). This difference leads to much smaller entanglement in the nontrivial phase. 

The entanglement in the present model is related to the free energy cost of twisted boundary conditions in the sigma model. The possibility of continuous twisting of the order parameter makes this smaller than in a system with discrete symmetry. In finite dimensions, the nontrivial phase has only a logarithmic violation of the area law, rather than volume law entanglement \cite{nahum2020entanglement}.

In the all-to-all setting we again characterize the phases by the operator entanglement between initial and final times. In the nontrivial phase the $RP^{n-1}$ field is ``ordered'' (on timescales much shorter than $\tau$ below) and so we use a nonlinear sigma model formulation for an $n\times n$ matrix $Q_{ab}$ that parameterizes $RP^{n-1}$. This has the form ${S = K \int \dd t \tr (\partial_t Q)^2}$, with a stiffness that scales as 
\be
K \propto (\delta r)^2 N 
\ee
close to the transition on the nontrivial side. 
A rescaling of $t$ shows that this leads to a characteristic timescale of order
\be\label{eq:appmajoranatimescale}
\tau \sim (\delta r)^2 N.
\ee
For $t\gg \tau$, the simple one-dimensional field theory has exponentially decaying correlations, and the operator entanglement (computed from the cost of imposing twisted boundary conditions between the initial and final time \cite{nahum2013loop}) decays exponentially. 
For $t\ll \tau$, the operator entanglement is extensive in $N$ and scales as $S\sim \delta r^2 N/t$.

Note the large difference between the timescale in Eq.~\ref{eq:appmajoranatimescale}, which is linear in $N$, and the timescale in the entangled phase (which is accessed in more generic dynamics) that is exponentially large in $N$. 
The model outlined in this Appendix has only pairwise Majorana correlations, and a very restricted entanglement structure.
However this particular feature of the present problem is likely to carry over to  larger set of models involving unitaries and measurements or projections for free fermions \cite{chen2020emergent, tobyforthcoming}, since the key feature is having continuous, rather than discrete, replica symmetry \cite{evers2008anderson,tobyforthcoming}.

\section{Calculations for quantum tree}

\subsection{Numerical recursion for quantum tree}
\label{app:quantumTree}

In this section we briefly describe the numerical procedure used to simulate the quantum tree. 
Due to the single-site Haar rotations the Schmidt basis of the tree Eq.~\ref{eq:Tsvd} becomes uniformly distributed  and we can thus characterize the wave function using only the two Schmidt values (that is, using a single real positive number between $0$ and 1/2, which is the minimal Schmidt value squared $Z = \lambda_{min}^2$). 

The recursive procedure to generate the tree at generation ${k+1}$ consists of using three singular values at generation $k$, $Z_{k}$, ${Z_{k}}'$ and ${Z_{k}}''$ and connecting them to a node using Eq.~\ref{eq:T_recursive_relation}. Thus, the number of eigenvalues required to describe a certain instance of the tree {\it exactly} grows exponentially as $3^k$, which is clearly not simulable at large $k$. Here we take advantage of the fact that in the case of the FMPT the nodes of the tree are statistically independent. Thus, at a certain level $k$ we can generate a large constant pool of $N$ singular values, where ${1\ll N\ll 3^k}$. Assuming the pool spans the distribution function of $Z_k$ faithfully we can then draw randomly three singular values from this pool to generate a member in the pool of the next generation. This is known as the ``pool method'' \cite{miller1994weak,monthus2008anderson,garcia-mata2017scaling}. 

\begin{figure}
    \centering
    \includegraphics[width=0.85\linewidth]{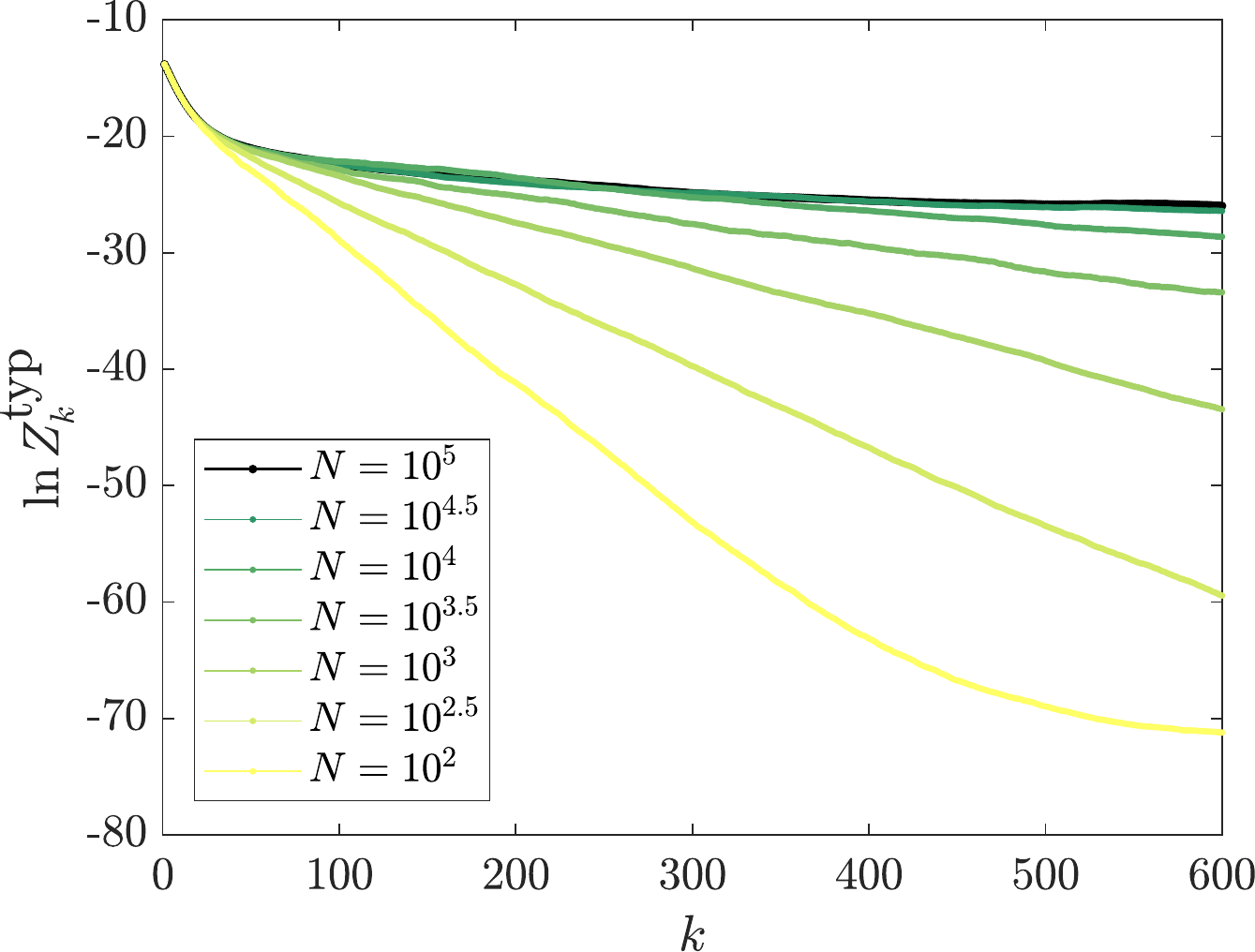}
    \caption{ $Z^{\textrm{typ}}_{k}$ vs. $k$ at $\ln (r_c-r) = -5.3$ for Haar ensemble and for various pool sizes $N$. This plot demonstrates the convergence with pool size $N$ and $k$ for the point closest to the phase transition in Fig.~\ref{fig:Z_critical}. The minimal $N$ and $k$ required for convergence diminish for points farther away from the critical point. In the case of half integer powers we round $N$ to the closest integer. }
    \label{fig:app:N_dep_tree_Haar}
\end{figure}

To verify that the pool spans the distribution of $Z_k$ faithfully we test the convergence of the evolution of $Z_k$ with the generation number $k$ as a function of $N$.
It is known that convergence in $N$ can be very slow for the pool method~\cite{miller1994weak,monthus2008anderson}.
For example, in Fig.~\ref{fig:app:N_dep_tree_Haar} we present $Z_{k}^{\text{typ}}$ at $\ln (r_c-r) = -5.3$ as a function of $k$ for different values of $N$, which is the point closest to the critical point in Fig.~\ref{fig:Z_critical}. 
The origin of the strong $N$-dependence lies in the exponent $\lambda$, which approaches $1/2$ at the transition causing the distribution of $Z$ to become broad.
Upon tuning farther away from the critical point the minimal $k$ and $N$ required for convergence is found to decay rapidly (not shown in the figure). 

Finally we also note that due to the forced measurements the distribution function of the singular values has a delta function at $Z =0$ with a known prefactor. Namely, the probability of a singular value at
generation $k$ to be exactly zero is given by the recursive relation
\be \label{eq:prob_of_exact_zero_f_k(p)}
f_k(p) = p+(1-p)[f_{k-1}(p)]^3\,.
\ee
In principle we can keep these zeros in our pool. However, this is highly inefficient, especially { when $p$ starts to get close to  the classical transition, where $f_\infty(p) = \left[\sqrt{1+(2-3p)p}/(1-p) -1\right]/2$ becomes unity}. Thus, in our simulations we keep only non-zero eigenvalues and, if needed, account for the zeros using Eq.~\ref{eq:prob_of_exact_zero_f_k(p)}.

\subsection{Averages of tree recursion constants}
\label{app:unitaryaverages}

\begin{figure}
    \includegraphics[width=\linewidth]{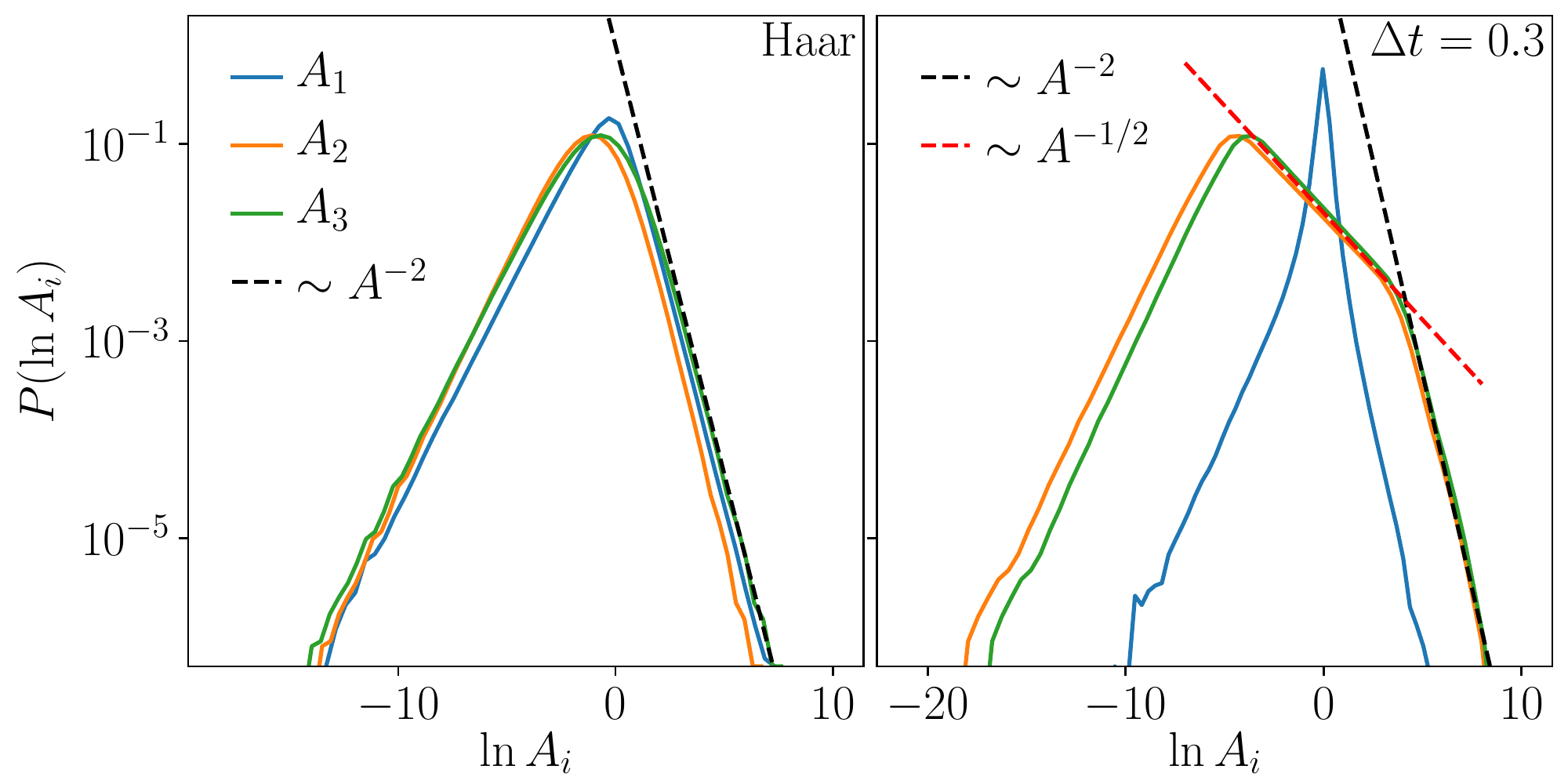}
    \caption{The distribution function of the parameters $V_j \equiv \ln A_j$ for  a Haar-random two-site unitary (Left) and the 
ensemble in Eqs.~\ref{eq:Ufixed},~\ref{eq:ufixeddistribution}   with $\Delta t=0.3$. The data here is collected from $N = 10^7$ random unitaries.}
    \label{fig:Adistrib}
\end{figure}

In Sec.~\ref{sec:recursionrelationfortree} we described the recursion relation for the singular values of the tree. The linearised recursion relation involves the random multiplicative constants in Eq.~\ref{sec:recursionrelationfortree}, which we repeat for convenience:
\ba\notag
A_1  &= \frac{ \left|U^{11}_{11} U^{21}_{21} - U^{11}_{21} U^{21}_{11}\right|^2}{\lf |U^{11}_{11}|^2 + |U^{21}_{11}|^2 \ri^2}, \\ \notag
A_2 &= \frac{ \left|U^{11}_{11} U^{21}_{12} - U^{11}_{12} U^{21}_{11}\right|^2}{\lf |U^{11}_{11}|^2 + |U^{21}_{11}|^2 \ri^2}\\
A_3 &  = \frac{ \left|
U^{11}_{11} U^{22}_{11} - U^{12}_{11} U^{21}_{11}
\right|^2}{\lf |U^{11}_{11}|^2 + |U^{21}_{11}|^2 \ri^2}.
\label{eq:AconstantsApp}
\end{align}
In this Appendix we derive the facts about the distribution of these quantities that were given in Sec.~\ref{sec:recursionrelationfortree}.
Some of these facts also hold for general node tensors, not necessarily expressed in terms of unitaries.

As in the text, we assume that the distribution of $U$ is invariant under multiplication by single-site $\mathrm{U}(2)$ matrices on any of its four legs. Initially however we do not assume that it is invariant under 2-site unitaries, i.e. we do not assume that $U$ is Haar-distributed in $\mathrm{U}(4)$.

First we show that, so long as the unitary $U$ is nontrivially entangling  (defined below) with probability 1, the average of any of the above quantities is exactly equal to one,
\be\label{eq:AunityApp}
\< A_i \> = 1,
\ee
and also
\be\label{eq:Ahalflog}
\<A_i^{1/2}\ln A_i \> = 0.
\ee
The argument is the same for any of the three $A_i$, so consider $A_3$ for definiteness. The argument only  relies on the property of $\mathrm{U}(2)$ invariance on a leg mentioned in Sec.~\ref{sec:treenetworkstructure} (together with the assumption that certain singular values are not fine-tuned to zero), so they hold for more general choices of $t$ satisfying this requirement. (We could also consider unitaries acting on more sites/trees with a larger branching number.)

The expression for $A_3$ involves only the matrix elements $U_{11}^{ad}$ for $a,d=1,2$. Regarding this as a ${2\times 2}$ matrix with row index $a$ and column index $d$, we make a singular value decomposition, with positive singular values $\eta_1$ and $\eta_2$:
\be
U_{11}^{ad} = \sum_{\mu=1,2} w_{a\mu} \eta_\mu v_{\mu d}.
\ee
Here, $w$ and $v$ are $\mathrm{U}(2)$ matrices. Now we note that, for any given $U$, the singular values of the tree we are considering are invariant under unitary basis transformations for the bond at the top of the tree. This implies that  $A_3$ must be invariant if $U_{11}^{ad}$ is multiplied by an arbitrary single-site unitary acting on the $a$ index (see Eq.~\ref{eq:4legtensordef}). We choose this unitary to be the inverse of $w$, so that $U_{11}^{ad}$ is replaced in Eq.~\ref{eq:AconstantsApp} by 
\be
U_{11}^{ad}  \longrightarrow \eta_a v_{a d}.
\ee
Together with $|\operatorname{det} v|^2 =1$, this gives the expression 
\be\label{eq:A3eta}
A_3 = \f{\eta_1^2 \eta_2^2 }{\lf \eta_1^2 |v_{11}|^2 + \eta_2^2 |v_{21}|^2 \ri^2}.
\ee
For some trivial, nonentangling two-site unitaries, such as the identity or swap, one of the singular values $\eta$ is exactly zero, and $A_3$ vanishes. We assume that the distribution of $U$ is such that, with probability 1, both singular values $\eta$ are nonzero. This is our definition of ``nontrivially entangling'' above. 

The above expression involves a single column, $v_{a1}$, of the $\mathrm{U}(2)$ matrix $v_{ad}$. Since we assumed that the distribution of $U$ is invariant under single-site rotations, $v_{ad}$ is Haar-distributed, and $v_{a1}$ is just a unit vector (with two complex or four real components) that must be averaged uniformly over the sphere $S^3$. 

This can be done in a standard way by relating the average over the sphere to a Gaussian average.  Let us write the four real components of the unit vector ${(v_{11},v_{21})}$ as ${V=(w,x,y,z)}$. If ${\<\ldots\>_\mu}$ is the Gaussian average with weight proportional to $e^{-\mu V^2}$, then
\be\label{eq:spherevsgaussian}
\<\ldots\>_\mu = 2\mu^2 \int \dd R \, R^3 \, e^{-\mu R^2} \< \ldots \>_{|V|=R},
\ee
as we see by splitting the Gaussian integral on the LHS into radial and angular parts. The latter gives the integral over a sphere of fixed radius, which is the last expression on the RHS. We are interested in $\<\ldots\>_{|V|=1}$, the average over the unit sphere.

The Gaussian average of $A_3$, which is of order $|V|^{-4}$, diverges at small $R$, so we instead first consider 
\ba
f(\mu) \equiv \< e^{-  ( \eta_1^2 (w^2+x^2) + \eta_2^2 (y^2 + z^2) )}\>_\mu  = \f{\mu^2}{(\mu+ \eta_1^2)(\mu+ \eta_2^2)}.\notag
\end{align}
Using Eq.~\ref{eq:spherevsgaussian} we may alternately write $f(\mu)$ as an average over the sphere. By scaling out a factor of $R$ from the components of $V$ we obtain an average over the unit sphere, and performing the $R$ integral gives 
\ba
f(\mu) =  \< \f{\mu^2}{\lf \mu + \left[ \eta_1^2 (w^2+x^2) + \eta_2^2 (y^2 + z^2) \right]  \ri^2 } \>_{|V|=1}.\notag
\end{align}
Equating the expressions for $f(\mu)$, and taking the limit $\mu\rightarrow 0$, the $\mathrm{U}(2)$ Haar average is 
\be
\< \f{1}{\lf \eta_1^2 |v_{11}|^2 + \eta_2^2 |v_{21}|^2 \ri^2} \>_v = 
\f{1}{\eta_1^2 \eta_2^2}
\ee
where we have restored the previous notation for the vector.
Plugging this into Eq.~\ref{eq:A3eta} gives  $\<A_3\>=1$, as stated above, regardless of the precise distribution of $\eta$. 

The same argument applies for $A_1$ and $A_2$, using the appropriate singular value decomposition. Note that each of the $A_i$ involves only a subset of the components of $U$, so that it effectively reduces to a ${2\times 2}$ matrix, as above for the matrix $U_{11}^{ad}$. That is, in each case two of the four legs of $U$ are set to index value ``1''. 

By multiplying $f(\mu)$ with $\mu^a$ and integrating over $\mu$, we find
\be\label{eq:Alambda}
\<A_i^\lambda\> = \f{1}{2\lambda-1} \< \f{1/H^{\lambda-1}- H^{\lambda}}{1-H} \>,
\ee
with $H = \eta_1^2/\eta_2^2$. 
The remaining average on the RHS is over these singular values, again for the appropriate singular value decomposition of $U$.

Differentiating with respect to $\lambda$ at $\lambda=1/2$ gives Eq.~\ref{eq:Ahalflog}, irrespective of the distribution of $H$.

While Eqs.~\ref{eq:AunityApp},~\ref{eq:Ahalflog} simplified, more general moments depend on the detailed distribution of $U$. For the 2-site Haar case we may write analytical formulas for $\<A_i^\lambda\>$ (given in Eqs.~\ref{eq:HaarA1A2},~\ref{eq:HaarA3} of the main text). 

First consider $A_3$, as above. This simplifies because, for a Haar-distributed unitary, $U_{11}^{ad}$ can be viewed as a normalized and uniformly random vector in a Hilbert space of dimension $2\times 2$ (a Page-random state). We are interested in the singular values when this state is split into two equal subsystems. Writing $s_i = \eta_i^2$, this distribution is \cite{nadal2011statistical},
\be\notag
P(s_1,s_2) = 3 \delta(s_1+s_2-1)(s_1-s_2)^2
\rightarrow
P(H) =  \f{3(1-H)^2}{(1+H)^4},
\ee
for $0<H<\infty$. Applying this to (\ref{eq:Alambda}) gives 
\be
\< A_3^\lambda\> 
=
\f{\pi \lambda(1-\lambda)}{\sin(\pi\lambda)}.
\ee
This is finite for $\lambda\in (-1,2)$.

By the right-invariance of the Haar measure, the distribution of $A_1$ is the same as that of $A_2$. We must consider the singular values for a decomposition of the matrix $U_{b1}^{a1}$, which is the upper left ${2\times 2}$ block of a ${4\times 4}$ Haar matrix. The distribution of singular values for such a sub-block of a Haar unitary may be found in Ref.~\cite{kieburg2016singular}. Writing again $s_i= \eta_i^2$, 
\be
P(s_1, s_2) = 6 (s_1 - s_2)^2
\ee
with the constraint $0<s_i<1$ but, unlike in the previous case, no constraint on ${s_1+s_2}$.
Since there is a relabelling symmetry under $s_1\leftrightarrow s_2$, or equivalently under $H\leftrightarrow 1/H$, we may insist $0<H<1$. Then
\be
P(H) = 3 (1-H)^2.
\ee
Applying this to (\ref{eq:Alambda}) gives
\be
\<A_1^\lambda\> = \<A_2^\lambda\> = 
\f{12}{12+8\lambda - 7 \lambda^2 - 2\lambda^3 + \lambda^4},
\ee
as stated in the text. Again this is finite for $-1<\lambda<2$.

Finally, let us discuss the asymptotics of the distributions of the $A_i$. 
This will clarify the following point. 
In the main text we described an ensemble of 2-site unitaries with an entangling strength parameterised by $\Delta t$. 
In the limit of small $\Delta t$, these unitaries become closer and closer to the identity. For the identity, $A_2=A_3=0$ exactly. But we have shown above that, for any nonzero value of $\Delta t$, no matter how small, ${\<A_2\>=\<A_3\>=1}$. Therefore the limit $\Delta t\rightarrow 0$ does not commute with the average over unitaries. This is because the distribution of $A_i$ develops a long tail when $\Delta t$ becomes small.

Define the 2-component vectors
\ba
\phi_a & 
= U^{a1}_{11} &
\psi_a & 
=U^{a1}_{21}, 
&
\chi'_a & 
= U^{a2}_{11} &
\psi'_a & 
= U^{a1}_{12}.
\end{align}
Then we may write Eqs.~\ref{eq:AconstantsApp} as
\ba\label{eq:Aiexpressionsphi}
A_1 & = 
\f{|\psi|^2|\phi|^2- |\psi^\dag \phi|^2} {|\phi|^4}, \\
A_2 & = 
\f{|\psi'|^2|\phi|^2- |\psi'^\dag \phi|^2} {|\phi|^4}, \\
A_3 & = 
\f{|\chi|^2|\phi|^2- |\chi^\dag \phi|^2} {|\phi|^4}.
\end{align}
We see that $A_i$ can become arbitrarily large if $|\phi|$ becomes small, with $A_i$ scaling like $|\phi|^{-2}$ in this limit.  Since $\phi$ has two complex (or four real) components, we expect that for a generic distribution of unitaries, the cumulative probability distribution of $|\phi|$ scales like $|\phi|^4$ at small $|\phi|$. This gives 
\ba\label{eq:Atailexponent}
P_{\ln} ( \ln A_i ) \, \dd \ln A_i 
&\sim \f{\dd \ln A_i}{A_i^2}, &
A_i & \gg 1
\end{align}
at large $A_i$, as stated in the text. 
At small $A$, similar considerations for the numerators in Eq.~\ref{eq:AconstantsApp} show that generically 
\ba\label{eq:Ailefttail}
P(\ln A_i) \dd \ln A_i & \sim A_i \, \dd \ln A_i, &
A_i & \ll 1.
\end{align}
These power laws are consistent with numerics and also with the fact that the moments $\<A_i^\lambda\>$ for the Haar case 
diverge at $\lambda = 2$ and at $\lambda = -1$.
The $A_i$ are of course correlated, but we do not consider their joint distribution here. 

Now consider the case of a weakly entangling unitary with random single-site scramblers,
\be
U = (u_1\otimes u_2) e^{- i \Delta t H } (u_3 \otimes u_4),
\ee
with $\Delta t$ small but fixed. ($H$ may either be  fixed, as in an ensemble discussed in the main text, or random.)
We focus on $A_2$ and $A_3$, whose distributions become broad at small $\Delta t$ (that of $A_1$ does not). The two cases are similar, so consider $A_3$, which is  given by Eq.~\ref{eq:A3eta}. 
We expect that for small $\Delta t$ we typically have $\eta_1^2 \sim \Delta t^2$ (here we keep only the scaling with $\Delta t$). Therefore so long as $|v_{22}|^2 \gg \Delta t^2$, i.e. in the regime $A_3 \ll \Delta t^{-2}$, we have 
\be
A_3 = \f{\eta_1^2 \eta_2^2}{(\eta_1^2 |v_{12}|^2 + \eta_2^2 |v_{22}|^2)^2 } \sim \f{\Delta t^2}{|v_{22}|^4}.
\ee
For small $|v_{22}|$, the cumulative distribution of $|v_{22}|$ scales like $|v_{22}|^2$, which gives
\ba
P(\ln A_3) \dd \ln A_3 & \sim \f{\Delta t}{\sqrt{A_3}} \dd \ln A_3,
& 
& \Delta t^{2} \ll A_3 \ll \Delta t^{-2}.
\end{align}
On the other hand for $A_3\gg \Delta t^{-2}$, we expect to recover the generic exponent $-2$ (Eq.~\ref{eq:Atailexponent}) for the distribution, suggesting
\ba
P(\ln A_3) \dd \ln A_3 & \sim \f{(\Delta t)^{-2}}{A_3^2} \dd \ln A_3,
& 
& A_3 \gg \Delta t^{-2}.
\end{align}
Similarly for $A_3\ll \Delta t^2$ we expect to recover the exponent 1 for the generic case (Eq.~\ref{eq:Ailefttail}). 
The existence  of three regimes with different power law exponents, 1, $-1/2$, and -2, when $\Delta t$ is relatively small, is in good agreement with the numerical data.
Note that the scaling above, with the $-1/2$ tail being cut off at $A_3\sim \Delta t^{-2}$, is also consistent with $\<A_3\>$ being of order $1$ at small $\Delta t$.

\begin{center}
\begin{table}[h!] 
  \begin{tabular}{ |c|c|c| }
    \hline
    Quantity &Haar & $\Delta t = 0.3$\\
      \hline
    \hline
    $\langle A_i \rangle$ & 1 & 1 \\ 
    \hline
    $\langle \sqrt{A_i} \log A_i\rangle$ & 0 & 0 \\
    \hline
    $\langle \sqrt {A_1}\rangle$ &64/75 & $0.981 $\\
      \hline
    $\langle \sqrt{ A_2}\rangle$ &64/75  & $0.420$ \\
      \hline
    $\langle \sqrt{A_3}\rangle$ & $\pi/4$ & $0.419$ \\
    
      \hline
        $\langle \sqrt{A_1} (\log A_1)^2\rangle$ &$ {17408 \over 16875}$ & 0.227 \\
      \hline
      $ \langle \sqrt{A_2} (\log A_2)^2\rangle$ &$ {17408 \over 16875}$ & 1.429 \\
      \hline
      $ \langle \sqrt{A_3} (\log A_3)^2\rangle$ & ${\pi^3/4}-2\pi$ & 1.428 \\
       \hline
  \end{tabular} 
  \caption{Some averages of the $A_i$ that are used in the text.}
  \label{Table:As_averages}
\end{table} 
\end{center}

\subsection{Continuum recursion relation at $\Delta>1$}
\label{app:FKPPcontinuum}

We start with the FKPP equation with the spatially varying diffusion constant ${D(x) =1+e^x}$,
\be\label{eq:htravelingwaveapp}
\partial_\tau H =
\partial_x  \Big( D(x) \partial_x  - a \Big) H 
+ \Delta H \big( 1-H \big),
\ee
in the regime $\Delta>1$, such that the linearized problem is in the paramagnetic phase. The entanglement transition is at ${a=-(\Delta+1)}$ so we write
\be
a = -(\Delta + 1) + \sigma
\ee
with ${0<\sigma\ll 1}$. When $\sigma>0$, we converge at late times to a stationary solution satisfying $\partial_\tau H=0$. We would like to determine the position $x_f$ of the front in this solution, or equivalently the value of $\Ztyp$ (recall that ${x_f \sim \ln \Ztyp}$).

Since $H$ varies by an exponentially large factor over the relevant range of $x$, it is useful to look instead at the local exponential decay rate, which is order 1:
\be
R(x) \equiv \partial_x \ln H(x).
\ee
From the stationary version of (\ref{eq:htravelingwaveapp}), this satisfies
\be
\partial_x R = -\Delta f(x) - \left[ 1- (\Delta-\sigma) f(x)\right] R - R^2 + \Delta f(x) H(x),
\ee 
where 
\be
f(x) = \f{1}{e^x+1}.
\ee
$R$ tends to 1 in the limit $x\rightarrow \infty$. (To see this note that in this limit Eq.~\ref{eq:htravelingwaveapp} becomes ${\partial_x e^x \partial_x H = 0}$. Together with $\lim_{x\rightarrow \infty} H=0$ this gives $H\propto e^{-x}$.) Let us define
\be
R(x) = - 1 + \sigma S(x).
\ee
Then
\be\label{eq:SDeltaApp}
\partial_x S = - f(x) + \left[ 1- (\Delta-\sigma)f(x)\right] S - \sigma S^2 + \f{\Delta}{\sigma} f(x) H(x).
\ee
We will see in a moment that we must treat the cases ${1<\Delta<2}$ and ${\Delta >2}$ separately. For now let us just note that if $x$ is sufficiently large, the final term above will be subleading since both $H$ and $f$ tend to zero at large $x$. Assuming we are in the range of $x$ where this term is negligible, and and  dropping terms that are subleading in $\sigma$,
\be\label{eq:Seqsimplified}
\partial_x S \simeq - f(x) + \left[ 1- \Delta f(x)\right] S.
\ee
Solving this equation, and fixing the integration constant by demanding that $S$ does not blow up as ${x\rightarrow\infty}$,
\be
S = \f{1}{\Delta(\Delta-1)}\left[
e^{-(\Delta-1)x}\lf e^x+1 \ri^\Delta - \lf e^x + \Delta \ri
\right].
\ee
Now consider this solution for large negative $x$
\be\label{eq:expansionofSapp}
S = \f{1}{\Delta(\Delta-1)} \left[ e^{(\Delta-1)|x|} + \ldots \right].
\ee
We will check below that for $1<\Delta<2$, there is a range of $x$ where this expansion is valid, i.e. where the final term in (\ref{eq:SDeltaApp}) can indeed be neglected.

According to this expansion, as we increase $-x$, the value of $R$ begins to increase significantly from $-1$ once 
\be
x \sim - \f{1}{\Delta-1} \ln \f{1}{\sigma}.
\ee
This suggests that the front is at ${x_f\sim -\f{1}{\Delta-1} \ln \f{1}{\sigma}}$, and that here we can match onto the stationary solution of the travelling wave equation Eq.~\ref{eq:htravelingwaveapp} with $\sigma=0$ and with a spatially constant diffusion coefficient.
(This matching makes sense since the forward part of this solution has $R=-1$ when $\sigma=0$.) 

However, we must  check the self-consistency of our neglect of the final term in Eq.~\ref{eq:SDeltaApp}. 
Assuming the above scaling for $x_f$, we have 
\be\label{eq:neglectedterm}
\f{1}{\sigma} f(x) H(x) \sim \sigma^{(2-\Delta)/(\Delta-1)} \f{e^{-2x}}{1+e^{-x}}
\ee
for $x\gg x_f$.  If $1<\Delta < 2$, this term is indeed much smaller than the RHS of Eq.~\ref{eq:Seqsimplified} for ${x\gg x_f}$. Therefore for this range of $\Delta$ the above analysis, giving 
\ba\label{eq:xfintermediateDeltaApp}
x_f & \sim -\f{1}{\Delta-1} \ln \f{1}{\sigma}
&
(1< &\Delta < 2),
\end{align}
is self-consistent, though not rigorous.

On the other hand, for $\Delta>2$, when the power of $\sigma$ in Eq.~\ref{eq:neglectedterm} is negative,  this term cannot be dropped from Eq.~\ref{eq:SDeltaApp} for ${x\gg x_f}$. 
Numerically solving Eq.~\ref{eq:SDeltaApp} suggests that instead the final term in Eq.~\ref{eq:SDeltaApp} contributes at leading order if we fix $x$ and take $\sigma\rightarrow 0$. 
Indeed, the alternative would be to have some $x_c$, with ${x_f \ll x_c \ll 0}$, such that the term  is negligible for $x\gg x_c$ but not for ${x_f < x < x_c}$, and this may be seen to be inconsistent by examining the ratio of this term to the right hand side of Eq.~\ref{eq:Seqsimplified}.
Using this fact, that $H/\sigma$ should be of order 1 when $x\sim 0$, and assuming that $H\sim e^{- (x-x_f)}$ for $x\gg x_f$, we find that 
\ba\label{eq:xflargeDeltaApp}
x_f & \sim - \ln \f{1}{\sigma}
&
(\Delta & > 2).
\end{align}
Eqs.~\ref{eq:xfintermediateDeltaApp},~\ref{eq:xflargeDeltaApp} give the power laws $\Ztyp \sim \sigma^{1/(\Delta-1)}$ and $\Ztyp\sim \sigma$ stated in Sec.~\ref{sec:othertree}.

These power laws can be checked directly by making a numerical solution of Eq.~\ref{eq:htravelingwaveapp} for $H(x, \tau)$ at different values of $\Delta$ and $a$. We use a numeric differential equation solver, solved over a wide domain of discrete values ${x \in (x_L, x_R)}$, with boundary conditions such that ${H(x_L) = 1}$ and ${H(x_R) = 0}$.  These are exponentially close in $x_L$, $x_R$ respectively to the true values for the solution on the infinite domain.  An initial guess is used for $H(x, 0)$ and then evolved until a very long time $\tau = \tau_f$ in order to arrive at the
 steady-state solution, from which we can read off the
position of the front.  We define this as the value of $x = x_\textrm{front}$ such that $H(x_\textrm{front}, \tau_f) = 1/2$.  For the data presented in Fig.~\ref{fig:FKPPslopes}, $x_L = -70$, $x_R = 40$, $\tau_f = 10^8$, and the domain of $x$ is discretized into $8001$ points.

Making a linear fit of  ${x_\textrm{front}}$
 against $\ln \sigma$ for a given value of $\Delta$, and recalling ${x_\textrm{front}\sim  \ln Z^\text{typ}}$, allows one to determine the value of the exponent $\gamma$, defined by ${Z^\textrm{typ} \propto \sigma^\gamma}$.  We make this linear fit over the range of $\sigma$ such that $-10 < \ln \sigma < -6$.  Smaller $\sigma$ requires very high numerical accuracy (a dense discretization of the domain of $x$), while at larger $\sigma$ the critical behavior may not be apparent.
  The results are shown in the main text in Fig.~\ref{fig:FKPPslopes}. The error bars in this figure are defined by the difference in slope obtained from fits using only the left half of this range, $-10 < \ln \sigma < -8$, as compared to only the right half of the range, $-8 < \ln \sigma < -6$.

\subsection{Minimal cut formula on tree}
\label{app:treemincut}

Assume that the minimal cut chops out $m$ subtrees from the full tree, with each subtree being cut only once, at its apex.
The singular values of a given subtree ${a\in \{1, \ldots, m\}}$ are $\{\lambda^{(a)}_1, \lambda^{(a)}_2\}$, with ${(\lambda^{(a)}_1)^2+(\lambda^{(a)}_2)^2=1}$ and $(\lambda_2^{(a)})^2 = Z^{(a)}$. The full state may be written
\be
\ket{\psi} = \sum_{i_1,\ldots, i_m}\hspace{-2mm}
\lambda^{(1)}_{i_1}\ldots \lambda^{(m)}_{i_m}
 \ket{i_1, \ldots, i_m}_\text{subtrees}
 \ket{i_1, \ldots, i_m}_\text{rest}.
\ee
The states $\ket{i_1, \ldots, i_m}_\text{subtrees}$ are products of the Schmidt states at the base of the subtrees, as in Sec.~\ref{sec:recursionrelationfortree}, and are orthonormal. The states $\ket{i_1, \ldots, i_m}_\text{rest}$, for the remaining spins, are neither normalized nor orthogonal. They are obtained by contracting the tensor network on the other side of the min-cut with Schmidt states at the tops of the subtrees.
The R\'enyi entropies are determined by the singular values of the (unnormalized) density matrix
\be
\rho = \sum_{i_1,\ldots, i_m}\hspace{-2mm}
(\lambda^{(1)}_{i_1})^2\ldots (\lambda^{(m)}_{i_m})^2
 \ket{i_1, \ldots, i_m}_\text{rest}
 \bra{i_1, \ldots, i_m}_\text{rest}.
\ee
In the limit of small $Z^{(a)}$, with everything else fixed,
\be
\rho \simeq \ket{e_0}\bra{e_0} + \sum_{a=1}^m Z^{(a)} \lf
\ket{e_a}\bra{e_a} -  \ket{e_0}\bra{e_0} \ri,
\ee
with $e_0= \ket{1, \ldots, 1}_\text{rest}$ and $e_a=\ket{1,\ldots,2,\ldots,1}_\text{rest}$, with the ``$2$'' in slot $a$. For $n>1$ this gives: 
\be\label{eq:simpletreemincutexample}
S_n \simeq \f{n}{n-1} \sum_{a=1}^m
\lf \f{|e_a|^2 |e_0|^2 - |e_0^\dag e_a|^2}{|e_0|^4}
\ri Z^{(a)} 
+ \mathcal{O}(Z^2).
\ee 
Each term is associated with a bond lying on the minimal cut. Let the height of this bond above the base be $k(a)$.

For a given term, the coefficient in brackets will vanish if the states $e_a$ and $e_0$ become parallel. 
Each of these is a state in the ``rest'' Hilbert space, given by the tensor network made up of the tensors on one side of the minimal cut. 
The bonds lying on the minimal cut have fixed states attached to them: one of these is changed in going from $e_0$ to $e_1$.
Exploiting the fact that these truncated tensor networks are still trees, we can write $e_0$ and $e_1$ in terms of the singular value decompositions of two subtrees. We see that the  the term in brackets is of the same order as the singular value for a tree of depth $k(a)$. (This assumes that the singular values are not \textit{growing} with $k$: this can occur if we fine-tune the boundary conditions, but is not relevant to the case we are discussing.) 
Therefore in the present limit of small $Z$ we confirm the minimal cut conjecture in the main text, according to which each bond $a$ on the minimal cut contributes an entanglement of order $Z^{(a)}  {Z'}^{(a)}$, where $Z^{(a)}$ and ${Z'}^{(a)}$ are two random variables each distributed like $Z_{k}$ for $k=k(a)$. 
(The $Z^{(a)}$ values are independent of each other and of the ${Z'}^{(a)}$ values, but the ${Z'}^{(a)}$ are correlated among themselves.)

\section{More on circuit simulations}
\label{app:circuit-numerics}

We present some additional details regarding the simulations of quantum circuits  in Sec.~\ref{sec:simulationscircuits}. The non-unitary time-evolution operator, $V$, for a given realisation of the circuit is built by choosing at every step a measurement with probability $r$ on a randomly chosen site or entangling two randomly chosen sites with the unitary with probability $1-r$. For a system of size $N$, time progresses by one unit for every $N$ unitaries applied. The Haar random unitaries are generated using Mezzadri's algorithm~\cite{mezzadri2006generate}.
\begin{figure}
    \includegraphics[width=\linewidth]{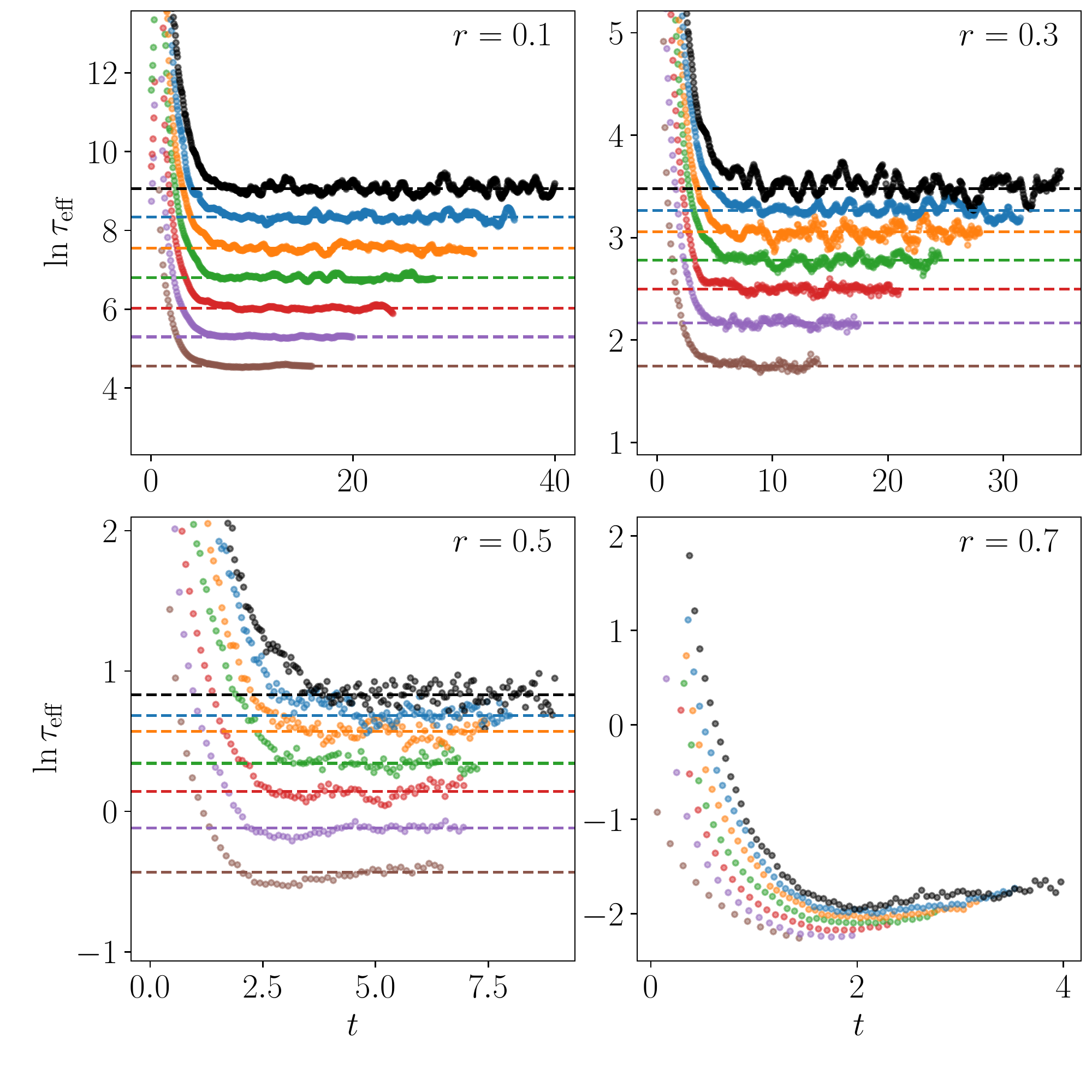}
    \caption{$\tau_\mathrm{eff}$ from time-derivatives of $\ln\mathcal{D}^\mathrm{typ}$. The plateaux denoted by the dashed lines show the $\tau(r,N)$ values which are used to extract $a(r)$ (see Fig.~\ref{fig:haar-fit-tau}). Different colours correspond to the different $N$ following the same convention as in Fig.~\ref{fig:overlap-linear}.}
    \label{fig:haar-log-derivs}
\end{figure}

\begin{figure}
    \includegraphics[width=\linewidth]{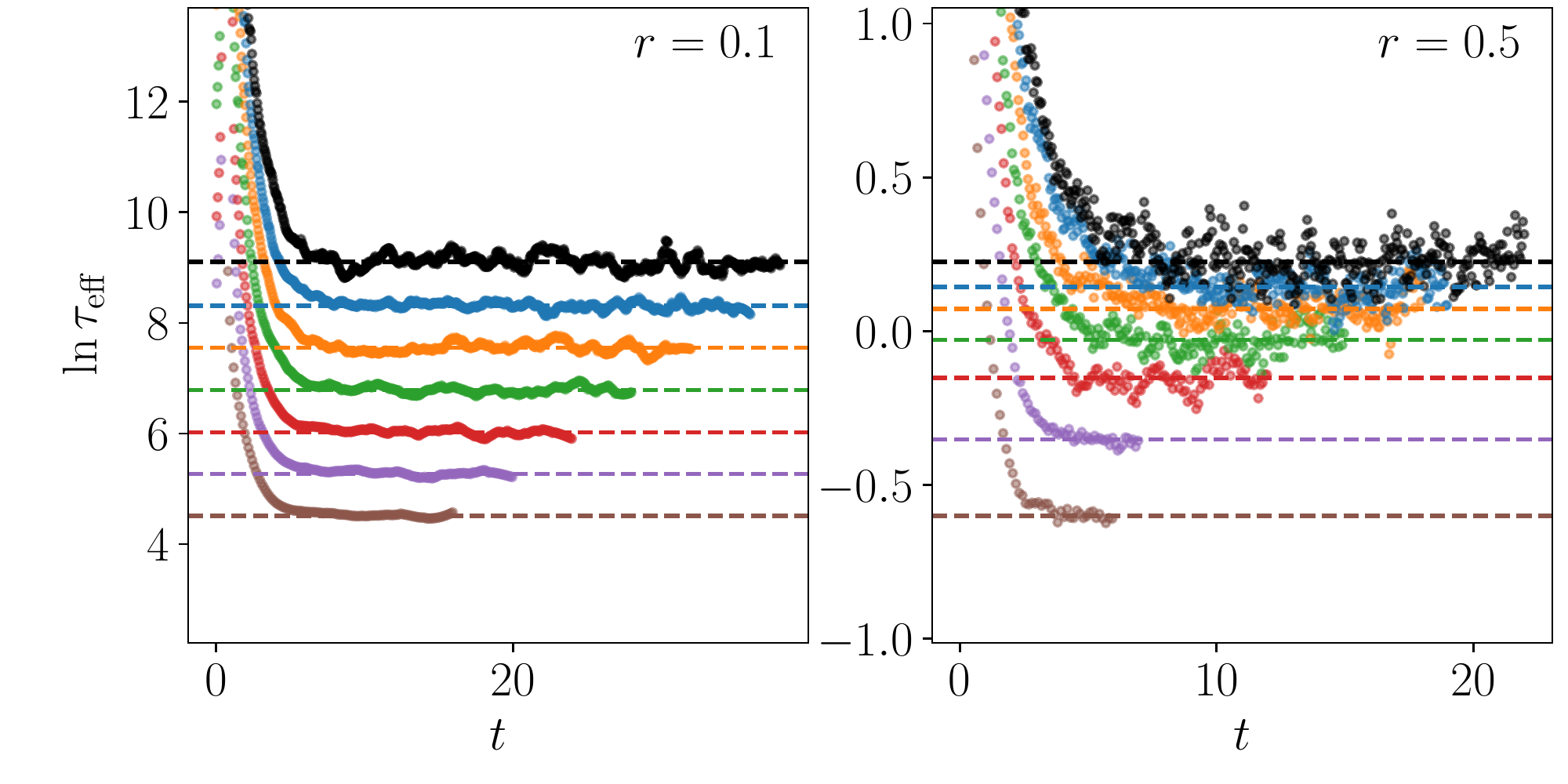}
    \caption{Same as in Fig.~\ref{fig:haar-log-derivs} but for the Haar circuit with measurements.}
    \label{fig:haar-mes-log-derivs}
\end{figure}

Note that the $\tau_\mathrm{eff}$ described in the main text is obtained by taking a log-derivative of $\mathcal{D}(t)$. Since, numerical deriatives are notoriously noisy, we smooth the data for $\mathcal{D}(t)$ using a Savitzky-Golay filter~\cite{savitzky1964smoothing} and then take the derivative. Representative examples of $\tau_\mathrm{eff}$ and the plateaux therein are shown in Figs.~\ref{fig:haar-log-derivs} and \ref{fig:haar-mes-log-derivs}.

\begin{figure}
    \includegraphics[width=\linewidth]{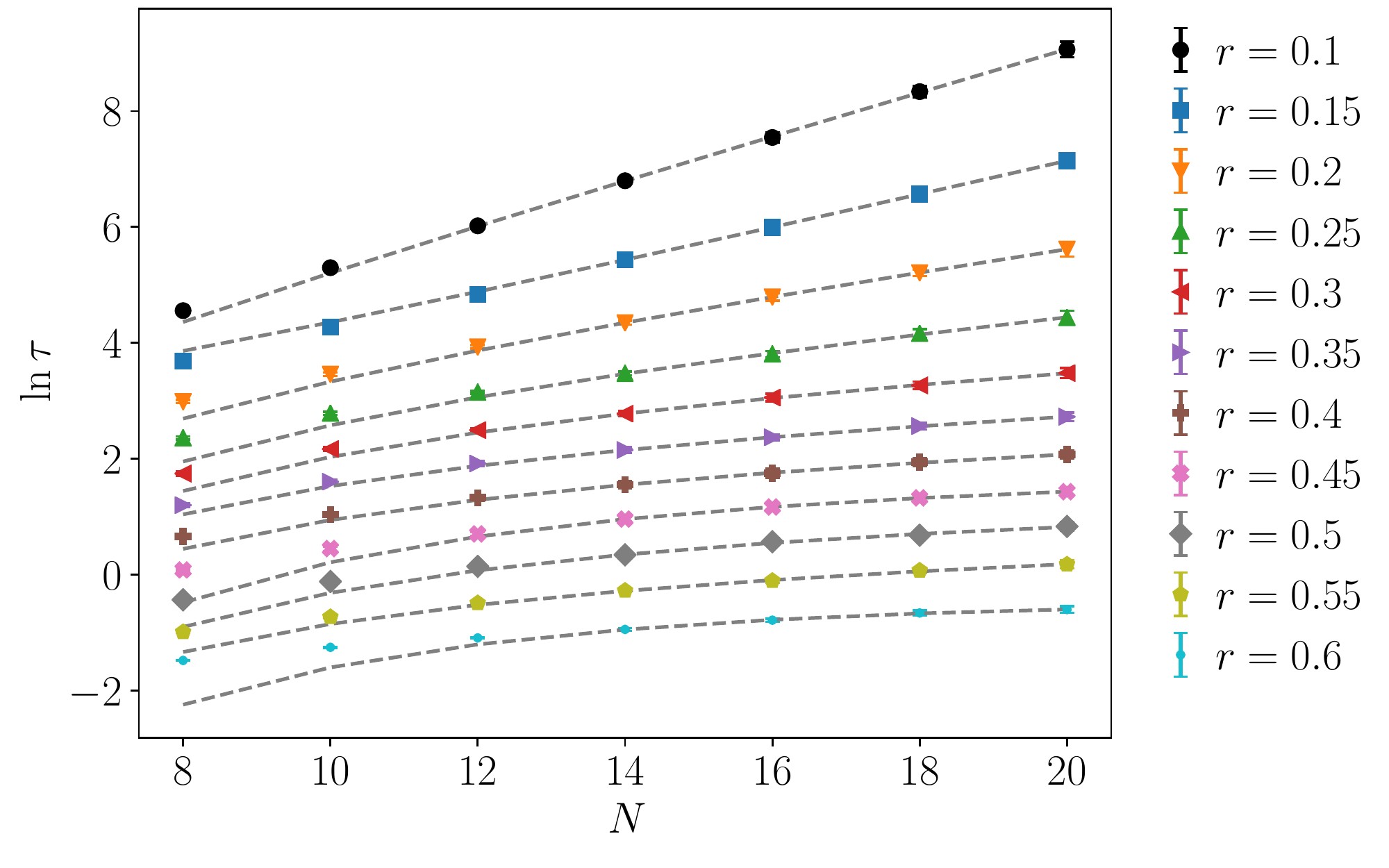}\\
    \includegraphics[width=\linewidth]{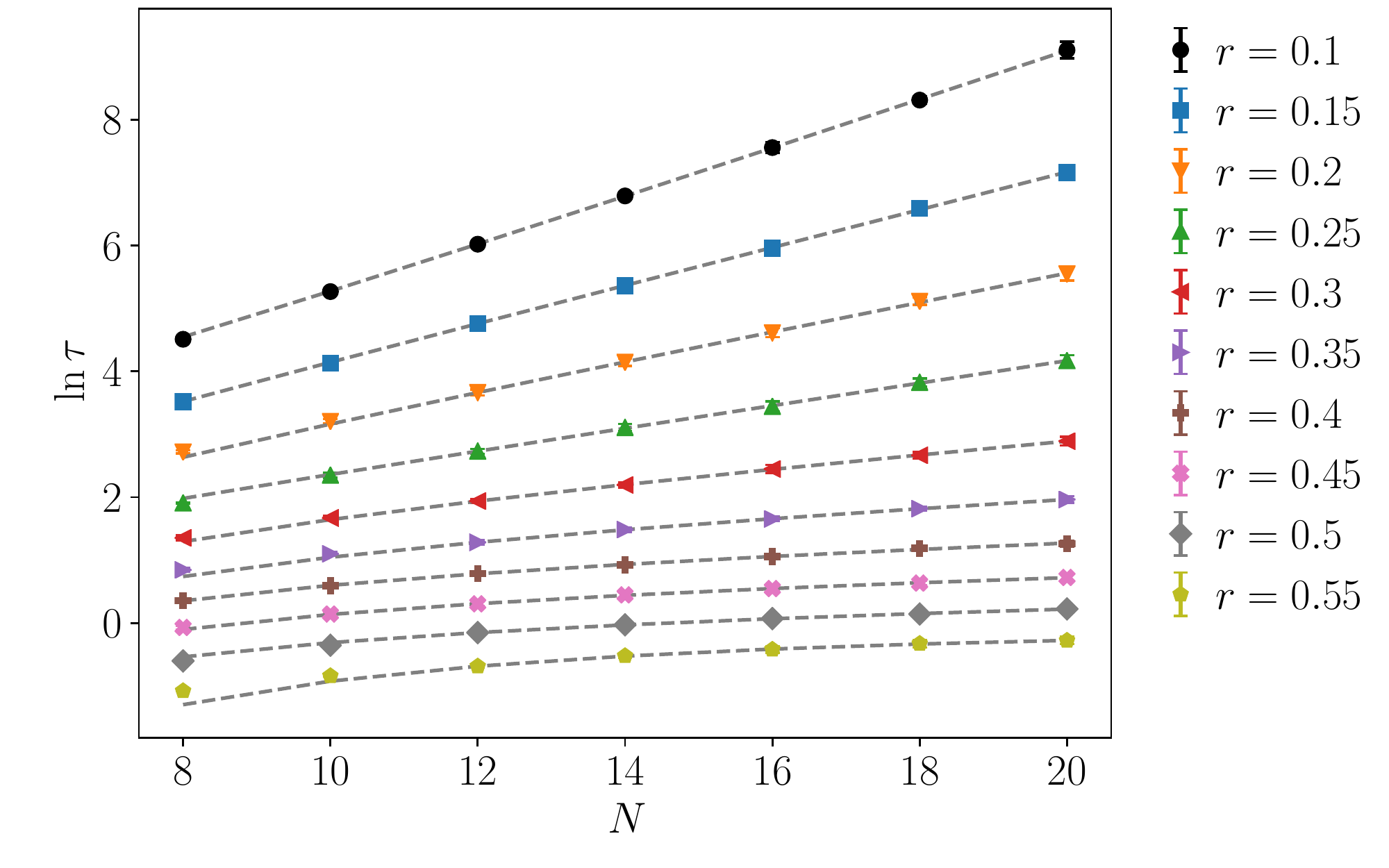}
    \caption{Fits of $\ln\tau$ to a function of the form $a(r)N + b(r)+c(r)/N$ for different values of $r$ for the Haar circuit with forced-measurements (top) and measurements(bottom). We use $N\ge 14$ for the fits shown which were used to extract $a(r)$ shown in Fig.~\ref{fig:ar-haar}. Fits including smaller $N$ (not shown) were used to estimate the errorbars on $a(r)$. }
    \label{fig:haar-fit-tau}
\end{figure}

From the $\tau$ so-obtained, we extract $a(r)$ by fitting the data to a form $a(r)N + b(r)+c(r)/N$. The fits are shown in Fig.~\ref{fig:haar-fit-tau}.

\section{Toy model for the slow decay of $S$}
\label{app:multiplyingmatrices}

In the entangled phase the operator entanglement has a plateau at an extensive value ${S_n /N \simeq s_n >0}$, which persists for a time that scales exponentially in the number of spins, $N$.
In this appendix we describe the crudest toy model for this plateau, which neglects both locality of the interactions and, for the most part, distinctions between different R\'enyi entropies. 
We consider the forced measurement case, where the circuit is made up of uncorrelated random pieces.

Imagine dividing up the circuit up into blocks of temporal duration $\Delta t$, which corresponds to writing the time evolution operator $V(t)$ as a product of random matrices $W_i$, with each matrix of size $2^N \times 2^N$ and $i=1,\ldots, t/\Delta t$.
$\Delta t$ is chosen to be much larger than 1 but much smaller than the timescale $\tau$ that will emerge below.
Each block has a singular value decomposition ${W_i  =  U^{(1)}_i D {U^{(2)}_i}^\dag}$.
As a toy model, we will treat the unitaries $U^{(1)}$ and $U^{(2)}$ as Haar random (neglecting locality) and we will make the simplest choice of $D$ that yields a given value of $S_1=sN$, which is a flat entanglement spectrum:
\ba\notag
D & = \f{1}{\sqrt{B}}
\left(
\begin{array}{cc}
 \mathbb{I}_{B\times B} &   0  \\
 0 &  0   
\end{array}
\right), 
& 
B & = e^{s N}, 
&
0& < s<\ln 2.
\end{align}
Note that the nonzero block is a small fraction of the size of the matrix for large $N$.
Finally, we will make an uncontrolled simplification by also treating the entanglement spectrum of $V(t)$ as flat:
\ba\notag
V(t) & =  U_L(t) D(t) {U_R(t)}^\dag,
&
D(t) & = \f{1}{\sqrt{B(t)}}
\left(
\begin{array}{cc}
 \mathbb{I}_{B(t)\times B(t)} &   0  \\
 0 &  0   
\end{array}
\right), 
\end{align}
with $B(t) = e^{S(t)}$. Note that $B$ and $D$ refer to quantities for a single slice of width $\Delta t$, while $B(t)$ and $D(t)$ refer to the complete evolution operator up to time $t$.

We are interested in the singular values of the new evolution matrix $V(t+\Delta t)$, which may be written
\ba
V(t+\Delta t) = U' D(t) U D {U''}^\dag.
\end{align}
for Haar unitaries $U'$, $U$, $U''$. These values are also the singular values of 
\be\label{eq:appwidetildeVdefn}
\widetilde V = D(t) U D.
\ee
Up to a normalization factor, $\widetilde V$ is just a rectangular block, of size $B(t)\times B$, taken from a Haar unitary of exponentially larger size ($2^N\times 2^N$).
Correlations between unitary matrix elements become weaker as the size of the matrix increases, so we expect that we can treat them as Gaussian, with ${\mathbb{E} \, U_{ab}  = 0}$ and ${\mathbb{E} \, U_{ab} (U_{a'b'})^*   = \f{1}{2^N} \delta_{aa'} \delta_{bb'}}$. (Higher cumulants are suppressed by powers of $2^N$.) 
The singular-values-squared of $\widetilde V$,  denoted $v_i = \eta_i^2$, are eigenvalues of the ${B(t)\times B(t)}$  matrix 
\be\label{eq:appMmatrixVVdag}
M = \widetilde V \widetilde V^\dag.
\ee
When the matrix elements are of $\widetilde V$ are complex Gaussian random numbers, this is as a Wishart random matrix (see e.g.\ Ref.~\cite{nadal2011statistical} for an application in a related context). 
The distribution of its eigenvalues depends on $B$ as well as on $B(t)$. We assume that ${1\ll B(t) \ll B}$. 
Normalizing the matrix so the $s_i$ sum to one,  the eigenvalue density for $\widetilde V \widetilde V^\dag$ is  the Marcenko-Pastur distribution,
\ba\notag
\rho(v) & = \f{8\sqrt{v - v_{-}}\sqrt{v_{+}-v}}{\pi (v_{+}-v_{-})^2 v} , 
&
v_{\pm} & = \f{\lf \sqrt{B}\pm \sqrt{B(t)} \ri^2}{B \, B(t)}.
\end{align}
From this distribution the operator R\'enyi entropies can be calculated as
\ba\notag
e^{-(n-1)S_n(t+\Delta t)} & = \int_{v_{-}}^{v_{+}} \dd v v^{n} \rho(v)
\\ \notag
&  \simeq B(t)^{-(n-1)} \lf 1 + \f{n(n-1)}{2} \f{B(t)}{B} + \ldots  \ri 
\\ \notag
S_n(t+\Delta t) - S(t) &  \simeq -  \exp \lf - \lf s N- S(t) - \ln \f{n}{2} \ri \ri.
\end{align}
As expected, the entanglement spectrum does not remain flat. In our crude approximation, however, we neglect this, and apply the above transformation iteratively, so that in the continuum limit for times $\gg \Delta t$:
\be
\partial_t S(t) \sim - C \exp\left[ - (s N - S(t)) \right].
\ee
Here $C$ is an order-1 constant.
At times larger than $\Delta t$, but short enough such that $S(t)\gg 1$ (which we assumed above), this equation gives a solution:
\be
S(t) = s N - \ln t + \ldots
\ee
independently of the value of $C$. Note that we have neglected random fluctuations (see Sec.~\ref{sec:replicatimescale}).

This analysis suggests a characteristic timescale $\tau$ with $\ln \tau \simeq s N$. 
We may verify this dependence directly in the opposite limit of asymptotically late times, where (as usual for a product of random matrices) there is a separation of scale between the largest singular value, the second largest, and so on.
This separation allows us to consider only the two largest singular values. At a given time $t$, let them be normalized as
\be
\{ \eta_1, \eta_2\}  = \{1 , \epsilon(t)\}.
\ee
Then in place of $M$ in Eq.~\ref{eq:appMmatrixVVdag} we have a $2\times 2$ matrix 
\be
M_{ik} = \epsilon^{i+k-2}\sum_{j=1}^{B} U_{ij} U_{kj}^*.
\ee
Each of the elements of $M_{ik}$ is a different sum of many random variables, so we assume that $M$ can be approximated as Gaussian:
\ba\notag
\langle M_{ik} \rangle  &\simeq   \f{B \epsilon^{2(i-1)}  \delta_{ik}}{2^N} , 
  &
\langle \langle M_{ik} M_{i'k'} \rangle\rangle 
 & \simeq 
\f{B  \epsilon^{2(i+i'-2)} \delta_{ik'} \delta_{i'k}  }{2^{2N}} .
\end{align}
After absorbing a normalization constant into~$M$, 
\be
M = 
\left(
\begin{array}{cc}
 1 & 0    \\
 0 &  \epsilon^2
\end{array}
\right)
+
\f{1}{\sqrt{B}}
\left(
\begin{array}{cc}
 a & \epsilon \beta    \\
  \epsilon  \beta^* &  \epsilon^2 b
\end{array}
\right),
\ee
where $a,b,\beta$ have mean zero and ${\<a^2\>=\<b^2\>=\< |\beta|^2\>=1}$. The new (small) singular value squared is
\be
\epsilon^2_\text{new} = \epsilon^2 \lf 
1 - \f{a-b}{\sqrt{B}}
+ \f{a^2-ab-|\beta|^2}{B}
+ \ldots 
\ri
\ee
Let us study the typical value of this exponentially small quantity, defined by ${(\epsilon^2_\text{new})_\text{typ}
=
\exp \< \ln \epsilon^2_\text{new} \>}$.
We have
\be
 \ln \epsilon^2_\text{new} =
  \ln \epsilon^2
 + \f{b-a}{\sqrt{B}}
+ \f{a^2-b^2-2|\beta|^2}{2B}  + \ldots
\ee
Here the average is taken over $a,b,\beta$.
Note that the leading fluctuation term, of order $1/\sqrt{B}$, averages to zero. The next term, however, gives a negative drift  under the recursion.
Recalling that $B=e^{sN}$, and applying this map iteratively with each added block,
\ba
\epsilon^2_\text{typ}(t) &\sim \exp \lf
- \f{t}{ \tau}
\ri,
&
\tau & =  \Delta t \times \exp ( s N).
\end{align}
Therefore  at the latest times,
\ba
S_n(t) & \sim 
\exp \lf -\f{t}{\tau} \ri
\times
\left\{
\begin{array}{cc}
[n/(n-1)]   & \quad n>1    \\
(2 t / \tau)   &\quad n=1
\end{array}
\right. .
\end{align}
Thus, the main conclusion is that in this toy model $s N$ sets both the value of the early time plateau and also the timescale for the late-time exponential decay, in agreement with the picture from the replica treatment.

\section{Field theory: further details}

\subsection{$N\rightarrow 1$ ordered phase}
\label{app:nto1masses}

In Sec.~\ref{sec:FTconsequences} we stated that for ${\mu^2<0}$ the field theory of Sec.~\ref{sec:MPTFT} has an ordered phase with ${X_{ab}  = f (\delta_{ab} - 1/N) + W_{ab}}$, where ${f=\f{-\mu^2}{3g}\f{N}{N-2}}$ is the order parameter, and
$W_{ab}$ represents fluctuations around the saddle-point value whose  quadratic Lagrangian is
\be\label{eq:appmasscheck}
\mathcal{L} = \f{1}{2} \sum_{ab} (\partial W_{ab})^2 +  {\f{-\mu^2 N}{2 (2-N)} \big( \sum_{a b} W_{ab}^2 -2  \sum_a W_{aa}^2 \big)}.
\ee
We would like to check that if we compute the masses of the fluctuation modes using this expression, and then take $N\rightarrow 1$, these masses remain positive.
Viewing $W_{ab}$ as a vector, the term  $\sum_a W_{aa}^2$ is  $W.M.W$ where the matrix $M_{ab,cd}$ (with row index $a,b$ and column index $c,d$) is 1 if ${a=b=c=d}$ and zero otherwise. We want the eigenvalues of $M$ when projected onto the subspace of $W$ satisfying $\sum_a W_{ab}=0$ and $\sum_b W_{ab}=0$, i.e. the eigenvalues of $\widetilde M_{ab,cd} = P_{aa'} P_{bb'}M_{a'b', c'd'} P_{c'c} P_{d'd}$ where $P$ is the projector ${P_{aa'} = \delta_{aa'} - 1/N}$. 
Since $M$ is nonzero only when its four indices are equal, drawing a diagram shows that ${\tr \widetilde M^k = \tr S^k}$, where ${S_{a,b} = (1-2/N) \delta_{ab} + 1/ N^2}$. This gives the nonzero eigenvalues of $\widetilde M$ as $(N-1)/N\rightarrow 0$ with multiplicity 1 and $(N-2)/N\rightarrow -1$ with multiplicity ${N-1}$.
Altogether, the eigenvalues of the matrix $(\mathbb{I} - 2 \widetilde M)$ which appears in (\ref{eq:appmasscheck}) are either $1$ or $3$ in the limit and are positive.

\subsection{Effect of $YFY$ coupling when ${N>0}$}
\label{app:mFreduction}

Here we show that for ${N>0}$ (for example in the replica limit ${N\rightarrow 1}$, 
 but \textit{not} in the limit ${N\rightarrow 0}$), and in the vicinity of the critical point $r=0$, the theory 
\be\label{eq:appRTNactionYlanguage}
\mathcal{L} =
 \sum_{ab} \left[
 \f{1}{2}(\partial Y_{ab})^2 
+ r  Y_{ab} 
+ g Y_{ab}^3
\right]
+ \f{m_F^2}{2} \sum_{abcd} Y_{ab} F_{ab,cd} Y_{cd}
\ee
can be reduced to the theory 
\be
\label{eq:appMPTlagrangian}
\mathcal{L} = \sum_{ab} \, \left[ \,
\f{1}{2} (\partial  X_{ab})^2 + \f{\mu^2}{2}  X_{ab}^2 + g  X_{ab}^3
\, \right],
\ee
for a matrix $X$ with vanishing row and column sums, by discarding massive modes. 

First, shifting the field by ${Y_{ab} \rightarrow Y_{ab}+c}$ with
${c=- r / (2Nm_F^2)+ \mathcal{O}(r^2)}$ removes the linear term and generates a mass term. The quadratic part of the Lagrangian is then 
\be\label{eq:appYFYL2}
\mathcal{L}_2 = \f{1}{2} \sum_{ab,cd} Y_{ab} \lf (k^2 + \mu^2) \delta_{ac}\delta_{bd} 
+ m_F^2 F_{ab,cd} \ri
Y_{cd}
\ee
with ${\mu^2=-  3 g r/ (N m_F^2)}+ \mathcal{O}(r^2)$. 
 As a matrix, ${F = \mathbb{I} \otimes E +    E\otimes  \mathbb{I}}$, where $E$ is the  ${N\times N}$ matrix with unit elements: ${E_{ab}=1}$, so 
the matrix appearing in the brackets in Eq.~\ref{eq:appYFYL2} is 
\ba
& (k^2 + \mu^2) \mathbb{I}\otimes \mathbb{I} + m_F^2 \lf  \mathbb{I} \otimes E +    E\otimes  \mathbb{I} \ri 
\end{align}
This can be decomposed in terms of the  projection matrices ${P_1 \equiv {N}^{-1} E}$ and 
${P_{N-1}  \equiv \mathbb{I} - P_1}$ as:
\ba
&(k^2 + \mu^2) \lf P_{N-1} \otimes P_{N-1} \ri 
\\ & + 
(k^2 + \mu^2+  m_F^2 N) \lf P_{1} \otimes P_{N-1} + P_{N-1} \otimes P_{1} \ri 
\\ & +
(k^2 + \mu^2+ 2  m_F^2 N) \lf P_{1} \otimes  P_{1} \ri.
\end{align}
This is a decomposition into three representations of $G_N$ of dimensions ${(N-1)^2}$,  
$2(N-1)$, and $1$.
We see that, at the critical point (where $r$ and therefore $\mu^2$ vanish) the second and the third representations remain massive and only the first becomes massless. 
Retaining only this representation is equivalent to fixing the row and column sums of $Y$ to zero. Doing so and renaming the resulting  field $X$ yields precisely Eq.~\ref{eq:appMPTlagrangian}. 
More precisely the two massive representations ought to be integrated out, renormalizing the values of the couplings in Eq.~\ref{eq:appMPTlagrangian}.

These manipulations manifestly require ${N>0}$: for example they are appropriate for the replica limit ${N\rightarrow 1}$ relevant to the MPT. The replica limit ${N\rightarrow 0}$ must be handled separately.

\subsection{Rewriting Lagrangian in $N\rightarrow 0$ limit}
\label{app:ntozerolimit}

We give details of field redefinitions in Sec.~\ref{sec:FTRTNFMPT}.
First define the $N$-component vectors $\vec{v}^+$, $\vec{v}^-$, and $\vec{v}^i$ for ${i=2,\ldots,N}$ \cite{cardy1985nonperturbative,cardy1985field,cardy2001exact,kaviraj2020random}:
\ba
\vec{v}^+ & = \f{1}{2} \lf   1,0,\ldots, 0 \ri + \f{1}{2} \f{\lf   0,1,\ldots, 1 \ri }{N-1}
\\
\vec{v}^- & = \f{1}{2} \lf   1,0,\ldots, 0 \ri - \f{1}{2} \f{\lf   0,1,\ldots, 1 \ri }{N-1}
\\
\vec{v}^i & =  \lf   0,\ldots,0, 1,0, \ldots, 0 \ri - \f{\lf   0,1,\ldots, 1 \ri }{N-1}
\end{align}
where the extra ``1'' in the third line is in the $i$th place. There are ${N+1}$ of these vectors but the $\vec{v}^i$ are not linearly independent: ${\sum_{i>1} \vec{v}^i=0}$. 

We use these vectors to rewrite $Y_{ab}$ in terms of  $y_{\alpha\beta}$:
\be\label{eq:appydef}
y_{\alpha\beta} = \vec{v}^\alpha. Y . \vec{v}^\beta =\sum_{ab} {v}^\alpha_a Y_{ab} {v}^\beta_b.
\ee
In this appendix we use $a,b,a',b',\ldots$ to denote indices that run from $1$ to $N$, and
$\alpha, \beta,\ldots$ to denote indices that take the $N+1$ values $\{ +, -, 2, \ldots, N\}$. 
We will use $i, j,k$ to denote indices that run only over $2$ to $N$.
Note that 
\ba\label{eq:appyrowcolsums}
\sum_{j=2}^N y_{j, \beta}&= 0,
&
\sum_{k=2}^N y_{\alpha, k}& = 0.
\end{align}
(Alternately we could define an ${N\times N}$ matrix without this redundancy.)

To invert the transformation defining $y$, define ${N+1}$-component vectors
\ba
\vec{x}^1 & = (1, 1, 0, \ldots, 0), \\
\vec{x}^{i>1} & = ( 1,-1, 0,\ldots, 1, \ldots, 0).
\end{align}
The final 1 in the second line is in $(i+1)$st place, which in our labelling convention corresponds to the component $x^i_\beta$ with ${\beta = i}$.
We have $\sum_\alpha x_\alpha^a v^\alpha_{a'} = \delta^a_{a'}$, so that
\be
Y_{ab} =
\vec{x}^a. y. \vec{x}^b =
 \sum_{\alpha\beta} x^a_\alpha \, y_{\alpha\beta} \, x^b_\beta.
\ee
Explicitly (for $j,k>1$):
\ba
Y_{11} = & \lf y_{++} + y_{+-} + y_{-+} + y_{--} \ri,
\\
Y_{1k} =& \lf y_{++} - y_{+-} + y_{-+}  - y_{--} \ri+ ( y_{+k} + y_{-k}  ),
\\
Y_{j1}  =&\lf y_{++} + y_{+-} - y_{-+}  - y_{--} \ri + ( y_{j+} + y_{j-}  ),
\\
Y_{jk}  =&  \lf y_{++} - y_{+-} - y_{-+}  + y_{--} \ri 
\\
& + ( y_{j+} - y_{j-}  ) 
+( y_{+k} - y_{-k}  ) 
+ y_{jk}.
\end{align}

Inserting these relations into the derivative term,
\ba \notag
\sum_{ab} (\partial Y_{ab})^2 = 
&\phantom{+} 
8 \lf \partial y_{+-} \partial y_{-+} + \partial y_{++} \partial y_{--} \ri 
\\ \notag
& +  4 \sum_j  \partial y_{j+}\partial y_{j-} 
\\ \notag &+ 4 \sum_k  \partial y_{+k}\partial y_{-k} 
\\  \notag&+ \sum_{jk} (\partial y_{jk})^2
\\ & + \mathcal{O}(N), \label{eq:appYykinetic}
\end{align}
where the final line contains terms whose coefficients contain an explicit factor of $N$, which we assume can be neglected in the limit $N\rightarrow 0$.

Next consider the $F$ term in the Lagrangian, which has the form
\ba
\sum_{abcd} Y_{ab} F_{ab,cd} Y_{cd} = \sum_a \bigg( \sum_b Y_{ab}  \bigg)^2 + \sum_b  \bigg( \sum_a Y_{ab}  \bigg)^2.
\end{align}
This simplifies to 
\ba \notag
\sum_{abcd} Y_{ab} F_{ab,cd} Y_{cd} 
= &\phantom{+} 
16 y_{--} \lf y_{+-} + y_{-+} \ri 
\\  \notag& + 4 \sum_j y_{j-}^2 
\\ \notag & + 4 \sum_k y_{-k}^2
\\ & + \mathcal{O}(N). \label{eq:appYyFterm}
\end{align}

The linear term in the Lagrangian is simply
\be\label{eq:appYylinearterm}
\sum_{ab} Y_{ab} = 4 y_{--} + \mathcal{O}(n).
\ee

Since the coefficient of the linear term is zero at the critical point, we assign ``engineering'' dimensions to the fields such that the quadratic terms in (\ref{eq:appYykinetic}) and (\ref{eq:appYyFterm}) are marginal. Denoting the spacetime dimension by $D$, the inverse length dimensions of the fields are, in order of increasing scaling dimension,
\ba
[y_{++}] & =({D-6})/{2},
\\
[y_{j+}]=[y_{+k}] &=({D-4})/{2},
\\
[y_{+-}]=[y_{-+}]=[y_{jk}] &= ({D-2})/{2},
\\
[y_{j-}]=[y_{-k}]&={(D+0)}/{2},
\\
[y_{--}] &=({D+2})/{2},
\end{align}
so that if the number of ``$+$'' indices for a component of $y$ is $n_+$, and the number of $-$ indices is $n_-$, the engineering dimension is 
\be\label{eq:appxplusesminuses}
x(n_+, n_-) = \f{(D - 2) - 2(n_+ - n_-) }{2}.
\ee 
This formula implies that, at a given order in $y$, terms with the largest number of $+$ indices and the smallest number of $-$ indices are most relevant.

Now consider the additional terms in the Lagrangian beyond those in Eqs.~\ref{eq:appYykinetic},~\ref{eq:appYylinearterm},~\ref{eq:appYyFterm}, order by order in $Y$. 

$G_N$ symmetry does not allow any linear terms other than (\ref{eq:appYylinearterm}). 
At quadratic order the remaining possibilities are $\sum_{ab} Y_{ab}^2$ and $(\sum_{ab} Y_{ab})^2$.
The former is  redundant --- it can be cancelled by a shift in $Y$ because of the presence of $\sum_{ab} Y_{ab}^3$. We will also check this below in the new parameterization. 
The latter, $(\sum_{ab} Y_{ab})^2$, is prortional to $(y_{--})^2$ by Eq.~\ref{eq:appYylinearterm}, so is less relevant than the terms on the RHS of Eq.~\ref{eq:appYyFterm} and can be neglected.

Cubic terms  are obtained by contracting indices in 
\be
Y_{ab} Y_{a'b'} Y_{a''b''},
\ee
i.e. by setting some indices equal to others and summing. 
Left indices may only be set equal to other left indices, and similarly for right indices.
This allows three different types of index contraction: 
an index (e.g. $a$) can be summed without being set equal to any other index; 
two indices can be set equal and then summed (e.g. $a=a'$); or three indices can be set equal and summed (${a=a'=a''}$). 
When we rewrite the contracted expression in terms of ${y_{\alpha\beta}y_{\alpha'\beta'}y_{\alpha''\beta''}}$, 
the single, double, and triple index contractions lead to contractions with the tensors
\ba
d^{(1)}_\alpha & =   \sum_a x^a_\alpha, & 
d^{(2)}_{\alpha \alpha'} & =   \sum_a x^a_\alpha x^a_{\alpha'}, &
d^{(3)}_{\alpha\alpha'\alpha''} & =   \sum_a x^a_\alpha x^a_{\alpha'} x^a_{\alpha''},
\end{align}
respectively, for the indices $\alpha, \alpha', \alpha'''\in {\{ +, -, 2, \ldots, N\}}$. 
For each of these, we may check the maximal value of 
\be
{\Delta \equiv n_+ - n_-},
\ee
 the difference in the number of $+$ and $-$ indices, that may appear on the right hand side.
 ($\Delta$ should not be confused with a scaling dimension.)
Since the difference in the total number of $+$ and $-$ indices is what determines the engineering dimension of a field or a product of fields (Eq.~\ref{eq:appxplusesminuses}), identifying $\Delta_\text{max}$ for each type of index contraction allows us to say which types of contraction will give the most relevant cubic terms: they are those for which the sum of $\Delta_\text{max}$, over all contractions, is largest. 

Explicitly, 
\ba
d^{(1)}_\alpha & =  
(N, 2-N, 1, \ldots, 1)_\alpha.
\end{align}
However $y$ vanishes when contracted with $(0, 1, \ldots, 1)$ (Eq.~\ref{eq:appyrowcolsums}). After dropping this part, and taking the limit $N\rightarrow 0$ directly in the coefficients,
\ba
d^{(1)}_\alpha & \rightarrow  2 \delta_{\alpha,- }.
\end{align}
Therefore this pattern of index contraction contributes $\Delta = -1$.
Using similar simplifications,
\be
d^{(2)}_{\alpha,\alpha'} \rightarrow
2 \lf  \delta_{+\alpha} \delta_{-\alpha'} + \delta_{-\alpha} \delta_{+\alpha'}    \ri + \sum_{j=2}^N \delta_{j \alpha}\delta_{j \alpha'}.
\ee
This pattern of index contraction contributes $\Delta = 0$. Finally, $d^{(3)}$ contains various patterns of index contraction with different values for $\Delta$:
\ba
d^{(3)}_{\alpha,\alpha', \alpha''} \rightarrow  
& \phantom{+} 2 \lf \delta_{- \alpha} \delta_{+\alpha'} \delta_{+\alpha''} + \ldots \ri 
\\ 
& + \sum_{i=2}^N \lf \delta_{+\alpha} \delta_{i \alpha'} \delta_{i \alpha''} + \ldots \ri
\\ 
& + \sum_i \delta_{i\alpha} \delta_{i\alpha'} \delta_{i \alpha''} 
\\
& - \sum_i \lf \delta_{- \alpha} \delta_{i \alpha'} \delta_{i \alpha''} + \ldots \ri 
\\
& + 2 \delta_{-\alpha} \delta_{-\alpha'} \delta_{-\alpha''}. 
\end{align}
(Ellipses indicate terms related to those shown by cyclic permutations of $\alpha$, $\alpha'$, $\alpha''$.)
The first two lines on the RHS have ${\Delta = 1}$; all the others have smaller values of $\Delta$.
If we are interested only in keeping the most relevant terms in  a given expression, we can truncate to only the first two lines:
\ba
\widetilde d^{(3)}_{\alpha,\alpha', \alpha''} \equiv   
& \phantom{+} 2 \lf \delta_{- \alpha} \delta_{+\alpha'} \delta_{+\alpha''} + \ldots \ri 
\\ 
& + \sum_{i=2}^N \lf \delta_{+\alpha} \delta_{i \alpha'} \delta_{i \alpha''} + \ldots \ri. 
\end{align}

Therefore the most relevant cubic terms allowed by $G_N$ symmetry are those with ${\Delta = 2}$
(in total, i.e. counting both row and column indices) 
that arise by discarding the less-relevant parts of $\sum_{ab} Y_{ab}^3$: 
\ba
\sum_{ab} Y_{ab}^3
\rightarrow &  
 \sum \widetilde d^{(3)}_{\alpha,\alpha', \alpha''} \widetilde d^{(3)}_{\beta,\beta', \beta''} y_{\alpha\beta} y_{\alpha'\beta'} y_{\alpha''\beta''}
\end{align}
which is equal to
 \ba \notag
\bigg(\sum_{ab} Y_{ab}^3\bigg)_{\Delta=2} & =  12 y_{++}  \lf  
  y_{++}  y_{--} + 2 y_{+-} y_{-+}  \ri \\   \notag
  & + 12 y_{++}  \lf    y_{-k} y_{+k} + y_{j-} y_{j+} + \f{y_{jk} y_{jk}}{4}   \ri 
  \\   \notag
  & +
  6 \lf 
y_{+-} y_{j+} y_{j+}+  y_{-+} y_{+k} y_{+k}
  \ri \\  
  & + 
  6 y_{j+} y_{+k} y_{jk}. 
  \label{eq:appcubictermexplicit1}
  \end{align}
(Repeated $j$ or $k$ indices are summed from $2$ to $N$.)
All other contractions of $YYY$, such as  $\sum_{aa'bb'} Y_{a b} Y_{a b'} Y_{a' b'}$, give terms that are strictly less relevant according to the engineering dimensions.

Let us check that, having dropped less relevant terms from the Langrangian in the process of rewriting it in terms of $y$, the  coupling of the quadratic term
\ba\label{eq:appredundantquadraticy}
& \f{1}{4} \sum_{ab} Y_{ab}^2= \\
& 2 (y_{+-}y_{-+} + y_{++} y_{--}) + y_{j+}y_{j-} + y_{+k} y_{-k} + \f{y_{jk} y_{jk}}{4},
\notag
\end{align}
 is still redundant.
The shift ${Y_{ab} \rightarrow Y_{ab} + c}$ for each element corresponds to shifting ${y_{++} \rightarrow y_{++} + c}$ and leaving other elements of $y$ unchanged (by Eq.~\ref{eq:appydef}).
Under this shift, the cubic term  (\ref{eq:appcubictermexplicit1}) generates precisely Eq.~\ref{eq:appredundantquadraticy}, with a  coefficient of order $c$ (and the linear term in Eq.~\ref{eq:appYylinearterm} with a coefficient of order $c^2$). Therefore by an appropriate shift of $y_{++}$ we may eliminate the quadratic term (\ref{eq:appredundantquadraticy}).

\bibliography{alltoallref}

\end{document}